\documentclass[11pt,a4paper]{article}
\usepackage{jheppub,tcolorbox}
\usepackage[T1]{fontenc}

\usepackage{tikz}
\usetikzlibrary {arrows.meta}
\usepackage{caption}
\usepackage{multirow}

\usetikzlibrary{decorations.pathmorphing}
\usetikzlibrary{decorations.markings}
\usetikzlibrary{positioning,shapes.misc}
\usetikzlibrary {graphs}

\tikzset{snake it/.style={decorate, decoration={snake,amplitude=.4mm,segment length=2mm,
                       post length=0mm,pre length=0mm}}}

\newcommand{\ord}[1]{{\scriptscriptstyle(#1)}}

\newcommand{\tr}{{\operatorname{tr}}}
\newcommand*{\flip}[1]{\scalebox{-1}[1]{\rotatebox[origin = c]{180}{#1}}}

\title{\boldmath{Binary Black Holes and \\Quantum Off-Shell Recursion}}

\preprint{APCTP Pre2023 - 012}

\author[a]{Kyoungho Cho}
\author[b]{Kwangeon Kim}
\author[a,c]{Kanghoon Lee}

\affiliation[a]{Asia Pacific Center for Theoretical Physics, Postech, Pohang 37673, Korea}
\affiliation[b]{Department of Physics, Yonsei University, Seoul 03722, Korea}
\affiliation[c]{Department of Physics, Postech, Pohang 37673, Korea}

\emailAdd{kyoungho.cho@apctp.org}
\emailAdd{kim64656@yonsei.ac.kr}
\emailAdd{kanghoon.lee1@gmail.com}

\abstract{
The quantum off-shell recursion provides an efficient and universal computational tool for loop-level scattering amplitudes. In this work, we present a new comprehensive computational framework based on the quantum off-shell recursion for binary black hole systems. Using the quantum perturbiner method, we derive the recursions and solve them explicitly up to two-loop order. We develop a power-counting prescription that enables the straightforward separation of classical diagrams. We also devise a classification scheme that optimizes the integration by parts (IBP) reduction process, which makes higher-loop calculations more tractable. By employing the soft expansion technique, we remove irrelevant terms from the loop integrands and express them in terms of master integrals. We classify the one-loop and the two-loop classical diagrams, and their loop integrands are represented by linear combinations of the master integrals. Finally, we explicitly calculate the classical scalar $2\to 2$ amplitudes in the potential region up to the 3PM order and reproduce the known results.
}

\begin{document}
\maketitle	
\flushbottom
%
\section{Introduction}

Recently, quantum field theory (QFT) techniques have been employed to develop a rigorous theoretical framework for studying binary black hole systems despite the significant difference in the scale between elementary particles and black holes. There has been substantial progress in this direction and established a comprehensive theoretical framework that can accurately describe the dynamics of binary black hole systems, especially in the inspiral phase where the black holes are sufficiently separated. In this limit, black holes can be considered as point particles up to a particular scale, allowing their interactions to be treated as perturbations. In this context, the field theory method \cite{Neill:2013wsa,Luna:2017dtq,Kosower:2018adc,Foffa:2016rgu,Bjerrum-Bohr:2018xdl,Chung:2018kqs,Cheung:2018wkq,Plefka:2018dpa,Bern:2019nnu,Bern:2019crd,Blumlein:2019zku,Brandhuber:2019qpg,Cristofoli:2019ewu,Cristofoli:2019neg,Damour:2019lcq,Foffa:2019hrb,Kalin:2019rwq,KoemansCollado:2019ggb,Bern:2020gjj,Blumlein:2020znm,Bini:2020wpo,Bini:2020uiq,Bini:2020rzn,Cheung:2020gyp,Cheung:2020sdj,Cristofoli:2020uzm,DiVecchia:2020ymx,Kalin:2020mvi,Kalin:2020fhe,Parra-Martinez:2020dzs,Bern:2021yeh,Bern:2021dqo,Bjerrum-Bohr:2021din,Bjerrum-Bohr:2021vuf,Bjerrum-Bohr:2021wwt,Britto:2021pud,DiVecchia:2021bdo,DiVecchia:2021ndb,Dlapa:2021npj,Dlapa:2021vgp,Herrmann:2021lqe,Herrmann:2021tct,Bellazzini:2022wzv,Bern:2022jvn,Blumlein:2022qjy,DiVecchia:2022piu,Dlapa:2022lmu,Heissenberg:2022tsn,Jones:2022aji,Barack:2023oqp,Bern:2023ccb,Bjerrum-Bohr:2023iey,Damgaard:2023ttc,Jones:2023ugm,Kosmopoulos:2023bwc,Rettegno:2023ghr,Wilson-Gerow:2023syq,Porto:2005ac,Vaidya:2014kza,Bini:2017xzy,Cachazo:2017jef,Vines:2017hyw,Guevara:2017csg,Bini:2018ywr,Guevara:2018wpp,Levi:2018nxp,Vines:2018gqi,Chung:2019duq,Damgaard:2019lfh,Guevara:2019fsj,Maybee:2019jus,Aoude:2020onz,Bern:2020buy,Guevara:2020xjx,Levi:2020kvb,Levi:2020uwu,Jakobsen:2022fcj,Aoude:2023dui,Aoude:2023vdk,DiVecchia:2023frv,Heissenberg:2023uvo,Jakobsen:2023ndj,Jakobsen:2023hig} along with the worldline QFT method \cite{Bel:1981be,Westpfahl:1985tsl,Goldberger:2004jt,Goldberger:2006bd,Ledvinka:2008tk,Goldberger:2009qd,Steinhoff:2014kwa,Damour:2016gwp,Blanchet:2018yvb,Comberiati:2022cpm,Jakobsen:2022psy,Mogull:2020sak,Saketh:2022xjb,Wang:2022ntx,Bhattacharyya:2023kbh,Bern:2023ity,Damgaard:2023vnx,Kopp:2023eas} has emerged as an efficient and powerful computational tool that provides a systematic procedure for converting loop corrections to classical perturbations, which are crucial for enhancing the accuracy of the description.

However, applying the QFT techniques to binary black hole systems is not straightforward and poses several challenges. Traditional methods based on the Feynman diagrams involve cumbersome calculations and do not provide a clear pathway for incorporating higher-order corrections. Moreover, the perturbative nature of QFT requires a careful analysis of loop diagrams, which can quickly become intractable for higher-loop calculations. Modern methods, primarily based on the double copy \cite{Kawai:1985xq,Bern:2008qj,BjerrumBohr:2009rd,Stieberger:2009hq,Bjerrum-Bohr:2010diw,Bjerrum-Bohr:2010mtb,Bern:2010ue,Bern:2010yg,Tye:2010dd,BjerrumBohr:2010zs,Feng:2010my,Mafra:2011kj,Monteiro:2011pc,BjerrumBohr:2012mg,Monteiro:2014cda,Luna:2016hge,Bern:2019prr,Bautista:2019tdr} and unitarity cut approach \cite{Bern:1994zx,Bern:1994cg,Bern:1995db,Britto:2004nc,Ossola:2006us,Forde:2007mi,Berger:2008sj,Bern:2011qt}, offer a more efficient computational framework that dispenses with explicit Feynman diagrams or Lagrangian.

In the present article, we establish an alternative computational framework based on the quantum off-shell recursion relations \cite{Lee:2022aiu}, which is the loop-level generalization of the Berends-Giele recursions \cite{Berends:1987me} for tree-level. The quantum off-shell recursion provides an efficient and robust tool for deriving loop integrands. Furthermore, this framework is universally applicable to any theory with Lagrangian.

We initiate by introducing the field theory of non-rotating binary black hole systems \cite{Goldberger:2004jt,Porto:2016pyg,Cheung:2018wkq}. In the field theory description, the non-rotating black holes are realized by massive scalar fields minimally coupled to gravity. We then construct the perturbative Lagrangian and its corresponding equations of motion (EoM) by adopting  Cheung and Remmen's convention \cite{Cheung:2016say,Cheung:2017kzx} of metric perturbations. This choice yields simpler expressions for the perturbative expansions than conventional perturbative General Relativity (GR) approaches. Next, we derive the Dyson-Schwinger (DS) equation, a quantum counterpart to the classical EoM, by deforming the perturbative EoM. This plays a crucial role in constructing quantum off-shell recursions.

Subsequently, we construct the quantum perturbiner method \cite{Lee:2022aiu} derived from the quantum effective action formalism. The perturbiner expansion is the generating function of the off-shell currents directly related to the scattering amplitude. In this context, the quantum off-shell recursions can be derived by substituting the perturbiner expansion into the DS equation. We need to perform the $\hbar$-expansion of off-shell currents, which is crucial for constructing self-consistent recursion relations. We solve the recursions up to two-loop order explicitly. Interestingly, the solutions of the off-shell recursions are loop integrands and include both quantum and classical contributions.

Next, we introduce a new prescription for extracting and classifying classical loop diagrams compatible with the off-shell recursion. Since the off-shell recursion generates loop integrands by solving the recursion relations without considering explicit diagrams, we need to invent a new prescription, a more algebraic method. We developed a power-counting prescription by defining an identifier for diagrams that can be easily implemented in the loop integrands. To this end, we introduce a set of coupling constants by dividing the Newton constant $\kappa$ according to the interaction types and a label for each propagator. We then define the identifier as power counting of the new couplings and labels of a diagram. It is straightforward to filter out the integrands corresponding to the classical diagrams from the total result.

Further, we devise a classification scheme for optimizing the IBP reduction \cite{Chetyrkin:1981qh,Kotikov:1990kg,Kotikov:1991pm,Remiddi:1997ny,Henn:2013pwa,Henn:2014qga} process. We first group the classical diagrams according to the power countings of the new coupling constants, and we call the groups \emph{classes}. We next categorize each class according to the diagram topologies. We first define the \emph{primary diagrams} containing the maximum number of internal lines. We then may generate diagrams by pinching the internal lines from a primary diagram. The set of diagrams generated from the same primary diagram is called \emph{sector}. The main advantage of this classification is that the diagrams included in the same class incorporate the same integral family. In the one-loop case, there is only one class with two sectors. In the two-loop case, there are four classes, and these classes have their own sectors.

Even for the classical diagrams, numerators of their loop integrands may include terms that are irrelevant to the classical long-range interaction. Utilizing the soft expansion that scales the loop momenta and momentum transfer \cite{Parra-Martinez:2020dzs}, we truncate the unnecessary terms in the loop integrands by the power counting rules, thereby obtaining the classical loop integrands. We also define the soft integral family by replacing the conventional scalar propagators with linear ones and derive the master integrals for each sector using IBP reduction. Using the IBP reduction once more, we may express the classical loop integrands for each sector in terms of these master integrals, which can be computed by solving the associated differential equations. We refer to the solutions of the differential equations derived in \cite{Parra-Martinez:2020dzs}. Finally, by summing contributions from each sector, we acquire the explicit expression of the classical scalar $2\to 2$ amplitudes up to 3PM order. These results are the same as the known results.

This article is structured as follows: Section 2 introduces our field theory and derives the perturbative EoMs. We also construct the DS equations using the deformation of the fields to operators. Section 3 introduces the perturbiner expansion at the loop level and the quantum off-shell recursion relations. Section 4 presents our power-counting prescription for filtering and classifying classical loop diagrams. Section 5 discusses the soft expansion and IBP reduction techniques with the one-loop example. Section 6 provides explicit computations of scattering amplitudes at two loop levels, and Section 7 concludes the paper with a summary and outlook for future research.

\section{Field Theory for Non-spinning Binary Black Holes} \label{Sec:2}

In this section, we discuss the field theory description of a non-rotating black hole merger in the inspiral phase. During this phase, black holes are sufficiently separated relative to the horizon scale, thereby allowing us to approximate the black holes as weakly interacting massive point particles. We first introduce the field theory, which consists of the Einstein-Hilbert (EH) action and the minimally coupled massive scalar fields. Next, we derive the perturbations of the field theory action and its equations of motion (EoM). Finally, we construct the Dyson-Schwinger (DS) equation, which is crucial in constructing the quantum off-shell recursion.

\subsection{Perturbations of the action and EoM} \label{Sec:2.1}

We consider the QFT framework of the non-rotating binary black hole merger in the inspiral phase. The field theory action consists of two parts: the EH action $S_{\rm EH}$ and the massive scalar fields $S_{\rm matter}$
\begin{equation}
  S_{\rm total} = S_{\rm EH} + S_{\rm matter}\,.
\label{}\end{equation}
The dynamics of gravitons are governed by the EH action, which is proportional to the Ricci scalar $R$
\begin{equation}
\begin{aligned}
  S_{\rm EH} &= \int \mathrm{d}^{4}x L_{\rm EH} \,,
  \qquad
  L_{\rm EH} = \frac{1}{2 \kappa^{2}} \sqrt{-g} R\,,
\end{aligned}\label{}
\end{equation}
where $\kappa^{2} = 8\pi G$ and $G$ is the Newton constant. The matter action is given by massive scalar fields, $\phi_{1}$ and $\phi_{2}$, interacting through only the minimally coupled gravitons. We also ignore the self-interaction terms due to the absence of pair-creations or pair-annihilations of black holes,
\begin{equation}
    S_{\rm matter} =\int_{x}\sqrt{-g}
  	\sum_{I=1}^{2} \left(- \frac{1}{2} g^{\mu\nu} \partial_{\mu} \phi_{I} \partial_{\nu} \phi_{I} -\frac{1}{2} m_{I}^{2} \phi^{2}_{I} \right)\,.
\label{}\end{equation}

For a concise perturbation formulation of GR, we transform the metric tensor to the tensor density in the Landau-Lifshitz book \cite{Landau:1975pou}. The tensor density $\sigma$ is defined as
\begin{equation}
  \sigma^{\mu\nu} = \sqrt{-g} g^{\mu\nu } \,,
  \qquad
  \sigma_{\mu\nu} = \frac{1}{\sqrt{-g}} g_{\mu\nu} \,,
\label{def_sigma1}\end{equation}
and $\sigma^{\mu\nu}$ is the inverse of $\sigma_{\mu\nu}$. We recast the EH action and the matter action in terms of the tensor density up to the boundary term
\begin{equation}
\begin{aligned}
  S_{\rm EH} &= \frac{1}{2 \kappa^{2}}\int \mathrm{d}^{D} x \left[\
  	  \frac{1}{4} \sigma^{\mu\nu} \partial_{\mu} \sigma^{\rho\sigma} \partial_{\nu} \sigma_{\rho\sigma}
  	- \frac{1}{2} \sigma^{\mu\nu} \partial_{\mu} \sigma^{\rho\sigma} \partial_{\rho} \sigma_{\nu\sigma}
   	+ (D-2) \sigma^{\mu\nu}\partial_{\mu}\hat{d} \partial_{\nu}\hat{d} \
  	\right]\,,
  \\
    S_{\rm matter} &= \int \mathrm{d}^{D}x \sum_{I=1}^2\left[-\frac{1}{2} \sigma^{\mu \nu} \partial_{\mu} \phi_I \partial_{\nu} \phi_{I}
    -\frac{1}{2}m_{I}^{2}|\operatorname{det} \sigma|^{-\frac{1}{2}} \phi_{I}^{2}\right]\,,
\end{aligned}\label{EH_action_sigma}
\end{equation}
where $\partial_{\mu}\hat{d} = - \frac{1}{4} \sigma^{\rho\sigma} \partial_{\mu} \sigma_{\rho\sigma}$.

Let us consider the perturbations of the field theory around a flat background. Although we could consider perturbations of the metric tensor in the usual way, we employ Cheung and Remmen's convention of metric perturbations \cite{Cheung:2016say,Cheung:2017kzx,Cho:2022faq}. This convention yields much simpler expressions than the usual metric perturbation, defined as a linear fluctuation of $g_{\mu\nu}$ around a trivial background $\eta_{\mu\nu}$. In contrast, we define metric perturbations as a linear fluctuation of the inverse tensor density $\sigma^{\mu\nu}$,
\begin{equation}
  \sigma^{\mu\nu} = \eta^{\mu\nu} - \kappa h^{\mu\nu}\,,
  \qquad
  \sigma_{\mu\nu} = \eta_{\mu\nu} + \sum_{n=1}^{\infty} \kappa^{n} (h^{n})_{\mu\nu} \,,
  \qquad
  |h^{\mu\nu}| \ll 1\,.
\label{perturb_sigma}\end{equation}
Here the indices are raised or lowered by the background metric $\eta$

Upon this convention, the all-order expression of the perturbative EH action \eqref{EH_action_sigma} and the Einstein equation in $\kappa$ are obtained in \cite{Cho:2022faq}, and the perturbations of the Lagrangian action is given by
\begin{equation}
\begin{aligned}
  L_{\rm EH} =&\frac{1}{2\kappa^{2}}
  \sum_{n=0}^{\infty} \bigg[
    - \frac{1}{4}\kappa^{n+2} (\eta-\kappa h)^{\mu\nu} \sum_{m=0}^{n} \tr\big[ h^{m} \partial_{\mu}h h^{n-m} \partial_{\nu} h\big]
  	+ \frac{1}{2} \kappa^{n+2} \big[\partial_{\nu}h h^{n} \partial_{\mu}h\big]^{\mu\nu}
  \\&\qquad\qquad
  	+ \frac{1}{4(D-2)}\kappa^{n+2}(\eta-\kappa h)^{\mu\nu} \sum_{m=0}^{n} \tr\big[h^{m}\partial_{\mu}h\big] \tr\big[h^{n-m}\partial_{\nu}h\big]
  \bigg] \,.
\end{aligned}\label{pert_action_1}
\end{equation}
We impose the de Donder gauge choice, $\partial_{\mu} h^{\mu\nu} = 0$, and the gauge fixing term is
\begin{equation}
  S_{\rm gf} = \int \mathrm{d}^{4}{x} \Big(-\frac{1}{4} \partial_{\mu} h^{\mu\rho} \partial^{\nu} h_{\nu\rho}\Big)\,.
\label{gauge_fixing_term}\end{equation}

For general two-loop scattering amplitude, including both quantum and classical contributions, we should expand the EH action to $\kappa^{3}$. However, we are considering the ``classical'' two-loop amplitude -- it will be clear the meaning of classical in Section \ref{Sec:4}  -- and it is sufficient to expand the action up to $\kappa^{2}$-order terms or four-point vertices,
\begin{equation}
  L_{\rm EH} + L_{\rm gf} = \int \mathrm{d}^{4}x \left[
  \frac{1}{2}L^{h}_{2} + \frac{1}{2}\kappa L^{h}_{3} + \frac{1}{2}\kappa^{2} L^{h}_{4} + j^{\mu\nu} h_{\mu\nu}
  \right]\,,
\label{perturbed_action1}\end{equation}
where we added the bare external source for graviton $j^{\mu\nu}$. Here $L^{h}_{2}$ is the graviton kinetic term in the presence of the gauge fixing term \eqref{gauge_fixing_term}
\begin{equation}
  L^{h}_{2} =  -\frac{1}{4} \partial^{\rho} h^{\mu\nu} \partial_{\rho} h_{\mu\nu}
  + \frac{1}{8} \partial^{\mu}h \partial_{\mu} h\,,
\label{}\end{equation}
where $h = h^{\mu}{}_{\mu}$. The graviton self-interaction terms $L^{h}_{3}$ and $L^{h}_{4}$ are given by
\begin{equation}
\begin{aligned}
  L^{h}_{3} &=
      \frac{1}{2} \big[\partial_{\nu}h h\partial_{\mu}h\big]^{\mu\nu}
  	- \frac{1}{2} \tr\big[\partial^{\mu}h h \partial_{\mu} h\big]
	+ \frac{1}{4}  \tr\big[h\partial_{\mu}h\big]\partial^{\mu}h
	+ \frac{1}{4} h^{\mu\nu} \tr\big[\partial_{\mu}h \partial_{\nu} h\big]
	\\&\quad
	- \frac{1}{8} h^{\mu\nu} \partial_{\mu}h \partial_{\nu}h \,,
  \\
  L^{h}_{4} &=
  - \frac{1}{2} \tr\big[h^{2} \partial^{\mu} h \partial_{\mu}h\big]
  - \frac{1}{4} \tr\big[h\partial^{\mu} h h\partial_{\mu}h\big]
  + \frac{1}{4} \tr\big[h^{2}\partial^{\mu}h\big] \partial_{\mu} h
  + \frac{1}{8} \tr\big[h\partial^{\mu}h\big]\tr\big[h\partial_{\mu}h\big]
  \\&\quad
  + \frac{1}{2} h^{\mu\nu} \tr\big[h\partial_{\mu}h \partial_{\nu}h\big]
  - \frac{1}{4} h^{\mu\nu} \tr\big[h\partial_{\mu}h\big] \partial_{\nu}h
  + \frac{1}{2} \big[\partial_{\nu}h h^{2}\partial_{\mu}h\big]^{\mu\nu} \,.
\end{aligned}\label{perturbed_EH_action}
\end{equation}
Here we have employed the matrix notation to avoid cumbersome dummy indices.\footnote{ For arbitrary matrices $A^{\mu}{}_{\nu}$, $B^{\mu}{}_{\nu}$ and $C^{\mu}{}_{\nu}$, we denote their matrix product as
\begin{equation}
  A^{\mu}{}_{\rho} B^{\rho}{}_{\nu} = \big[A B\big]^{\mu}{}_{\nu}
  \qquad
  A^{\mu}{}_{\rho} B^{\rho}{}_{\sigma} C^{\sigma}{}_{\nu}= \big[A B C\big]^{\mu}{}_{\nu}\,,
\label{}\end{equation}
and the trace of matrices as
\begin{equation}
  A^{\mu}{}_{\rho} B^{\rho}{}_{\mu} := \tr\big[A B\big]\,,
  \qquad
  A^{\mu}{}_{\rho} B^{\rho}{}_{\sigma} C^{\sigma}{}_{\mu} := \tr\big[A B C\big]\,.
\label{}\end{equation}
}
One remark is that we do not need to introduce the ghost fields because the classical diagrams do not allow the graviton loops.

We may rewrite the kinetic term $L^{h}_{2}$ using the kinetic operator $K^{\mu\nu,\rho\sigma}[x,y]$
\begin{equation}
  S^{h}_{2} = \int \mathrm{d}^{4} x \frac{1}{2} L^{h}_{2} =  \int \mathrm{d}^{4}x\mathrm{d}^{4}y \bigg[ - \frac{1}{2} h_{\mu\nu}(x) K^{\mu\nu,\rho\sigma}[x,y] h_{\rho\sigma}(y) \bigg]\,,
\label{}\end{equation}
where
\begin{equation}
  K^{\mu\nu,\rho\sigma}[x,y] = - \frac{1}{4} \bigg( \eta^{(\mu|\rho|} \eta^{\nu)\sigma} - \frac{1}{2}\eta^{\mu\nu} \eta^{\rho\sigma}\bigg) \Box_{y}\delta^{D} (x-y) \,.
\label{}\end{equation}
Then the graviton propagator $D^{\mu\nu,\rho\sigma}[x,y]$ is the inverse of $K^{\mu\nu,\rho\sigma}[x,y]$
\begin{equation}
  D^{\mu\nu,\rho\sigma}[x,y] = 2 P^{\mu\nu,\rho\sigma} D[x,y]  \,,
\label{}\end{equation}
where
\begin{equation}
  D[x,y] = \int \frac{\mathrm{d}^{4} p}{(2\pi)^{4}} \frac{1}{p^{2}} e^{ip\cdot x}\,,
  \qquad
  P^{\mu\nu,\rho\sigma} = \eta^{(\mu|\rho|} \eta^{\nu)\sigma} - \frac{1}{2}\eta^{\mu\nu} \eta^{\rho\sigma}\,.
\label{}\end{equation}

Let us consider the perturbative action of the minimally coupled massive scalars. We should keep the terms up to the $\kappa^{3}$-orders, which is one order higher than the EH action,
\begin{equation}
\begin{aligned}
    L_{\rm matter} &= \sum_{I=1}^2\left[-\frac{1}{2} \sigma^{\mu \nu} \partial_{\mu} \phi_I \partial_{\nu} \phi_{I}
    -\frac{1}{2}m_{I}^{2}|\operatorname{det} \sigma|^{-\frac{1}{2}} \phi_{I}^{2}\right]
  \\
   &= \sum_{I=1}^{2} \Bigg[- \frac{1}{2} \big(\eta^{\mu\nu}- \kappa h^{\mu\nu}\big) \partial_{\mu} \phi_{I} \partial_{\nu} \phi_{I}
    - \frac{m_{I}^{2}}{2}\bigg(1-\frac{\kappa h}{2}+\frac{\kappa^{2}h^{2}}{8} - \frac{\kappa^{2}h_{\mu\nu}h^{\mu\nu}}{4} \bigg) \phi_{I}^{2}
    \\&\qquad\quad
    +\kappa^{3}(\cdots) \Bigg] \,.
\end{aligned}\label{perturbed_matter_action}
\end{equation}

For later convenience, we introduce the capital Roman indices $M, N, P\cdots$ to denote a pair of symmetric indices, such as
\begin{equation}
  h^{M} = h^{\mu\nu}\,, \qquad P^{M,N} = P^{\mu\nu,\rho\sigma}\,,
\label{}\end{equation}
and in general
\begin{equation}
  v^{MNP \cdots} = v^{(\mu\nu)(\rho\sigma)(\kappa\lambda)\cdots}\,.
\label{}\end{equation}
To avoid the complicated structures of contractions of dummy indices, we define tensors that consist of $\eta^{\mu\nu}$ and $\delta_{\mu}{}^{\nu}$. We will call them as $\mathcal{I}$-tensors. For instance, let us consider a matrix product of three $h^{\mu\nu}$,
\begin{equation}
\begin{aligned}
  {}[hhh]^{\mu\nu} &= h^{\mu\rho} h^{\rho\sigma} h^{\sigma\nu}
  \\&
  = \big(\eta^{\mu\kappa_{1}} \eta^{\nu\lambda_{3}} \eta^{\lambda_{1}\kappa_{2}} \eta^{\lambda_{2}\kappa_{3}} \big)h^{\kappa_{1}\lambda_{1}} h^{\kappa_{2}\lambda_{2}} h^{\kappa_{3}\lambda_{3}}\,,
  \\&
  = \mathcal{I}^{\mu\nu,\kappa_{1}\lambda_{1},\kappa_{2}\lambda_{2},\kappa_{3}\lambda_{3}} h^{\kappa_{1}\lambda_{1}} h^{\kappa_{2}\lambda_{2}} h^{\kappa_{3}\lambda_{3}}\,,
  \\&
  =  \mathcal{I}^{\mu\nu,N,P,Q} h^{N} h^{P} h^{Q}\,.
\end{aligned}\label{}
\end{equation}
where $\mathcal{I}^{\mu\nu,\kappa_{1}\lambda_{1},\kappa_{2}\lambda_{2},\kappa_{3}\lambda_{3}} = \eta^{\mu(\kappa_{1}} \eta^{\lambda_{1})(\kappa_{2}} \eta^{\lambda_{2})(\kappa_{3}} \eta^{\lambda_{3})\nu}$. Hereafter, we will ignore the position of the indices while keeping the summation convention for repeated indices. In Appendix \ref{App:A}, we present the explicit expressions of the $\mathcal{I}$-tensors and the general structure of the EoMs.

Equipped with the $\mathcal{I}$-tensors, we now represent the EoMs of $h^{\mu\nu}$ and $\phi^{I}$. The EoM of $h^{\mu\nu}$ is divided into two parts: the perturbations of the curvature part $\mathcal{G}^{M}_{n}$ and the energy-momentum (EM) tensor $\mathcal{T}^{M}_{n}$
\begin{equation}
  \Box h^{M} = 2P^{M,N} \Biggl[
   \sum_{n=1}^{2} \kappa^{n}\mathcal{G}^{N}_{n} - \sum_{n=1}^{3}\kappa^{n}\mathcal{T}^{N}_{n}  - 2 j^{N}\Bigg]\,.
\label{perturbed_Eistein_eq}\end{equation}
Here, curvature perturbations $\mathcal{G}^{M}_{1}$ and $\mathcal{G}^{M}_{2}$ are obtained by variations of $L^{h}_{3}$ and $L^{h}_{4}$ in \eqref{perturbed_EH_action} respectively
\begin{equation}
\begin{aligned}
  \mathcal{G}^{M}_{1} &=
	\mathcal{I}^{M,\rho N,\sigma P}_{1,1} \partial_{\rho} h^{N} \partial_{\sigma} h^{P}
  + \mathcal{I}^{M,N,\rho\sigma P}_{1,2} h^{N} \partial_{\rho} \partial_{\sigma} h^{P}\,,
  \\
  \mathcal{G}^{M}_{2} &=
   \mathcal{I}^{M,N,\rho P,\sigma Q}_{2,1} h^{N} \partial_{\rho} h^{P} \partial_{\sigma} h^{Q}
  + \mathcal{I}^{M,N,P,\rho \sigma Q}_{2,2} h^{N} h^{P} \partial_{\rho} \partial_{\sigma} h^{Q}\,,
\end{aligned}\label{G_tensor}
\end{equation}
where $\mathcal{I}_{m,n}$ are defined in \eqref{I_12_length3} and \eqref{I_12_length4}.
The EM tensor parts, $\mathcal{T}^{\mu\nu}_{1}$, $\mathcal{T}^{\mu\nu}_{2}$ and $\mathcal{T}^{\mu\nu}_{3}$, are given by variations of the perturbed matter action \eqref{perturbed_matter_action},
\begin{equation}
\begin{aligned}
  \mathcal{T}^{\mu\nu}_{1} &=  \sum_{I=1}^{2} \left[
  	\partial^{\mu} \phi_{I} \partial^{\nu} \phi_{I}
  + \frac{m_{I}^{2}}{2} \eta^{\mu\nu} \phi_{I} \phi_{I}\right]\,,
  \\
  \mathcal{T}^{M}_{2} &= \mathcal{I}_{\rm EM}^{M,N} \sum_{I=1}^{2} m^{2}_{I} h^{N} \phi_{I} \phi_{I} \,,
  \qquad
  \mathcal{T}^{M}_{3} = \mathcal{I}_{\rm EM}^{M,N,P} \sum_{I=1}^{2} m^{2}_{I} h^{N} h^{P} \phi_{I} \phi_{I} \,,
\end{aligned}\label{}
\end{equation}
where we present $\mathcal{I}_{{\rm EM}}$ in \eqref{I_EM1} and \eqref{I_EM2}.

Similarly, perturbations of EoM of the scalar fields can be represented as
\begin{equation}
\begin{aligned}
  \left(\square-m_{I}^{2}\right) \phi_{I}
  &= - \rho_{I}
	+ \kappa \bigg[\partial_{\kappa} \big(h^{\kappa\lambda} \partial_{\lambda} \phi_{I}\big)
	- \frac{\kappa m_{I}^{2}}{2} h^{\kappa\kappa} \phi_{I}\bigg]
  \\&\quad
	+ \kappa^{2} m^{2}_{I} \mathcal{I}^{M,N}_{\phi}h^{M} h^{N} \phi_{I}
  	+ \kappa^{3} m^{2}_{I} \mathcal{I}^{M,N,P}_{\phi} h^{M} h^{N} h^{P} \phi_{I}\,,
\end{aligned}
\label{EoM_phi}\end{equation}
where $\mathcal{I}_{\phi}$ are defined in \eqref{I_phi}.

\subsection{Dyson-Schwinger equations}

The Dyson-Schwinger (DS) equation, a quantum counterpart of the classical EoM, is crucial in constructing the quantum off-shell recursions. It can be obtained simply by deforming the fields of the classical EoMs. In this context, the classical fields are defined by the vacuum expectation value of the fields
\begin{equation}
\begin{aligned}
  \varphi^{I}_{x} = \frac{\delta W[j,\rho]}{\delta \rho^{I}_{x}}\,,
  \qquad
  h^{\mu\nu}_{x} = \frac{\delta W[j,\rho]}{\delta j^{\mu\nu}_{x}}\,,
\end{aligned}\label{classical_fields}
\end{equation}
where $\rho^{I}_{x}$ and $j^{\mu\nu}_{x}$ are the bare external sources for $\phi^{I}$ and $h^{\mu\nu}$, and $W[j,\rho]$ is the generating functional for connected diagrams
\begin{equation}
  \exp\big[W[j,\rho]\big] = \int \mathcal{D}\phi^{I} \mathcal{D} h_{\mu\nu} \exp\bigg[\frac{i}{\hbar}\Big(S_{\rm EH} +S_{\rm matter} + \rho^{I}_{x} \phi^{I}_{x} + j^{\mu\nu}_{x} h_{x}^{\mu\nu}\Big)\bigg]\,.
\label{}\end{equation}
We will encounter the dressed external sources in the next section. Here, the subscripts denote coordinates such as
\begin{equation}
  \varphi^{I}_{x} = \varphi^{I} (x)\,, \qquad \rho^{I}_{x} = \rho^{I}(x)\,, \qquad h^{\mu\nu}_{x} =h^{\mu\nu}(x)\,, \qquad j^{\mu\nu}_{x} = j^{\mu\nu}(x)\,.
\label{}\end{equation}
We will adopt this notation throughout this paper.

The DS equations are generated by the following deformations of the fields from the classical EoM \cite{Lee:2022aiu}
\begin{equation}
\begin{aligned}
  \phi^{I}_{x} \to \varphi^{I} + \frac{\hbar	}{i}\frac{\delta}{\delta \rho^{I}_{x}} \,,
  \qquad
  h^{M}_{x} \to h^{\mu\nu}_{x} + \frac{\hbar}{i}\frac{\delta}{\delta j^{\mu\nu}_{x}} \,.
\end{aligned}\label{}
\end{equation}
We summarized the general structure of the deformation rule in Appendix \ref{Sec:B.1}.

Thus, the DS equation includes several functional derivatives of the $\varphi^{I}$ and $h^{\mu\nu}$, which makes it difficult to solve. We follow the strategy that treats the functional derivatives as independent degrees of freedom, called \emph{descendant fields} \cite{Lee:2022aiu}. Considering up to two-loop order, the first and the second descendant fields arise in the DS equations. The first descendant fields are the single functional derivatives of the classical fields
\begin{equation}
\begin{aligned}
  \psi^{MN}_{xy} &= \frac{\delta h_{x}^{M}}{\delta j^{N}_{y}} \,, &\quad \psi^{MI}_{xy}&= \frac{\delta h_{x}^{M}}{\delta \rho^{I}_{y}}\,,
  &\quad
  \psi^{IM}_{xy} &= \frac{\delta \varphi_{x}^{I}}{\delta j^{M}_{y}} \,, &\quad \psi^{IJ}_{xy} &= \frac{\delta \varphi_{x}^{I}}{\delta \rho^{J}_{y}} \,,
\end{aligned}\label{first_descendant_fields}
\end{equation}
and the second descendant fields are the double-functional derivatives of the classical fields
\begin{equation}
\begin{aligned}
    \psi^{MNP}_{xyz} &= \frac{\delta^{2} h_{x}^{M}}{\delta j^{N}_{y}\delta j^{P}_{z}} \,, &\quad \psi^{MNI}_{xyz}&= \frac{\delta^{2} h_{x}^{M}}{\delta j^{N}_{y} \delta\rho^{I}_{z}}\,,
  &\quad
    \psi^{MIN}_{xyz} &= \frac{\delta^{2} h_{x}^{M}}{\delta \rho^{I}_{y}\delta j^{N}_{z}} \,, &\quad \psi^{MIJ}_{xyz}&= \frac{\delta^{2} h_{x}^{M}}{\delta \rho^{I}_{y} \delta\rho^{J}_{z}}\,,
  \\
  \psi^{IMN}_{xyz} &= \frac{\delta^{2} \varphi_{x}^{I}}{\delta j^{M}_{y} \delta j^{N}_{z}} \,, &\quad \psi^{IMJ}_{xyz} &= \frac{\delta^{2} \varphi_{x}^{I}}{\delta j^{M}_{y} \delta \rho^{J}_{z}} \,,
  &\quad
  \psi^{IJM}_{xyz} &= \frac{\delta^{2} \varphi_{x}^{I}}{\delta \rho^{J}_{y} \delta j^{M}_{z}} \,, &\quad \psi^{IJK}_{xyz} &= \frac{\delta^{2} \varphi_{x}^{I}}{\delta \rho^{J}_{y}\rho^{K}_{z}} \,.
\end{aligned}\label{second_descendant_fields}
\end{equation}

Again, for concise expressions, we introduce collective indices $A,B,C, \cdots$ by combining $M, N, \cdots $ and $I,J, \cdots$,
\begin{equation}
  A = \{M, I\}\,, \quad B = \{N, J\}\,,\quad \cdots \,.
\label{}\end{equation}
Using the notations, we may define the collective external sources $\mathbf{j}^{A} = \{j^{\mu\nu}, \rho^{I}\} = \{j^{M}, \rho^{I}\}$. The collective classical fields are given by
\begin{equation}
  \mathbf{h}^{A}_{x} = \frac{\delta W[\mathbf{j}]}{\delta \mathbf{j}^{A}_{x}} = \big\{h^{M}_{x},\varphi^{I}_{x}\big\}\,,
\label{collective_classical_field}\end{equation}
and their descendants defined in \eqref{first_descendant_fields} and \eqref{second_descendant_fields} are represented in a simple form
\begin{equation}
\begin{aligned}
  \psi^{IA}_{xy} = \frac{\delta \varphi_{x}^{I}}{\delta \mathbf{j}_{y}^{A}} \,,
  \qquad
  \psi^{M A}_{xy} = \frac{\delta h^{M}_{x}}{\delta \mathbf{j}_{y}^{A}}\,,
  \qquad
  \psi^{IAB}_{xyz} = \frac{\delta^{2} \varphi_{x}^{I}}{\delta \mathbf{j}^{A}_{y} \delta \mathbf{j}^{B}_{z}}\,,
  \qquad
  \psi^{MAB}_{xyz} = \frac{\delta h^{M}_{x}}{\delta \mathbf{j}^{A}_{y} \delta \mathbf{j}^{B}_{z}}\,.
\end{aligned}\label{}
\end{equation}

In solving the DS equations, the $\hbar$-expansion is crucial. Since the descendant fields are treated by independent degrees of freedom, their field equations should be imposed for solving the DS equation. We may derive the EoM for the descendant fields by acting on additional functional derivatives to the DS equation. However, this process never stops, and one encounters an infinite sequence of new descendant fields upon each functional derivative. To bypass this issue, we should expand the DS equation to a particular loop order. Specifically, we can expand the DS equation in $\hbar$, directly related to the loop order. Notably, all the classical fields and their descendant fields are derived from the full generating functional $W[\mathbf{j}]$ without any approximation; these contain all order expansions in $\hbar$
\begin{equation}
\begin{aligned}
  \mathbf{h}^{A}_{x} = \sum_{n=0}^{\infty} \mathbf{h}^{\ord{n}} |^{A}_{x} \hbar^{n}\,,
  \qquad
  \psi^{A,B}_{x,y} = \sum_{n=0}^{\infty} \psi^{\ord{n}} |^{A,B}_{x,y} \hbar^{n}\,,
  \qquad
  \psi^{A,B,C}_{x,y,z} = \sum_{n=0}^{\infty} \psi^{\ord{n}} |^{A,B,C}_{x,y,z} \hbar^{n}\,.
\end{aligned}\label{hbar_expansion_classical_fields}
\end{equation}
If we substitute these $\hbar$-expansions and keep the terms at a fixed order in $\hbar$, we can truncate the new descendant fields because these are higher $\hbar$-order terms.

The DS equation for the scalar fields is represented by the $\mathcal{I}_{\phi}$ tensors as
\begin{equation}
\begin{aligned}
  &\left(\square-m_{I}^{2}\right) \varphi^{I}_{x} =
  	- \rho_{x}^{I}
  	+ \kappa \Omega^{I}_{1}
  	+ \kappa^{2} \mathcal{I}^{I,M,N}_{\phi} \Omega^{IMN}_{2}
  	+ \kappa^{3} \mathcal{I}^{I,M,N,P}_{\phi} \Omega^{IMNP}_{3}\,,
\end{aligned}\label{DSeq_scalar}
\end{equation}
where the explicit form of the $\Omega$-tensors are listed in \eqref{Omega_fields}. It is straightforward to derive the field equations for the descendant fields by acting functional derivative $\frac{\delta}{\delta \mathbf{j}^{A}_{x}}$ on the DS equation \eqref{DSeq_scalar}. One can find their explicit form in \eqref{PsiIA} and \eqref{PsiIAB}.

The DS equation for $h^{M}$ is given by
\begin{equation}
  \Box h_{x}^{M} = 2 P^{M,N}\bigg(\sum_{n=1}^{2}\kappa^{n}\mathcal{G}^{M}_{n} - \sum_{n=1}^{3} \kappa^{n} \mathcal{T}^{N}_{n} -2 j^{N}_{x}\bigg)\,,
\label{DSeq_h}\end{equation}
where $\mathcal{G}^{M}_{x}$ represents the curvature perturbations
\begin{equation}
\begin{aligned}
  \mathcal{G}^{M}_{1} &=
  	  \mathcal{I}^{M,\rho N,\sigma P}_{1,1} \Lambda^{\rho N,\sigma P}_{1,1}
  	+ \mathcal{I}^{M,N,\rho \sigma P}_{1,2} \Lambda^{N,\rho \sigma P}_{1,2}\,,
  	\\
    \mathcal{G}^{M}_{2} &=
      \mathcal{I}^{M,N,\rho P,\sigma Q}_{2,1} \Lambda^{N,\rho P,\sigma Q}_{2,1}
    + \mathcal{I}^{M,N,P,\rho \sigma Q}_{2,2} \Lambda^{N,P,\rho \sigma Q}_{2,2} \,,
\end{aligned}\label{}
\end{equation}
and
\begin{equation}
\begin{aligned}
  \mathcal{T}^{M}_{1} = \Theta^{M}_{1}\,,
  \qquad
  \mathcal{T}^{M}_{2} = \mathcal{I}_{\rm EM}^{M,N} \Theta^{N}_{2}\,,
  \qquad
  \mathcal{T}^{M}_{3} = \mathcal{I}_{\rm EM}^{M,N,P} \Theta^{N,P}_{3}\,.
\end{aligned}\label{}
\end{equation}
The explicit form of the $\Lambda$ and $\Theta$ fields are listed in \eqref{Lambda_fields} and \eqref{Theta_fields}.

As we have stated, the field equations for the descendant fields are generated through the successive applications of the functional derivatives $\frac{\delta}{\delta \mathbf{j}^{A}_{x}}$ to the DS equations. The comprehensive formulations of these equations are presented in \eqref{PsiMA} and \eqref{PsiMAB}.

\section{Quantum Perturbiner methods for Binary Black Hole System}

In the present section, we will derive the quantum off-shell recursion relations up to the two-loop level. The central object in the off-shell recursion is the off-shell current, which is given by a correlation function with at least an off-shell leg and amputated on-shell external legs. We first construct the quantum perturbiner expansion for $\mathbf{h}^{A}_{x}$ and its descendant fields, employing the quantum effective action formalism. We then derive the recursions by substituting these expansions into the DS equations derived in the previous section. We also discuss the initial conditions of the recursions.

\subsection{Quantum perturbiner expansion}
The perturbiner method was introduced for deriving the off-shell recursions at tree level using the classical EoM \cite{Selivanov:1997an, Selivanov:1997aq, Rosly:1996vr, Rosly:1997ap, Selivanov:1997ts} instead of patterns in interaction vertices. It has also been applied to gravity \cite{Gomez:2021shh,Cho:2022faq,Tao:2023yxy}. Recently, the perturbiner method has been generalized to the quantum perturbiner method by including loop corrections \cite{Lee:2022aiu,Gomez:2022dzk,Chen:2023bji}. Here we will employ the approach based on the DS equation \cite{Lee:2022aiu}, which is available for higher loops. In this subsection, we derive the quantum perturbiner expansion from the definition of the classical fields \eqref{collective_classical_field} and their descendants.

Let us consider the formal series expansion of the generating functional for connected diagrams $W[\mathbf{j}]$ respective to the collective dressed external source $\mathbf{j}^{A}_{x}$
\begin{equation}
\begin{aligned}
  W[\mathbf{j}] &= \sum_{n>1} \frac{1}{n!} \int_{x_{1} x_{2} \cdots x_{n}} \frac{\delta^{n} W[\mathbf{j}]}{\delta \mathbf{j}^{A_{1}}_{x_{1}} \delta \mathbf{j}^{A_{2}}_{x_{2}} \cdots \delta \mathbf{j}^{A_{n}}_{x_{n}}} \Big|_{\mathbf{j} = 0}  \mathbf{j}^{A_{1}}_{x_{1}} \mathbf{j}^{A_{2}}_{x_{2}} \cdots \mathbf{j}^{A_{n}}_{x_{n}}\,,
  \\
  &= \frac{\hbar}{i} \sum_{n>1} \frac{1}{n!} \int_{x_{1} x_{2} \cdots x_{n}}
  \langle 0\! \left| T\big[\mathbf{h}^{A_{1}}_{x_{1}} \mathbf{h}^{A_{2}}_{x_{2}} \cdots \mathbf{h}^{A_{n}}_{x_{n}}\big] \right|\! 0 \rangle_{c}
  \frac{i \mathbf{j}^{A_{1}}_{x_{1}}}{\hbar} \frac{i \mathbf{j}^{A_{2}}_{x_{2}}}{\hbar} \cdots \frac{i \mathbf{j}^{A_{n}}_{x_{n}}}{\hbar}\,,
\end{aligned}\label{}
\end{equation}
where $\langle 0\! \left| T\big[\mathbf{h}^{A_{1}}_{x_{1}} \mathbf{h}^{A_{2}}_{x_{2}} \cdots \mathbf{h}^{A_{n}}_{x_{n}}\big] \right|\! 0 \rangle_{c}$ is a connected time-ordered $n$-point correlator and
\begin{equation}
  \int_{x, y \cdots}=\int \mathrm{d}^{4} x \mathrm{~d}^{4} y \cdots \quad \text { and } \quad \int_{p, q, \cdots}=\int \frac{\mathrm{d}^{4} p}{(2 \pi)^{4}} \frac{\mathrm{d}^{4} q}{(2 \pi)^{4}} \cdots\,.
\label{}\end{equation}
The corresponding one-point function $\frac{\delta W[\mathbf{j}]}{\delta \mathbf{j}^{A}_{x}}$ in the presence of the external source is given by
\begin{equation}
\begin{aligned}
  \mathbf{h}^{A}_{x}
  = \sum_{n=1}^{\infty} \frac{1}{n !} \int_{y_{1}, y_{2}, \cdots, y_{n}} \langle 0 | T\big[\mathbf{h}^{A}_{x} \mathbf{h}^{B_{1}}_{y_{1}} \cdots \mathbf{h}^{B_{n}}_{y_{n}}\big] | 0 \rangle_{c}
    \frac{i \mathbf{j}^{B_{1}}_{y_{1}}}{\hbar} \cdots \frac{i \mathbf{j}^{B_{n}}_{y_{n}}}{\hbar}\,.
\end{aligned}\label{classical_field}
\end{equation}

We next identify the external source $\mathbf{j}^{A}_{x}$ to make contact with the scattering amplitude. According to the LSZ reduction formula, $W[\mathbf{j}]$ can be written as the summation of scattering amplitudes by choosing the external source $\mathbf{j}^{A}_{x}$ as follows:
\begin{equation}
  \mathbf{j}^{A}_{x}=\sum_{i=1}^{N-1} \int_{y_{i}} \mathbf{K}^{AB_{i}}_{x y_{i}} \boldsymbol{\epsilon}^{B_{i}}_{i} e^{-i k_{i} \cdot y_{i}} = \sum_{i=1}^{N-1} \tilde{\mathbf{K}}^{A B_{i}}_{k_{i}} \boldsymbol{\epsilon}^{B_{i}}_{i} e^{-i k_{i} \cdot x}\,,
\label{def_j}\end{equation}
where $N$ is the total number of external particles, and $\mathbf{K}^{AB}_{x y}$ is the inverse of the dressed propagator $\mathbf{D}^{AB}_{xy}$ in the position space ($\tilde{\mathbf{K}}^{A B_{i}}_{k_{i}}$ and $\tilde{\mathbf{D}}^{AB}_{k_{i}}$ in momentum space) including all-loop corrections, which amputates the external lines. These satisfy the inverse relation:
\begin{equation}
\begin{aligned}
  \int_{y} \mathbf{D}^{AC}_{x, y} \mathbf{K}^{CB}_{y, z}= \delta^{AB} \delta_{x, z}\,,
  \qquad \text{or} \qquad
  \tilde{\mathbf{K}}^{AC}(p) \tilde{\mathbf{D}}^{CB}(p)=\delta^{AB}\,.
\end{aligned}\label{}
\end{equation}
Here $\boldsymbol{\epsilon}^{A}_{i}$ is the collective polarization tensor that consists of the graviton polarization tensor $\epsilon^{\mu\nu}_{i}$ and scalar polarization $\epsilon^{I}_{i}$
\begin{equation}
  \boldsymbol{\epsilon}^{A}_{i} = \big\{\epsilon^{\mu\nu}_{i}, \epsilon^{I}_{i}\big\}\,.
\label{inversePropagator}\end{equation}
The scalar polarization decides the type of the scalar fields $\varphi^{1}$ or $\varphi^{2}$ for each external scalar state. The choice of the polarization depends on the problems in which we are interested. Since we are considering $2\to 2$ scattering between two scalar particles, we choose the polarizations as follows:
\begin{equation}
\begin{aligned}
  \epsilon^{1}_{1} &= 1\,,
  &\qquad
  \epsilon^{1}_{2} &= 0\,,
  &\qquad
  \epsilon^{1}_{3} &= 0\,,
  \\
  \epsilon^{2}_{1} &= 0\,,
  &\qquad
  \epsilon^{2}_{2} &= 1\,,
  &\qquad
  \epsilon^{2}_{3} &= 1\,,
  \\
  \epsilon^{\mu\nu}_{1} &= 0\,,
  &\qquad
  \epsilon^{\mu\nu}_{2} &= 0\,,
  &\qquad
  \epsilon^{\mu\nu}_{3} &= 0\,.
\end{aligned}\label{}
\end{equation}
This is compatible with the kinematics depicted in Figure \ref{kinematics_figure}.

One remark is that we should distinguish the dressed external sources from the bare ones in the action and DS equations. The bare sources are written in the inverse propagators or the kinetic operators,
\begin{equation}
  j^{M}_{x} = \sum_{i=1}^{N-1} \tilde{K}^{MP}_{k_{i}} \epsilon^{P}_{i} e^{-ik_{i}\cdot x}\,,
  \qquad
  \rho^{I}_{x} = \sum_{i=1}^{N-1} \tilde{K}^{IJ}_{k_{i}} \epsilon^{J}_{i} e^{-ik_{i}\cdot x}\,,
\label{}\end{equation}
where $\tilde{K}^{\mu\nu,\rho\sigma}_{k_{i}}$ and $\tilde{K}^{IJ}_{k_{i}}$ are the kinetic operators defined by
\begin{equation}
\begin{aligned}
  \tilde{K}^{\mu\nu,\rho\sigma}_{k_{i}} &= \frac{1}{2} \bigg( \eta^{\mu\rho} \eta^{\nu\sigma} - \frac{1}{2}\eta^{\mu\nu} \eta^{\rho\sigma}\bigg) k_{i}^{2}\,,
  \\
  \tilde{K}^{IJ}_{k_{i}} &= \delta^{IJ} \big(k_{i}^{2} + m_{I}^{2} \big)\,.
\end{aligned}\label{}
\end{equation}

After substituting the definition of the dressed external source in \eqref{def_j} into the expansion of $\mathbf{h}^{A}_{x}$ \eqref{classical_field}, we convert to the momentum space by Fourier transformation
\begin{equation}
\begin{aligned}
  \mathbf{h}^{A}_{x} = \sum_{n=1}^{\infty} \sum_{i_{1},\cdots i_{n}=1}^{n} \int_{\ell} &(2\pi)^{4}\delta^{4}\big(l+k_{i_{1}\cdots i_{n}}\big) \langle 0 | \tilde{\mathbf{h}}^{A}(l) \tilde{\mathbf{h}}^{B_{i_1}} (k_{i_1}) \cdots \tilde{\mathbf{h}}^{B_{i_n}}(k_{i_n}) | 0 \rangle_{c}
  \\&
   \times \Big(\frac{i}{\hbar}\Big)^{n} \big(\tilde{\mathbf{K}}^{B_{i_1} C_{i_1}}_{k_{i_1}} \boldsymbol{\epsilon}^{C_{i_{1}}}\big)
   \cdots \big(\tilde{\mathbf{K}}^{B_{i_{n}}C_{i_{n}}}_{k_{i_n}} \boldsymbol{\epsilon}^{C_{i_{n}}}\big) e^{i l \cdot x} \,,
\end{aligned}\label{h_expansion}
\end{equation}
where $k_{i_{1}\cdots i_{n}} = k_{i_{1}} + k_{i_{2}} + \cdots + k_{i_{n}}$. We may recast the expansion and define the \emph{quantum perturbiner expansion} \cite{Lee:2022aiu} for $\mathbf{h}^{A}_{x}$
\begin{equation}
\begin{aligned}
  \mathbf{h}^{A}_{x} &= \sum_{n=1}^{\infty} \sum_{i_{1},\cdots, i_{n}=1}^{n} \mathbf{J}^{A}_{i_{1}\cdots i_{n}} e^{-ik_{ i_{1} i_{2}\cdots i_{n}}\cdot x}\,,
\end{aligned}\label{h_perturbiner0}
\end{equation}
where the coefficients $\mathbf{J}^{A}_{i_{1}\cdots i_{n}}$ is the \emph{quantum off-shell currents} and its explicit form is
\begin{equation}
\begin{aligned}
  \mathbf{J}^{A}_{i_{1}\cdots i_{n}} = \big\langle 0 \big| \tilde{\mathbf{h}}^{A}(-k_{i_{1}i_{2}\cdots i_{n}}) \tilde{\mathbf{h}}^{B_{1}} (k_{i_1}) \tilde{\mathbf{h}}^{B_{2}}(k_{i_2}) \cdots \tilde{\mathbf{h}}^{B_{n}}(k_{i_n}) \big| 0 \big\rangle_{c}
  \\
  \times\Big(\frac{i}{\hbar}\Big)^{n} \big(\tilde{\mathbf{K}}^{B_{i_1} C_{i_1}}_{k_{i_1}} \boldsymbol{\epsilon}^{C_{i_{1}}}\big)
   \cdots \big(\tilde{\mathbf{K}}^{B_{i_{n}}C_{i_{n}}}_{k_{i_n}} \boldsymbol{\epsilon}^{C_{i_{n}}}\big) \,.
\end{aligned}\label{rel_current_correl}
\end{equation}
This implies the quantum perturbiner expansion is a generating function of the off-shell currents $\mathbf{J}^{A}_{i_{1}\cdots i_{n}}$. Apparently, the definition of the quantum perturbiner expansion looks the same as the classical one, but the quantum case involves all order corrections in $\hbar$.

It is convenient to denote the quantum perturbiner expansion \eqref{h_perturbiner0} in terms of the so-called \emph{words}.
\begin{equation}
  \mathbf{h}^{A}_{x} = \sum_{\mathcal{P}} \mathbf{J}^{A}_{\mathcal{P}} e^{-ik_{\mathcal{P}}\cdot x}\,,
\label{perturbiner_expansion}\end{equation}
where $\mathcal{P}$ is the word labelling multi-particle states and $\sum_{\mathcal{P}}$ is a summation over all possible words. In the perturbiner formalism, single-particle states are labeled by what we refer to as ``letters,'' denoted by $i,j,k,l,\cdots$, whereas combinations of these states, or ``words,'' are represented as $\mathcal{P}, \mathcal{Q}, \mathcal{R} \cdots$, for instance, $\mathcal{P}= ijk$. Notably, within our field theory description, the order of letters within a given word is meaningless as single-particle states are represented solely by their momenta $k_{\mathcal{P}}$ in the plane wave basis. For instance, $\mathbf{J}^{A}_{123} = \mathbf{J}^{A}_{132} = \cdots$. To eliminate such redundancy, we impose a canonical ordering on the letters within each word, such that for an arbitrary word $\mathcal{P} = ijkl \cdots $, we have $i<j < k < l \cdots $.

It is evident that off-shell currents associated with the empty word, denoted by $\mathcal{P}=\emptyset$, are trivial, leading to
\begin{equation}
  \mathbf{J}^{A}_{\emptyset} = 0\,.
\label{}\end{equation}
On the other hand, as we will elaborate below, descendant currents do admit zero modes. The length of a word is called its ``rank'' denoted as $|\mathcal{P}|, |\mathcal{Q}|$, $|\mathcal{R}|$. We further require that the off-shell currents with repeated momenta vanish,
\begin{equation}
  \mathbf{h}^{A}_{i_{1}\cdots j \cdots j\cdots i_{n}} = 0\,.
\label{}\end{equation}
Whenever we want, we can separate the perturbiner expansion for the collective currents $\mathbf{h}^{A}_{x}$ into its components $h^{\mu\nu}_{x}$ and $\varphi^{I}_{x}$
\begin{equation}
\begin{aligned}
  h^{\mu\nu}_{x} = \sum_{\mathcal{P}} J^{\mu\nu}_{\mathcal{P}} e^{-ik_{\mathcal{P}}\cdot x}\,,
  \qquad
  \varphi^{I}_{x} &= \sum_{\mathcal{P}} \Phi^{I}_{\mathcal{P}} e^{-ik_{\mathcal{P}}\cdot x}\,,
\end{aligned}\label{perturbiner_h_scalar}
\end{equation}
where $\Phi^{I}_{\mathcal{P}}$ and $J^{\mu\nu}_{\mathcal{P}}$ are the scalars and graviton off-shell currents respectively.

According to the LSZ reduction formula, the off-shell current is directly related to the scattering amplitudes after amputating the off-shell leg and taking the on-shell limit. For instance, the $n$-loop four-point scattering amplitude between the scalar particles is given by
\begin{equation}
  \left(\frac{\hbar}{i}\right)^{n}\mathcal{M}^{(n)}\left(k_{1}, k_{2}, k_{3}, k_{4}\right)= \lim _{ k_{123}^{2} \rightarrow  -m^{2}_{1}} \sum_{p=0}^{n} \tilde{\mathbf{K}}^{(p)}_{123} \Phi^{(n-p)} \big|^{1}_{123}\,,
\label{amplitude_from_current}\end{equation}
where $k_{i}$ are external momenta.

We now consider the quantum perturbiner expansion for the descendant fields. The definition of the descendant fields can be rewritten by multiple functional derivatives on $W[\mathbf{j}]$, which is a $n$-point function in the presence of the external source,
\begin{equation}
  \psi^{A_{1}A_{2} \cdots A_{n}}_{x_{1}x_{2}\cdots x_{n}} = \frac{\delta^{n} W[\mathbf{j}]}{\delta \mathbf{j}^{A_{1}}_{x_{1}}\delta \mathbf{j}^{A_{2}}_{x_{2}}\cdots \delta \mathbf{j}^{A_{n}}_{x_{n}}}\,.
\label{}\end{equation}
Similar to \eqref{h_expansion}, descendant fields can be represented in terms of the correlation function with inverse two-point functions such as
\begin{equation}
\begin{aligned}
  \psi^{A_{1}A_{2} \cdots A_{n}}_{x_{1}x_{2}\cdots x_{n}} =& \sum_{m=1}^{\infty} \sum_{j_{1},\cdots j_{m} = 1}^{m} \int_{\ell_{1} \ell_{2}\cdots \ell_{n}} (2\pi)^{4}\delta^{4}\big(\ell_{1}+\ell_{2}+\cdots+\ell_{n}+k_{j_{1}\cdots j_{m}}\big)
  \\&
  \times\langle 0 | \tilde{\mathbf{h}}^{A_{1}}(\ell_{1})\cdots \tilde{\mathbf{h}}^{A_{n}}(\ell_{n}) \tilde{\mathbf{h}}^{B_{j_1}} (k_{j_1}) \cdots \tilde{\mathbf{h}}^{B_{j_m}}(k_{j_m}) | 0 \rangle_{c}
  \\&
   \times \Big(\frac{i}{\hbar}\Big)^{m} \big(\tilde{\mathbf{K}}^{B_{j_1} C_{j_1}}_{k_{j_1}} \boldsymbol{\epsilon}^{C_{j_{1}}}\big)
   \cdots \big(\tilde{\mathbf{K}}^{B_{j_{m}}C_{j_{m}}}_{k_{j_m}} \boldsymbol{\epsilon}^{C_{j_{m}}}\big) e^{i (\ell_{1} \cdot x_{1}+\cdots+\ell_{n} \cdot x_{n})} \,.
\end{aligned}\label{}
\end{equation}
We define the $n$-th descendant currents $\Psi^{A_{1}A_{2} \cdots A_{n+1}}_{\ell_{1}\ell_{2}\cdots \ell_{n}|\hat{\mathcal{P}}}$ in terms of the connected correlation functions
\begin{equation}
\begin{aligned}
  &\Psi^{A_{1}A_{2} \cdots A_{n+1}}_{\ell_{1}\ell_{2}\cdots \ell_{n}|\hat{\mathcal{P}}}
  = \sum_{m=1}^{\infty} \sum_{i_{1},\cdots i_{m} = 1}^{m} \int_{\ell_{1} \ell_{2}\cdots \ell_{n}}
	  \Big(\frac{i}{\hbar}\Big)^{m} \big(\tilde{\mathbf{K}}^{B_{i_1} C_{i_1}}_{k_{i_1}} \boldsymbol{\epsilon}^{C_{i_{1}}}\big)
   	  \cdots \big(\tilde{\mathbf{K}}^{B_{i_{m}}C_{i_{m}}}_{k_{i_m}} \boldsymbol{\epsilon}^{C_{i_{m}}}\big)
  \\&\quad
  \times\langle 0 | \tilde{\mathbf{h}}^{A_{1}}(\ell_{1})\cdots \tilde{\mathbf{h}}^{A_{n}}(\ell_{n}) \tilde{\mathbf{h}}^{A_{n+1}}(-\ell_{1\cdots n}-k_{1\cdots n}) \tilde{\mathbf{h}}^{A_{n}}(\ell_{n}) \tilde{\mathbf{h}}^{B_{i_1}} (k_{i_1}) \cdots \tilde{\mathbf{h}}^{B_{i_m}}(k_{i_m}) | 0 \rangle_{c}   \,,
\end{aligned}\label{}
\end{equation}
where $\hat{\mathcal{P}}$ denotes a word including the emptyset such as $\hat{\mathcal{P}}=\mathcal{P} \cup \emptyset$.
Upon this definition, we derive the perturbiner expansion
\begin{equation}
  \psi^{A_{1}A_{2} \cdots A_{n}}_{x_{1}x_{2}\cdots x_{n}} = \sum_{\hat{\mathcal{P}}} \int_{\ell_{2}\cdots \ell_{n}} \Psi^{A_{1}A_{2} \cdots A_{n}}_{\ell_{2}\ell_{3}\cdots \ell_{n}|\hat{\mathcal{P}}} e^{-i(k_{\mathcal{P}}-\ell_{2\cdots n)}) \cdot x_{1}-\sum_{i=2}^{n}\ell_{i}\cdot x_{i}}\,,
\label{perturbiner_descendant}\end{equation}
where $\ell_{12\cdots n} = \ell_{1}+ \ell_{2} + \cdots + \ell_{n}$. For the two-loop computations, we need up to second-descendant fields
\begin{equation}
\begin{aligned}
    \psi^{AB}_{xy} &= \sum_{\hat{\mathcal{P}}} \int_{\ell} \Psi_{\ell|\hat{\mathcal{P}}}^{AB} e^{-i(k_{\mathcal{P}}-p)\cdot x-ip\cdot y } \,,
  \\
  \psi^{ABC}_{xyz} &= \sum_{\hat{\mathcal{P}}} \int_{\ell_{1}\ell_{2}} \Psi_{\ell_{1}\ell_{2}|\hat{\mathcal{P}}}^{ABC} e^{-i(k_{\mathcal{P}}-\ell_{12})\cdot x -i \ell_{1}\cdot y -i \ell_{2}\cdot z } \,.
\end{aligned}\label{}
\end{equation}

As we have defined $\hbar$-expansion of the classical fields in \eqref{hbar_expansion_classical_fields}, we may expand the off-shell currents as
\begin{equation}
\begin{aligned}
  \Phi^{I}_{\mathcal{P}} &= \sum_{n=0}^{\infty} \int_{\ell_{1}\ell_{2}\cdots \ell_{n}}\Phi^{\ord{n}}_{\ell_{1}\cdots \ell_{n}} \big|_{\mathcal{P}}^{I}\bigg(\frac{\hbar}{i}\bigg)^{n}\,,
  \\
  J^{M}_{\mathcal{P}} &= \sum_{n=0}^{\infty} \int_{\ell_{1}\ell_{2}\cdots \ell_{n}} J^{\ord{n}}_{\ell_{1}\cdots \ell_{n}}\big|_{\mathcal{P}}^{M}\bigg(\frac{\hbar}{i}\bigg)^{n}\,,
  \\
  \Psi^{A B_{1}\cdots B_{m}}_{\ell_{1}\ell_{2}\cdots \ell_{m}} &=\sum_{n=0}^{\infty} \int_{\ell'_{1}\ell'_{2}\cdots \ell'_{n}} \Psi^{\ord{n}}_{\ell'_{1} \cdots \ell'_{n}}\big|^{A,B_{1},\cdots ,B_{m}}_{\ell_{1}\cdots \ell_{m}|\hat{\mathcal{P}}} \bigg(\frac{\hbar}{i}\bigg)^{n}\,,
\end{aligned}\label{current_loop_expansion}
\end{equation}
where $\ell_{1}, \ell_{2} \cdots$ are the loop momenta. This $\hbar$-expansion is crucial in constructing a self-consistent recursion relation by truncating unnecessary higher descendant currents.

\subsection{Bracket and Descend operator}

Let us consider a classical field $A_{x}$ that is expressed as the product of other classical fields $B^{1}_{x}, B^{2}_{x}, \cdots, B^{n}_{x}$,
\begin{equation}
  A_{x} = B^{1}_{x} B^{2}_{x} \cdots B^{n}_{x}\,,
\label{ex1}\end{equation}
and assume that their perturbiner expansions are given by
\begin{equation}
  A_{x} = \sum_{\mathcal{P}}\mathcal{A}_{\mathcal{P}} e^{-ik_{\mathcal{P}}\cdot x}\,,
  \qquad
  B^{i}_{x} = \sum_{\mathcal{P}} \mathcal{B}^{i}_{\mathcal{P}} e^{-ik_{\mathcal{P}}\cdot x}\,,
\label{}\end{equation}
where $\mathcal{A}_{\mathcal{P}}$ and $\mathcal{B}^{i}_{\mathcal{P}}$ denote the off-shell currents. By substituting these perturbiner expansion into \eqref{ex1}, we can express $\mathcal{A}_{\mathcal{P}}$ in terms of $\mathcal{B}^{i}$
\begin{equation}
  \mathcal{A}_{\mathcal{P}} = \sum_{\mathcal{P}=\mathcal{Q}_{1}\cup \mathcal{Q}_{2}\cdots \cup \mathcal{Q}_{n}} \mathcal{B}^{1}_{\mathcal{Q}_{1}} \mathcal{B}^{2}_{\mathcal{Q}_{2}} \cdots \mathcal{B}^{n}_{\mathcal{Q}_{n}} \,.
\label{ex2}\end{equation}

A typical form of off-shell recursions involves cumbersome products and summations over the off-shell currents. For concise expressions, we introduce a multilinear bracket denoted as $\lceil \bullet \rfloor_{\mathcal{P}}$, which effectively replaces the summations over products of the currents,
\begin{equation}
\begin{aligned}
  \big\lceil \mathcal{A}, \mathcal{B} \big\rfloor_{\mathcal{P}} = \sum_{\mathcal{P}=\mathcal{Q}\cup \mathcal{R}} \mathcal{A}_{\mathcal{Q}} \mathcal{B}_{\mathcal{R}}\,,
  \qquad
  \big\lceil A, B, C\big\rfloor_{\mathcal{P}} = \sum_{\mathcal{P}=\mathcal{Q}\cup \mathcal{R}\cup \mathcal{S}} A_{\mathcal{Q}} B_{\mathcal{R}} C_{\mathcal{S}} \,.
\end{aligned}\label{}
\end{equation}
If descendant currents are positioned in one of the entries of the bracket, the summation over the words is replaced with the hatted words, including the zero modes. By definition, the bracket is orderless, $\big\lceil \mathcal{A}, \mathcal{B} \big\rfloor_{\mathcal{P}} = \big\lceil \mathcal{B}, \mathcal{A} \big\rfloor_{\mathcal{P}}$, and satisfies the multilinear property, $\big\lceil a \mathcal{A} +b \mathcal{B}, \mathcal{C} \big\rfloor_{\mathcal{P}} = a \big\lceil \mathcal{A}, \mathcal{C} \big\rfloor_{\mathcal{P}} + b \big\lceil \mathcal{B}, \mathcal{C}\big\rfloor_{\mathcal{P}}$. Using this bracket, \eqref{ex2} can be rewritten by
\begin{equation}
   \mathcal{A}_{\mathcal{P}} = \big\lceil \mathcal{B}^{1}, \mathcal{B}^{2}, \cdots \mathcal{B}^{n} \big\rfloor_{\mathcal{P}} \,.
\label{}\end{equation}
One may insert momenta at any slot with an off-shell current, and the momenta inherits the word of the current in the same slot, for instance,
\begin{equation}
\begin{aligned}
  \big\lceil k^{\rho}J^{M},k^{\sigma}J^{N}\big\rfloor_{\mathcal{P}} &= \sum_{\mathcal{P}=\mathcal{Q}\cup \mathcal{R}} k^{\rho}_{\mathcal{Q}}J^{M}_{\mathcal{Q}} k^{\sigma}_{\mathcal{R}}J^{N}_{\mathcal{R}}\,,
  \\
  \big\lceil J^{M},k^{\rho}k^{\sigma}J^{N}\big\rfloor_{\mathcal{P}} &= \sum_{\mathcal{P}=\mathcal{Q}\cup \mathcal{R}}J^{M}_{\mathcal{Q}} k^{\rho}_{\mathcal{R}} k^{\sigma}_{\mathcal{R}}J^{N}_{\mathcal{R}} \,.
\end{aligned}\label{}
\end{equation}
We may also add a superscript labeling the $\hbar$-order to the bracket, which is the total $\hbar$-order of the currents in the bracket
\begin{equation}
\begin{aligned}
  \big\lceil \mathcal{A}, \mathcal{B}\big\rfloor^{\ord{n}}_{\mathcal{P}} &=
  \sum_{m=0}^{n} \sum_{\mathcal{P}=\mathcal{Q}\cup \mathcal{R}} \mathcal{A}^{\ord{n-m}}_{\mathcal{Q}} \mathcal{B}^{\ord{m}}_{\mathcal{R}}\,,
  \\
  \big\lceil \mathcal{A}, \mathcal{B}, \mathcal{C} \big\rfloor^{\ord{n}}_{\mathcal{P}} &= \sum_{m,p=0\atop n\geq m+p}^{n} \sum_{\mathcal{P}=\mathcal{Q}\cup \mathcal{R} \cup \mathcal{S}} \mathcal{A}^{\ord{n-m-p}}_{\mathcal{Q}} \mathcal{B}^{\ord{m}}_{\mathcal{R}} \mathcal{C}^{\ord{p}}_{\mathcal{S}}\,.
\end{aligned}\label{bracket_hbar}
\end{equation}

We next define a differential operator $\Delta^{A}_{\mathcal{P}}$ and call it the \textit{descend operator}. It maps an off-shell current and descendant currents as follows:
\begin{equation}
\begin{aligned}
  \Delta^{A}_{p} \Phi^{I}_{\mathcal{P}} &= \Psi^{IA}_{p|\mathcal{P}} \,,
  \\
  \Delta^{A}_{p} J^{M}_{\mathcal{P}} &= \Psi^{MA}_{p|\mathcal{P}} \,,
  \\
  \Delta^{A}_{p} \Psi^{BC_{1}\cdots C_{n}}_{q_{1}\cdots q_{n}|\hat{\mathcal{P}}} &= \Psi^{BC_{1}\cdots C_{n}A}_{q_{1}\cdots q_{n}p|\hat{\mathcal{P}}}\,.
\end{aligned}\label{Delta_operator}
\end{equation}
We also denote the successive applications of the descend operator as
\begin{equation}
  \Delta^{A_{1},A_{2},\cdots, A_{n}}_{\ell_{1},\ell_{2},\cdots,\ell_{n} }\big[\bullet\big] = \Delta^{A_{n}}_{\ell_{n}}\bigg[\cdots \Delta^{A_{2}}_{\ell_{2}}\Big[\Delta^{A_{1}}_{\ell_{1}}\big[\bullet\big]\Big]\cdots\bigg]\,.
\label{}\end{equation}
Since $\Delta^{A}_{p}$ is a differential operator, it should satisfy the Leibniz rule inside the bracket
\begin{equation}
\begin{aligned}
  \Delta^{A}_{p} \big\lceil J^{M}, J^{N}\big\rfloor_{\mathcal{P}} &= \big\lceil \Delta^{A}_{p} J^{M}, J^{N}\big\rfloor_{\mathcal{P}} + \big\lceil J^{M}, \Delta^{A}_{p} J^{N}\big\rfloor_{\mathcal{P}}\,,
  \\&
  = \big\lceil  \Psi^{MA}_{p}, J^{N}\big\rfloor_{\mathcal{P}}
  + \big\lceil J^{M}, \Psi^{NA}_{p}\big\rfloor_{\mathcal{P}}\,,
  \\
  &=\sum_{\mathcal{P}=\hat{\mathcal{Q}}\cup \hat{\mathcal{R}}} \Big( \Psi^{MA}_{p|\hat{\mathcal{Q}}} J^{N}_{\hat{\mathcal{R}}} + J^{M}_{\hat{\mathcal{Q}}} \Psi^{NA}_{p|\hat{\mathcal{R}}} \Big)\,.
\end{aligned}\label{}
\end{equation}
Note that the summation in the last line should be for the hatted words.
It also acts on the external on-shell momenta $k^{\mu}_{\mathcal{P}}$, and one can show that the external momenta are shifted
\begin{equation}
\begin{aligned}
  \Delta^{A}_{p}\big[k^{\mu}_{\mathcal{P}}\big] = k^{\mu}_{\mathcal{P}}-p^{\mu} := k^{\mu}_{\mathcal{P},p}\,.
\end{aligned}\label{Delta_operator2}
\end{equation}
%


\subsection{Quantum off-shell recursions up to two loops}

We now derive the off-shell recursion relations by combining the DS equation with the quantum perturbiner expansion up to two-loop order, following the steps established in \cite{Lee:2022aiu}. Initially, we substitute the perturbiner expansions of the classical fields and their descendant fields into the DS equations. After this substitution, the series is truncated to isolate contributions up to the two-loop order. We then extract the coefficients for an arbitrary rank $\mathcal{P}$ corresponding to the off-shell recursions under investigation. The descendant currents that contribute to the two-loop recursions are
\begin{equation}
  \Psi^{\ord{0}}\big|^{AB}_{\ell|\hat{\mathcal{P}}}\,,
  \qquad
  \Psi^{\ord{1}}\big|^{AB}_{\ell|\hat{\mathcal{P}}}\,,
  \qquad
  \Psi^{\ord{0}}\big|^{ABC}_{\ell_{1}\ell_{2}|\hat{\mathcal{P}}}\,.
\label{}\end{equation}
It is straightforward to derive the off-shell recursions for these descendant currents by acting on the $\Delta$ operator.

\subsubsection{Scalar field sector}

The DS equation and the perturbiner expansions for the scalar fields are derived in \eqref{DSeq_scalar} and \eqref{perturbiner_h_scalar} respectively. Substituting the expansion and truncating the DS equation at rank $|\mathcal{P}|$, in our case $\mathcal{P}=123$ and $|\mathcal{P}|=3$, we present the recursions for the off-shell currents for the massive scalars up to two loops,
\begin{equation}
\begin{aligned}
    \big(k_{\mathcal{P}}^{2}+m_{I}^{2}\big) \Phi^{\ord{n}}\big|^{I}_{\mathcal{P}}
  &= \kappa\Omega^{\ord{n}}_{1}\big|^{I}_{\mathcal{P}}
  	- \kappa^{2}\mathcal{I}^{M,N}_{\phi} \Omega^{\ord{n}}_{2}\big|^{MN}_{\mathcal{P}}
	- \kappa^{3}\mathcal{I}^{M,N,P}_{\phi} \Omega^{\ord{n}}_{3}\big|^{MNP}_{\mathcal{P}} \,, \quad n\leq2\,,
\end{aligned}
\label{Phi_recursion}\end{equation}
where $\Omega^{\ord{n}}_{i}$ are denoted in \eqref{Omega_currents}.

For consistency, we should require that each $\Omega^{\ord{n}}_{i}$ satisfies
\begin{equation}
  \Omega^{\ord{n}}_{i} = 0 \,, \qquad \text{for}~ n<0\,.
\label{}\end{equation}
One can derive the off-shell recursions for the descendant currents for $\Phi^{I}_{\mathcal{P}}$ by acting the $\Delta$-operators on \eqref{Phi_recursion} by using the relations in \eqref{Delta_operator} and \eqref{Delta_operator2}.

\subsubsection{Graviton sector}
Let us now consider the recursions for graviton current $J^{M}_{\mathcal{P}}$ and its descendants. We derive the recursions by substituting the perturbiner expansion into the DS equation for $h^{\mu\nu}$ in \eqref{DSeq_h}. Collecting the $n$-th order terms in $\hbar$, where $n\leq2$, and the coefficients of $e^{ik_{\mathcal{P}}\cdot x}$ provides the off-shell recursion relation. The structure of the graviton recursion relation is the same as the DS equation \eqref{DSeq_h}. The recursion at the $n$-loop order is given by
\begin{equation}
\begin{aligned}
  k^{2}_{\mathcal{P}} J^\ord{n}\big|_{\mathcal{P}}^{M} &=
  2P^{M,N} \Bigg[
  	  \sum_{i=1}^{2} \kappa^{i} \mathcal{G}^{\ord{n}}_{i}\big|^{N}_{\mathcal{P}}
  	+ \sum_{i=1}^{3} \kappa^{i} \mathcal{T}^{\ord{n}}_{i}\big|^{N}_{\mathcal{P}}
  \Bigg]\,.
\end{aligned}\label{recursion_h}
\end{equation}
Here the curvature part $\mathcal{G}^{\ord{n}}_{i}\big|^{M}_{\mathcal{P}}$ are defined as
\begin{equation}
\begin{aligned}
  \mathcal{G}^{\ord{n}}_{1}\big|^{N}_{\mathcal{P}} &=
    \tilde{\mathcal{I}}^{M,\rho N,\sigma P}_{1} \Lambda^\ord{n}_{1,1}\big|^{\rho N, \sigma P}_{\mathcal{P}}
  + \tilde{\mathcal{I}}^{M,N,\rho\sigma, P}_{2} \Lambda^\ord{n}_{1,2}\big|^{N, \rho \sigma, P}_{\mathcal{P}}\,
  \\
  \mathcal{G}^{\ord{n}}_{2}\big|^{N}_{\mathcal{P}} &=
    \tilde{\mathcal{I}}^{M,N,\rho P,\sigma Q}_{1} \Lambda^\ord{n}_{2,1}\big|^{N, \rho P, \sigma Q}_{\mathcal{P}}
  + \tilde{\mathcal{I}}_{2}^{M,N,P, \rho \sigma, Q} \Lambda^\ord{n}_{2,2}\big|^{N,P, \rho \sigma, Q}_{\mathcal{P}}\,,
\end{aligned}\label{}
\end{equation}
and the explicit form of the $\Lambda^{\ord{n}}_{n,1}$ and $\Lambda^{\ord{n}}_{n,2}$ are listed in \eqref{Lambda_currents}. The EM tensor part $\mathcal{T}^{\ord{n}}_{i}\big|^{N}_{\mathcal{P}}$ are defined as
\begin{equation}
\begin{aligned}
  \mathcal{T}^{\ord{n}}_{1}\big|^{M}_{\mathcal{P}} &= \Theta_{1}^\ord{n}\big|^{M}_{\mathcal{P}}\,,
  \\
  \mathcal{T}^{\ord{n}}_{2}\big|^{M}_{\mathcal{P}} &= \mathcal{I}_{\rm EM}^{M,N} \Theta_{2}^\ord{n}\big|^{N}_{\mathcal{P}}\,,
  \\
  \mathcal{T}^{\ord{n}}_{3}\big|^{M}_{\mathcal{P}} &= \mathcal{I}_{\rm EM}^{M,N,P} \Theta_{3}^\ord{n}\big|^{N,P}_{\mathcal{P}} \,,
\end{aligned}\label{}
\end{equation}
and the explicit form of $\Theta^\ord{n}_{i}$ are listed in \eqref{Theta_currents}.

Again, we require that $\Lambda^{\ord n}$ and $\Theta^{\ord n}$ vanish for $n<0$ cases
\begin{equation}
  \Lambda_{i,j}^{\ord n} = 0\,, \qquad \Theta_{i}^{\ord n}= 0 \,, \qquad \text{for}~n<0\,,
\label{}\end{equation}
and the associated descendent recursions can be derived by acting the $\Delta$-operators successively to \eqref{recursion_h}.
\subsection{Initial conditions}

We should impose an appropriate initial condition to address the solutions of the off-shell recursions. For $2\to 2$ scattering amplitudes for scalar particles, the external states consist of scalar fields only, while gravitons are restricted to the internal lines. The initial condition for the recursions is realized by the rank-1 off-shell currents for the scalars $\Phi^{I}_{i}$ and for graviton $J^{\mu\nu}_{i}$, as well as the descendant fields.

First, let us consider $\Phi^{I}_{i}$ and $J^{\mu\nu}_{i}$. By definition of the off-shell currents and the dressed external source, \eqref{rel_current_correl} and \eqref{def_j}, we may derive the initial condition
\begin{equation}
\begin{aligned}
  \Phi^{I}_{i} &= \big\langle 0 \big| \Phi^{I}(-k_{i}) \Phi^{J}(k_{i}) \big| 0 \big\rangle \tilde{\mathbf{K}}^{J,K}_{k_{i}} \epsilon^{K}_{i}
  = \tilde{\mathbf{D}}^{I,J}_{k_{i}} \tilde{\mathbf{K}}^{J,K}_{k_{i}} \epsilon^{K}_{i}=\epsilon^{I}_{i}\,,
  \\
  J^{\mu\nu}_{i} &= 0\,.
\end{aligned}\label{rel_current_correl}
\end{equation}
Note that this initial condition is for the all-loop order, and we should distribute the result to each loop order. In fact, one can show that the initial condition is only contributed from the rank-1 currents at the tree level, and the higher loops vanish. The tree-level two-point function, which is equivalent to the propagator $\tilde{D}^{I, J}_{k_{i}}$, and the inverse propagator $\tilde{K}^{I, J}_{k_{i}}$ satisfy the inverse relation $\tilde{D}^{I, J}_{k_{i}} \tilde{K}^{J, K}_{k_{i}} = \delta^{IK}$.

Consequently, the first equation of \eqref{rel_current_correl} can be expanded in $\hbar$ as
\begin{equation}
  \tilde{\mathbf{D}}^{I,J}_{k_{i}} \tilde{\mathbf{K}}^{J,K}_{k_{i}} = \delta^{IK} + \sum_{n=1}^{\infty} \sum_{m=1}^{n} \Big(\frac{\hbar}{i}\Big)^{n} \tilde{\mathbf{D}}^{\ord{n-m}}\big|^{I,J}_{k_{i}} \tilde{\mathbf{K}}^{\ord{m}}\big|^{J,K}_{k_{i}} = \delta^{IK}\,.
\label{}\end{equation}
As this equation holds for each order in $\hbar$, contributions from higher loop orders vanish
\begin{equation}
  \Phi^{\ord{n}}|^{I}_{i} = \sum_{m=1}^{n} \Big(\frac{\hbar}{i}\Big)^{n} \tilde{\mathbf{D}}^{\ord{n-m}}\big|^{I,J}_{k_{i}} \tilde{\mathbf{K}}^{\ord{m}}\big|^{J,K}_{k_{i}} \epsilon^{K}= 0 \,, \qquad \text{for}~ n>0\,.
\label{}\end{equation}
Thus our initial condition for $\Phi^{I}_{\mathcal{P}}$ and $J^{\mu\nu}_{\mathcal{P}}$ is
\begin{equation}
\begin{aligned}
  \Phi^{\ord{0}}\big|^{I}_{i} &= \epsilon^{I}_{i}\,,
  &\qquad
  J^{\ord{0}}\big|^{I}_{i} &= 0\,,
  \\
  \Phi^{\ord{n}}\big|^{I}_{i} &= 0\,,
  &\qquad
  J^{\ord{n}}\big|^{I}_{i} &= 0\,, \qquad \text{for}~n>0\,.
\end{aligned}\label{initial_condition_Phi_J}
\end{equation}
The DS equations with the bare external sources are reduced to
\begin{equation}
\begin{aligned}
  \sum_{i} k_{i}^{2} \big(J^{\ord{0}}|^{\mu\nu}_{i} e^{-ik_{i}\cdot x}\big) &= 4 P^{\mu\nu,\rho\sigma} j^{\rho\sigma}_{x} = \sum_{i}k^{2} \epsilon^{\mu\nu}_{i}e^{ik_{i}\cdot x}\,,
  \\
  \sum_{i}\big(k_{i}^{2} + m^{2}_{I}\big) \big(\Phi^{I}_{i}e^{-ik_{i}\cdot x} \big) &= \rho^{I}_{x} =  \big(k_{i}^{2} + m^{2}_{I}\big) \epsilon^{I}_{i} e^{-ik_{i}\cdot x}\,,
\end{aligned}\label{}
\end{equation}
and the previous initial conditions satisfy them.

Finally, let us consider the initial condition for the descendant currents case. As before, we derive the initial condition from their definition. One can show that it is enough to determine the case of the first descendant currents because the recursion relations generate the initial conditions for the other descendant currents. Unlike the $J^{\mu\nu}_{\mathcal{P}}$ and $\Phi^{I}_{\mathcal{P}}$, the descendant currents may have the zero-mode contribution
\begin{equation}
\begin{aligned}
  \Psi^{AB}_{\ell,\emptyset} = \langle 0 | \tilde{\mathbf{h}}^{A}(-\ell) \tilde{\mathbf{h}}^{B}(\ell) | 0 \rangle_{c} = \tilde{\mathbf{D}}^{AB}_{\ell}\,.
\end{aligned}\label{}
\end{equation}

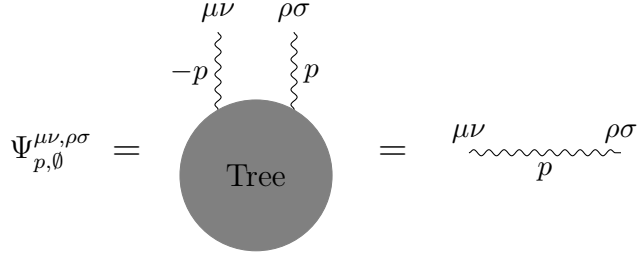
\begin{figure}[t]
\begin{center}
\begin{tikzpicture}
	\path (-2.2,-0.5) node {\large $\Psi^{\mu\nu,\rho\sigma}_{p,\emptyset}$};
  	\path (-1.2,-0.5) node {\Large $=$};
  	\draw[snake it] (0,1.1) -- (0,0) node[anchor=east,pos=1/2]{$-p$} node[anchor=south,pos=0]{$\mu\nu$};
  	\draw[snake it] (1,1.1) -- (1,0) node[anchor=west,pos=1/2]{$p$} node[anchor=south,pos=0]{$\rho\sigma$};
  	\filldraw[gray] (0.5,-0.8) circle (1) ;
  	\path (0.5,-0.8) node {\large Tree};
  	\path (2.3,-0.5) node {\Large $=$};
  	\draw[snake it] (3.3,-0.5) -- (5.3,-0.5) node[anchor=north,pos=1/2]{$p$} node[anchor=south,pos=0]{$\mu\nu$} node[anchor=south,pos=1]{$\rho\sigma$};
\end{tikzpicture}
\end{center}
  \caption{the wiggly lines represent off-shell gravitons}\label{figure:initial_condition_Psi}
\end{figure}

However, the higher loop currents $\Psi^{\ord{n>0}}\big|^{AB}_{\ell,\emptyset}$ contribute to the self-energy or the vacuum polarizations, which are classified as quantum contributions according to the prescription discussed in Section \ref{Sec:4.2}. Thus only the tree-level current $\Psi^{\ord{0}}\big|^{AB}_{\ell,\emptyset}$ contributes to the classical amplitudes. Thus, our initial condition is
\begin{equation}
\begin{aligned}
  \Psi^{\ord{0}}\big|^{IJ}_{\ell,\emptyset} &= \frac{\delta^{IJ}}{\ell^{2}+m^{2}_{I}}\,,
  &\qquad
  \Psi^{\ord{n}}\big|^{IJ}_{\ell,\emptyset} &= 0\,, \quad \text{for}~ n>0\,,
  \\
  \Psi^{\ord{0}}\big|^{MN}_{\ell,\emptyset} &= \frac{2P^{M,N}}{\ell^{2}}\,,
  &\qquad
  \Psi^{\ord{n}}\big|^{MN}_{\ell,\emptyset} &= 0\,, \quad \text{for}~ n>0\,,
\end{aligned}\label{initial_condition_Psi}
\end{equation}
and these are propagators and represented diagrammatically in Figure \ref{figure:initial_condition_Psi}.

One can show that $\Psi^{\ord{0}}\big|^{AB}_{\ell,\emptyset}$ satisfies the DS equation as we expect:
\begin{equation}
\begin{aligned}
  \int_{\ell}\left(\square_{x}-m_{I}^{2}\right) \big(\Psi^{\ord{0}}\big|^{IJ}_{\ell|\emptyset} e^{i\ell\cdot(x-y)}\big) &= - \int_{\ell}\delta^{IJ} e^{i\ell \cdot(x-y)}\,,
  \\
  \int_{\ell} \square_{x}\big(\Psi^{\ord{0}}\big|^{MN}_{\ell|\emptyset} e^{i\ell\cdot(x-y)}\big) &= - \int_{\ell}2P^{MN} e^{i\ell \cdot(x-y)}\,.
\end{aligned}\label{}
\end{equation}
Interestingly, all the other descendant currents are determined by the $\Phi^{I}_{i}$, $J^{\mu\nu}_{i}$ and $\Psi^{AB}_{\ell,\emptyset}$ by the recursion relations. Consequently, the total initial condition is given by \eqref{initial_condition_Phi_J} and \eqref{initial_condition_Psi}.


\section{Classifying Classical Diagrams} \label{Sec:4}

So far, we have constructed the quantum off-shell recursions for the scattering amplitude of scalar particles up to the two-loop order. We may solve the recursions using symbolic computational tools, such as Mathematica, and the solutions of the off-shell recursions serve as loop integrands. The loop integrals include not only quantum but also classical contributions. Since the quantum effects are negligible in the binary black hole systems, we will focus on establishing a method for identifying the classical diagrams pertinent to the off-shell recursions.

Notably, the off-shell recursion method offers an algebraic approach that contrasts with the conventional unitarity cut method -- it does not require any explicit diagram topologies for solving the recursions. Consequently, we intend to develop a novel prescription to identify and extract the classical diagrams by employing power counting rules, which do not rely on explicit diagrammatic representation.

\subsection{Kinematics}
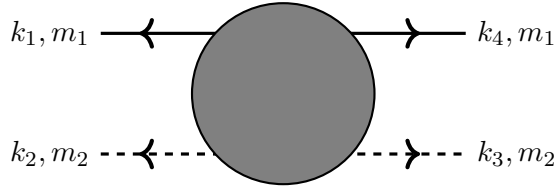
\begin{figure}[t]
  \begin{center}
\begin{tikzpicture}[scale=0.8,decoration={markings,
    mark= at position 0.7cm with {\arrow[scale=1.3]{<}},
    mark= at position -0.6cm with {\arrow[scale=1.3]{>}}},]

  	\draw[very thick,dashed,postaction={decorate}] (0,0) node[anchor=east]{$k_{2},m_{2}$} -- (6,0) node[anchor=west]{$k_{3},m_{2}$};
  	\draw[very thick,postaction={decorate}] (0,2) node[anchor=east]{$k_{1},m_{1}$} -- (6,2) node[anchor=west]{$k_{4},m_{1}$} ;
  	\filldraw[fill=gray!,thick](3,1) circle (1.5) ;
\end{tikzpicture}
\end{center}
\caption{The kinematics of binary BH systems}
\label{kinematics_figure}
\end{figure}

Let us consider the classical $2 \to 2$ scattering amplitude $\mathcal{M}_{\rm cl}[k_{1},k_{2},k_{3},k_{4}]$ of two massive scalar particles, where $k_{1}$, $k_{2}$, $k_{3}$ and $k_{4}$ are external momenta for the scalar particles as depicted in Figure \ref{kinematics_figure}. From the relation between the off-shell current and the amplitude \eqref{amplitude_from_current}, the amplitude is associated with the rank-3 scalar off-shell current $\Phi^{1}_{123}$ depicted in Figure \ref{scalar_current}. Note that $\Phi^{1}_{234}$ or $\Phi^{2}_{124}$ etc would give the same result, but here we will focus on $\Phi^{1}_{123}$ only.

The external momenta satisfy the momentum conservation, $k_{1}+k_{2}+k_{3}+k_{4} = 0$, and the mass-shell condition, $k_{1}^{2}+m_{1}^{2} = 0$, $k_{2}^{2}+m_{2}^{2} = 0$, $k_{3}^{2}+m_{2}^{2} = 0$, $k_{4}^{2}+m_{1}^{2} = 0$. For the soft expansion, which we will discuss in the next section, it is useful to parametrize the external momenta \cite{Parra-Martinez:2020dzs} as follows:
\begin{equation}
  k^{\mu}_1= -\bar{k}^{\mu}_{1} +\frac{q^{\mu}}{2}\,,
  \qquad
  k^{\mu}_2= -\bar{k}^{\mu}_{2} -\frac{q^{\mu}}{2}\,,
  \qquad
  k^{\mu}_3= \bar{k}^{\mu}_{2} -\frac{q^{\mu}}{2}\,,
  \qquad
  k^{\mu}_4= \bar{k}^{\mu}_{1} +\frac{q^{\mu}}{2}\,,
\label{}\end{equation}
where $q^{\mu}$ is the momentum transfer between the scalar particles. The mass-shell conditions require $\bar{k}^{\mu}_{1}$ and $\bar{k}^{\mu}_{2}$ are orthogonal to $q^{\mu}$
\begin{equation}
  k^{2}_{1} - k^{2}_{4} = -2 \bar{k}_{1} \cdot q = 0\,, \qquad k^{2}_{2} - k^{2}_{3} = 2 \bar{k}_{2} \cdot q = 0\,.
\label{}\end{equation}
We also introduce the dimensionless momenta $u^{\mu}_{1}$ and $u^{\mu}_{2}$
\begin{equation}
  u^{\mu}_{1} = \frac{\bar{k}^{\mu}_{1}}{\bar{m}_{1}} \,,
  \qquad
  u^{\mu}_{2} = \frac{\bar{k}^{\mu}_{2}}{\bar{m}_{2}} \,,
\label{}\end{equation}
where
\begin{equation}
  \bar{m}^{2}_{1} = -\bar{k}^{2}_{1} = m^{2}_{1} + \frac{q^{2}}{4} \,,
  \qquad
  \bar{m}^{2}_{2} = -\bar{k}^{2}_{2} = m^{2}_{2} + \frac{q^{2}}{4}\,.
\label{}\end{equation}

It is useful to introduce parameters $y$ and $\sigma$ as
\begin{equation}
  y = - u_{1} \cdot u_{2}\,,
  \qquad
  \sigma = - \frac{k_{1}\cdot k_{2}}{m_{1}m_{2}}\,,
\label{y_variable}\end{equation}
and these are related to
\begin{equation}
  y = \sigma - \frac{2 m_{1} m_{2} + \sigma (m_{1}^{2}+ m_{2}^{2})}{8 m_{1}^{2} m_{2}^{2}} q^{2} + \cdots \,.
\label{}\end{equation}
We further introduce a parameter $x$ to rationalize $\sqrt{y^{2}-1}$ which appears in the middle of computations
\begin{equation}
  x = y - \sqrt{y^{2}-1} \,, \qquad \sqrt{y^{2}-1} = \frac{1-x^{2}}{x}\,, \qquad y \geq 1 \,,
  \qquad  0<x\leq 1\,.
\label{x_variable}\end{equation}
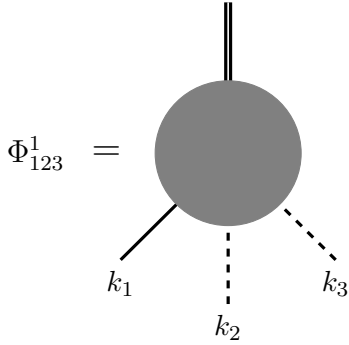
\begin{figure}[t]
\begin{center}
\begin{tikzpicture}[scale=0.8]
    \path (-2,0) node {\Large $=$} node[anchor=east]{\large $\Phi^{1}_{123}$  ~~ };
  	\draw[very thick, double] (0,0) -- (0,2.5);
  	\draw[very thick] (0,0) -- (-135:2.5)node[anchor=north]{$k_{1}$};
  	\draw[very thick,dashed] (0,0) -- (0,-2.5) node[anchor=north]{$k_{2}$};
  	\draw[very thick,dashed] (0,0) -- (-45:2.5) node[anchor=north]{$k_{3}$};
	\filldraw[gray] (0,0) circle (1.2);
\end{tikzpicture}
\end{center} \caption{Diagrammatical representation of the currents associated with the tree and one-loop scalar four-point amplitude. The double lines represent the off-shell lines}
\label{scalar_current}
\end{figure}
%

\subsection{Classical loop integrands} \label{Sec:4.2}

Solving the quantum off-shell recursions, we can derive the off-shell current $\Phi^{1}_{123}$ associated with the $2\to 2$ scattering amplitude for the scalar particles at a given loop order. Interestingly, even at the loop levels, the amplitude includes both classical and quantum contributions. For describing the classical conservative dynamics, we only focus on the classical part.

For a consistent field theory limit, we assume that the black holes are well separated to ignore the internal structure of the black holes, and the Compton length scale must be much smaller than the impact parameter $|b|$. The Fourier dual of $|b|$ is given by the momentum transfer $|q|$ between two black holes, $|b| \sim \frac{1}{|q|}$, and this indicates $|q|$ is much smaller than the mass and the momenta of the black holes
\begin{equation}
  m_{1},m_{2}, |k_{i}|\gg |q|\,.
\label{}\end{equation}
This implies that we may characterize the amplitude according to the expansion in small $|q|$. We further consider the conservative dynamics by assuming the small relative velocity limit $\vec{v}\ll 1$, or a scale of the components of $q^{\mu} = (q^{0},\vec{q})$
\begin{equation}
  q^{0} \gg \vec{q}\,.
\label{}\end{equation}
Thus we consider two limits -- near-static classical limit, $|q|\ll 1$ and  $\vec{v}\ll 1$.

To isolate the classical contributions from the total scattering amplitude, it is necessary to identify the classical loop integrands contributing to the classical amplitudes. A simple and efficient prescription was introduced in \cite{Bern:2019crd}. The established criteria for a classical loop integrands are:
\begin{itemize}
\item Every independent loop must contain at least one matter line
\item Both ends of a graviton line cannot be connected to the same matter line
\item Truncate the higher power of loop momenta and momentum transfer
\end{itemize}
The first two are rules on the topology of the diagrams, and the diagrams satisfying the criteria are called classical diagrams. The last one is a condition on the numerators of loop integrands, which is realized by the soft expansion in Section \ref{Sec:5}. This prescription is particularly useful in the context of the unitarity cut method because cut topologies should be imposed before calculation thereby one can focus on the classical diagrams only.

In contrast, the quantum off-shell recursion formalism generates the loop integrands without concerning the diagram topologies by solving the recursions algebraically. However, it is challenging to figure out the diagram topologies from the total loop integrands. Thus we need an alternative strategy for identifying classical diagrams that do not rely on diagram topologies.

To address this issue, we propose a novel prescription that employs power counting rules to separate classical diagrams. To this end, we introduce additional fictitious coupling constants and labeling for each propagator. The only coupling constant in the field theory is the Newton constant, $G_{N}$ or $\kappa$, and the loop integrands at $L$-loop order have the same couplings dependence, $\kappa^{2(L+1)}$. For a finer separation, we introduce new formal coupling constants, $\kappa_{1}$, $\kappa_{2}$ and $\kappa_{g}$, obtained by dividing $\kappa$ according to the types of interaction vertice as depicted in Figure \ref{new_couplings}. These are useful bookkeeping devices that convey more information than $\kappa$.

\begin{figure}[t]
\begin{center}
\begin{tikzpicture}[scale=0.7]
	\path (-1,1) node[anchor=east]{\bf{matter 1:}};
  	\draw[very thick] (0,2) -- (3,2) ;
  	\draw[snake it] (1.5,0) -- (1.5,2) node[anchor= south ]{ $(\kappa_{1})^{1}$};
  	\filldraw[black] (1.5,2) circle (2pt);
\end{tikzpicture}
\qquad
\begin{tikzpicture}[scale=0.7]
  	\draw[very thick] (0,2) -- (3,2) ;
  	\draw[snake it] (0,0) -- (1.5,2);
  	\draw[snake it] (1.5,2) -- (3,0);
  	\filldraw[black] (1.5,2) circle (2pt) node[anchor= south ]{ $(\kappa_{1})^{2}$};
\end{tikzpicture}
\qquad
\begin{tikzpicture}[scale=0.7]
  	\draw[very thick] (0,2) -- (3,2) ;
  	\draw[snake it] (0,0) -- (1.5,2);
  	\draw[snake it] (1.5,2) -- (3,0);
  	\draw[snake it] (1.5,0) -- (1.5,2);
  	\filldraw[black] (1.5,2) circle (2pt) node[anchor= south ]{ $(\kappa_{1})^{3}$};
\end{tikzpicture}
\\~\\
\begin{tikzpicture}[scale=0.7]
	\path (-1,1) node[anchor=east]{\bf{matter 2:}};
  	\draw[very thick,dashed] (0,0) -- (3,0) ;
  	\draw[snake it] (1.5,0) -- (1.5,2);
  	\filldraw[black] (1.5,0) circle (2pt) node[anchor= north ]{ $(\kappa_{2})^{1}$};
\end{tikzpicture}
\qquad
\begin{tikzpicture}[scale=0.7]
  	\draw[very thick,dashed] (0,0) -- (3,0) ;
  	\draw[snake it] (0,2) -- (1.5,0);
  	\draw[snake it] (1.5,0) -- (3,2);
  	\filldraw[black] (1.5,0) circle (2pt) node[anchor= north ]{ $(\kappa_{2})^{2}$};
\end{tikzpicture}
\qquad
\begin{tikzpicture}[scale=0.7]
  	\draw[very thick,dashed] (0,0) -- (3,0) ;
  	\draw[snake it] (0,2) -- (1.5,0);
  	\draw[snake it] (1.5,0) -- (1.5,2);
  	\draw[snake it] (1.5,0) -- (3,2);
  	\filldraw[black] (1.5,0) circle (2pt) node[anchor= north ]{ $(\kappa_{2})^{3}$};
\end{tikzpicture}
\\~\\
\begin{tikzpicture}[scale=0.8]
	\path (-1,1) node[anchor=east]{\bf{graviton:}};
  	\draw[snake it] (0,0) -- (1,1);
  	\draw[snake it] (2,0) -- (1,1);
  	\filldraw[black] (1,1) circle (2pt) node[anchor=west]{~~~$(\kappa_{g})^{1} $};
  	\draw[snake it] (1,1) -- (1,2);
\end{tikzpicture}
\qquad
\begin{tikzpicture}[scale=0.8]
  	\draw[snake it] (0,0) -- (2,2);
  	\filldraw[black] (1,1) circle (2pt) node[anchor=west]{~~~$(\kappa_{g})^{2}$};
  	\draw[snake it] (2,0) -- (0,2);
\end{tikzpicture}
\end{center}
\caption{We replace the Newton constant $\kappa$ to new formal coupling constants $\kappa_{1}$, $\kappa_{2}$ and $\kappa_{g}$ according to the interaction types. }
\label{new_couplings}\end{figure}

Next, we introduce labels for each propagator, $n_{1}$, $n_{2}$ and $n_{g}$, for the $\phi_{1}$, $\phi_{2}$ and the graviton respectively
\begin{equation}
  \begin{tikzpicture}
  \draw[snake it] (0,3) node[anchor=east]{ graviton ~~} -- (2,3) node[anchor=west]{~:~$n_{g}$} ;
  \draw[very thick] (0,2) node[anchor=east]{matter 1 ~~} -- (2,2) node[anchor=west]{~:~$n_{1}$};
  \draw[very thick,dashed] (0,1) node[anchor=east]{ matter 2 ~~}  -- (2,1) node[anchor=west]{~:~$n_{2}$};
\end{tikzpicture}\label{labels_propagators}
\end{equation}

Collecting the couplings and labelings, we can define an identifier of diagrams as follows:
\begin{equation}
  (\kappa_{g})^{N_{1}}(\kappa_{1})^{N_{2}} (\kappa_{2})^{N_{3}} (n_{g})^{N_{4}} (n_{1})^{N_{5}} (n_{2})^{N_{6}} \,.
\label{diagram_identifier}\end{equation}
We may assign a set of the six numbers $(N_{1}, \cdots N_{6})$ for each diagram. These numbers can be facilitated to define conditions on the identifiers for filtering the classical diagrams efficiently. However, this assignment is not an one-to-one correspondence between identifiers and diagrams. Multiple diagrams can correspond to a set of exponents in \eqref{diagram_identifier}; even quantum diagrams can satisfy the conditions for classical diagrams. Nevertheless, such quantum diagrams can be effectively removed because of their distinctive propagator structure compared to classical diagrams. This is sufficient for our purpose.

There is a subtle issue in assigning the identifier. Let us consider a diagram having graviton propagators without a loop momentum, which involves $\frac{1}{q^{2}}$ factor. Since the factor does not affect the loop integrals, we may consider them as diagrams without such a graviton propagator. While these diagrams may exhibit distinct power-counting behaviors, we do not treat them as independent diagrams as in the following examples,
\begin{equation}
\begin{aligned}
  \begin{tikzpicture}[scale=0.7]
  	\draw[very thick,dashed] (0,0) -- (3,0);
  	\draw[very thick] (0,2) -- (3,2);
  	\draw[snake it] (0.5,2) -- (1.5,1);
  	\draw[snake it] (2.5,2) -- (1.5,1);
  	\draw[snake it] (1.5,0) -- (1.5,1);
	\path (4,1) node {\Large $= \frac{1}{q^{2}}$};
\end{tikzpicture}\
\begin{tikzpicture}[scale=0.7]
  	\draw[very thick,dashed] (0,0) -- (3,0);
  	\draw[very thick] (0,2) -- (3,2);
  	\draw[snake it] (0.5,2) -- (1.5,0);
  	\draw[snake it] (2.5,2) -- (1.5,0);
\end{tikzpicture}
\qquad \qquad
\begin{tikzpicture}[scale=0.7]
  	\draw[very thick,dashed] (0,0) -- (3,0);
  	\draw[very thick] (0,2) -- (3,2);
  	\draw[snake it] (0.5,2) -- (1.5,0.7);
  	\draw[snake it] (2.5,2) -- (1.5,0.7);
  	\draw[snake it] (1.5,0) -- (1.5,0.7);
  	\draw[snake it] (1.5,2) -- (2,1.3);
	\path (4,1) node {\Large $= \frac{1}{q^{2}}$};
\end{tikzpicture}
\begin{tikzpicture}[scale=0.7]
  	\draw[very thick,dashed] (0,0) -- (3,0);
  	\draw[very thick] (0,2) -- (3,2);
  	\draw[snake it] (0.5,2) -- (1.5,0);
  	\draw[snake it] (2.5,2) -- (1.5,0);
  	\draw[snake it] (1.5,2) -- (2,1);
\end{tikzpicture}
\end{aligned}\label{trivial_graviton_propagator}
\end{equation}
This feature simplifies the classification scheme discussed in the following subsection.

In practice, we do not have to impose those numbers by hand. Instead, we can assign the numbers at the level of recursions. It is achieved simply by replacing $\kappa$ to $\{\kappa_{1}, \kappa_{1}, \kappa_{g}\}$ and scaling each propagator in the recursions such as
\begin{equation}
  \frac{1}{(\ell\pm k_{\mathcal{P}})^{2}} \to \frac{n_{g}}{(\ell\pm k^{2}_{\mathcal{P}})^{2}}\,,
  \qquad
  \frac{1}{(\ell\pm k_{\mathcal{P}})^{2}+m^{2}_{1,2}} \to \frac{n_{1,2}}{(\ell\pm k_{\mathcal{P}})^{2}+m^{2}_{1,2}}\,.
\label{}\end{equation}
For instance, we may rewrite the recursion of $\Phi^{I}$ in \eqref{Phi_recursion} as
\begin{equation}
\begin{aligned}
   \Phi^{{\scriptscriptstyle(2)}}_{\ell_{1}\ell_{2}}\big|^{I}_{\mathcal{P}}
  = \frac{n_{I}}{k_{\mathcal{P}}+m^{2}_{I}} \Big(\kappa_{I} \Omega^{{\scriptscriptstyle(2)}}_{\ell_{1}\ell_{2}}\big|^{I}_{\mathcal{P}}
  	- \kappa_{I}^{2}\mathcal{I}^{M,N}_{\varphi,I} \Omega^{{\scriptscriptstyle(2)}}_{\ell_{1}\ell_{2}}\big|^{I,M,N}_{\mathcal{P}}
	- \kappa_{I}^{3}\mathcal{I}^{M,N,P}_{\varphi, I} \Omega^{{\scriptscriptstyle(2)}}_{\ell_{1}\ell_{2}}\big|^{I,M,N,P}_{\mathcal{P}}\Big)\,.
\end{aligned}
\label{Phi_recursion2}\end{equation}
If we solve the modified recursions, each term in the off-shell currents contains the polynomial of the couplings and the labeling of the propagators automatically.

\subsection{Classifying the classical diagrams}

We now introduce a new classification scheme for efficient evaluations of loop integrals, especially optimizing the integration-by-parts (IBP) reduction procedure. Before discussing IBP reduction in the next section, we need to introduce integral family for a diagram.

Typically, one may associate an integral family for each diagram. An integral family $F^{\Gamma_{l}}_{i_{1},i_{2},\cdots,i_{n}}$ at $l$ loops associated with a diagram $\Gamma_{l}$ is represented by
\begin{equation}
  F^{\Gamma_{l}}_{i_{1},i_{2},\cdots,i_{n}} = \int \frac{\mathrm{d}^{D}\ell_{1}}{(2\pi)^{D}} \cdots \frac{\mathrm{d}^{D}\ell_{l}}{(2\pi)^{D}}
  \frac{1}{\big(\rho_{1}\big)^{i_{1}} \big(\rho_{2}\big)^{i_{2}} \cdots \big(\rho_{n}\big)^{i_{n}}}\,, \qquad i_{1}, \cdots, i_{n} \in \mathbb{Z}\,,
\label{integral_family}\end{equation}
where $\rho_{i}$ represent the propagators that constitute the diagram $\Gamma_{l}$. Since the exponents can be zero or negative integers, the above is the minimum form -- one may add several propagators not included in the loop integrand associated with the diagram $\Gamma_{l}$. Note that the assignment of integral families for each diagram is computationally redundant.

Instead, our method offers a more efficient alternative: grouping diagrams characterized by the integral family. After clustering the diagrams, the IBP reduction can be executed on the groups collectively rather than individually. This collective approach significantly enhances the computational efficiency.


\subsubsection{Classes and Sectors} \label{sec:4.2.1}
Our classification scheme occurs through a two-step process. The first step is grouping the classical diagrams according to the power counting of the new couplings $\kappa_{1}, \kappa_{2}$ and $\kappa_{g}$. To facilitate this, we introduce a process called \textit{pinching}, which removes either an internal matter or a graviton line (propagator) from a diagram. While this process reduces the total number of internal lines, it leaves the powers of the coupling constants $\kappa_{1}, \kappa_{2}$ and $\kappa_{g}$ unaltered. We may illustrate the pinching process as follows:
\begin{equation}
\begin{aligned}
  \begin{tikzpicture}[scale=0.8]
  	\draw[very thick] (0,0) -- (3,0) ;
  	\draw[snake it] (0.7,0) -- (0.7,1.5);
  	\draw[snake it] (2.2,0) -- (2.2,1.5);
	\draw[thick, ->] (3.5,1) -- (5,1) node[anchor=north,pos=1/2,align=center]{\small pinching};
\end{tikzpicture}
\begin{tikzpicture}

\end{tikzpicture}
\begin{tikzpicture}[scale=0.8]
  	\draw[very thick] (0,0) -- (3,0) ;
  	\draw[snake it] (1.5,0) -- (0.7,1.5);
  	\draw[snake it] (1.5,0) -- (2.2,1.5);
\end{tikzpicture}
\\~\\
\begin{tikzpicture}[scale=0.8]
  	\draw[snake it] (0,0) -- (3,0) ;
  	\draw[snake it] (0.7,0) -- (0.7,1.5);
  	\draw[snake it] (2.2,0) -- (2.2,1.5);
	\draw[thick, ->] (3.5,1) -- (5,1) node[anchor=north,pos=1/2,align=center]{\small pinching};
\end{tikzpicture}
\begin{tikzpicture}

\end{tikzpicture}
\begin{tikzpicture}[scale=0.8]
  	\draw[snake it] (0,0) -- (3,0) ;
  	\draw[snake it] (1.5,0) -- (0.7,1.5);
  	\draw[snake it] (1.5,0) -- (2.2,1.5);
\end{tikzpicture}
\\
\end{aligned}\label{}
\end{equation}
Thus, any integrands related through the pinching share the same powers of coupling constants. This invariance enables us to define a specific set of diagrams, which we shall call this collection as a \textit{class}. All the integrands constituting a given class carry the same power-counting of the coupling constants regardless of the number of internal lines. Thus, the powers of the couplings can be used for markers of classes:
\begin{equation}
  \text{class} \iff \big(\kappa_{g}\big)^{N_{1}} \big(\kappa_{1}\big)^{N_{2}} \big(\kappa_{2}\big)^{N_{3}}\,.
\label{classes_powercounting}\end{equation}

The next step is classifying according to the structure of propagators, in other words, the diagram topologies. The power-counting criterion alone is insufficient for distinguishing the difference among all diagram topologies. This implies that a class allows a further decomposition into multiple subclasses according to the available integral families, which depend on the structure of propagators. We shall call such subclasses \textit{sector}. Obviously, the choice of the subclasses is not unique. We want a systematic construction that minimizes the number of subclasses.

Let us consider specific diagrams with maximal internal lines, which maximize the sum of $n_{1} + n_{2} + n_g$ within a class and refer to these as \textit{primary diagrams}. In organizing sectors within a class, we use these primary diagrams as a reference point. More precisely, a sector comprises a collection of diagrams directly related to a designated primary diagram through pinching while keeping the structure of the interaction vertices. Generally, the primary diagram is defined uniquely for a given class, and we may use the primary diagram to label the class.

\begin{figure}[t]
\begin{center}
\begin{tikzpicture}[
roundnode/.style={circle, draw=green!60, very thick, minimum size=7mm},
squarednode/.style={rectangle, draw=red!60,	 very thick, minimum size=7mm},
squarednode2/.style={rectangle, draw=blue!60,	 very thick, minimum size=7mm}
]
\node[roundnode]    (class)                           {Class};
\node[squarednode]	(sector2)       [right=of class] {Sector 2};
\node[squarednode]  (sector1)       [above=of sector2] {Sector 1};
\node[squarednode]  (sector3)       [below=of sector2] {Sector 3};

\node[squarednode2]	(InFa2)       [right=of sector2] {Integral Family 2};
\node[squarednode2]  (InFa1)       [right=of sector1] {Integral Family 1};
\node[squarednode2]  (InFa3)       [right=of sector3] {Integral Family 3};

\draw[->,thick] (class) -- (sector2);
\draw[->,thick] (class) -- (sector1.west);
\draw[->,thick] (class) -- (sector3.west);
\draw[->,thick] (sector2) -- (InFa2);
\draw[->,thick] (sector1) -- (InFa1);
\draw[->,thick] (sector3) -- (InFa3);
\end{tikzpicture}
\end{center}\caption{A class is decomposed by each sector, generated from a primary diagram by pinching. An integral family is associated with each sector. }\label{sector_integral_family}
\end{figure}
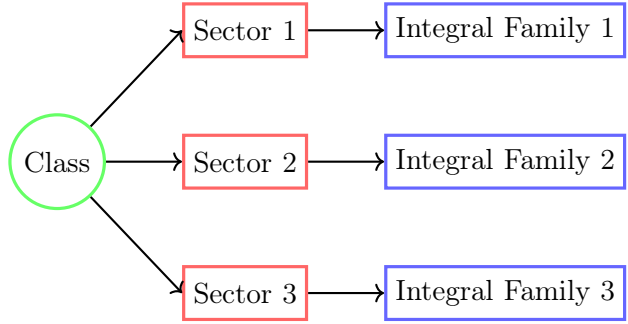

A salient feature of the sector is that every integrand within a given sector belongs to the same integral family. This consistency arises from the property of pinching: it removes an internal propagator by setting the power of the propagator to zero, yet the resultant integrand remains in the original integral family. Consequently, each primary diagram defines a sector, and each sector is associated with a specific integral family. In Figure \ref{sector_integral_family}, we summarized the sector.

One important remark is that constructing a sector is not unique. For instance, consider the process of obtaining a four-point vertex through pinching as below,
\begin{equation}
\begin{aligned}
  \begin{tikzpicture}[scale=0.7]
  	\draw[very thick] (-0.5,2) -- (2.5,2);
  	\draw[snake it] (0,0) -- (0,2);
  	\draw[snake it] (2,0) -- (2,2);

  	\draw[very thick] (-0.5,-0.5) -- (2.5,-0.5);
  	\draw[snake it] (0,-2.5) -- (0.8,-1.7);
  	\draw[snake it] (2,-0.5) -- (1.2,-1.3);
  	\draw[snake it] (0,-0.5) -- (2,-2.5);
  	
  	\path (4.75,-0.25) node {\small pinching};
  	\draw[thick, ->] (3.5,1) -- (6, 0.5);
  	\draw[thick, ->] (3.5,-1.5) -- (6, -1 ) ;

  	\draw[very thick] (7,0.75) -- (10,0.75);
  	\draw[snake it] (8.5,0.75) -- (9.5,-1.25);
  	\draw[snake it] (8.5,0.75) -- (7.5,-1.25);
  \end{tikzpicture}
\end{aligned}\label{nonunique_sector}
\end{equation}
This implies that a pinched diagram can have multiple origins, and this feature creates ambiguity in defining a sector. However, this ambiguity does not affect the final results. As such, the choice of a sector remains flexible and can be adapted depending on our purpose.


\subsubsection{Crossing and mirror duals} \label{sec:4.2.2}

Let us consider a set of all possible diagrams associated with an amplitude. For a given diagram, one can always identify its crossing and mirror dual counterparts within the set. Combining these diagrams, we can define a quartet. The loop integrands associated with the elements of a quartet are related to each other in a specific way. This structure enables us to reduce the number of loop integrals to be performed.
\begin{figure}[t]
\begin{center}
\begin{tikzpicture}[scale=0.7]
  	\draw[very thick,dashed] (0,0) -- (3,0) node[anchor=north, pos=1/2]{IY};
  	\draw[very thick] (0,2) -- (3,2);
  	\draw[snake it] (0.5,2) -- (0.5,0);
  	\draw[snake it] (1,2) -- (1.75,1);
  	\draw[snake it] (2.5,2) -- (1.75,1);
  	\draw[snake it] (1.75,1) -- (1.75,0);
\end{tikzpicture}
\qquad
\begin{tikzpicture}[scale=0.7]
  	\draw[very thick,dashed] (0,0) -- (3,0) node[anchor=north, pos=1/2]{$\overline{\rm IY}$};
  	\draw[very thick] (0,2) -- (3,2);
  	\draw[snake it] (0.5,2) -- (1.75,0);
  	\draw[snake it] (1,2) -- (1.75,1);
  	\draw[snake it] (2.5,2) -- (1.75,1);
  	\draw[snake it] (1.1,0.55) -- (0.5,0);
  	\draw[snake it] (1.75,1) -- (1.45,0.8);
\end{tikzpicture}
\qquad
\begin{tikzpicture}[scale=0.7]
  	\draw[very thick,dashed,xscale=-1] (0,0) -- (3,0) node[anchor=north, pos=1/2]{YI};
  	\draw[very thick,xscale=-1] (0,2) -- (3,2);
  	\draw[snake it,xscale=-1] (0.5,2) -- (0.5,0);
  	\draw[snake it,xscale=-1] (1,2) -- (1.75,1);
  	\draw[snake it,xscale=-1] (2.5,2) -- (1.75,1);
  	\draw[snake it,xscale=-1] (1.75,1) -- (1.75,0);
\end{tikzpicture}
\qquad
\flip{\begin{tikzpicture}[scale=0.7]
  	\draw[very thick] (0,0) -- (3,0);
  	\draw[very thick,dashed] (0,2) -- (3,2)node[anchor=south, pos=1/2]{$\rm IY$};
  	\draw[snake it] (0.5,2) -- (0.5,0);
  	\draw[snake it] (1,2) -- (1.75,1);
  	\draw[snake it] (2.5,2) -- (1.75,1);
  	\draw[snake it] (1.75,1) -- (1.75,0);
\end{tikzpicture}}
\end{center}\caption{A quartet including the IY diagram, which includes its crossed, the horizontal mirror dual and the vertical mirror dual diagrams are shown in order.}
\label{example_crossing_mirror}
\end{figure}
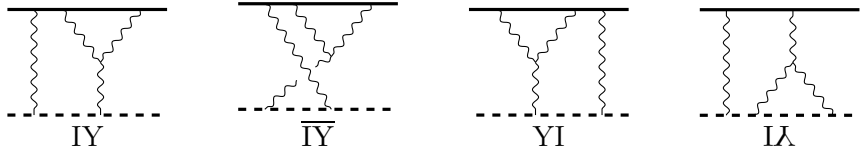

To begin with, we explore the concept of crossing. An integrand for a crossed diagram is generated by swapping the momenta $k_{2}$ and $k_{3}$ for a given diagram. This exchange is represented as follows:
\begin{equation}
\begin{aligned}
  \text{\bf crossing}: ~ k_{2} \leftrightarrow k_{3} \Longrightarrow ~
  &u_{1} \to u_{1}\,,~ u_{2} \to -u_{2}\,,~ y \to - y\,, ~ q^{\mu} \to q^{\mu}\,,
  \\
  & \sigma \to -\sigma + \frac{q^{2}}{2m_{1}m_{2}}\,,~
 \sqrt{y^{2}-1} \to - \sqrt{y^{2}-1} \,, \quad x \to -x\,.
\end{aligned}\label{eq:crossed_diagram}
\end{equation}
We denote the crossed diagram as $\overline{\Gamma}$ for a diagram $\Gamma$. We illustrate an example in Figure \ref{example_crossing_mirror}. The second diagram ($\overline{\rm IY}$) is the crossed diagram of the first diagram (IY). Note that the crossing map modifies the pole structure of loop integrals at the potential region, and we cannot be analytically continued from $y>1 \to y<-1$ \cite{Parra-Martinez:2020dzs}. Thus, this relation should be applied at the integrand level rather than the loop integral values.

Next, let us consider the mirror dual diagrams generated by reflecting the spatial orientation of the original diagram. We distinguish between two types of mirror duals, namely horizontal and vertical, formally defined as:
\begin{equation}
\begin{aligned}
  \text{\bf horizontal mirror dual}: &~ k_{1} \leftrightarrow k_{4} ~ \text{and} ~ k_{2} \leftrightarrow k_{3} \Longrightarrow y \to y\,, ~ q^{\mu} \to q^{\mu}\,,
	\\
  \text{\bf vertical mirror dual}: & ~ k_{1} \leftrightarrow k_{2} ~ \text{and} ~ k_{3} \leftrightarrow k_{4} \Longrightarrow y \to y\,,~ q^{\mu} \to -q^{\mu}\,,~ m_{1} \leftrightarrow m_{2}\,.
\end{aligned}\label{eq:mirror_duals}
\end{equation}
An example of the horizontal and vertical mirror duals is illustrated in Figure \ref{example_crossing_mirror}.
Unlike the crossing, these mirror duals preserve the sign of $y$, circumventing any concerns related to the pole structure. The mirror dual relations can be directly applied to the integral values, and we do not have to evaluate loop integrals for mirror dual diagrams.

The total loop integrands, the solution of the recursions at a specific loop order, contain all possible diagram topologies. Within the set of total integrands, one may find the crossing and mirror dual pairs for a given diagram, including the self-dual diagrams. According to our classification scheme, we may extend the notion of crossing and mirror duals to the level of sectors. This structure significantly enhances computational efficiency because the symmetry properties inherent in the set of diagrams render it unnecessary to evaluate the loop integrals corresponding to each mirror dual independently. Furthermore, such relations among the sectors serve as a robust consistency check for the loop integrands.


\subsubsection{One-loop example} \label{Sec:4.3.3}
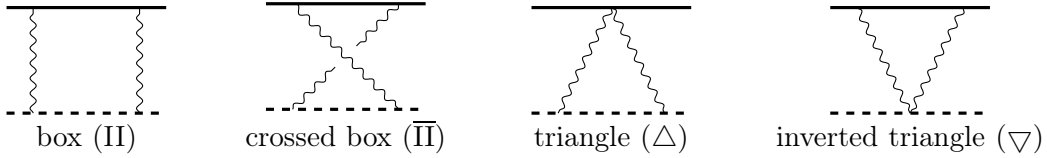
\begin{figure}[t]
\begin{center}
	\begin{tikzpicture}[scale=0.7]
  	\draw[very thick,dashed] (0,0) -- (3,0)node[anchor=north,pos=1/2,align=center]{box (II)};
  	\draw[very thick] (0,2) -- (3,2);
  	\draw[snake it] (0.5,0) -- (0.5,2);
  	\draw[snake it] (2.5,0) -- (2.5,2);
\end{tikzpicture}
\qquad
\begin{tikzpicture}[scale=0.7]
  	\draw[very thick,dashed] (0,0) -- (3,0)node[anchor=north,pos=1/2,align=center]{crossed box ($\overline{\rm II}$)};
  	\draw[very thick] (0,2) -- (3,2);
  	\draw[snake it] (0.5,0) -- (1.3,0.8);
  	\draw[snake it] (2.5,2) -- (1.7,1.2);
  	\draw[snake it] (0.5,2) -- (2.5,0);
\end{tikzpicture}
\qquad
\begin{tikzpicture}[scale=0.7]
	\draw[very thick,dashed] (0,0) -- (3,0) node[anchor=north,pos=1/2,align=center]{ triangle ($\triangle$)};
  	\draw[very thick] (0,2) -- (3,2);
  	\draw[snake it] (1.5,2) -- (2.5,0);
  	\draw[snake it] (1.5,2) -- (0.5,0);
  	\end{tikzpicture}
\qquad
\begin{tikzpicture}[scale=0.7]
  	\draw[very thick,dashed] (6,0) -- (9,0) node[anchor=north,pos=1/2,align=center]{inverted triangle ($\flip{$\triangle$}$)};
  	\draw[very thick] (6,2) -- (9,2);
  	\draw[snake it] (8.5,2) -- (7.5,0);
  	\draw[snake it] (6.5,2) -- (7.5,0);
\end{tikzpicture}
\end{center}
\caption{One loop diagrams}
\label{one_loop_diagrams}\end{figure}

To elucidate the definitions previously outlined, such as the class, sector and integral family, let us examine the classical $2 \to 2$ amplitude at one loop. It is well-known that the classical one-loop contribution arises from both box and triangle diagrams \cite{Bern:2019crd,Cristofoli:2020uzm}. There exist two kinds of box diagrams: the box (II) diagram and the crossed box ($\overline{\rm II}$), which are illustrated in figure \ref{one_loop_diagrams}. These have the same power counting,
\begin{equation}
  \text{II, $\overline{\rm II}$}: ~(\kappa_{1})^{2} (\kappa_{2})^{2} (\kappa_{g})^{0} (n_{1})^{1} (n_{2})^{1} (n_{g})^{2}\,.
\label{box_power}\end{equation}

Similarly, there are two kinds of triangle diagrams: the triangle ($\triangle$) and the inverted triangle (\flip{$\triangle$}). Their power countings are given by\footnote{Note that we omitted the diagrams containing a graviton propagator without loop momenta, $\frac{1}{q^{2}}$ as in \eqref{trivial_graviton_propagator}.}
\begin{equation}
\begin{aligned}
    \triangle: ~&
  (\kappa_{1})^{2} (\kappa_{2})^{2} (\kappa_{g})^{0} (n_{1})^{0} (n_{2})^{1} (n_{g})^{2}\,,
  \\
  \bigtriangledown:~&
  (\kappa_{1})^{2} (\kappa_{2})^{2} (\kappa_{g})^{0} (n_{1})^{1} (n_{2})^{0} (n_{g})^{2}\,.
\end{aligned}\label{triangle_power}
\end{equation}
One can check that the box and the triangle diagrams have the same power-counting of the couplings, $\kappa^{2}_{1} \kappa^{2}_{2} \kappa^{0}_{g}$. According to the relation between a class and power-counting of the couplings \eqref{classes_powercounting}, all the classical diagrams at one-loop form a single class, which we shall refer to $\Box$-class. Furthermore, the triangle diagrams can be generated by pinching a matter line from the box diagrams.

The primary diagrams in $\Box$-class are the II and $\overline{\rm II}$ diagrams. These have four internal lines, whereas the triangle diagrams have three lines. Each primary diagram defines a sector: the II-sector and the $\overline{\rm II}$-sector. Thus, the entire one-loop integrands are decomposed into the two sectors. As described in \eqref{nonunique_sector}, there is no canonical choice for defining a sector. We should determine how to allocate the triangle diagrams into the sectors. Our convention is to equally distribute the triangles into each sector as follows:
\begin{equation}
\begin{aligned}
  \text{II-sector}: &~ \mbox{II} + \frac{1}{2} \big(\triangle+\flip{$\triangle$}\big)\,,
  \\
  \text{$\overline{\rm II}$-sector}: &~ \mbox{X} + \frac{1}{2} \big(\triangle+\flip{$\triangle$}\big)\,.
\end{aligned}\label{sectors_one_loop}
\end{equation}
Since the triangle diagrams are invariant under the crossing relation \eqref{eq:crossed_diagram}, this choice ensures a manifest invariance under the crossing at the level of integrands. We may exploit this property as a consistency check of the solutions of the off-shell recursion.

There are two integral families associated with the II and the $\overline{\rm II}$ sectors. First, the integral family for the II sector is defined by
\begin{equation}
\begin{aligned}
  F^{\rm II}_{i_{1},i_{2},i_{3},i_{4}} &= \int_{\ell} \frac{1}{\big(\rho^{\rm II,1}_{g}\big)^{i_{1}} \big(\rho^{\rm II,2}_{g}\big)^{i_{2}} \big(\rho^{\rm II,3}_{1}\big)^{i_{3}} \big(\rho^{\rm II,4}_{2}\big)^{i_{4}}}\,,
  \qquad
  i_{1}, \cdots i_{4} \in \mathbb{Z}\,,
\end{aligned}\label{II_integral_family}
\end{equation}
where the loop integral is defined by \cite{Smirnov:2012gma,Parra-Martinez:2020dzs}
\begin{equation}
  \int_{\ell} \equiv \int \frac{d^{D} \ell \,e^{\gamma_{E} \epsilon}}{i \pi^{D / 2}}\,,
\label{}\end{equation}
which is different from the usual convention of the Fourier transformation, $\int\frac{\mathrm{d}^{4}\ell_{i}}{(2\pi)^{4}}$. We should multiply the following factor for each loop integral to restore the original convention and compute the scattering amplitude,
\begin{equation}
  \frac{i}{\big(4\pi)^{2}}(4\pi e^{-\gamma_{E}}\mu^{2}\big)^{\epsilon}\,.
\label{Normalization_factor_loop_integral}\end{equation}
Here the propagators $\rho^{\rm II}_{g}$ and $\rho^{\rm II}_{1,2}$ are graviton and matter propagators respectively defined by
\begin{equation}
\begin{aligned}
  \rho^{\rm II,1}_{g} &=  \ell^{2}\,,
  \\
  \rho^{\rm II,2}_{g} &= \big(\ell-k_{23} \big)^{2} = \big(\ell + q\big)^{2}\,,
  \\
  \rho^{\rm II,3}_{1} &= \big(\ell -k_{123} \big)^{2} + m_{1}^{2} = \big(\ell + \bar{m}_{1} u_{1} +\frac{q}{2} \big)^{2} + m_{1}^{2}\,,
  \\
  \rho^{\rm II,4}_{2} & = \big(\ell-k_{3} \big)^{2} + m_{2}^{2} = \big(\ell- \bar{m}_{2} u_{2} + \frac{q}{2} \big)^{2} + m_{2}^{2}\,.
\end{aligned}\label{propagators_II}
\end{equation}
We omitted the $-i\varepsilon$ factors in the propagators.

Similarly, the integral family for the $\overline{\rm II}$ sector is given by
\begin{equation}
  F^{\overline{\rm II}}_{i_{1},i_{2},i_{3},i_{4}} = \int_{\ell} \frac{1}{\big(\rho^{\overline{\rm II},1}_{g}\big)^{i_{1}} \big( \rho^{\overline{\rm II},2}_{g} \big)^{i_{2}} \big( \rho^{\overline{\rm II},3}_{1} \big)^{i_{3}} \big( \rho^{\overline{\rm II},4}_{2} \big)^{i_{4}}}\,,
  \qquad
  i_{1}, \cdots i_{4} \in \mathbb{Z}\,,
\label{crossed_II_integral_family}\end{equation}
where the propagators are defined by
\begin{equation}
\begin{aligned}
  \rho^{\overline{\rm II},1}_{g} &=  \ell^{2}\,,
  \\
  \rho^{\overline{\rm II},2}_{g} &= \big(\ell-k_{23} \big)^{2} = \big(\ell + q\big)^{2}\,,
  \\
  \rho^{\overline{\rm II},3}_{1} &= \big(\ell -k_{123} \big)^{2} + m_{1}^{2} = \big(\ell + \bar{m}_{1} u_{1} +\frac{q}{2} \big)^{2} + m_{1}^{2}\,,
  \\
  \rho^{\overline{\rm II},4}_{2} & = \big(\ell-k_{2} \big)^{2} + m_{2}^{2} = \big(\ell + \bar{m}_{2} u_{2} + \frac{q}{2} \big)^{2} + m_{2}^{2}\,.
\end{aligned}\label{propagators_crossed_II}
\end{equation}
One can check that the integral family of the $\overline{\rm II}$-sector is related to the II-sector by crossing relation \eqref{eq:crossed_diagram}.

The loop integrals for the II and the $\overline{\rm II}$ diagrams are represented by
\begin{equation}
  F^{\rm II}_{1,1,1,1} \,,
  \qquad
  F^{\overline{\rm II}}_{1,1,1,1} \,.
\label{box_integrals}\end{equation}
The triangle diagrams can be generated from the box diagram by pinching the box (II) diagram. The corresponding loop integrals are given by
\begin{equation}
\begin{aligned}
  F^{\rm II}_{1,1,0,1} = F^{\overline{\rm II}}_{1,1,0,1}\,,
  \qquad
  F^{\rm II}_{1,1,1,0} = F^{\overline{\rm II}}_{1,1,1,0}\,.
\end{aligned}\label{triangle_integrals}
\end{equation}
%

\subsection{Two loops} \label{Sec:4.4}
\begin{table}[b]
\begin{center}
\begin{tabular}{ |c|c|c|}
\hline
Class & couplings & power-countings \\
  \hline\hline
  $\Box\!\Box$ & $\kappa_{1}^{3} \kappa_{2}^{3} \kappa_{g}^{0}$ & $\kappa_{1}^{3} \kappa_{2}^{3} \kappa_{g}^{0} n_{1}^{2} n_{2}^{2} n_{g}^{3} + \kappa_{1}^{3} \kappa_{2}^{3} \kappa_{g}^{0} n_{1}^{2} n_{2}^{1} n_{g}^{3}+\kappa_{1}^{3} \kappa_{2}^{3} \kappa_{g}^{0} n_{1}^{1} n_{2}^{2} n_{g}^{3} $
  \\& &
  $ +\kappa_{1}^{3} \kappa_{2}^{3} \kappa_{g}^{0} n_{1}^{1} n_{2}^{1} n_{g}^{3}+\kappa_{1}^{3} \kappa_{2}^{3} \kappa_{g}^{0} n_{1}^{0} n_{2}^{2} n_{g}^{3} +\kappa_{1}^{3} \kappa_{2}^{3} \kappa_{g}^{0} n_{1}^{2} n_{2}^{0} n_{g}^{3} $
\\
\hline
  $\mathbf{H}$& $\kappa_{1}^{2} \kappa_{2}^{2} \kappa_{g}^{2}$ & $\kappa_{1}^{2} \kappa_{2}^{2} \kappa_{g}^{2} n_{1}^{1} n_{2}^{1} n_{g}^{5}+\kappa_{1}^{2} \kappa_{2}^{2} \kappa_{g}^{2} n_{1}^{1} n_{2}^{1} n_{g}^{4} $
\\
\hline
  $\mathbf{IY}$ & $\kappa_{1}^{3} \kappa_{2}^{2} \kappa_{g}^{1}$ & $\kappa_{1}^{3} \kappa_{2}^{2} \kappa_{g}^{1} n_{1}^{2} n_{2}^{1} n_{g}^{4} + \kappa_{1}^{3} \kappa_{2}^{2} \kappa_{g}^{1} n_{1}^{1} n_{2}^{1} n_{g}^{4}+\kappa_{1}^{3} \kappa_{2}^{2} \kappa_{g}^{1} n_{1}^{2} n_{2}^{0} n_{g}^{4} $
\\
\hline
$\textbf{\flip{IY}}$ & $\kappa_{1}^{2} \kappa_{2}^{3} \kappa_{g}^{1}$ & $\kappa_{1}^{2} \kappa_{2}^{3} \kappa_{g}^{1} n_{1}^{1} n_{2}^{2} n_{g}^{4} +\kappa_{1}^{2} \kappa_{2}^{3} \kappa_{g}^{1} n_{1}^{1} n_{2}^{1} n_{g}^{4} +\kappa_{1}^{2} \kappa_{2}^{3} \kappa_{g}^{1} n_{1}^{0} n_{2}^{2} n_{g}^{4}$
\\
\hline
\end{tabular}	
\end{center}
\caption{Power countings for classical diagrams at two loops}
\label{2loop_types}\end{table}
Let us turn to our main interest, the classification of the classical two-loop diagrams. At the two-loop level, there exist 14 distinct power-countings, along with four different kinds of coupling constants. According to the relation between the classes and power countings of couplings \eqref{classes_powercounting}, we are led to define four classes at two loops, which are denoted as follows: $\Box\!\Box$, $\mathbf{H}$, $\mathbf{IY}$ and $\textbf{\flip{IY}}$. The results are summarized in Table \ref{2loop_types}.

\begin{figure}[t]
  \begin{tikzpicture}[scale=0.7]
  	\draw[very thick,dashed] (0,0) -- (3,0) node[anchor=north,pos=1/2]{(III)};
  	\draw[very thick] (0,1.5) -- (3,1.5);
  	\draw[snake it] (0.5,1.5) -- (0.5,0);
  	\draw[snake it] (1.5,1.5) -- (1.5,0);
  	\draw[snake it] (2.5,1.5) -- (2.5,0);
\end{tikzpicture}
~
\begin{tikzpicture}[scale=0.7]
  	\draw[very thick,dashed] (0,0) -- (3,0) node[anchor=north,pos=1/2]{(IX)};
  	\draw[very thick] (0,1.5) -- (3,1.5);
  	\draw[snake it] (0.5,1.5) -- (0.5,0);
  	\draw[snake it] (1.5,1.5) -- (1.9,0.9);
  	\draw[snake it] (2.1,0.6) -- (2.5,0);
  	\draw[snake it] (2.5,1.5) -- (1.5,0);
\end{tikzpicture}
~
\begin{tikzpicture}[scale=0.7]
  	\draw[very thick,dashed] (0,0) -- (3,0) node[anchor=north,pos=1/2]{(XI)};
  	\draw[very thick] (0,1.5) -- (3,1.5);
  	\draw[snake it] (2.5,1.5) -- (2.5,0);
  	\draw[snake it] (0.5,1.5) -- (0.9,0.9);
  	\draw[snake it] (1.1,0.6) -- (1.5,0);
  	\draw[snake it] (1.5,1.5) -- (0.5,0);
\end{tikzpicture}
~
\begin{tikzpicture}[scale=0.7]
  	\draw[very thick,dashed] (0,0) -- (3,0) node[anchor=north,pos=1/2]{$(\overline{\rm III})$} ;
  	\draw[very thick] (0,1.5) -- (3,1.5);
  	\draw[snake it] (0.5,1.5) -- (1.35,0.85);
  	\draw[snake it] (1.65,0.65) -- (2.5,0);
  	\draw[snake it] (1.5,1.5) -- (1.5,0);
  	\draw[snake it] (2.5,1.5) -- (1.65,0.80);
  	\draw[snake it] (1.35,0.65) -- (0.5,0);
\end{tikzpicture}
~
\begin{tikzpicture}[scale=0.7]
  	\draw[very thick,dashed] (0,0) -- (3,0) node[anchor=north,pos=1/2]{$(\overline{\rm IX})$};
  	\draw[very thick] (0,1.5) -- (3,1.5);
  	\draw[snake it] (0.5,1.5) -- (2.5,0);
  	\draw[snake it] (1,0.8) -- (0.5,0);
  	\draw[snake it] (1.5,1.5) -- (1.3,1.2);
  	\draw[snake it] (2.5,1.5) -- (1.9,0.7);
  	\draw[snake it] (1.7,0.36) -- (1.5,0);
\end{tikzpicture}
~
\begin{tikzpicture}[scale=0.7]
  	\draw[very thick,dashed] (0,0) -- (3,0) node[anchor=north,pos=1/2]{$(\overline{\rm XI})$};
  	\draw[very thick] (0,1.5) -- (3,1.5);
  	\draw[snake it] (2.5,1.5) -- (0.5,0);
  	\draw[snake it] (0.5,1.5) -- (1.05,0.7);
  	\draw[snake it] (1.3,0.35) -- (1.5,0);
  	\draw[snake it] (1.5,1.5) -- (1.75,1.2);
  	\draw[snake it] (2,0.8) -- (2.5,0);
\end{tikzpicture}
  \caption{Primary diagrams in $\Box\!\Box$-class. These define their own sectors.}
\label{doublebox_diagrams}\end{figure}
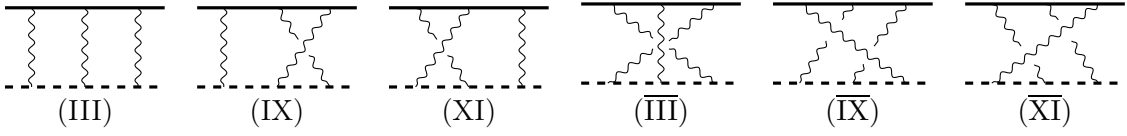

We define primary diagrams and sectors for each class. We start from $\Box\!\Box$-class. In this class, there are 6 primary diagrams, ${\rm III}\,, {\rm IX}\,, {\rm XI}$ and their crossed diagrams $\overline{\rm III}\,, \overline{\rm IX}\,, \overline{\rm XI}$ as depicted in Figure \ref{doublebox_diagrams}. Since each primary diagram defines a sector, $\Box\!\Box$-class consists of six sectors
\begin{equation}
  \Box\!\Box = {\rm III} + {\rm IX} +  {\rm XI} + \overline{\rm III} + \overline{\rm IX}+\overline{\rm XI}\,,
\label{classes_double_box}\end{equation}
Note that the IX and XI ($\overline{\rm IX}$ and $\overline{\rm XI}$) are a pair of the horizontal mirror dual. Since their loop integrals are identical, we thus omit them.

For each sector in $\Box\!\Box$-class, an integral family is associated as follows:
\begin{equation}
\begin{aligned}
  F^{\rm III}_{i_{1},i_{2},\cdots i_{9}} &= \int_{\ell_{1},\ell_{2}}
  \frac{\big(\rho^{\rm III,8}_{g}\big)^{-i_{8}} \big(\rho^{\rm III,9}_{g}\big)^{-i_{9}}}{\big(\rho^{\rm III,1}_{g}\big)^{i_{1}} \big(\rho^{\rm III,2}_{g}\big)^{i_{2}}\big(\rho^{\rm III,3}_{g}\big)^{i_{3}}\big(\rho^{\rm III,4}_{1}\big)^{i_{4}} \big(\rho^{\rm III,5}_{1}\big)^{i_{5}} \big(\rho^{\rm III,6}_{2}\big)^{i_{6}}\big(\rho^{\rm III,7}_{2}\big)^{i_{7}}}\,,
\end{aligned}\label{III_integral_family_full_propa}
\end{equation}
where
\begin{equation}
\begin{aligned}
  \rho^{\rm III,1}_{g} &= \ell_{1}^{2}\,,
  &\quad
  \rho^{\rm III,2}_{g} &= \ell_{2}^{2}\,,
  &\quad
  \rho^{\rm III,3}_{g} &= (\ell_{1}+\ell_{2}+q)^{2}\,,
  \\
  \rho^{\rm III,4}_{1} &= (\ell_{1} - k_{123})^{2}+m^{2}_{1}\,,
  &\quad
  \rho^{\rm III,5}_{1} &= (\ell_{1} + k_{1})^{2}+m^{2}_{1}\,,
  &\quad
  \rho^{\rm III,6}_{2} &= (\ell_{1} - k_{3})^{2}+m^{2}_{2}\,,
  \\
  \rho^{\rm III,7}_{2} &= (\ell_{1} - k_{2})^{2}+m^{2}_{2}\,,
  &\qquad
  \rho^{\rm III,8}_{g} &= (\ell_{1}+q)^{2},
  &\qquad
  \rho^{\rm III,9}_{g} &= (\ell_{2}+q)^{2}\;.
\end{aligned}\label{}
\end{equation}
and
\begin{equation}
  F^{\rm IX}_{i_{1},i_{2},\cdots i_{9}} = \int_{\ell_{1},\ell_{2}} \frac{\big(\rho^{\rm IX,8}_{g}\big)^{-i_{8}} \big(\rho^{\rm IX,9}_{g}\big)^{-i_{9}}}{\big(\rho^{\rm IX,1}_{g}\big)^{i_{1}} \big(\rho^{\rm IX,2}_{g}\big)^{i_{2}} \big(\rho^{\rm IX,3}_{g}\big)^{i_{3}}\big(\rho^{\rm IX,4}_{1}\big)^{i_{4}} \big(\rho^{\rm IX,5}_{1}\big)^{i_{5}}\big(\rho^{\rm IX,6}_{2}\big)^{i_{6}}\big(\rho^{\rm IX,7}_{2}\big)^{i_{7}}} \,,
\label{IX_integral_family_full_propa}\end{equation}
where
\begin{equation}
\begin{aligned}
  \rho^{\rm IX,1}_{g} &= \ell_{1}^{2}\,,
  &\quad
  \rho^{\rm IX,2}_{g} &=\ell_{2}^{2}\,,
  &\quad
  \rho^{\rm IX,3}_{g} &=(\ell_{1}+\ell_{2}+q)^{2},
  \\
  \rho^{\rm IX,4}_{1} &= (\ell_{1}-k_{123})^{2}+m_{1}^{2}\,,
  &\quad
  \rho^{\rm IX,5}_{1} &= (\ell_{2}+k_{1})^{2}+m_{1}^{2}\,,
  &\quad
  \rho^{\rm IX,6}_{2}&= (\ell_{2}-k_{2})^{2}+m_{2}^{2}\,,
  \\
  \rho^{\rm IX,7}_{2} &= (\ell_{1}+\ell_{2}-k_{2})^{2}+m_{2}^{2}\,,
  &\quad
  \rho^{\rm IX,8}_{g} &=(\ell_{1}+q)^{2}\,,
  &\quad
  \rho^{\rm IX,9}_{g}&=(\ell_{2}+q)^{2}\,,
\end{aligned}\label{}
\end{equation}
Here the $\rho^{\rm III,8}_{g}, \rho^{\rm III,9}_{g},\rho^{\rm IX,8}_{g}$ and $\rho^{\rm IX,9}_{g}$ are auxiliary propagators that are not included in the integrands of the primary diagrams. These are crucial for consistent IBP reduction. The integral families of the crossed diagrams, $F^{\overline{\rm III}}_{n_{1},n_{2},\cdots n_{9}}$ and $F^{\overline{\rm IX}}_{n_{1},n_{2},\cdots n_{9}}$, can be simply derived by the crossing relation defined in \eqref{eq:crossed_diagram}.

The $\mathbf{H}$-class consists of two sectors, H and $\overline{\rm H}$
\begin{equation}
  \mathbf{H} = {\rm H} + \overline{\rm H}\,.
\label{}\end{equation}
The H and the $\overline{\rm H}$ sectors are in crossing relation and depicted by
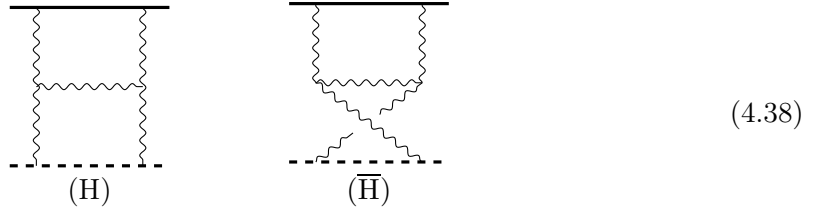
\begin{equation}
\begin{aligned}
\begin{tikzpicture}[scale=0.7]
  	\draw[very thick,dashed] (0,0) -- (3,0) node[anchor=north,pos=1/2]{(H)};
  	\draw[very thick] (0,3) -- (3,3);
  	\draw[snake it] (2.5,3) -- (2.5,0);
  	\draw[snake it] (0.5,3) -- (0.5,0);
  	\draw[snake it] (0.5,1.5) -- (2.5,1.5);
\end{tikzpicture}
\qquad\qquad
\begin{tikzpicture}[scale=0.7]
  	\draw[very thick,dashed] (0,0) -- (3,0) node[anchor=north,pos=1/2]{$(\overline{\rm H})$};
  	\draw[very thick] (0,3) -- (3,3);
  	\draw[snake it] (2.5,1.5) -- (2.5,3);
  	\draw[snake it] (0.5,1.5) -- (0.5,3);
  	\draw[snake it] (0.5,1.5) -- (2.5,1.5);
  	\draw[snake it] (0.5,0) -- (1.2,0.55);
  	\draw[snake it] (1.7,0.9) -- (2.5,1.5);
  	\draw[snake it] (0.5,1.5) -- (2.5,0);
\end{tikzpicture}
\end{aligned}\label{}
\end{equation}
Their integral families $F^{\rm H}_{i}$ and $F^{\overline{\rm H}}_{i}$ are defined by
\begin{equation}
\begin{aligned}
  F^{\rm H}_{i_{1},\cdots, i_{9}} &= \int_{\ell_{1},\ell_{2}}  \frac{ \big(\rho^{\rm H,8}_{1}\big)^{-i_{8}}\big(\rho^{\rm H,9}_{2}\big)^{-i_{9}}}{\big(\rho^{\rm H,1}_{g}\big)^{i_{1}} \big(\rho^{\rm H,2}_{g}\big)^{i_{2}}\big(\rho^{\rm H,3}_{g}\big)^{i_{3}}\big(\rho^{\rm H,4}_{g}\big)^{i_{4}} \big(\rho^{\rm H,5}_{g}\big)^{i_{5}}\big(\rho^{\rm H,6}_{1}\big)^{i_{6}}\big(\rho^{\rm H,7}_{2}\big)^{i_{7}}}\,,
\end{aligned}\label{H_integral_family_full_propa}
\end{equation}
where
\begin{equation}
\begin{aligned}
  \rho^{\rm H,1}_{g} &= (\ell_{1}+q)^{2}\,,
  &\qquad
  \rho^{\rm H,2}_{g} &= (\ell_{2}+q)^{2}\,,
  &\qquad
  \rho^{\rm H,3}_{g} &= \ell_{1}^{2}\,,
  \\
  \rho^{\rm H,4}_{g} &= \ell_{2}^{2}\,,
  &\qquad
  \rho^{\rm H,5}_{g} &= (\ell_{1}+\ell_{2}+q)^{2}\,,
  &\qquad
  \rho^{\rm H,6}_{1} &= (\ell_{1}-k_{123})^{2}+m_{1}^{2}\,,
  \\
  \rho^{\rm H,7}_{2} &= (\ell_{2}-k_{2})^{2}+m_{2}^{2}\,,
  &\qquad
  \rho^{\rm H,8}_{1} &= (\ell_{1} + k_{1})^{2}+m^{2}_{1}\,,
  &\qquad
  \rho^{\rm H,9}_{2} &= (\ell_{1} - k_{3})^{2}+m^{2}_{2}\,.
\end{aligned}\label{}
\end{equation}
The integral family of $\overline{\rm H}$ can be derived by the crossing relation.

The $\mathbf{IY}$-class consists of 6 integral families: $\rm IY$, $/\!\!\!\rm{Y}$, $\rm YI$, $\overline{\rm IY}$, $\overline{/\!\!\!\rm{Y}}$ and $\overline{\rm YI}$, which are depicted as follows:
\begin{equation}
\begin{aligned}
\begin{tikzpicture}[scale=0.7]
  	\draw[very thick,dashed] (0,0) -- (3,0) node[anchor=north, pos=1/2]{(IY)};
  	\draw[very thick] (0,1.5) -- (3,1.5);
  	\draw[snake it] (0.5,1.5) -- (0.5,0);
  	\draw[snake it] (1,1.5) -- (1.75,0.6);
  	\draw[snake it] (2.5,1.5) -- (1.75,0.6);
  	\draw[snake it] (1.75,0.6) -- (1.75,0);
\end{tikzpicture}
\quad
\begin{tikzpicture}[scale=0.7]
  	\draw[very thick,dashed] (0,0) -- (3,0) node[anchor=north, pos=1/2]{({\rm \slash \!\!\!\!\! Y})};
  	\draw[very thick] (0,1.5) -- (3,1.5);
  	\draw[snake it] (0.5,0) -- (2,1.5);
  	\draw[snake it] (1,1.5) -- (1.75,0.6);
  	\draw[snake it] (2.5,1.5) -- (1.75,0.6);
  	\draw[snake it] (1.75,0.6) -- (1.75,0);
\end{tikzpicture}
\quad
\begin{tikzpicture}[scale=0.7]
  	\draw[very thick,dashed,xscale=-1] (0,0) -- (3,0) node[anchor=north, pos=1/2]{(YI)};
  	\draw[very thick,xscale=-1] (0,1.5) -- (3,1.5);
  	\draw[snake it,xscale=-1] (0.5,1.5) -- (0.5,0);
  	\draw[snake it,xscale=-1] (1,1.5) -- (1.75,0.6);
  	\draw[snake it,xscale=-1] (2.5,1.5) -- (1.75,0.6);
  	\draw[snake it,xscale=-1] (1.75,0.6) -- (1.75,0);
\end{tikzpicture}
\quad
\begin{tikzpicture}[scale=0.7,xscale=-1]
  	\draw[very thick,dashed] (0,0) -- (3,0) node[anchor=north, pos=1/2]{($\overline{\rm IY}$)};
  	\draw[very thick] (0,1.5) -- (3,1.5);
  	\draw[snake it] (0.25,0) -- (2.75,1.5);
  	\draw[snake it] (0.5,1.5) -- (1.25,0.6);
  	\draw[snake it] (2,1.5) -- (1.25,0.6);
  	\draw[snake it] (1.25,0.6) -- (1.25,0);
\end{tikzpicture}
\quad
\begin{tikzpicture}[scale=0.7]
  	\draw[very thick,dashed,xscale=-1] (0,0) -- (3,0) node[anchor=north, pos=1/2]{($\overline{\rm \slash \!\!\! Y}$)};
  	\draw[very thick,xscale=-1] (0,1.5) -- (3,1.5);
  	\draw[snake it,xscale=-1] (0.5,0) -- (2,1.5);
  	\draw[snake it,xscale=-1] (1,1.5) -- (1.75,0.6);
  	\draw[snake it,xscale=-1] (2.5,1.5) -- (1.75,0.6);
  	\draw[snake it,xscale=-1] (1.75,0.6) -- (1.75,0);
\end{tikzpicture}
\quad
\begin{tikzpicture}[scale=0.7]
  	\draw[very thick,dashed] (0,0) -- (3,0) node[anchor=north, pos=1/2]{($\overline{\rm YI}$)};
  	\draw[very thick] (0,1.5) -- (3,1.5);
  	\draw[snake it] (0.25,0) -- (2.75,1.5);
  	\draw[snake it] (0.5,1.5) -- (1.25,0.6);
  	\draw[snake it] (2,1.5) -- (1.25,0.6);
  	\draw[snake it] (1.25,0.6) -- (1.25,0);
\end{tikzpicture}
\end{aligned}\label{}
\end{equation}
The IY and YI ($\overline{\rm IY}$ and $\overline{\rm YI}$) sectors are related by the horizontal mirror dual. Thus, their integral values are the same, and we shall omit the YI and $\overline{\rm YI}$ sectors. The integral families for the IY and the $\rm \slash \!\!\! Y$ sectors are given by as follows:
\begin{equation}
\begin{aligned}
  F^{\rm IY}_{i_{1},i_{2},\cdots,i_{9}} &= \int_{\ell_{1},\ell_{2}} \frac{\big(\rho^{\rm IY,8}_{2}\big)^{-i_{8}}\big(\rho^{\rm IY,9}_{g}\big)^{-i_{9}}}{\big(\rho^{\rm IY,1}_{g}\big)^{i_{1}} \big(\rho^{\rm IY,2}_{g}\big)^{i_{2}}\big(\rho^{\rm IY,3}_{g}\big)^{i_{3}}\big(\rho^{\rm IY,4}_{g}\big)^{i_{4}} \big(\rho^{\rm IY,5}_{1}\big)^{i_{5}}\big(\rho^{\rm IY,6}_{1}\big)^{i_{6}}\big(\rho^{IY,7}_{2}\big)^{i_{7}}}\,,
\end{aligned}\label{IY_integral_family_full_propa}
\end{equation}
where
\begin{equation}
\begin{aligned}
  \rho^{\rm IY,1}_{g} &= (\ell_{1}+q)^{2}\,,
  &\quad
  \rho^{\rm IY,2}_{g} &= \ell_{1}^{2}\,,
  &\quad
  \rho^{\rm IY,3}_{g} &= \ell_{2}^{2}\,,
  \\
  \rho^{\rm IY,4}_{g} &= (\ell_{1}+\ell_{2}+q)^{2}\,,
  &\quad
  \tilde{\rho}^{\rm IY,5}_{1} &= (\ell_{2}-k_{123})^{2}+m_{1}^{2}\,,
  &\quad
  \tilde{\rho}^{\rm IY,6}_{1} &= (\ell_{1}+k_{1})^{2}+m_{1}^{2}\,,
  \\
  \tilde{\rho}^{\rm IY,7}_{2} &= (\ell_{1} - k_{2})^{2}+m_{2}^{2}\,,
  &\quad
  \tilde{\rho}^{\rm IY,8}_{2} &= (\ell_{2} - k_{3})^{2}+m_{2}^{2}\;,
  &\quad
  \rho^{\rm IY,9}_{g} &= (\ell_{2}+q)^{2}\,.
\end{aligned}\label{}
\end{equation}
and
\begin{equation}
  F^{\rm \slash \!\!\! Y}_{i_{1},i_{2},\cdots,i_{9}} =
  \int_{\ell_{1},\ell_{2}} \frac{\big(\rho^{\rm \slash \!\!\! Y,8}_{2}\big)^{-i_{8}}\big(\rho^{\rm \slash \!\!\! Y,9}_{g}\big)^{-i_{9}}}{\big(\rho^{\rm \slash \!\!\! Y,1}_{g}\big)^{i_{1}} \big(\rho^{\rm \slash \!\!\! Y,2}_{g}\big)^{i_{2}}\big(\rho^{\rm \slash \!\!\! Y,3}_{g}\big)^{i_{3}}\big(\rho^{\rm \slash \!\!\! Y,4}_{g}\big)^{i_{4}}\big(\rho^{\rm \slash \!\!\! Y,5}_{1}\big)^{i_{5}}\big(\tilde{\rho}^{\rm \slash \!\!\! Y,6}_{1}\big)^{i_{6}}\big(\rho^{\rm \slash \!\!\! Y,7}_{2}\big)^{i_{7}}}\,,
\label{slashedIY_integral_family_full_propa}\end{equation}
where
\begin{equation}
\begin{aligned}
  \rho^{\text{/\!\!\!\rm{Y}},1}_{g} &= (\ell_{1}+q)^{2}\,,
  &\quad
  \rho^{\text{/\!\!\!\rm{Y}},2}_{g} &= \ell_{1}^{2}\,,
  &\quad
  \rho^{\text{/\!\!\!\rm{Y}},3}_{g} &= \ell_{2}^{2}\,,
  \\
  \rho^{\text{/\!\!\!\rm{Y}},4}_{g} &= (\ell_{1}+\ell_{2}+q)^{2}\,,
  &\quad
  \rho^{\text{/\!\!\!\rm{Y}},5}_{1} &= (\ell_{2}-k_{123})^{2}+m_{1}^{2}\,,
  &\quad
  \rho^{\text{/\!\!\!\rm{Y}},6}_{1} &= (\ell_{1}+\ell_{2}-k_{123})^{2}+m_{1}^{2}\,,
  \\
  \rho^{\text{/\!\!\!\rm{Y}},7}_{2} &= (\ell_{1}-k_{2})^{2}+m_{2}^{2}\,,
  &\quad
  \rho^{\text{/\!\!\!\rm{Y}},8}_{2} &= (\ell_{1}-k_{3})^{2}+m_{2}^{2}\;,
  &\quad
  \rho^{\text{/\!\!\!\rm{Y}},9}_{g} &= (\ell_{2}+q)^{2}\,.
\end{aligned}\label{}
\end{equation}

Finally, the $\textbf{\flip{IY}}$-class consists of six sectors, $\backslash \!\!\!\!\!\flip{ \rm{Y}}$, $\flip{\rm YI}$, $\overline{\flip{\rm IY}}$, $\overline{\backslash\!\!\!\flip{\rm{Y}}}$ and $\overline{\flip{\rm YI}}$, which are illustrated as follows:
\begin{equation}
  \begin{tikzpicture}[scale=0.7,yscale=-1]
  	\draw[very thick] (0,0) -- (3,0);
  	\draw[very thick,dashed] (0,1.5) -- (3,1.5) node[anchor=north, pos=1/2]{($\flip{IY}$)};
  	\draw[snake it] (0.5,1.5) -- (0.5,0);
  	\draw[snake it] (1,1.5) -- (1.75,0.6);
  	\draw[snake it] (2.5,1.5) -- (1.75,0.6);
  	\draw[snake it] (1.75,0.6) -- (1.75,0);
\end{tikzpicture}
\quad
\begin{tikzpicture}[scale=0.7,yscale=-1]
  	\draw[very thick] (0,0) -- (3,0);
  	\draw[very thick,dashed] (0,1.5) -- (3,1.5) node[anchor=north, pos=1/2]{($\flip{\rm \slash \!\!\!\!\! Y}$)};
  	\draw[snake it] (0.5,0) -- (2,1.5);
  	\draw[snake it] (1,1.5) -- (1.75,0.6);
  	\draw[snake it] (2.5,1.5) -- (1.75,0.6);
  	\draw[snake it] (1.75,0.6) -- (1.75,0);
\end{tikzpicture}
\quad
\begin{tikzpicture}[scale=0.7,yscale=-1,xscale=-1]
  	\draw[very thick] (0,0) -- (3,0);
  	\draw[very thick,dashed] (0,1.5) -- (3,1.5) node[anchor=north, pos=1/2]{$(\flip{YI})$};
  	\draw[snake it] (0.5,1.5) -- (0.5,0);
  	\draw[snake it] (1,1.5) -- (1.75,0.6);
  	\draw[snake it] (2.5,1.5) -- (1.75,0.6);
  	\draw[snake it] (1.75,0.6) -- (1.75,0);
\end{tikzpicture}
\quad
\begin{tikzpicture}[scale=0.7,yscale=-1]
  	\draw[very thick] (0,0) -- (3,0);
  	\draw[very thick,dashed] (0,1.5) -- (3,1.5) node[anchor=north, pos=1/2]{($\overline{\flip{\rm IY}}$)};
  	\draw[snake it] (0.25,0) -- (2.75,1.5);
  	\draw[snake it] (0.5,1.5) -- (1.25,0.6);
  	\draw[snake it] (2,1.5) -- (1.25,0.6);
  	\draw[snake it] (1.25,0.6) -- (1.25,0);
\end{tikzpicture}
\quad
\begin{tikzpicture}[scale=0.7,yscale=-1,xscale=-1]
  	\draw[very thick] (0,0) -- (3,0);
  	\draw[very thick,dashed] (0,1.5) -- (3,1.5) node[anchor=north, pos=1/2]{($\overline{ \backslash
  	\!\!\! \flip{Y}}$)};
  	\draw[snake it] (0.5,0) -- (2,1.5);
  	\draw[snake it] (1,1.5) -- (1.75,0.6);
  	\draw[snake it] (2.5,1.5) -- (1.75,0.6);
  	\draw[snake it] (1.75,0.6) -- (1.75,0);
\end{tikzpicture}
\quad
\begin{tikzpicture}[scale=0.7,yscale=-1,xscale=-1]
  	\draw[very thick] (0,0) -- (3,0);
  	\draw[very thick,dashed] (0,1.5) -- (3,1.5) node[anchor=north, pos=1/2]{($\overline{\flip{\rm YI}}$)};
  	\draw[snake it] (0.25,0) -- (2.75,1.5);
  	\draw[snake it] (0.5,1.5) -- (1.25,0.6);
  	\draw[snake it] (2,1.5) -- (1.25,0.6);
  	\draw[snake it] (1.25,0.6) -- (1.25,0);
\end{tikzpicture}
\label{}\end{equation}
One can easily notice that the $\textbf{\flip{IY}}$-class is the vertical mirror image of the $\mathbf{IY}$-class. This implies that the loop integrals of $\textbf{\flip{IY}}$-class can be derived by the mapping for the vertical mirror dual in \eqref{eq:mirror_duals} without explicit computation from the results of $\mathbf{IY}$-class.

\section{Soft Expansion and IBP reduction}\label{Sec:5}

In the previous section, we introduced the new framework for separating the classical diagrams and their classification. Nonetheless, one must also account for the residual quantum contributions present in the numerators of loop integrands, even for the classical diagrams. These quantum contributions arise from the terms with higher powers of momentum transfer $q^{\mu}$ or loop momenta $\ell_{i}$ and do not contribute to the classical long-range interaction -- these correspond to UV properties of GR \cite{Bern:2019crd, Bern:2019nnu}.

This section briefly reviews the soft expansion and IBP reduction framework established in \cite{Parra-Martinez:2020dzs} in the context of the off-shell recursions. We also apply the framework to the one-loop classical amplitude.

\subsection{Brief review on the soft expansion}

To obtain the classical contributions for a given loop integrand for a classical diagram, one typically employs the soft expansion in the context of the method of regions \cite{Beneke:1997zp}. There are four different regions for the internal graviton loop momenta: hard, soft, potential and radiation. The potential region corresponds to the conservative dynamics. In this region the loop momenta $\ell = (\omega,\vec{\ell})$ scales as
\begin{equation}
  (\omega,\vec{\ell}) \sim (|\vec{q}||\vec{v}|,|\vec{q}|)\,.
\label{}\end{equation}

The scales in the soft region satisfy the hierarchy $m_{1}, m_{2}, \bar{k}_{i} \gg q^{\mu}, \ell_{i}$. We implement a scaling transformation on the variables in $\hbar$ as
\begin{equation}
\begin{aligned}
  \ell_{i} \to \hbar \ell_{i}\,, \qquad q \to \hbar q\,, \qquad u_{i} \to u_{i} \,, \qquad m_{i} \to m_{i}\,.
\end{aligned}\label{s_scaling}
\end{equation}
Upon this scaling, a loop integral $I$ for a classical diagram is expanded in $\hbar$
\begin{equation}
  I = \cdots + \hbar^{-2} I^{(-2)} + \hbar^{-1} I^{(-1)} + \hbar^{0} I^{(0)} + \sum_{n>0} \hbar^{n} I^{(n)}\,.
\label{}\end{equation}
Obviously, the terms with positive powers in $\hbar$ are the quantum contributions, and we should truncate them. The remaining $\hbar^{0}$ and $\hbar^{n<0}$ order terms are the so-called classical and superclassical terms, respectively.

Let us consider in more detail the scaling of the integrands. Each integration measure scales as $\hbar^{4}$ in $4$-dimensions
\begin{equation}
  \int_{\ell_{i}} \to \hbar^{4} \int_{\ell_{i}} \,.
\label{}\end{equation}
Regarding the propagators, the graviton propagators scale homogeneously as $\hbar^{-2}$
\begin{equation}
  \frac{1}{\ell_{i}^{2}} \to \frac{\hbar^{-2}}{\ell_{i}^{2}} \,,
  \qquad
  \frac{1}{(\ell_{i}\pm q)^{2}} \to \frac{\hbar^{-2}}{(\ell_{i}\pm q)^{2}}\,.
\label{}\end{equation}
On the other hand, the matter propagators scale inhomogeneously and can be expanded in $\hbar$ as
\begin{equation}
\begin{aligned}
  \frac{1}{\ell^{2} + 2\ell\cdot(\frac{q}{2} \pm \bar{m}_{I} u_{I} )}
  \to \frac{\pm 1}{2\bar{m}_{I}(u_{I} \cdot \ell) }
  \left[\hbar^{-1} \mp \frac{\ell^2 + \ell\cdot q}{2 \bar{m}_{I} (\ell \cdot u_{I})}
  	+\hbar\left( \frac{\ell^2 + \ell\cdot q}{2 \bar{m}_{I} (\ell \cdot u_{I})}\right)^2 + \cdots \right]\,,
\end{aligned}\label{expansion_matter_propagator}
\end{equation}
and the scaling of the matter propagators start from the $\hbar^{-1}$-order.

The expansion of the matter propagator is conveniently represented by introducing a linear matter propagator, which scales precisely as $\hbar^{-1}$,
\begin{equation}
  \frac{\pm 1}{2 \bar{m}_{I}(u_{I} \cdot \ell) }\,,
\label{linearPropa}\end{equation}
and the remainder in the expansion \eqref{expansion_matter_propagator} is absorbed into the numerator. Let us consider a loop integrand $I$ for a classical diagram with a numerator $\mathcal{N}$
\begin{equation}
    I = \int_{\ell_{1},\cdots, \ell_{l}} \frac{
    \mathcal{N}[\ell_{1},\cdots, \ell_{l};u_{1},u_{2},q^{\mu}]
    }{
    \big(\rho^{1}_{g}\big)^{i_{1}} \cdots \big(\rho^{m}_{g}\big)^{i_{m}}
    \big(\rho^{m+1}_{1}\big)^{i_{m+1}} \cdots \big(\rho_{1}^{n}\big)^{i_{n}}
    \big(\rho_{2}^{i_{n+1}}\big)^{i_{n+1}}\cdots \big(\rho_{2}^{p}\big)^{i_{p}}
    }\,,
\label{}\end{equation}
where $1<m<n<p$ and $\rho^{i}_{g}$, and $\rho^{i}_{1,2}$ are the gravitons and the full matter propagators respectively.

After the soft expansion on the loop integrands, we may recast the integrand using the modified numerator $\tilde{\mathcal{N}}$, which includes all the terms arising from the expansion of the matter propagators \eqref{expansion_matter_propagator}
\begin{equation}
  I = \int_{\ell_{1},\cdots, \ell_{l}} \frac{
  \tilde{\mathcal{N}}[\ell_{1},\cdots, \ell_{l};u_{1},u_{2},q^{\mu}]
  }{
    \big(\rho^{1}_{g}\big)^{i_{1}} \cdots \big(\rho^{m}_{g}\big)^{i_{m}}
    \big(\tilde{\rho}^{m+1}_{1}\big)^{i_{m+1}} \cdots \big(\tilde{\rho}_{1}^{n}\big)^{i_{n}}
    \big(\tilde{\rho}_{2}^{i_{n+1}}\big)^{i_{n+1}}\cdots \big(\tilde{\rho}_{2}^{p}\big)^{i_{p}}
  }\,,
\label{}\end{equation}
where $\tilde{\rho}^{i}_{1}$ and $\tilde{\rho}^{i}_{2}$ are the linearized matter propagators defined in \eqref{linearPropa}. To get a consistent classical contribution, we truncate the numerators $\tilde{\mathcal{N}}$ by imposing the power counting constraint after the $\hbar$ expansion \eqref{s_scaling}. If $\tilde{\mathcal{N}}$ scales as $\tilde{\mathcal{N}} \to \sum_{N} \hbar^{N} \tilde{\mathcal{N}}^{\ord{N}}$, we keep the terms satisfying the following relation to truncate the quantum part:
\begin{equation}
   4l-2n_{g}-n_{1} - n_{2} + N \leq 0 \,,
\label{}\end{equation}
where $n_{g}$ and $n_{1,2}$ are the number of graviton and the scalar fields respectively.
 Accordingly, we have $\tilde{\mathcal{N}}_{\rm classical}$ after the truncation as
\begin{equation}
  \tilde{\mathcal{N}}_{\rm classical} = \sum_{N\leq 2n_{g} +n_{1} +n_{2}-4l} \hbar^{N} \tilde{\mathcal{N}}^{\ord{N}}\,.
\label{}\end{equation}

As the matter propagators are replaced with the linear ones, we have to modify the integral families in \eqref{integral_family} by accommodating the linearized matter propagators, which we shall call the \textit{soft integral family},
\begin{equation}
\begin{aligned}
  j_{i_{1},\cdots,i_{p}} = \int_{\ell_{1},\ell_{2},\cdots, \ell_{l}}
  \frac{1}{
    \big(\rho^{1}_{g}\big)^{i_{1}} \cdots \big(\rho^{m}_{g}\big)^{i_{m}}
    \big(\tilde{\rho}^{m+1}_{1}\big)^{i_{m+1}} \cdots \big(\tilde{\rho}_{1}^{n}\big)^{i_{n}}
    \big(\tilde{\rho}_{2}^{i_{n+1}}\big)^{i_{n+1}}\cdots \big(\tilde{\rho}_{2}^{p}\big)^{i_{p}}
  }\,.
\end{aligned}\label{}
\end{equation}
Nonetheless, the intrinsic relationship between the sectors and the integral families remains intact even after the soft expansion. Accordingly, we will exploit IBP reduction with respect to the soft integral families.

The IBP reduction allows us to represent a loop integrand as a linear combination of a set of basis integrals called master integrals. A complete set of master integrals may generate a system of differential equations by acting on the following differential operator to the master integrals:
\begin{equation}
  \frac{\mathrm{d}}{\mathrm{d}y} = \frac{2 x^{2}}{x^{2}-1} \frac{\mathrm{d}}{\mathrm{d}x}\,.
\label{diff_operator}\end{equation}
Here $x$ and $y$ are defined in \eqref{y_variable} and \eqref{x_variable}, and master integrals are functions of these variables. We can evaluate the loop integrals by solving the differential equations with a proper boundary condition. We may adopt the results in \cite{Parra-Martinez:2020dzs}.

\subsection{One loop example}

Let us consider the one-loop example for elucidating our approach. As discussed in Section \ref{Sec:4.3.3}, the only contribution to the classical amplitude at one loop is from the box and triangle diagrams. We shall collect the integrands associated with the box and triangle diagrams using the power counting rules in \eqref{box_power} and \eqref{triangle_power}. At one loop level, there is only $\Box$ class, which is decomposed into two sectors, the II and $\overline{\rm II}$ sectors defined in \eqref{sectors_one_loop}. The soft integral families for each sector, \eqref{II_integral_family} and \eqref{crossed_II_integral_family}, are given by
\begin{equation}
\begin{aligned}
  j^{\rm II}_{i_{1},i_{2},i_{3},i_{4}} &= \int_{\ell} \frac{1}{\mathcal{D}^{\rm II}_{i_{1},i_{2},i_{3},i_{4}}}\,,
  \qquad
  \mathcal{D}^{\rm II}_{i_{1},i_{2},i_{3},i_{4}} = \big({\rho}^{\rm II,1}_{g}\big)^{i_{1}} \big({\rho}^{\rm II,2}_{g}\big)^{i_{2}} \big(\tilde{\rho}^{\rm II,3}_{1}\big)^{i_{3}} \big(\tilde{\rho}^{\rm II,4}_{2}\big)^{i_{4}}\,,
  \\
  j^{\overline{\rm II}}_{i_{1},i_{2},i_{3},i_{4}} &= \int_{\ell} \frac{1}{\mathcal{D}^{\overline{\rm II}}_{i_{1},i_{2},i_{3},i_{4}}}\,,
  \qquad
  \mathcal{D}^{\overline{\rm II}}_{i_{1},i_{2},i_{3},i_{4}} = \big({\rho}^{\overline{\rm II},1}_{g}\big)^{i_{1}} \big({\rho}^{\overline{\rm II},2}_{g}\big)^{i_{2}} \big(\tilde{\rho}^{\overline{\rm II},3}_{1}\big)^{i_{3}} \big(\tilde{\rho}^{\overline{\rm II},4}_{2}\big)^{i_{4}}\,,
\end{aligned}\label{integral_family_one_loop}
\end{equation}
where $\rho^{{\rm II}}_{g}$ and ${\rho}^{\overline{\rm II}}_{g}$ are the graviton propagators in each sector, and $\tilde{\rho}^{{\rm II}}_{1}$ and $\tilde{\rho}^{{\rm II}}_{2}$ ($\tilde{\rho}^{\overline{\rm II}}_{1}$ and $\tilde{\rho}^{\overline{\rm II}}_{2}$) are the linear matter propagators in the II sector ($\overline{\rm II}$ sector) for the scalar field 1 and 2 respectively
\begin{equation}
\begin{aligned}
  \tilde{\rho}^{\rm II,3}_{1} &= 2 \ell \cdot u_{1}\,,
  &\qquad
  \tilde{\rho}^{\rm II,4}_{2} &= -2 \ell \cdot u_{2}\,,
  \\
  \tilde{\rho}^{\overline{\rm II},3}_{1} &= 2 \ell \cdot u_{1}\,,
  &\qquad
  \tilde{\rho}^{\overline{\rm II},4}_{2} &= 2 \ell \cdot u_{2}\,.
\end{aligned}\label{}
\end{equation}
One can check that the integral families \eqref{integral_family_one_loop} are related to each other by the crossing relation \eqref{eq:crossed_diagram}.

\subsubsection{II sector}
We first consider the II sector. From the IBP reduction of the soft integral family, there are three master integrals $f^{\rm II}_{1}$, $f^{\rm II}_{2}$ and $f^{\rm II}_{3}$ defined as
\begin{equation}
\begin{aligned}
  f^{\rm II}_{1}=\epsilon q^{2} j_{0,0,2,1}\,,
  \qquad
  f^{\rm II}_{2}=-\epsilon |q| j_{1,0,1,1}\,,
  \qquad
  f^{\rm II}_{3}=\epsilon^{2} \sqrt{y^{2}-1} q^{2} j_{1,1,1,1}\,,
\end{aligned}\label{}
\end{equation}
where $\epsilon$ is the dimensional regularization parameter, $d=4-2\epsilon$. Acting on the differential operator \eqref{diff_operator}, we derive a differential equation
\begin{equation}
  \frac{\mathrm{d} \vec{f}_{\rm II}}{\mathrm{~d} x}=\epsilon \frac{A}{x} \vec{f}_{\rm II}\,, \qquad \vec{f}_{\rm II} = \big\{ f^{\rm II}_{1} \,, f^{\rm II}_{2} \,, f^{\rm II}_{3}\big\}\,,
\label{diff_eq_II}\end{equation}
where $A$ is a $3\times3$ constant matrix
\begin{equation}
  A= \begin{pmatrix} 0 & 0 & 0 \\ 0 & 0 & 0 \\ 1 & 0 & 0 \end{pmatrix}\,.
\label{}\end{equation}

In \cite{Parra-Martinez:2020dzs,Bern:2019crd,Beenakker:1988jr}, the solution of the differential equation \eqref{diff_eq_II} is constructed in the potential region with a boundary condition at the static limit $y\to 1$
\begin{equation}
\begin{aligned}
	f_{1}^{\rm II} &=0 \,,
	\\
	f_{2}^{\rm II} &=\epsilon^{2}\left(q^{2}\right)^{-\epsilon} e^{\gamma_{\mathrm{E}} \epsilon} \frac{\sqrt{\pi} \Gamma\left(\frac{1}{2}-\epsilon\right)^{2} \Gamma\left(\epsilon+\frac{1}{2}\right)}{2 \Gamma(1-2 \epsilon)} \,,
	\\
	f_{3}^{\rm II} &=\epsilon^{2}\left(q^{2}\right)^{-\epsilon} e^{\gamma_{\mathrm{E}} \epsilon} \frac{\mathrm{i} \pi}{2} \frac{\Gamma(-\epsilon)^{2} \Gamma(1+\epsilon)}{\Gamma(-2 \epsilon)} \,.
\end{aligned}
\label{sol_diff_eq_II}\end{equation}
Since $f^{\rm II}_{1}$ does not contribute to the result at the potential region, we ignore it from now on.

From the solution of the one-loop recursions, we can derive the explicit form of the one-loop soft integral for the II sector, which is decomposed into three parts:
\begin{equation}
\begin{aligned}
  \mathcal{M}^{\rm II} &=
  	\int_{\ell} \frac{\tilde{\mathcal{N}}^{\rm II}_{1,1,1,1}}{\mathcal{D}^{\rm II}_{1,1,1,1}}
  + \int_{\ell} \frac{\tilde{\mathcal{N}}^{\rm II}_{1,1,0,1}}{\mathcal{D}^{\rm II}_{1,1,0,1}}
  + \int_{\ell} \frac{\tilde{\mathcal{N}}^{\rm II}_{1,1,1,0}}{\mathcal{D}^{\rm II}_{1,1,1,0}}\,,
\end{aligned}\label{II_integrand}
\end{equation}
where $\mathcal{D}^{\rm II}_{i_{1},i_{2},i_{3},i_{4}}$ is the propagators defined in \eqref{integral_family_one_loop}. The first term corresponds to the II diagram, and the second and third correspond to the $\triangle$ and \flip{$\triangle$} diagrams. The numerators $\tilde{\mathcal{N}}^{\rm II}$ are given by
\begin{equation}
\begin{aligned}
  \tilde{\mathcal{N}}^{\rm II}_{1,1,1,1} &=
  \frac{4 m_1^4 m_2^4 \left(2 \sigma^2-1\right)^2}{\bar{m}_{1} \bar{m}_{2}}
  \\&\quad
  +2 m_1^2 m_2^2 \left(2 \sigma ^2-1\right) \Big(\frac{m_1}{\ell\cdot u_2} -\frac{m_2}{\ell\cdot u_1} \Big) \left[L +8 \sigma (\ell\cdot u_1) (\ell\cdot u_2) \right]
  \\&\quad
  + m_1 m_2 \bigg[
    \frac{\big(m_1 L-4 m_2 \sigma (\ell\cdot u_2)^2\big)^2}{\left(\ell\cdot u_2\right){}^2}
  + \frac{m_2 L \left(8 m_2 \sigma  \left(\ell\cdot u_2\right){}^2 - m_1 L\right)}{(\ell\cdot u_1) (\ell\cdot u_2)}
  \\&\qquad\qquad\quad
  + \frac{m_2^2 L^{2}}{\left(\ell\cdot u_1\right)^2}
  + \frac{8 m_1 \sigma (\ell\cdot u_1) \left(m_1 L-4 m_2 \sigma (\ell\cdot u_2)^2\right)}{\ell\cdot u_2}
  \\&\qquad\qquad\quad
  +16 m_1^2 \sigma ^2 \left(\ell\cdot u_1\right){}^2+4 m_1 m_2 q^2 \sigma \left(1-2 \sigma ^2\right)\bigg]\,,
  \\
  \tilde{\mathcal{N}}^{\rm II}_{1,1,0,1} &=
	\frac{2 m_2^2 m_1^4}{q^2 \bar{m}_1} \bigg[ \sigma^2 (2\ell+q)^{2}+q^2 \left(5 \sigma ^2-1\right)\bigg]
	\\&\quad
	- \frac{ m_{1}^{2}m_2^2(\ell\cdot u_1)}{q^{2}} \bigg[ 16 \left(\ell\cdot u_2\right){}^2
	+ (2\ell+q)^{2} + q^2 \left(5-4 \sigma ^2\right) \bigg]
	\\&\quad
	-\frac{m_1^2 m_2^2 }{(\ell\cdot u_1)q^2} \Big(\ell\cdot (\ell+ q)\Big)
	  \bigg[ \sigma^2 (2\ell + q)^{2} + q^2 \left(5 \sigma ^2-1\right) +2 \left(\ell\cdot u_2\right)^2\bigg]
	\\&\quad
	+\frac{4 m_1^{2} m_2^2}{q^2} \bigg[
	  4 m_{1} \sigma^2\left(\ell\cdot u_1\right)^2
	+ m_{1}\left(\ell\cdot u_2\right)^2
	- \sigma \Big(2 \ell\cdot(\ell+q)+5 q^2\Big) (\ell\cdot u_2)
	\bigg] \,,
  \\
  \tilde{\mathcal{N}}^{\rm II}_{1,1,1,0} &= \mathcal{N}^{\rm II}_{1,1,0,1}\Big|_{u_{1}\leftrightarrow u_{2}, m_{1}\leftrightarrow m_{2}, q\to -q}\,.
\end{aligned}\label{}
\end{equation}
where $L = \left(2 \sigma^{2} -1\right) (\ell^{2}+\ell\cdot q)$.

Applying IBP reduction for tensor reduction, we can represent $\mathcal{M}^{\rm II}$ in terms of the master integrals
\begin{equation}
  \mathcal{M}^{\rm II} = \sum_{i=2}^{3} c^{\rm II}_{i} f^{\rm II}_{i} \big|_{\rm non-analytic}\,.
\label{lin_comb_amp_II}\end{equation}
We should keep only non-analytic terms in $|q|$ or $\epsilon$ since the analytic terms are irrelevant for the classical long-range interaction \cite{Parra-Martinez:2020dzs}.
The coefficients $c^{\rm II}_{i}$, which are independent on the loop momentum $\ell$, are
\begin{equation}
\begin{aligned}
  c^{\rm II}_{2} &= \frac{4 m_1^{2} m_2^{2} \left(m_1{+}m_2\right)}{ |q|} \Bigg[
  \frac{\left(2 \sigma^2 - 1\right)^{2}}{\sigma -1} -\frac{2\left(2 \sigma^2-1\right) }{\epsilon}
  + \frac{(15 \sigma^2+3)+4 \left(4 \sigma ^2-1\right) \epsilon}{8\epsilon(\epsilon -1)}
  \Bigg]\,,
  \\
  c^{\rm II}_{3} &= \frac{4 m_1^3 m_2^3 \left(1-2 \sigma ^2\right)^2}{\sqrt{\sigma ^2-1} \epsilon ^2 q^{2}}
  -\frac{m_1 m_2 \left(2 \sigma ^2-1\right)^2 \left(2 m_2 m_1 \sigma +m_1^2+m_2^2\right)}{\left(\sigma ^2-1\right)^{3/2} \epsilon} \,.
\end{aligned}\label{}
\end{equation}
Substituting the coefficients and the master integrals into \eqref{lin_comb_amp_II}, we have
\begin{equation}
\begin{aligned}
  \mathcal{M}^{\rm II} &=
  \frac{4 \pi^2 m_{1}^2 m_{2}^2 \sigma  \left(2 \sigma ^2-1\right) (m_{1}+m_{2})}{ (q^{2})^{\epsilon +\frac{1}{2}}}
  - \frac{4i \pi m_{1}^3 m_{2}^3 (2 \sigma^2-1)^2}{\epsilon \sqrt{\sigma ^2-1} (q^{2})^{\epsilon +1}}
  \\&\quad
  + \frac{3 \pi ^2 m_{1}^2 m_{2}^2 (m_{1}+m_{2})\left(5 \sigma ^2-1\right)}{4(q^{2})^{\frac{1}{2}+\epsilon}}\,,
\end{aligned}\label{amp_II}
\end{equation}
where the first line is from the box and the crossed box diagrams, and the second line is the contribution from the triangle diagrams.

\subsubsection{$\overline{\rm II}$ sector}

We next consider the $\overline{\rm II}$ sector. According to the crossing relation \eqref{eq:crossed_diagram}, the definition of the master integrals is replaced by
\begin{equation}
  f^{\overline{\rm II}}_{1}=\epsilon q^{2} j^{\overline{\rm II}}_{0,0,2,1}\,,
  \qquad
  f^{\overline{\rm II}}_{2}=-\epsilon |q| j^{\overline{\rm II}}_{1,0,1,1}\,,
  \qquad
  f^{\overline{\rm II}}_{3}= - \epsilon^{2} \sqrt{y^{2}-1} q^{2} j^{\overline{\rm II}}_{1,1,1,1}\,.
\label{}\end{equation}
On the other hand, the differential equation is not altered under the crossing rule. The difference arises from the boundary condition at the static limit, $y \to 1$. Solving the differential equation, we have the master integrals \cite{Parra-Martinez:2020dzs}
\begin{equation}
\begin{aligned}
  f_{1}^{\overline{\rm II}}&=0 \,,
  \\
  f_{2}^{\overline{\rm II}}&= \epsilon^{2}\left(q^{2}\right)^{-\epsilon} e^{\gamma_{\mathrm{E}} \epsilon} \frac{\sqrt{\pi} \Gamma\left(\frac{1}{2}-\epsilon\right)^{2} \Gamma\left(\epsilon+\frac{1}{2}\right)}{2 \Gamma(1-2 \epsilon)} \,,
  \\
  f_{3}^{\overline{\rm II}}&=0 \,.
\end{aligned}
\label{sol_diff_beq_II}\end{equation}
The explicit form of the one-loop integrand for the $\overline{\rm II}$ sector is also derived by solving the one-loop recursions and divided into three parts:
\begin{equation}
\begin{aligned}
  \mathcal{M}^{\overline{\rm II}} &=
  	\int_{\ell} \frac{\mathcal{N}^{\overline{\rm II}}_{1,1,1,1}}{\mathcal{D}^{\overline{\rm II}}_{1,1,1,1}}
  + \int_{\ell} \frac{\mathcal{N}^{\overline{\rm II}}_{1,1,0,1}}{\mathcal{D}^{\overline{\rm II}}_{1,1,0,1}}
  + \int_{\ell} \frac{\mathcal{N}^{\overline{\rm II}}_{1,1,1,0}}{\mathcal{D}^{\overline{\rm II}}_{1,1,1,0}}\,,
\end{aligned}\label{bII_integrand}
\end{equation}
where $\mathcal{D}^{\overline{\rm II}}_{i_{1},i_{2},i_{3},i_{4}}$ is defined in \eqref{integral_family_one_loop} and the numerators are given by
\begin{equation}
\begin{aligned}
  \mathcal{N}^{\overline{\rm II}}_{1,1,1,1} &= \mathcal{N}^{{\rm II}}_{1,1,1,1}\Big|_{\rm crossing}\,,
  \\
  \mathcal{N}^{\overline{\rm II}}_{1,1,0,1} &= \mathcal{N}^{{\rm II}}_{1,1,0,1}\,,
  \\
  \mathcal{N}^{\overline{\rm II}}_{1,1,1,0} &= \mathcal{N}^{{\rm II}}_{1,1,1,0}\,,
\end{aligned}\label{}
\end{equation}

Again, applying IBP reduction to \eqref{bII_integrand} for tensor reduction, $\mathcal{M}^{\overline{\rm II}}$ reduces to
\begin{equation}
  \mathcal{M}^{\overline{\rm II}} = c^{\overline{\rm II}}_{2} f^{\overline{\rm II}}_{2} \Big|_{\rm non-analytc}\,,
\label{lin_comb_amp_bII}\end{equation}
where the coefficients are
\begin{equation}
\begin{aligned}
  c^{\overline{\rm II}}_{2} &= \frac{4 m_1^{2} m_2^{2} \left(m_1{+}m_2\right)}{ |q|}  \Bigg[
  \frac{\left(2 \sigma^2 - 1\right)^{2}}{\sigma -1} +\frac{2\left(2 \sigma^2-1\right) }{\epsilon}
  + \frac{15 \sigma^2+3}{8\epsilon(\epsilon -1)}+\frac{4 \sigma ^2-1}{2(\epsilon -1)}
  \Bigg]\,.
\end{aligned}\label{sol_diff_eq_bII}
\end{equation}
By substituting the result into \eqref{lin_comb_amp_bII}, we have the amplitude from the $\overline{\rm II}$ sector
\begin{equation}
\begin{aligned}
  \mathcal{M}^{\overline{\rm II}} =
  - \frac{4 \pi^2 m_{1}^2 m_{2}^2 \sigma  \left(2 \sigma ^2-1\right) (m_{1}+m_{2})}{(q^{2})^{\epsilon +\frac{1}{2}}}
  + \frac{3 \pi ^2 m_{1}^2 m_{2}^2 (m_{1}+m_{2})\left(5 \sigma ^2-1\right) }{4 (q^{2})^{\epsilon +\frac{1}{2}}}\,.
\end{aligned}\label{amp_bII}
\end{equation}
%

\subsubsection{Total classical one-loop amplitude}

The total classical one-loop amplitude is given by combining the results from the II and $\overline{\rm II}$ sectors,
\begin{equation}
  \mathcal{M}^{(1)}_{\rm pot} = 4 G^{2} \Big(\mathcal{M}^{\rm II} + \mathcal{M}^{\overline{\rm II}}\Big)\,.
\label{Total_1loop_amp}\end{equation}
Here the overall factor arises from the $(-i)$ factor in the LSZ reduction formula \eqref{amplitude_from_current}, 1-loop coupling $\kappa^{2} = 8\pi G$ and the normalization of the loop integral \eqref{Normalization_factor_loop_integral}, $-i\frac{i(8\pi G)^{2}}{(4\pi)^{2}} = 4G^{2}$.

Adding the contributions from the II and $\overline{\rm II}$ sectors, \eqref{amp_II} and \eqref{amp_bII}, and taking $\epsilon\to 0$ limit, we obtain total classical one-loop amplitude in the potential region using \eqref{Total_1loop_amp}
\begin{equation}
\begin{aligned}
  \mathcal{M}^{\ord{1}}_{\rm pot} &=
  - \frac{16i G^{2} \pi  m_{1}^3 m_{2}^3 \left(2 \sigma ^2 -1\right)^2 }{\sqrt{\sigma ^2-1} q^{2}}
  + \frac{6G^{2}\pi^2  m_{1}^2 m_{2}^2(m_{1}+m_{2}) \left(5 \sigma ^2-1\right)}{|q|} \,.
\end{aligned}\label{}
\end{equation}
The first term on the RHS is from the box and crossed box diagrams, and the last is from the triangles. The box diagram contribution is superclssical and does not contribute to the conservative potential -- it vanishes by the iteration. This result is consistent with the known result \cite{Bjerrum-Bohr:2021vuf}.

\section{Two Loop Amplitude} \label{Sec:6}

In this section, we calculate the classical loop amplitude at two loops, combining all the ingredients we developed so far. According to the power-countings of the couplings, we identified the four classes, $\Box\!\Box$, $\mathbf{H}$, $\mathbf{IY}$ and $\textbf{\flip{IY}}$, and the corresponding sectors for each class in Section \ref{Sec:4.4}. We represent each loop integrand as a linear combination of the master integrals using IBP reduction. We employed the LiteRed \cite{Lee:2012cn,Lee:2013mka,Smirnov:2019qkx} based on the Laporta program \cite{Laporta:2000dsw}.

For each sector, we introduce the soft integral family and the master integrals. We also present the differential equations for the soft integral families and solutions at the potential region. The solutions are obtained in \cite{Smirnov:2001cm,Henn:2013woa,Bianchi:2016yiq,Parra-Martinez:2020dzs}, and we convert the result in our convention. Finally, we derive the total classical amplitude by collecting all the amplitude contributions for each sector as follows:
\begin{equation}
  \mathcal{M}^{\ord{2}}_{\rm pot} = \frac{2G^{3}}{\pi}\Big(\mathcal{M}^{\Box\!\Box} + \mathcal{M}^{\mathbf{H}} + \mathcal{M}^{\mathbf{IY}}+ \mathcal{M}^{\mathbf{IY}}\big|_{m_{1}\leftrightarrow m_{2}}\Big)\,.
\label{Sum_all_two_loop_classes}\end{equation}
%

\subsection{Double box ($\Box\!\Box$) class} \label{Sec:6.1}

Six sectors are identified within the $\Box\!\Box$ class: III, IX, XI, and their crossings, as described in \eqref{classes_double_box}. Since the IX and XI sectors are in the horizontal mirror dual, as well as their crossed counterparts $\overline{\rm IX}$ and $\overline{\rm XI}$, we will consider the IX and $\overline{\rm IX}$ sectors only. The overall contribution to the amplitude from the $\Box\!\Box$ class is
\begin{equation}
  \mathcal{M}^{\Box\!\Box} = \mathcal{M}^{\rm III} + 2 \mathcal{M}^{\rm IX} + \mathcal{M}^{\overline{\rm III}} + 2\mathcal{M}^{\overline{\rm IX}}\,.
\label{}\end{equation}
%
\subsubsection{$\rm III$ and $\overline{\rm III}$ sectors}

The integral family associated with the III sector is introduced in \eqref{III_integral_family_full_propa}. The corresponding soft integral family is obtained by replacing the full matter propagators $\rho^{\rm III}_{1}$ and $\rho^{\rm III}_{1}$ with the linearized propagators, $\tilde{\rho}^{\rm III}_{1}$ and $\tilde{\rho}^{\rm III}_{2}$,
\begin{equation}
\begin{aligned}
  j^{\text{III}}_{i_{1},i_{2},\cdots,i_{9}} = \int_{\ell_{1},\ell_{2}} \frac{\big(\rho^{\rm III,8}_{g}\big)^{-i_{8}} \big(\rho^{\rm III,9}_{g}\big)^{-i_{9}}}{\big(\rho^{\rm III,1}_{g}\big)^{i_{1}} \big(\rho^{\rm III,2}_{g}\big)^{i_{2}} \big(\rho^{\rm III,3}_{g}\big)^{i_{3}}
  \big(\tilde{\rho}^{\rm III,4}_{1}\big)^{i_{4}} \big(\tilde{\rho}^{\rm III,5}_{1}\big)^{i_{5}} \big(\tilde{\rho}^{\rm III,6}_{2}\big)^{i_{6}} \big(\tilde{\rho}^{\rm III,7}_{2}\big)^{i_{7}}}\,,
\end{aligned}\label{integral_family_III}
\end{equation}
where the propagators are defined by
\begin{equation}
\begin{aligned}
  \rho^{\rm III,1}_{g} &= \ell_{1}^{2}\,,
  &\quad
  \rho^{\rm III,2}_{g} &= \ell_{2}^{2}\,,
  &\quad
  \rho^{\rm III,3}_{g} &= (\ell_{1}+\ell_{2}+q)^{2}\,,
  \\
  \tilde{\rho}^{\rm III,4}_{1} &= 2 \ell_{1} \cdot u_{1}\,,
  &\quad
  \tilde{\rho}^{\rm III,5}_{1} &= -2 \ell_{2} \cdot u_{1}\,,
  &\quad
  \tilde{\rho}^{\rm III,6}_{2} &= -2\ell_{1} \cdot u_{2}\,,
  \\
  \tilde{\rho}^{\rm III,7}_{2} &= 2\ell_{2}\cdot u_{2}\,,
  &\qquad
  \rho^{\rm III,8}_{g} &= (\ell_{1}+q)^{2},
  &\qquad
  \rho^{\rm III,9}_{g} &= (\ell_{2}+q)^{2}\;.
\end{aligned}\label{}
\end{equation}

After the IBP reduction, we have ten master integrals defined by
\begin{equation}
\begin{aligned}
  f^{\text{III}}_{1}&=-q^2 \epsilon ^2 j^{\text{III}}_{2,2,1,0,0,0,0,0,0}\,,
  \\
  f^{\text{III}}_{2}&=-\sqrt{y^2-1} \epsilon ^4 j^{\text{III}}_{1,1,1,1,0,0,1,0,0}\,,
  \\
  f^{\text{III}}_{3}&=q^2 \sqrt{y^2-1} \epsilon ^3 j^{\text{III}}_{1,1,2,1,0,0,1,0,0}\,,
  \\
  f^{\text{III}}_{4}&=q^2 y \epsilon ^3 j^{\text{III}}_{1,1,2,1,0,0,1,0,0}+q^2 \epsilon ^2 j^{\text{III}}_{1,1,1,2,0,0,2,0,0}\,,
  \\
  f^{\text{III}}_{5}&=q^2 \sqrt{y^2-1} \epsilon ^3 j^{\text{III}}_{1,1,2,0,1,0,1,0,0}\,,
  \\
  f^{\text{III}}_{6}&=-(1-6 \epsilon ) \epsilon ^3 j^{\text{III}}_{1,1,1,1,1,0,0,0,0}\,,
  \\
  f^{\text{III}}_{7}&=-q^2 \left(y^2-1\right) \epsilon ^4 j^{\text{III}}_{1,1,1,1,1,1,1,0,0}\,,
  \\
  f^{\text{III}}_{8}&=-|q| \epsilon ^3 j^{\text{III}}_{1,1,2,0,1,0,0,0,0}\,,
  \\
  f^{\text{III}}_{9}&=|q| \epsilon ^3 j^{\text{III}}_{1,1,1,1,0,0,2,0,0}\,,
  \\
  f^{\text{III}}_{10}&=|q| \sqrt{y^2-1} \epsilon ^4 j^{\text{III}}_{1,1,1,1,1,0,1,0,0}\,.
\end{aligned}\label{master_integral_III}
\end{equation}
Differentiating $f^{\rm III}_{i}$ with respect to $y$ and changing the variable to $x$, one can derive a set of differential equations in the canonical form
\begin{equation}
  \frac{\mathrm{d} \vec{f}_{\rm III}}{\mathrm{d} x} = \epsilon \bigg(\frac{A_{\rm III,0}}{x} + \frac{A_{\rm III,1}}{x-1} + \frac{A_{\rm III,-1}}{x+1}\bigg) \vec{f}_{\rm III}\,,
  \qquad
  \vec{f}_{\rm III} = \big\{f^{\rm III}_{1},f^{\rm III}_{2},\cdots,f^{\rm III}_{10}\}\,,
\label{}\end{equation}
where $A_{\rm III}$ are $10 \times 10$ constant matrices, which are divided into two parts depending on even and odd powers in $|q|$
\begin{equation}
  A_{\mathrm{III}, i}=\left(\begin{array}{cc}A_{\mathrm{III}, i}^{(\mathrm{e})} & 0 \\ 0 & A_{\mathrm{III}, i}^{(\mathrm{o})}\end{array}\right)\,,
\label{}\end{equation}
and
{\small
\begin{equation}
\begin{aligned}
  A_{\mathrm{III}, 0}^{(\mathrm{e})}=\left(\begin{array}{ccccccc}0 & 0 & 0 & 0 & 0 & 0 & 0 \\ -\frac{1}{2} & -6 & 0 & -1 & 0 & 0 & 0 \\ -\frac{3}{2} & 0 & 2 & -2 & 0 & 0 & 0 \\ 0 & 12 & 2 & 0 & 0 & 0 & 0 \\ -\frac{3}{4} & 0 & 0 & 0 & 0 & 0 & 0 \\ 0 & 0 & 0 & 0 & 0 & 0 & 0 \\ 0 & 0 & 1 & 0 & -2 & 0 & 0\end{array}\right)\,,
  \qquad
  A_{\mathrm{III}, \pm 1}^{(\mathrm{e})}= \begin{pmatrix} 0 & 0 & 0 & 0 & 0 & 0 & 0 \\ 0 & 6 & 0 & 0 & 0 & 0 & 0 \\ 0 & 0 & -2 & 0 & 0 & 0 & 0 \\ 0 & 0 & 0 & 0 & 0 & 0 & 0 \\ 0 & 0 & 0 & 0 & 0 & 0 & 0 \\ 0 & 0 & 0 & 0 & 0 & 0 & 0 \\ 0 & 0 & 0 & 0 & 0 & 0 & 0 \end{pmatrix} \,,
\end{aligned}\label{}
\end{equation}
}
and
{\small
\begin{equation}
  A_{\mathrm{III}, 0}^{(\mathrm{o})}=\left(\begin{array}{ccc}0 & 0 & 0 \\ 0 & -2 & 0 \\ 0 & 1 & 0\end{array}\right), \quad A_{\mathrm{II},+1}^{(\mathrm{o})}=\left(\begin{array}{ccc}0 & 0 & 0 \\ 3 & 6 & 0 \\ 0 & 0 & 0\end{array}\right), \quad A_{\mathrm{III},-1}^{(\mathrm{o})}=\left(\begin{array}{ccc}0 & 0 & 0 \\ -3 & -2 & 0 \\ 0 & 0 & 0\end{array}\right)\,.
\label{}\end{equation} }
We present the solution of the differential equation under our convention in \eqref{MasterIII}.

We perform IBP reduction to the loop integrands for the $\rm III$ sector, which are obtained by solving the off-shell recursions after the soft expansion. Consequently, the amplitude $\mathcal{M}^{\rm III}$ is expanded in terms of the master integrals \eqref{master_integral_III}
\begin{equation}
  \mathcal{M}^{\rm III} = \sum_{i=1}^{10} c^{\rm III}_{i} f^{\rm III}_{i}\Big|_{\text{non-analytic terms} }\,.
\label{lin_comb_amplitude_III}\end{equation}
The explicit form of the coefficients $c^{\rm III}_{i}$ and the value of the master integrals $f^{\rm III}_{i}$ are listed in Appendix \ref{App:C.1.1} and \eqref{MasterIII} respectively.

The classical amplitude $\mathcal{M}^{\rm III}$ can be obtained as
\begin{equation}
\begin{aligned}
  \mathcal{M}^{\rm III} &=\frac{2 \pi^2 m_1^4 m_2^4 (2 \sigma^2-1)^3 }{3 q^{2}(\sigma^2-1)}\Bigg[\pi ^2-4 \log^2 x-\frac{6}{ \epsilon ^2}\Bigg]
  \\&\
  +\frac{4 i \pi^3 m_1^3 m_2^3 (m_1{+}m_2)(2 \sigma^2-1)}{|q|\sqrt{\sigma^2-1}} \Bigg[
  	 (4\sigma^3 -2\sigma+1)\bigg(\frac{1}{\epsilon}
  	- 2\log2\bigg)
  	- \frac{ (1-2 \sigma ^2)^2 }{\sigma -1}\Bigg]
  \\&\
  +\frac{\pi ^2 m_1^2 m_2^2}{15  \epsilon } \Bigg[
    10 m_1 m_2 \left( \frac{72 \sigma^7-112 \sigma^5+44 \sigma^3-\sigma}{(\sigma ^2-1)^{2}} -\frac{8\left(6 \sigma ^4-5 \sigma ^2+1\right) \log x}{\sqrt{\sigma^{2}-1}}\right)
  \\&\qquad\qquad\quad
  +\left(m_1^2{+}m_2^2\right) \left( \frac{520 \sigma^8{-}1272 \sigma^6{+}1216 \sigma^4{-}516 \sigma ^2{+}67}{(\sigma ^2-1)^{2}}-\frac{80 \sigma \log x}{\sqrt{\sigma^{2}-1}}\right)\Bigg]\,.
\end{aligned}\label{amplitude_III}
\end{equation}

Next, we consider the $\overline{\rm III}$, the crossed counterpart of the III-sector. The integral family and its master integrals of the $\overline{\rm III}$ sector can be derived by the crossing relation from the III sector in \eqref{integral_family_III} and \eqref{master_integral_III}. The differential equations are also modified by the same crossing relation. The explicit form of the master integrals is presented in \eqref{MasterxIII}.

Using IBP reduction, we may represent the classical amplitude $\mathcal{M}^{\overline{\rm III}}$ as a linear combination of the master integrals $f^{\overline{\rm III}}_{i}$
\begin{equation}
  \mathcal{M}^{\overline{\rm III}} = \sum_{i=1}^{10} c^{\overline{\rm III}}_{i} f^{\overline{\rm III}}_{i}\Big|_{\text{non-analytic terms}} \,.
\label{expansion_amp_bIII}\end{equation}
Similar in \eqref{lin_comb_amplitude_III}, we keep the non-analytic terms in $q$ or $\epsilon$, which are relevant in the classical long-range interaction. The coefficients $c^{\overline{\rm III}}_{i}$ are denoted in Appendix \ref{App:C.1.2}. Note that the coefficients for the III and $\overline{\rm III}$ sectors are in the crossing relation,
\begin{equation}
  c^{\rm III}_{i} = c^{\overline{\rm III}}_{i}\Big|_{\eqref{eq:crossed_diagram}}\,.
\label{}\end{equation}
On the other hand, we can derive the coefficients from the IBP reduction of the loop integrands that are the solution of the off-shell recursions instead of the crossing relation. This relation provides a consistency check.

Substituting the values of the coefficients and the master integrals into \eqref{expansion_amp_bIII}, we have
\begin{equation}
\begin{aligned}
    \mathcal{M}^{\overline{\rm III}} &= \frac{4 \pi ^2 m_1^4 m_2^4 (2 \sigma ^2-1)^3 \log^2x}{3 \left(\sigma ^2-1\right) q^{2}}
  \\&\quad
  +\frac{4 \pi ^2 m_1^2 m_2^2}{15 \epsilon } \Bigg[\left(m_1^2+m_2^2\right) \left(40 \sigma ^4-73 \sigma ^2+13\right)-20 m_1 m_2 \sigma
  \\&\qquad\qquad\qquad\quad
  + \Big(\big(m_1^2+m_2^2\big) \sigma  - m_1 m_2 (6 \sigma^4 -5 \sigma^2 +1 )\Big)\frac{10\log x}{\sqrt{\sigma ^2-1}}\Bigg]\,.
\end{aligned}\label{amplitude_bIII}
\end{equation}
%

\subsubsection{$\rm IX$ and $\overline{\rm IX}$ sectors}

The integral family for the IX sector is defined in \eqref{IX_integral_family_full_propa} in terms of the full matter propagators. The corresponding soft integral family is derived by replacing the full matter propagators $\rho^{\rm IX}_{1}$ and $\rho^{\rm IX}_{2}$ with the linear propagators $\tilde{\rho}^{\rm IX}_{1}$ and $\tilde{\rho}^{\rm IX}_{2}$ respectively
\begin{equation}
\begin{aligned}
  j^{\text{IX}}_{i_{1},i_{2},\cdots,i_{9}} = \int_{\ell_{1},\ell_{2}} \frac{\big(\rho^{\rm IX,8}_{g}\big)^{-i_{8}} \big(\rho^{\rm IX,9}_{g}\big)^{-i_{9}}}{\big(\rho^{\rm IX,1}_{g}\big)^{i_{1}} \big(\rho^{\rm IX,2}_{g}\big)^{i_{2}} \big(\rho^{\rm IX,3}_{g}\big)^{i_{3}}
  \big(\tilde{\rho}^{\rm IX,4}_{1}\big)^{i_{4}} \big(\tilde{\rho}^{\rm IX,5}_{1}\big)^{i_{5}} \big(\tilde{\rho}^{\rm IX,6}_{2}\big)^{i_{6}} \big(\tilde{\rho}^{\rm IX,7}_{2}\big)^{i_{7}}}\,,
\end{aligned}\label{integral_family_III}
\end{equation}
where the set of propagators is given by
\begin{equation}
\begin{aligned}
  \rho^{\rm IX,1}_{g} &= \ell_{1}^{2}\,,
  &\quad
  \rho^{\rm IX,2}_{g} &=\ell_{2}^{2}\,,
  &\quad
  \rho^{\rm IX,3}_{g} &=(\ell_{1}+\ell_{2}+q)^{2},
  \\
  \tilde{\rho}^{\rm IX,4}_{1} &= 2\ell_{1}\cdot u_{1}\,,
  &\quad
  \tilde{\rho}^{\rm IX,5}_{1} &= -2\ell_{2}\cdot u_{1}\,,
  &\quad
  \tilde{\rho}^{\rm IX,6}_{2} &= 2\ell_{2}\cdot u_{2}\,,
  \\
  \tilde{\rho}^{\rm IX,7}_{2} &= 2(\ell_{1}+\ell_{2})\cdot u_{2} \,,
  &\quad
  \rho^{\rm IX,8}_{g} &=(\ell_{1}+q)^{2}\,,
  &\quad
  \rho^{\rm IX,9}_{g}&=(\ell_{2}+q)^{2}\,.
\end{aligned}\label{}
\end{equation}

There are 15 master integrals for the $j^{\rm IX}_{i_{1}\cdots i_{9}}$:
\begin{equation}
\begin{aligned}
  f^{\text{IX}}_{1}&=-q^2 \epsilon^2 j^{\text{IX}}_{2,2,1,0,0,0,0,0,0}\,,
  \\
  f^{\text{IX}}_{2}&=-\sqrt{y^2-1} \epsilon ^4 j^{\text{IX}}_{1,1,1,1,0,0,1,0,0}\,,
  \\
  f^{\text{IX}}_{3}&=q^2 \sqrt{y^2-1} \epsilon ^3 j^{\text{IX}}_{1,2,1,1,0,0,1,0,0}\,,
  \\
  f^{\text{IX}}_{4}&=q^2 y \epsilon ^3 j^{\text{IX}}_{1,2,1,1,0,0,1,0,0}-q^2 \epsilon ^2 j^{\text{IX}}_{1,1,1,2,0,0,2,0,0}\,,
  \\
  f^{\text{IX}}_{5}&=-\sqrt{y^2-1} \epsilon ^4 j^{\text{IX}}_{1,1,1,1,0,1,0,0,0}\,,
  \\
  f^{\text{IX}}_{6}&=q^2 \sqrt{y^2-1} \epsilon ^3 j^{\text{IX}}_{1,1,2,1,0,1,0,0,0}\,,
  \\
  f^{\text{IX}}_{7}&= -q^2 y \epsilon ^3 j^{\text{IX}}_{1,1,2,1,0,1,0,0,0}-q^2 \epsilon ^2 j^{\text{IX}}_{1,1,1,2,0,2,0,0,0}\,,
  \\
  f^{\text{IX}}_{8}&= -(1-6 \epsilon ) \epsilon ^3 j^{\text{IX}}_{1,1,1,1,1,0,0,0,0}\,,
  \\
  f^{\text{IX}}_{9}&= q^2 \sqrt{y^2-1} \epsilon ^3 j^{\text{IX}}_{1,1,2,0,1,1,0,0,0}\,,
  \\
  f^{\text{IX}}_{10}&=-q^2 \left(y^2-1\right) \epsilon ^4 j^{\text{IX}}_{1,1,1,1,1,1,1,0,0}\,,
  \\
  f^{\text{IX}}_{11}&= -|q| \epsilon ^3 j^{\text{IX}}_{1,1,2,0,1,0,0,0,0}\,,
  \\
  f^{\text{IX}}_{12}&= |q| \epsilon ^3 j^{\text{IX}}_{1,1,1,1,0,2,0,0,0}\,,
  \\
  f^{\text{IX}}_{13}&= |q| \epsilon ^3 j^{\text{IX}}_{1,1,1,2,0,0,1,0,0}\,,
  \\
  f^{\text{IX}}_{14}&= |q| \sqrt{y^2-1} \epsilon ^4 j^{\text{IX}}_{1,1,1,1,1,0,1,0,0}\,,
  \\
  f^{\text{IX}}_{15}&= |q| \sqrt{y^2-1} \epsilon ^4 j^{\text{IX}}_{1,1,1,1,1,1,0,0,0}\,.
\end{aligned}\label{}
\end{equation}
Differentiating the master integrals with respect to $y$ and changing the variable to $x$, we have a set of differential equations
\begin{equation}
  \frac{\mathrm{d} \vec{f}_{\rm IX}}{\mathrm{d} x} = \epsilon \bigg(
  	\frac{A_{\rm IX,0}}{x} + \frac{A_{\rm IX,1}}{x-1} + \frac{A_{\rm IX,-1}}{x+1} \bigg) \vec{f}_{\rm IX}\,,
\label{}\end{equation}
where the $A_{\rm IX}$ are $15\times 15$ constant matrices decomposed as
\begin{equation}
  A_{\mathrm{IX}, i}=\left(\begin{array}{cc}A_{\mathrm{IX}, i}^{(\mathrm{e})} & 0 \\ 0 & A_{\mathrm{IX}, i}^{(\mathrm{o})}\end{array}\right)\,,
\label{}\end{equation}
and
{\scriptsize \begin{equation}
\begin{aligned}
  A_{\mathrm{IX}, 0}^{(\mathrm{e})}=\left(\begin{array}{cccccccccc}0 & 0 & 0 & 0 & 0 & 0 & 0 & 0 & 0 & 0 \\ \frac{1}{2} & -6 & 0 & -1 & 0 & 0 & 0 & 0 & 0 & 0 \\ \frac{3}{2} & 0 & 2 & -2 & 0 & 0 & 0 & 0 & 0 & 0 \\ 0 & 12 & 2 & 0 & 0 & 0 & 0 & 0 & 0 & 0 \\ -\frac{1}{2} & 0 & 0 & 0 & -6 & 0 & 1 & 0 & 0 & 0 \\ -\frac{3}{2} & 0 & 0 & 0 & 0 & 2 & 2 & 0 & 0 & 0 \\ 0 & 0 & 0 & 0 & -12 & -2 & 0 & 0 & 0 & 0 \\ 0 & 0 & 0 & 0 & 0 & 0 & 0 & 0 & 0 & 0 \\ -\frac{3}{4} & 0 & 0 & 0 & 0 & 0 & 0 & 0 & 0 & 0 \\ 0 & 0 & -\frac{1}{2} & 0 & 0 & -1 & 0 & 0 & 1 & 0\end{array}\right)\,,
  \qquad
  A_{\mathrm{IX}, \pm 1}^{(\mathrm{e})}=\left(\begin{array}{cccccccccc}0 & 0 & 0 & 0 & 0 & 0 & 0 & 0 & 0 & 0 \\ 0 & 6 & 0 & 0 & 0 & 0 & 0 & 0 & 0 & 0 \\ 0 & 0 & -2 & 0 & 0 & 0 & 0 & 0 & 0 & 0 \\ 0 & 0 & 0 & 0 & 0 & 0 & 0 & 0 & 0 & 0 \\ 0 & 0 & 0 & 0 & 6 & 0 & 0 & 0 & 0 & 0 \\ 0 & 0 & 0 & 0 & 0 & -2 & 0 & 0 & 0 & 0 \\ 0 & 0 & 0 & 0 & 0 & 0 & 0 & 0 & 0 & 0 \\ 0 & 0 & 0 & 0 & 0 & 0 & 0 & 0 & 0 & 0 \\ 0 & 0 & 0 & 0 & 0 & 0 & 0 & 0 & 0 & 0 \\ 0 & 0 & 0 & 0 & 0 & 0 & 0 & 0 & 0 & 0\end{array}\right)
\end{aligned}\label{}
\end{equation}
\begin{equation}
  A_{\mathrm{IX}, 0}^{(\mathrm{o})}=\left(\begin{array}{ccccc}0 & 0 & 0 & 0 & 0 \\ 0 & -2 & 0 & 0 & 0 \\ 0 & 0 & -2 & 0 & 0 \\ 0 & -1 & 1 & 0 & 0 \\ 0 & 1 & 0 & 0 & 0\end{array}\right)\,,
  \quad
  A_{\mathrm{IX},+1}^{(\mathrm{o})}=\left(\begin{array}{ccccc}0 & 0 & 0 & 0 & 0 \\ 3 & 6 & 0 & 0 & 0 \\ -3 & 0 & -2 & 0 & 0 \\ 0 & 0 & 0 & 0 & 0 \\ 0 & 0 & 0 & 0 & 0\end{array}\right)\,,
  \quad
  A_{\mathrm{IX},-1}^{(\mathrm{o})}=\left(\begin{array}{ccccc}0 & 0 & 0 & 0 & 0 \\ -3 & -2 & 0 & 0 & 0 \\ 3 & 0 & 6 & 0 & 0 \\ 0 & 0 & 0 & 0 & 0 \\ 0 & 0 & 0 & 0 & 0 \\ 0 & 0 & 0 & 0 & 0\end{array}\right)\,.
\label{}\end{equation}
}
The solution of the differential equation is presented in \eqref{MasterIX}.

After the tensor reduction for the integrands of the IX sector, $\mathcal{M}^{\rm IX}$ is represented by a linear combination of the master integrals
\begin{equation}
  \mathcal{M}^{\rm IX} = \sum_{i=1}^{15} c^{\rm IX}_{i} f^{\rm IX}_{i}\Big|_{\text{non-analytic}}\,,
\label{expansion_amp_IX}\end{equation}
and the explicit form of the coefficients is presented in Appendix \ref{App:C.1.3}. Substituting all the results into \eqref{expansion_amp_IX}, we have
\begin{equation}
\begin{aligned}
  \mathcal{M}^{\rm IX} &= \frac{10 \pi ^2 m_1^4 m_2^4 \left(2 \sigma ^2-1\right)^3 \log^2 x}{3 q^{2}\left(\sigma ^2-1\right)}
  \\&\quad
  +\frac{2 i \pi^3 m_1^3 m_2^3 \left(m_1+m_2\right)(2 \sigma^2-1)^2}{|q| \sqrt{\sigma ^2-1}} \Bigg[
  	-\frac{2 \sigma }{\epsilon} +4 \sigma \log x+\frac{ 2\sigma^{2} -1}{\sigma +1}
  \Bigg]
  \\&\quad
  +\frac{2 \pi ^2 m_1^2 m_2^2}{3 \epsilon } \Bigg[ 6 m_1 m_2 \sigma ^2 \left(\frac{ 3\sigma-4 \sigma ^3}{\sigma ^2-1} + \frac{5 \left(2 \sigma ^4-3 \sigma ^2+1\right) \log x}{\left(\sigma ^2-1\right)^{\frac{3}{2}}}\right)
  \\&\qquad\qquad\qquad\quad
  -(m_1^2+m_1^2) \left(\frac{62 \sigma^6-91 \sigma ^4+34 \sigma ^2-2}{\sigma ^2-1} -\frac{\sigma \log x}{\sqrt{\sigma ^2-1}} \right)\Bigg]\,.
\end{aligned}\label{amplitude_IX}
\end{equation}

Let us consider the $\overline{\rm IX}$ sector. As the $\overline{\rm III}$ sector, the master integrals and the differential equations can be derived by the crossing relation \eqref{eq:crossed_diagram}. The classical amplitude from the $\overline{\rm IX}$ sector $\mathcal{M}^{\overline{\rm IX}}$ is expanded as
\begin{equation}
  \mathcal{M}^{\overline{\rm IX}} = \sum_{i=1}^{15} c^{\overline{\rm IX}}_{i} f^{\overline{\rm IX}}_{i}\Big|_{\rm non-analytic} \,,
\label{expansion_amp_bIX}\end{equation}
and the explicit form of the coefficients and values of the master integrals are denoted in Appendix \ref{App:C.1.4}  and \eqref{MasterxIX}. Again, the coefficients in the IX and $\overline{\rm IX}$ sectors are related by the crossing relation \eqref{eq:crossed_diagram}
\begin{equation}
  c^{\rm IX}_{i} = c^{\overline{\rm IX}}_{i}\Big|_{\rm crossing}\,.
\label{}\end{equation}
As before, this relation provides a consistency check for the solutions of the off-shell recursions, and we showed that our result satisfies the relation.

Substituting the results into \eqref{expansion_amp_bIX}, we have
\begin{equation}
\begin{aligned}
  \mathcal{M}^{\overline{{\rm IX}}} &=
  -\frac{8 \pi ^2 m_1^4 m_2^4 \left(2 \sigma ^2-1\right)^3 \log^2 x}{3 \left(\sigma ^2-1\right) q^{2}}
  \\&\quad
  +\frac{2 \pi ^2 m_1^2 m_2^2 }{3 \epsilon }\Bigg[\big(m_1^2 +m_1^2\big) \left(28 \sigma ^4-16 \sigma ^2+1\right) -6 m_2 m_1 \sigma
  \\&\qquad\qquad\qquad\quad
  +\left(\big(m_1^2 +m_2^2\big) -6 m_2 m_1 \left(8 \sigma ^4-6 \sigma ^2+1\right) \sigma \right) \frac{\log x}{\sqrt{\sigma ^2-1}}\Bigg]\,.
\end{aligned}\label{amplitude_bIX}
\end{equation}
%

\subsubsection{Total amplitude from the $\Box\!\Box$ class}

Adding up the results for each sector, \eqref{amplitude_III}, \eqref{amplitude_bIII}, \eqref{amplitude_IX} and \eqref{amplitude_bIX}, we have the total two-loop amplitude
\begin{equation}
\begin{aligned}
  \mathcal{M}^{\Box\!\Box} &=
  \frac{2 \pi ^4 m_1^4 m_2^4 \left(2 \sigma^2-1\right)^3}{3 q^{2}\left(\sigma ^2-1\right)}
  	\Bigg[1-\frac{6}{\epsilon^{2}}\Bigg]
  \\&\quad
  +\frac{4 i \pi ^3 m_1^3 m_2^3 \left(m_1+m_2\right) \left(2 \sigma ^2-1\right)}{\sqrt{\sigma ^2-1} |q|} \Bigg[\frac{1}{\epsilon} -2 \log2 -\frac{ 2(2\sigma^2-1)^{2}}{\sigma ^2-1}\Bigg]
  \\&\quad
  +\frac{\pi ^2 m_1^3 m_2^3}{ \epsilon } \Bigg[\frac{2 \sigma  \left(8 \sigma ^2 \left(\sigma ^4-2 \sigma ^2+2\right)-7\right) }{\left(\sigma ^2-1\right)^2}-16 \left(2 \sigma ^2-1\right) \sqrt{\sigma^{2}-1}\log x
  \\&\qquad\qquad\qquad\quad
  + \frac{(m_1^2 + m_2^2) \left(-88 \sigma ^6+244 \sigma ^4-184 \sigma ^2+33\right)}{5m_{1}m_{2}\left(\sigma ^2-1\right)^{2} }\Bigg]\,.
\end{aligned}\label{total_double_box_class}
\end{equation}
%

\subsection{$\mathbf{H}$ class} \label{Sec:6.2}
The $\mathbf{H}$ class consists of the two sectors: H and $\overline{\rm H}$. The total amplitude contribution from the $\mathbf{H}$ class is given by
\begin{equation}
  \mathcal{M}^{\mathbf{H}} = \mathcal{M}^{\rm H}+ \mathcal{M}^{\overline{\rm H}}\,.
\label{}\end{equation}
For the H-sector, the integral family is defined in \eqref{H_integral_family_full_propa}, using the full matter propagators $\rho^{\rm H}_{1,2}$. The corresponding soft integral family that employs the linearized matter propagators $\tilde{\rho}^{\rm H}_{1,2}$ is given by
\begin{equation}
\begin{aligned}
  j^{\text{\rm H}}_{i_{1},i_{2},\cdots,i_{9}} = \int_{\ell_{1},\ell_{2}} \frac{\big(\tilde{\rho}^{\rm H,8}_{1}\big)^{-i_{8}} \big(\tilde{\rho}^{\rm H,9}_{2}\big)^{-i_{9}}}{\big(\tilde{\rho}^{\rm H,1}_{g}\big)^{i_{1}} \big(\tilde{\rho}^{\rm H,2}_{g}\big)^{i_{2}} \big(\tilde{\rho}^{\rm H,3}_{g}\big)^{i_{3}}
  \big(\tilde{\rho}^{\rm H,4}_{g}\big)^{i_{4}} \big(\tilde{\rho}^{\rm H,5}_{g}\big)^{i_{5}} \big(\tilde{\rho}^{\rm H,6}_{1}\big)^{i_{6}} \big(\tilde{\rho}^{\rm H,7}_{2}\big)^{i_{7}}}\,,
\end{aligned}\label{integral_family_H}
\end{equation}
where the propagators are defined by
\begin{equation}
\begin{aligned}
  \rho^{\rm H,1}_{g} &= (\ell_{1}+q)^{2}\,,
  &\qquad
  \rho^{\rm H,2}_{g} &= (\ell_{2}+q)^{2}\,,
  &\qquad
  \rho^{\rm H,3}_{g} &= \ell_{1}^{2}\,,
  \\
  \rho^{\rm H,4}_{g} &= \ell_{2}^{2}\,,
  &\qquad
  \rho^{\rm H,5}_{g} &= (\ell_{1}+\ell_{2}+q)^{2}\,,
  &\qquad
  \tilde{\rho}^{\rm H,6}_{1} &= 2 \ell_{1} \cdot u_{1}\,,
  \\
  \tilde{\rho}^{\rm H,7}_{2} &= 2\ell_{2}\cdot u_{2}\,,
  &\qquad
  \tilde{\rho}^{\rm H,8}_{1} &= -2 \ell_{2} \cdot u_{1}\,,
  &\qquad
  \tilde{\rho}^{\rm H,9}_{2} &= -2\ell_{1} \cdot u_{2}\,.
\end{aligned}\label{}
\end{equation}
Here $\tilde{\rho}^{\rm H,8}_{1}$ and $\tilde{\rho}^{\rm H,9}_{2}$ are the auxiliary propagators, which are not included in the propagator set of the H-diagram.

Taking the IBP reduction, we identify 16 master integrals for $j^{\text{\rm H}}_{i_{1},i_{2},\cdots,i_{9}}$. We may divide the master integrals into the even and odd powers in $|q|$ terms. Although the H-sector amplitude at the potential region requires the even terms only, we present all the master integrals for completeness. Note that, as we will see later, the odd $|q|$ terms appear in the \textbf{IY} class, even though these are irrelevant to the total amplitude. The master integrals corresponding to the even $|q|$ terms are as follows:
\begin{equation}
\begin{aligned}
  f^{\text{H}}_{1}&=-q^2 \epsilon ^2 j^{\text{H}}_{2,2,0,0,1,0,0,0,0}\,,
  \\
  f^{\text{H}}_{2}&=-(1-4 \epsilon ) \epsilon ^2 j^{\text{H}}_{0,1,0,1,1,2,0,0,0}\,,
  \\
  f^{\text{H}}_{3}&=q^4 \epsilon ^2 j^{\text{H}}_{2,1,1,2,0,0,0,0,0}\,,
  \\
  f^{\text{H}}_{4}&=q^2 \epsilon ^4 j^{\text{H}}_{1,1,1,1,0,1,1,0,0}\,,
  \\
  f^{\text{H}}_{5}&=-\sqrt{y^2-1} \epsilon ^4 j^{\text{H}}_{1,1,0,0,1,1,1,0,0}\,,
  \\
  f^{\text{H}}_{6}&=q^2 \sqrt{y^2-1} \epsilon ^3 j^{\text{H}}_{1,1,0,0,2,1,1,0,0}\,,
  \\
  f^{\text{H}}_{7}&=q^2 y \epsilon ^3 j^{\text{H}}_{1,1,0,0,2,1,1,0,0}+q^2 \epsilon ^2 j^{\text{H}}_{1,1,0,0,1,2,2,0,0}\,,
  \\
  f^{\text{H}}_{8}&=-\frac{\epsilon ^2 (4 \epsilon -1) }{\sqrt{y^2-1}}\left(y j^{\text{H}}_{1,0,1,0,1,0,2,0,0}+(2 \epsilon -1) j^{\text{H}}_{1,0,1,0,1,1,1,0,0}\right)\,,
  \\
  f^{\text{H}}_{9}&=-q^4 \sqrt{y^2-1} \epsilon ^4 j^{\text{H}}_{1,1,1,1,1,1,1,0,0}\,,
  \\
  f^{\text{H}}_{10}&=2 q^2 y \epsilon ^4 j^{\text{H}}_{1,1,1,1,0,1,1,0,0}+q^2 \epsilon ^4 j^{\text{H}}_{1,1,1,1,1,1,1,-1,-1}
  \\
  &\quad
  -\frac{(3 \epsilon -2) (3 \epsilon -1) \epsilon  j^{\text{H}}_{0,0,1,1,1,0,0,0,0}}{q^2}+\frac{1}{2} (2 \epsilon -1) \epsilon ^2 j^{\text{H}}_{1,1,1,1,0,0,0,0,0}\,,
\end{aligned}\label{master_integral_H}
\end{equation}
and the odd $|q|$ terms are
\begin{equation}
\begin{aligned}
  f^{\rm H}_{11} &= \frac{\epsilon^2 (3\epsilon -1)(2\epsilon-1) j^{\rm H}_{1,1,0,0,1,1,0,0,0}}{|q|}\,,
  \\
  f^{\rm H}_{12} &= \epsilon^3 |q| j^{\rm H}_{1,1,0,0,1,1,2,0,0}\,,
  \\
  f^{\rm H}_{13} &= \epsilon^3 (2 \epsilon -1) |q| j^{\rm H}_{1,1,1,1,0,1,0,0,0}\,,
  \\
  f^{\rm H}_{14} &= \frac{(1-2 \epsilon )^2 \epsilon ^2 j^{\rm H}_{,0,1,0,1,1,1,0,0,0}}{|q|}\,,
  \\
  f^{\rm H}_{15} &= \epsilon^3 |q| j^{\rm H}_{1,0,1,0,1,1,2,0,0}\,,
  \\
  f^{\rm H}_{16} &= \frac{16 \epsilon ^4  |q| j^{\rm H}_{1,0,1,0,1,1,2,0,0} }{(2 \epsilon +1)(y+1)}
  	-\frac{4 (y+1) \epsilon^4 |q| j^{\rm H}_{1,1,0,0,1,1,2,0,0}}{2 \epsilon +1}
  	\\&\quad
  	+(y-1) \epsilon ^3 |q|^{5/2} j^{\rm H}_{1,1,1,1,1,1,2,0,0}
    +\frac{16 y \epsilon ^3 (1-2 \epsilon)^2 j^{\rm H}_{0,1,0,1,1,1,0,0,0}}{(y+1) (2 \epsilon +1) |q|}
    \\&\quad
    +\frac{12 \epsilon^3 (3\epsilon-1)(2\epsilon-1) j^{\rm H}_{1,1,0,0,1,1,0,0,0}}{(2 \epsilon +1) |q|}\,.
\end{aligned}\label{}
\end{equation}
Acting on the differential operator \eqref{diff_operator}, we can obtain a set of differential equations
\begin{equation}
  \frac{\mathrm{d} \vec{f}_{\rm H}}{\mathrm{d} x} = \epsilon \bigg(
  	\frac{A_{\rm H,0}}{x} + \frac{A_{\rm H,1}}{x-1} + \frac{A_{\rm H,-1}}{x+1} \bigg) \vec{f}_{\rm H}\,,
\label{}\end{equation}
where the constant matrices $A_{\mathrm{H}, i}$ are decomposed into the even and the odd terms
\begin{equation}
  A_{\mathrm{H}, i}=\left(\begin{array}{cc}A_{\mathrm{H}, i}^{(\mathrm{e})} & 0 \\ 0 & A_{\mathrm{H}, i}^{(\mathrm{o})}\end{array}\right)\,,
\label{}\end{equation}
and the explicit expressions of the even terms are given by
{\footnotesize
\begin{equation}
  A_{\mathrm{H}, 0}^{(\mathrm{e})}=\left(\begin{array}{cccccccccc}0 & 0 & 0 & 0 & 0 & 0 & 0 & 0 & 0 & 0 \\ 0 & 0 & 0 & 0 & 0 & 0 & 0 & 0 & 0 & 0 \\ 0 & 0 & 0 & 0 & 0 & 0 & 0 & 0 & 0 & 0 \\ 0 & 0 & 0 & 0 & 0 & 0 & 0 & 0 & 0 & 0 \\ -\frac{1}{2} & 0 & 0 & 0 & -6 & 0 & -1 & 0 & 0 & 0 \\ -\frac{3}{2} & 0 & 0 & 0 & 0 & 2 & -2 & 0 & 0 & 0 \\ 0 & 0 & 0 & 0 & 12 & 2 & 0 & 0 & 0 & 0 \\ 0 & 2 & 0 & 0 & 0 & 0 & 0 & 2 & 0 & 0 \\ 2 & -4 & 0 & 0 & 0 & 4 & 2 & 4 & 2 & -2 \\ -1 & 0 & -1 & 0 & 12 & 8 & 0 & 8 & 2 & -2\end{array}\right)\,,
  \qquad
  A_{\mathrm{H}, \pm 1}^{(\mathrm{e})} =\left(\begin{array}{cccccccccc}0 & 0 & 0 & 0 & 0 & 0 & 0 & 0 & 0 & 0 \\ 0 & 0 & 0 & 0 & 0 & 0 & 0 & 0 & 0 & 0 \\ 0 & 0 & 0 & 0 & 0 & 0 & 0 & 0 & 0 & 0 \\ 0 & 0 & 0 & 0 & 0 & 0 & 0 & 0 & 0 & 0 \\ 0 & 0 & 0 & 0 & 6 & 0 & 0 & 0 & 0 & 0 \\ 0 & 0 & 0 & 0 & 0 & -2 & 0 & 0 & 0 & 0 \\ 0 & 0 & 0 & 0 & 0 & 0 & 0 & 0 & 0 & 0 \\ 0 & 0 & 0 & 0 & 0 & 0 & 0 & -2 & 0 & 0 \\ 0 & 0 & 0 & 0 & 0 & -4 & 0 & -4 & -2 & 0 \\ 1 & 0 & 1 & \pm 4 & 0 & 0 & 0 & 0 & 0 & 2\end{array}\right)\,.
\label{}\end{equation}
}
We omit the odd terms since these do not contribute to the amplitude from the $\mathbf{H}$ class.

As before, we can represent $\mathcal{M}^{\rm H}$ in terms of the master integrals through the IBP reduction to the H sector
\begin{equation}
  \mathcal{M}^{\rm H} = \sum_{i=1}^{10} c^{\rm H}_{i} f^{\rm H}_{i}\,.
\label{IBP_coeff_H}\end{equation}
The explicit form of the coefficients $c^{\rm H}_{i}$ and the solutions of the differential equation are listed in Appendix \ref{App:C.2.1} and \eqref{MasterH}. If we substitute the values of the master integrals, we have the amplitude
\begin{equation}
\begin{aligned}
  \mathcal{M}^{\rm H} &=  -\frac{8 \pi ^2 m_{1}^3 m_{2}^3}{3 \epsilon} \bigg(8 \sigma +  \frac{ 12 \sigma ^4-8 \sigma ^2+9}{\sqrt{\sigma ^2-1}}  \log x \bigg) \,.
\end{aligned}\label{amp_H}
\end{equation}

Similarly, we derive the master integral and its master integrals for the $\overline{\rm H}$ sector using the crossing relation \eqref{integral_family_H} and \eqref{master_integral_H}. The amplitude from the $\overline{\rm H}$ sector $\mathcal{M}^{\overline{\rm H}}$ is expanded as
\begin{equation}
  \mathcal{M}^{\overline{\rm H}} = \sum_{i=1}^{10} c^{\overline{\rm H}}_{i} f^{\overline{\rm H}}_{i} \,,
\label{IBP_coeff_bH}\end{equation}
and the explicit form of the coefficients and values of the master integrals are denoted in Appendix \ref{App:C.2.2} and \eqref{MasterxH}. Combining all the results, we have
\begin{equation}
\begin{aligned}
    \mathcal{M}^{\overline{\rm H}} &=
    -\frac{4 \pi ^2 m_{1}^3 m_{2}^3}{3 \epsilon} \bigg(8 \sigma +  \frac{ 12 \sigma ^4-8 \sigma ^2+9}{\sqrt{\sigma ^2-1}}  \log x \bigg) \,.
\end{aligned}\label{amp_bH}
\end{equation}

Finally, we derive the amplitudes from the total $\mathbf{H}$ class by adding the results of the H and $\overline{\rm H}$ sectors in \eqref{amp_H} and \eqref{amp_bH} respectively
\begin{equation}
\begin{aligned}
  \mathcal{M}^{\mathbf{H}} &= -\frac{4 \pi ^2 m_{1}^3 m_{2}^3}{\epsilon} \bigg(8 \sigma +  \frac{ 12 \sigma ^4-8 \sigma ^2+9}{\sqrt{\sigma ^2-1}}  \log x \bigg)\,.
\end{aligned}\label{total_H_class}
\end{equation}
%

\subsection{ $\mathbf{IY}$ and inverted $\mathbf{IY}$ classes} \label{Sec:6.3}

Finally, let us consider the \textbf{IY} and \textbf{\flip{IY}} classes. The $\mathbf{IY}$ class consists of the 6 sectors, $\rm IY$, $\rm YI$, $\rm /\!\!\!Y$, $\overline{\rm IY}$, $\overline{\rm YI}$, and $\overline{\rm /\!\!\!Y}$. Within this class, one can find three pairs under the horizontal mirror duals, which are connected through the dual mapping in \eqref{eq:mirror_duals},
\begin{equation}
  {\rm IY} \iff {\rm YI} \,,
  \qquad
  \overline{\rm IY} \iff \overline{\rm YI} \,,
  \qquad
  {\rm /\!\!\!Y} \iff \overline{\rm /\!\!\!Y} \,.
\label{}\end{equation}
Given these relations, it suffices to consider the IY, $\overline{\rm IY}$ and $\rm /\!\!\!Y$ sectors alone. Furthermore, the \textbf{\flip{IY}} class exhibits vertical mirror duality to the \textbf{IY} class. This allows us to derive the corresponding results through the simple mapping in \eqref{eq:mirror_duals}, without additional computations. As a result, our ensuing discussion will focus on IBP reduction and the amplitudes for the IY, $\overline{\rm IY}$ and $\rm /\!\!\!Y$ sectors. The amplitudes from the $\mathbf{IY}$ and $\textbf{\flip{IY}}$ classes are
\begin{equation}
\begin{aligned}
  \mathcal{M}^{\textbf{IY}} &=
  	2\mathcal{M}^{\rm IY} +2\mathcal{M}^{\rm IY} +2\mathcal{M}^{\text{/\!\!\!\rm{Y}}}\,,
  	\\
  \mathcal{M}^{\textbf{\flip{IY}}} &= \mathcal{M}^{\textbf{IY}} \big|_{m_{1}\leftrightarrow m_{2}}
\end{aligned}\label{sum_IY_class}
\end{equation}
One remarkable property of the master integrals in \textbf{IY} class is that these can be represented solely by master integrals in the other classes. This means that solving the differential equations arising in the $\mathbf{IY}$ and $\textbf{\flip{IY}}$ classes is unnecessary.

\subsubsection{IY and $\overline{\rm IY}$ sectors}
The soft integral family for the IY sector is derived by replacing the full matter propagators $\rho^{\rm IY}_{1}$ and $\rho^{\rm IY}_{2}$ in \eqref{IY_integral_family_full_propa} with their linearized counterparts $\tilde{\rho}^{\rm IY}_{1}$ and $\tilde{\rho}^{\rm IY}_{2}$ respectively
\begin{equation}
\begin{aligned}
j^{\text{\rm IY}}_{i_{1},i_{2},\cdots,i_{9}} = \int_{\ell_{1},\ell_{2}} \frac{\big(\tilde{\rho}^{\rm IY,8}_{2}\big)^{-i_{8}} \big(\rho^{\rm IY,9}_{g}\big)^{-i_{9}}}{\big(\rho^{\rm IY,1}_{g}\big)^{i_{1}} \big(\rho^{\rm IY,2}_{g}\big)^{i_{2}} \big(\rho^{\rm IY,3}_{g}\big)^{i_{3}}
\big(\rho^{\rm IY,4}_{g}\big)^{i_{4}} \big(\tilde{\rho}^{\rm IY,5}_{1}\big)^{i_{5}} \big(\tilde{\rho}^{\rm IY,6}_{1}\big)^{i_{6}} \big(\tilde{\rho}^{\rm IY,7}_{2}\big)^{i_{7}}}\,,
\end{aligned}\label{integral_family_IY}
\end{equation}
where the set of the propagators are defined by
\begin{equation}
\begin{aligned}
\rho^{\rm IY,1}_{g} &= (\ell_{1}+q)^{2}\,,
&\quad
\rho^{\rm IY,2}_{g} &= \ell_{1}^{2}\,,
&\quad
\rho^{\rm IY,3}_{g} &= \ell_{2}^{2}\,,
\\
\rho^{\rm IY,4}_{g} &= (\ell_{1}+\ell_{2}+q)^{2}\,,
&\quad
\tilde{\rho}^{\rm IY,5}_{1} &= 2 \ell_{2} \cdot u_{1}\,,
&\quad
\tilde{\rho}^{\rm IY,6}_{1} &= -2\ell_{1}\cdot u_{1}\,,
\\
\tilde{\rho}^{\rm IY,7}_{2} &= 2 \ell_{1} \cdot u_{2}\,,
&\quad
\tilde{\rho}^{\rm IY,8}_{2} &= -2\ell_{2} \cdot u_{2}\;,
&\quad
\rho^{\rm IY,9}_{g} &= (\ell_{2}+q)^{2}\,.
\end{aligned}\label{}
\end{equation}
Here $\tilde{\rho}^{\rm IY,8}_{2}$ and $\rho^{\rm IY,9}_{g}$ are the auxiliary propagators.

Interestingly, the soft integral family for the IY sector can be represented by the integral families for the III and H sectors,
\begin{equation}
\begin{aligned}
j^{\text{IY}}_{0,i_{2},i_{3},i_{4},i_{5},i_{6},i_{7},0,0}&=j^{\text{III}}_{i_{3},i_{2},i_{4},i_{5},i_{6},0,i_{7},0,0}\,,
\\
j^{\text{IY}}_{i_{1},i_{2},0,i_{4},i_{5},0,0,0,0}&=j^{\text{H}}_{0,i_{1},0,i_{2},i_{4},i_{5},0,0,0}\,,
\\
j^{\text{IY}}_{i_{1},i_{2},0,i_{4},i_{5},0,i_{7},0,0}&=j^{\text{H}}_{0,i_{1},0,i_{2},i_{4},i_{5},i_{7},0,0}\,.
\end{aligned}\label{}
\end{equation}
This implies that we do not have to define the master integrals for the IY sector; we exploit the master integrals defined in the III and H sectors. There are a total of 13 master integrals in this sector:
\begin{equation}
\begin{aligned}
  \text{III}: \quad
  f^{\rm IY}_{1} &= f^{\rm III}_{1}\,,
  &\quad
  f^{\rm IY}_{2} &= f^{\rm III}_{2}\,,
  &\quad
  f^{\rm IY}_{3} &= f^{\rm III}_{3}\,,
  &\quad
  f^{\rm IY}_{4} &= f^{\rm III}_{4}\,,
  \\
  f^{\rm IY}_{5} &= f^{\rm III}_{5}\,,
  &\quad
  f^{\rm IY}_{6} &= f^{\rm III}_{6}\,,
  &\quad
  f^{\rm IY}_{7} &= f^{\rm III}_{8}\,,
  &\quad
  f^{\rm IY}_{8} &= f^{\rm III}_{9}\,,
  \\
  f^{\rm IY}_{9} &= f^{\rm III}_{10}\,,
  \\
  \text{H even}:\quad f^{\rm IY}_{10} &= f^{\rm H}_{2}\,,
  &\quad
  f^{\rm IY}_{11} &= f^{\rm H}_{8}\,,
  \\
  \text{H odd}:\quad f^{\rm IY}_{12} &= f^{\rm H}_{14}\,,
  &\quad
  f^{\rm IY}_{13} &= f^{\rm H}_{15}\,.
\end{aligned}\label{master_integral_IY}
\end{equation}
Consequently, we do not have to solve the differential equations to determine these master integrals -- we can use the previous results.

After the IBP reduction, the amplitude from the IY sector is represented by a linear combination of the master integrals
\begin{equation}
  \mathcal{M}^{\rm IY} = \sum_{i=1}^{13} c^{\rm IY}_{i} f^{\rm IY}_{i}\Big|_{\rm non-analytic} \,.
\label{IBP_coeff_IY}\end{equation}
Again, we keep the non-analytic terms only for the same reason in the other classes. The explicit form of the coefficients $c^{\rm IY}_{i}$ are listed in Appendix \ref{App:C.3.1}. If we substitute the values of the master integrals, we have the amplitude
\begin{equation}
\begin{aligned}
  \mathcal{M}^{\rm IY} &=-\frac{i \pi ^3 m_{1}^4 m_{2}^3 \left(2 \sigma ^2-1\right) }{2 \sqrt{\sigma ^2-1} |q|}\Bigg[\sigma ^2-1+2\left(15 \sigma ^2-7\right) \log 2+\frac{7-15 \sigma ^2}{\epsilon }\Bigg]
\\&\
 +\frac{\pi ^2 m_{1}^3 m_{2}^2 }{45 \left(\sigma ^2-1\right) \epsilon } \Bigg[6 m_{1} \left(120 \sigma ^6-98 \sigma ^4-19 \sigma ^2+12\right)+5 m_{2} \sigma  \left(146 \sigma ^4-157 \sigma ^2+29\right)\Bigg]
\\&\
 -\frac{4 \pi ^2 m_{1}^3 m_{2}^2 \left(2 \sigma ^2-1\right)}{3 \epsilon\sqrt{\sigma ^2-1}} \left(4 m_{1} \sigma +m_{2} \left(2 \sigma ^2+3\right)\right)  \log x
\\&\
 -\frac{4 m_{1}^4 m_{2}^3 \sigma  \left(2 \sigma ^2-1\right) }{\epsilon^2|q|} \bigg[
    \bigg(13-\frac{15}{\epsilon }+\frac{4}{\epsilon ^2}\bigg) \frac{f^{\rm H}_{14}}{(1-2 \epsilon )^2 }
  + \bigg(\frac{6 \sigma ^2-1}{\epsilon}+\frac{1-2 \sigma ^2}{\epsilon^{2} }\bigg)f^{\rm H}_{15}\bigg]\,.
\end{aligned}\label{IY_amplitude}
\end{equation}

We now turn to the crossed one. The integral family and its master integrals for the $\overline{\rm IY}$ sector can be derived by the crossing relation from \eqref{integral_family_IY} and \eqref{master_integral_IY}. The master integrals are given by
\begin{equation}
\begin{aligned}
  \text{III}: \quad
  f^{\overline{\rm IY}}_{1} &= f^{\overline{\rm III}}_{1}\,,
  &\quad
  f^{\overline{\rm IY}}_{2} &= f^{\overline{\rm III}}_{2}\,,
  &\quad
  f^{\overline{\rm IY}}_{3} &= f^{\overline{\rm III}}_{3}\,,
  &\quad
  f^{\overline{\rm IY}}_{4} &= f^{\overline{\rm III}}_{4}\,,
  \\
  f^{\overline{\rm IY}}_{5} &= f^{\overline{\rm III}}_{5}\,,
  &\quad
  f^{\overline{\rm IY}}_{6} &= f^{\overline{\rm III}}_{6}\,,
  &\quad
  f^{\overline{\rm IY}}_{7} &= f^{\overline{\rm III}}_{8}\,,
  &\quad
  f^{\overline{\rm IY}}_{8} &= f^{\overline{\rm III}}_{9}\,,
  \\
  f^{\overline{\rm IY}}_{9} &= f^{\overline{\rm III}}_{10}\,,
  \\
  \text{H even}:\quad f^{\overline{\rm IY}}_{10} &= f^{\overline{\rm H}}_{2}\,,
  &\quad
  f^{\overline{\rm IY}}_{11} &= f^{\overline{\rm H}}_{8}\,,
  \\
  \text{H odd}:\quad f^{\overline{\rm IY}}_{12} &= f^{\overline{\rm H}}_{14}\,,
  &\quad
  f^{\overline{\rm IY}}_{13} &= f^{\overline{\rm H}}_{15}\,.
\end{aligned}\label{master_integral_IY}
\end{equation}
Note that $f^{\rm H}_{2} = f^{\overline{\rm H}}_{2}$ and $f^{\rm H}_{14} = f^{\overline{\rm H}}_{14}$ because these do not have the $u_{2}$ dependence. Applying the IBP reduction, the amplitude $\mathcal{M}^{\overline{\rm IY}}$ is represented by a linear combination of the above master integrals
\begin{equation}
  \mathcal{M}^{\overline{\rm IY}} = \sum_{i=1}^{13} c^{\overline{\rm IY}}_{i} f^{\overline{\rm IY}}_{i}\Big|_{\rm non-analytic}\,,
\label{IBP_coeff_bIY}\end{equation}
and the explicit form of the coefficients is denoted in Appendix \ref{App:C.3.2}. Combining all the results, we have
\begin{equation}
\begin{aligned}
    \mathcal{M}^{\overline{\rm IY}} &=-\frac{4\pi^2 m_{1}^3 m_{2}^2}{45}  \Bigg[\frac{3 m_{1} \left(30 \sigma ^4-41 \sigma ^2+6\right)+5 m_{2} \sigma  \left(23 \sigma ^2-14\right)}{\epsilon } \Bigg]
\\&\quad
 -\frac{2 \pi ^2 m_{1}^3 m_{2}^2 \left(2 \sigma ^2-1\right)}{3 \epsilon\sqrt{\sigma ^2-1}}  \left(m_{2} \left(2 \sigma ^2+3\right)-4 m_{1} \sigma \right) \log x
\\&\quad
 +\frac{4 m_{1}^4 m_{2}^3 \sigma \left(2 \sigma ^2-1\right) }{(1-2 \epsilon )^2 \epsilon ^2|q|}\left(13-\frac{15}{\epsilon }+\frac{4}{\epsilon ^2}\right) f^{\overline{\text{H}}}_{14}
\\&\quad
  -\frac{4 m_{1}^4 m_{2}^3 \left(2 \sigma ^2-1\right)}{\epsilon^{3}|q|} \left(
	6\sigma ^2-1+\frac{1-2 \sigma ^2}{\epsilon }\right) f^{\overline{\text{\rm H}}}_{15}\,.
\end{aligned}\label{bIY_amplitude}
\end{equation}
%

\subsubsection{$/\!\!\!\rm{Y}$ sector}
The soft integral family associated with the $/\!\!\!\rm{Y}$-sector is defined from \eqref{slashedIY_integral_family_full_propa} by replacing the full matter propagators $\rho_{1,2}$ with the linearized propagators $\tilde{\rho}_{1,2}$
\begin{equation}
\begin{aligned}
j^{\text{/\!\!\!\rm{Y}}}_{i_{1},i_{2},\cdots,i_{9}} = \int_{\ell_{1},\ell_{2}} \frac{\big(\tilde{\rho}^{\text{/\!\!\!\rm{Y}},8}_{2}\big)^{-i_{8}} \big(\rho^{\text{/\!\!\!\rm{Y}},9}_{g}\big)^{-i_{9}}}{\big(\rho^{\text{/\!\!\!\rm{Y}},1}_{g}\big)^{i_{1}} \big(\rho^{\text{/\!\!\!\rm{Y}},2}_{g}\big)^{i_{2}} \big(\rho^{\text{/\!\!\!\rm{Y}},3}_{g}\big)^{i_{3}}
\big(\rho^{\text{/\!\!\!\rm{Y}},4}_{g}\big)^{i_{4}} \big(\tilde{\rho}^{\text{/\!\!\!\rm{Y}},5}_{1}\big)^{i_{5}} \big(\tilde{\rho}^{\text{/\!\!\!\rm{Y}},6}_{1}\big)^{i_{6}} \big(\tilde{\rho}^{\text{/\!\!\!\rm{Y}},7}_{2}\big)^{i_{7}}}\,,
\end{aligned}\label{integral_family_/Y}
\end{equation}
where the propagators are defined by
\begin{equation}
\begin{aligned}
  \rho^{\text{/\!\!\!\rm{Y}},1}_{g} &= (\ell_{1}+q)^{2}\,,
  &\quad
  \rho^{\text{/\!\!\!\rm{Y}},2}_{g} &= \ell_{1}^{2}\,,
  &\quad
  \rho^{\text{/\!\!\!\rm{Y}},3}_{g} &= \ell_{2}^{2}\,,
  \\
  \tilde{\rho}^{\text{/\!\!\!\rm{Y}},5}_{1} &= 2 \ell_{2} \cdot u_{1}\,,
  &\quad
  \rho^{\text{/\!\!\!\rm{Y}},4}_{g} &= (\ell_{1}+\ell_{2}+q)^{2}\,,
  &\quad
  \tilde{\rho}^{\text{/\!\!\!\rm{Y}},6}_{1} &= 2(\ell_{1}+\ell_{2})\cdot u_{1}\,,
  \\
  \tilde{\rho}^{\text{/\!\!\!\rm{Y}},7}_{2} &= 2 \ell_{1} \cdot u_{2}\,,
  &\quad
  \tilde{\rho}^{\text{/\!\!\!\rm{Y}},8}_{2} &= -2\ell_{2} \cdot u_{2}\;,
  &\quad
  \rho^{\text{/\!\!\!\rm{Y}},9}_{g} &= (\ell_{2}+q)^{2}\,.
\end{aligned}\label{}
\end{equation}
Here $\tilde{\rho}^{\text{/\!\!\!\rm{Y}},8}_{2}$ and $\rho^{\text{/\!\!\!\rm{Y}},9}_{g}$ are auxiliary propagators for consistent IBP reduction.

As for the IY sector, the soft integral family of the $\text{/\!\!\!\rm{Y}}$ class can be represented by the soft integral families for the IX, H, and $\overline{\text{H}}$ sectors as follows:
\begin{equation}
\begin{aligned}
  j^{\text{/\!\!\!\rm{Y}}}_{0,i_{2},i_{3},i_{4},i_{5},i_{6},i_{7},0,0}&=j^{\text{IX}}_{i_{2},i_{3},i_{4},i_{7},0,i_{5},i_{6},0,0}\,,
  \\
  j^{\text{/\!\!\!\rm{Y}}}_{i_{1},i_{2},0,i_{4},i_{5},0,i_{7},0,0}&=j^{\text{H}}_{i_{1},0,i_{2},0,i_{4},i_{7},i_{5},0,0}\,,
  \\
  j^{\text{/\!\!\!\rm{Y}}}_{i_{1},i_{2},i_{3},0,0,i_{6},i_{7},0,0}&=j^{\overline{\text{H}}}_{i_{1},0,i_{2},0,i_{3},i_{7},i_{6},0,0}\,.
\end{aligned}\label{}
\end{equation}
This implies that we can construct the master integrals for the $\text{/\!\!\!\rm{Y}}$ sector by employing the previous results in the IX and H sectors. There are a total of 18 master integrals in this sector:
\begin{equation}
\begin{aligned}
  \text{IX}: \quad f^{\text{/\!\!\!\rm{Y}}}_{1} &= f^{\rm IX}_{1}\,,
  &\quad
  f^{\text{/\!\!\!\rm{Y}}}_{2} &= f^{\rm IX}_{2}\,,
  &\quad
  f^{\text{/\!\!\!\rm{Y}}}_{3} &= f^{\rm IX}_{3}\,,
  &\quad
  f^{\text{/\!\!\!\rm{Y}}}_{4} &= f^{\rm IX}_{4}\,,
  \\
  f^{\text{/\!\!\!\rm{Y}}}_{5} &= f^{\rm IX}_{5}\,,
  &\quad
  f^{\text{/\!\!\!\rm{Y}}}_{6} &= f^{\rm IX}_{6}\,,
  &\quad
  f^{\text{/\!\!\!\rm{Y}}}_{7} &= f^{\rm IX}_{7}\,,
  &\quad
  f^{\text{/\!\!\!\rm{Y}}}_{8} &= f^{\rm IX}_{8}\,,
  \\
  f^{\text{/\!\!\!\rm{Y}}}_{9} &= f^{\rm IX}_{11}\,,
  &\quad
  f^{\text{/\!\!\!\rm{Y}}}_{10} &= f^{\rm IX}_{12}\,,
  &\quad
  f^{\text{/\!\!\!\rm{Y}}}_{11} &= f^{\rm IX}_{13}\,,
  &\quad
  f^{\text{/\!\!\!\rm{Y}}}_{12} &= f^{\rm IX}_{14}\,,
  \\
  \text{H even}:\quad f^{\text{/\!\!\!\rm{Y}}}_{13} &= f^{\rm H}_{2}\,,
  &\quad
  f^{\text{/\!\!\!\rm{Y}}}_{14} &= f^{\rm H}_{8}\,,
  \\
  \text{$\overline{\rm H}$ even}:\quad f^{\text{/\!\!\!\rm{Y}}}_{15} &= f^{\overline{\rm H}}_{8}\,,
  \\
  \text{H odd}:\quad f^{\text{/\!\!\!\rm{Y}}}_{16} &= f^{\rm H}_{14}\,,
  &\quad
  f^{\text{/\!\!\!\rm{Y}}}_{17} &= f^{\rm H}_{15}\,,
  \\
  \text{$\overline{\rm H}$ odd}:\quad f^{\text{/\!\!\!\rm{Y}}}_{18} &= f^{\overline{\rm H}}_{15}\,.
\end{aligned}\label{}
\end{equation}
After the IBP reduction, $\mathcal{M}^{\text{/\!\!\!\rm{Y}}}$ is represented by the following linear combination
\begin{equation}
  \mathcal{M}^{\text{/\!\!\!\rm{Y}}} = \sum_{i=1}^{18} c^{\text{/\!\!\!\rm{Y}}}_{i} f^{\text{/\!\!\!\rm{Y}}}_{i}\,,
\label{IBP_coeff_Yslash}\end{equation}
and the explicit form of the coefficients $c^{\text{/\!\!\!\rm{Y}}}_{i}$ are listed in Appendix \ref{App:C.3.3}.

If we substitute the values of the master integrals for the IX, H and $\overline{\rm H}$ sectors, we have
\begin{equation}
\begin{aligned}
  \mathcal{M}^{\text{/\!\!\!\rm{Y}}} &=-\frac{2\pi ^2 m_{1}^3 m_{2}^2}{225}  \Bigg[\frac{15 \left(m_{1} \left(30 \sigma ^4-41 \sigma ^2+6\right)-5 m_{2} \sigma  \left(\sigma ^2-1\right)\right)}{\epsilon }
\\&\qquad\qquad\qquad\quad
 +m_{1} \left(600 \sigma ^4-2353 \sigma ^2+353\right)+25 m_{2} \sigma  \left(29-50 \sigma ^2\right)\Bigg]
\\&\
 +\frac{4 \pi ^2 m_{1}^3 m_{2}^2 }{\sqrt{\sigma ^2-1}} \Bigg[\frac{\left(2 \sigma ^2-1\right) }{3 \epsilon } \left(2 m_{1} \sigma +3 m_{2} \left(4 \sigma ^2-1\right)\right) \log (x)
\Bigg]
\\&\
+\frac{4 m_{1}^4 m_{2}^3 \left(2 \sigma ^2-1\right)}{\epsilon^3 |q|} \left(6 \sigma ^2-1+\frac{1-2 \sigma ^2}{\epsilon }\right) \left(f^{\text{\rm H}}_{15}+f^{\overline{\text{\rm H}}}_{15}\right)\,.
\end{aligned}\label{Yslash_amplitude}
\end{equation}
Here we do not substitute the explicit expressions of $f^{\text{\rm H}}_{15}$ and $f^{\overline{\text{\rm H}}}_{15}$, because these do not contribute to the total amplitude.

\subsubsection{Total $\mathbf{IY}$}

The total contribution from the $\mathbf{IY}$ class can be derived by substituting the results in each sector, \eqref{IY_amplitude}, \eqref{bIY_amplitude} and \eqref{Yslash_amplitude}, into \eqref{sum_IY_class}
\begin{equation}
\begin{aligned}
  \mathcal{M}^{\mathbf{IY}} &= -\frac{i \pi ^3 m_{1}^4 m_{2}^3 \left(2 \sigma ^2-1\right) }{\sqrt{\sigma ^2-1} |q|}\Bigg[\sigma^2-1 -\big(15 \sigma ^2-7\big)\bigg(\frac{1}{\epsilon } -2\log2\bigg)\Bigg]
\\&\quad
 +\frac{2\pi ^2 m_{1}^3 m_{2}^2 }{15 \left(\sigma ^2-1\right) \epsilon } \Bigg[6 m_{1} \left(62 \sigma ^4-69 \sigma ^2+12\right)+5 m_{2} \sigma  \left(22 \sigma ^4-11 \sigma ^2-5\right)\Bigg]
\\&\quad
 +\frac{4 \pi ^2 m_{1}^3 m_{2}^3 }{\epsilon (\sigma ^2-1)^{\frac{3}{2}}} \big(12 \sigma^4-16 \sigma ^2+5\big) \log x \,.
\end{aligned}\label{total_IY_class}
\end{equation}
As we have mentioned, $f^{\text{\rm H}}_{14}$, $f^{\overline{\text{\rm H}}}_{15}$, $f^{\text{\rm H}}_{15}$ and $f^{\overline{\text{\rm H}}}_{15}$ are cancled as we expected.

\subsection{Total two-loop amplitude}
The total two-loop amplitude in the potential region is derived by substituting the results for each class, \eqref{total_double_box_class}, \eqref{total_H_class} and \eqref{total_IY_class}, into \eqref{Sum_all_two_loop_classes}
\begin{equation}
\begin{aligned}
  &\mathcal{M}_{\rm pot}^{\ord{2}}(k_{1},k_{2},k_{3},k_{4})
  \\
  &= \frac{4G^{3} \pi m_{1}^4 m_{2}^4 \left(2 \sigma ^2-1\right)^3 \left(\pi ^2 \epsilon ^2-6\right)}{3 \epsilon^{2}\left(\sigma ^2-1\right)  q^2}
  +\frac{6G^{3} i \pi^2 m_{1}^3 m_{2}^3 (5\sigma^{2}-1)(2\sigma^{2}-1) (m_{1}+m_{2})}{\epsilon\sqrt{\sigma^2-1} |q|}
  \\&\quad
  -2G^{3}\log |q|^{2} \Bigg[ m_{1}^{2}m_{2}^{2}\pi\left(m_{1}^2 +m_{2}^2 + 2m_{1} m_{2} \sigma\right) \left[\frac{\left(2 \sigma ^2-1\right)^3}{(\sigma^{2}-1)^{2}} + \frac{3\left(5 \sigma ^2-1\right) \left(2 \sigma ^2-1\right)}{2(\sigma^{2}-1)}\right]
  \\&\qquad\qquad\qquad\quad
  + \frac{\pi m_{1}^{2}m_{2}^{2}}{6}\bigg[\left(3-54 \sigma ^2\right) \left(m_{1}^2+m_{2}^2\right)+m_{1}m_{2} \sigma  \left(4 \sigma ^2+206\right)\bigg]
  \\&\qquad\qquad\qquad\quad
  + \frac{8\pi m_{1}^{3} m_{2}^{3} \left(4 \sigma^4 -12\sigma^{2}-3\right)}{\sqrt{\sigma ^2-1}}\operatorname{arcsinh}\left(\sqrt{\frac{\sigma-1}{2}}\right)\Bigg]\,.
\end{aligned}\label{}
\end{equation}
This result exactly matches with the classical two-loop amplitude at potential region \cite{Bern:2019crd}.

\section{Conclusion}

This article introduces a novel calculation framework for the efficient and systematic computations of the classical scattering amplitude. The focus is on elucidating the conservative dynamics of non-rotating binary black hole systems, especially during the inspiral phase absent of gravitational wave emission. Our framework builds upon quantum off-shell recursion techniques and offers a systematic method for classifying classical diagrams through specific power-counting rules. On top of this classification, we employ IBP reduction on the loop integrands, deriving the $2\to 2$ classical amplitude for massive scalar particles within a potential region.

Initially, we introduce the field theory description that treats black holes as scalar point particles. The action comprises the EH action, along with two massive scalar theories interacting through the gravitational forces only. Adopting Cheung and Remmen's convention of metric perturbations, we developed the perturbative Einstein equation and the scalar equations of motion. Notably, the DS equation plays a crucial role in deriving the quantum off-shell recursions, which serve as a quantum counterpart to the classical equations.

Next, we derived the quantum off-shell recursion for the field theory. This was achieved by substituting the quantum perturbiner expansions for the graviton and scalar fields and their descendant fields into the DS equation up to two-loop order. The perturbiner formalism directly relates the gap between scattering amplitudes and solutions of EoMs or DS equations. We explicitly derived the recursions and solved them up to the two loops.

We then introduced a new power-counting prescription for identifying the classical diagrams. For optimizing the IBP reduction, we provided a systematic classification scheme of the classical diagrams by introducing the classes and sectors directly associated with integral families. For consistent truncation of the quantum contributions in the loop integrands for the classical diagrams, we applied the soft expansion established in \cite{Parra-Martinez:2020dzs}.

We provided a concrete example of our prescription by analyzing the classical one-loop amplitude, demonstrating its consistency with known results. Applying the classification scheme, we showed that one class and two sectors exist in this class. We then performed the soft expansion and the IBP reduction for the one-loop integrands. The classical one-loop integrand can be represented by the linear combinations of the one-loop master integrals. Referring to the differential equations and their solutions corresponding to the master integrals, we derived the one-loop classical amplitude in the potential region consistent with the known results.

Similarly, the classical two-loop diagrams are classified -- the four classes, $\Box\!\Box$, \textbf{H}, \textbf{IY} and \flip{\textbf{IY}}, and their sectors. We identified the mirror dual pairs for each sector and showed that the \flip{\textbf{IY}}-class can be obtained from the results of the \textbf{IY}-class by the simple relations. Again, we performed the soft expansion and the IBP reduction to the two-loop integrands for each sector to represent the soft two-loop integrands in terms of linear combinations of the master integrals. Interestingly, the entire master integrals of the \textbf{IY}-class can be represented by the master integrals of the $\Box\!\Box$ and \textbf{H} classes. Thus, we do not have to derive and solve the differential equations. Referring to the solutions of the differential equations obtained by \cite{Parra-Martinez:2020dzs}, we derived the two-loop classical amplitudes in the potential region. The amplitude exactly agrees with the known results in \cite{Bern:2019crd}.

This paper applies our formalism to the 3PM amplitude for the non-rotating binary black hole system. However, many directions remain open for future research. One of the immediate challenges is to derive the 4PM results for the same setup by three-loop computations \cite{Dlapa:2021npj,Dlapa:2021vgp,Bern:2021yeh,Bern:2021dqo,Bern:2022jvn}. The extension of our framework to include rotating black holes \cite{Porto:2005ac,Vaidya:2014kza,Bini:2017xzy,Cachazo:2017jef,Vines:2017hyw,Guevara:2017csg,Bini:2018ywr,Guevara:2018wpp,Levi:2018nxp,Vines:2018gqi,Chung:2019duq,Damgaard:2019lfh,Guevara:2019fsj,Maybee:2019jus,Aoude:2020onz,Bern:2020buy,Guevara:2020xjx,Levi:2020kvb,Levi:2020uwu,Jakobsen:2022fcj,Aoude:2023dui,Aoude:2023vdk,DiVecchia:2023frv,Heissenberg:2023uvo,Jakobsen:2023ndj,Jakobsen:2023hig} and the finite size correction for describing other celestial systems, such as neutron star binaries or black hole-neutron star systems, by incorporating the higher dimensional operators \cite{Bini:2020flp,Cheung:2020gbf,Aoude:2020ygw,Bern:2020uwk,Kalin:2020lmz,Haddad:2020que,Cheung:2020sdj,Mandal:2023hqa}. Furthermore, it would be interesting to apply the KMOC formulation \cite{Kosower:2018adc,Maybee:2019jus,Herrmann:2021lqe,Herrmann:2021tct,Cristofoli:2021vyo,Cristofoli:2021jas} for describing gravitational wave radiation.

\acknowledgments
We thank Jung-Wook Kim, Seok Kim, Sangmin Lee, and Mao Zeng for their useful comments and discussion. We would like to express our special gratitude to Kimyeong Lee and Jeong-Hyuck Park for suggesting to initiate this work and encouraging us. 
This work is supported by appointment to the JRG Program at the APCTP through the Science and Technology Promotion Fund and Lottery Fund of the Korean Government. KL is also supported by the National Research Foundation of Korea(NRF) grant funded by the Korean government(MSIT) RS-2023-00249451 and the Korean Local Governments of Gyeongsangbuk-do Province and Pohang City. KC is supported by the NRF grant funded by the Korean government (MSIT) 2022R1F1A1068489.

\newpage
\appendix

\section{Equations of Motions and $\mathcal{I}$-tensors} \label{App:A}

The EoM for the field theory is substantially complicated due to the graviton sector. This makes the derivation of the DS equation difficult. As we have discussed in Section \ref{Sec:2.1}, it is efficient to introduce the $\mathcal{I}$-tensors. Recall that the perturbed Einstein equation in \eqref{perturbed_Eistein_eq} is decomposed into two parts, $\mathcal{G}^{\mu\nu}$ and $\mathcal{T}^{\mu\nu}$. The general form of the curvature perturbations $\mathcal{G}^{\mu\nu}_{n}$ is given by
\begin{equation}
\begin{aligned}
  \mathcal{G}^{\mu\nu}_{n} &=
  \mathcal{I}^{\mu\nu,M_{1},\cdots, M_{n-1},\kappa N,\lambda P}_{n,1} h^{M_{1}} \cdots h^{M_{n-1}} \partial_{\kappa} h^{N} \partial_{\lambda} h^{P}
  \\&\quad
  +\mathcal{I}^{\mu\nu,M_{1},\cdots, M_{n-1}, N,\kappa \lambda P}_{n,2} h^{M_{1}} \cdots h^{M_{n-1}} h^{N} \partial_{\kappa} \partial_{\lambda} h^{P}\,.
\end{aligned}\label{}
\end{equation}
The general form of the $n$-th order of the energy-momentum perturbations $\mathcal{T}^{\mu\nu}_{n}$ are given by
\begin{equation}
\begin{aligned}
  \mathcal{T}^{\mu\nu}_{n} &= \frac{1}{2} \sum_{I=1}^{2} \left[\mathcal{I}_{\rm EM}^{\mu\nu,N_{1},\cdots N_{n-1},I} h^{N_{1}} \cdots h^{N_{n-1}} \varphi_{I}^{2} \right]\,, \qquad n>1\,.
\end{aligned}\label{}
\end{equation}
The EoM of the scalar fields are defined in \eqref{EoM_phi}. The general $n$-th order term (for $n>1$) of the EoM is also denoted by
\begin{equation}
   \mathcal{I}^{I,M_{1},M_{2},\cdots, M_{n}}_{\phi} h^{M_{1}} h^{M_{2}} \cdots h^{M_{n}} \phi^{I}\,.
\label{}\end{equation}

Here we denote the $\mathcal{I}$-tensors explicitly associated with the classical two-loop amplitudes. The relevant $\mathcal{I}$-tensors are as follows:
\begin{equation}
\begin{aligned}
  &\mathcal{I}_{1,1}^{\mu\nu,\rho_{1}\kappa_{1}\lambda_{1},\rho_{2}\kappa_{2}\lambda_{2}} \,,
  &\quad
  &\mathcal{I}_{1,2}^{\mu\nu,\kappa_{1}\lambda_{1},\rho_{1}\rho_{2}\kappa_{2}\lambda_{2}}\,,
  \\
  &\mathcal{I}^{\mu\nu,\kappa_{1} \lambda_{1},\rho_{1} \kappa_{2} \lambda_{2},\rho_{2} \kappa_{3} \lambda_{3}}_{2,1}\,,
  &\quad
  &\mathcal{I}^{\mu\nu,\kappa_{1} \lambda_{1},\kappa_{2} \lambda_{2},\rho_{1} \rho_{2} \kappa_{3} \lambda_{3}}_{2,2}\,,
  \\
  &\mathcal{I}^{\mu \nu, \kappa_{1} \lambda_{1},I}_{\rm EM}\,,
  &\quad
  &\mathcal{I}^{\mu \nu, \kappa_{1} \lambda_{1},\kappa_{2},\lambda_{2},I}_{\rm EM}\,,
  \\
  &\mathcal{I}^{I,\kappa_{1} \lambda_{1}, \kappa_{2} \lambda_{2}}_{\phi}\,,
  &\quad
  &\mathcal{I}^{I,\kappa_{1} \lambda_{1}, \kappa_{2} \lambda_{2}}_{\phi}\,.
\end{aligned}\label{}
\end{equation}
%

\subsection{Graviton sector}
Let us consider the curvature tensor part, $\mathcal{G}^{\mu\nu}_{1}$ and $\mathcal{G}^{\mu\nu}_{2}$
using the $\mathcal{I}$-tensors. First $\mathcal{G}^{\mu\nu}_{1}$ is defined in \eqref{G_tensor} as
\begin{equation}
  \mathcal{G}_{1}^{\mu\nu} = \mathcal{I}_{1,1}^{\mu\nu,\rho_{1}\kappa_{1}\lambda_{1},\rho_{2}\kappa_{2}\lambda_{2}} \partial_{\rho_{1}}h_{\kappa_{1}\lambda_{1}} \partial_{\rho_{2}}h_{\kappa_{2}\lambda_{2}}
  + \mathcal{I}_{1,2}^{\mu\nu,\kappa_{1}\lambda_{1},\rho_{1}\rho_{2}\kappa_{2}\lambda_{2}} h_{\kappa_{1}\lambda_{1}} \partial_{\rho_{1}} \partial_{\rho_{2}}h_{\kappa_{2}\lambda_{2}}\,.
\label{}\end{equation}
We want to present the explicit form of the $\mathcal{I}^{\mu\nu,\rho_{1}\kappa_{1}\lambda_{1},\rho_{2}\kappa_{2}\lambda_{2}}_{1,1}$ and $\mathcal{I}^{\mu\nu,\kappa_{1}\lambda_{1},\rho_{1}\rho_{2}\kappa_{2}\lambda_{2}}_{1,2}$. For simplicity, we decompose the $\mathcal{I}$-tensors according to the following tensor structure:
\begin{equation}
\begin{aligned}
  \mathcal{I}_{1,1}^{\mu\nu,\rho_{1}\kappa_{1}\lambda_{1},\rho_{2}\kappa_{2}\lambda_{2}} &=
    \bar{\mathcal{I}}_{1,1}^{\mu\nu,\rho_{1}\kappa_{1}\lambda_{1},\rho_{2}\kappa_{2}\lambda_{2}}
  + \bar{\mathcal{I}}_{1,1}^{\mu\nu,\rho_{2}\kappa_{2}\lambda_{2},\rho_{1}\kappa_{1}\lambda_{1}}\,,
  \\
  \mathcal{I}_{1,2}^{\mu\nu,\kappa_{1}\lambda_{1},\rho_{1}\rho_{2}\kappa_{2}\lambda_{2}} &=
    \bar{\mathcal{I}}_{1,2}^{\mu\nu,\kappa_{1}\lambda_{1},\rho_{1}\rho_{2}\kappa_{2}\lambda_{2}}
  + \bar{\mathcal{I}}_{1,2}^{\nu\mu,\kappa_{1}\lambda_{1},\rho_{1}\rho_{2}\kappa_{2}\lambda_{2}} \,,
\end{aligned}\label{I_12_length3}
\end{equation}
where
{\small
\begin{equation}
\begin{aligned}
&{\bar{\mathcal{I}}_{1,1}}^{\mu \nu, \rho_{1} \kappa_1 \lambda_1, \rho_{2} \kappa_2 \lambda_2}
\\&=
\frac{1}{8} \eta^{\mu \nu} \eta^{\rho_{1} \rho_{2}} \eta^{\kappa_1 (\kappa_2} \eta^{\lambda_2) \lambda_1}
+\frac{1}{16} \eta^{\rho_{1} (\mu} \eta^{\nu) \rho_{2}} \eta^{\kappa_1 \lambda_1} \eta^{\kappa_2 \lambda_2}
- \frac{1}{8} \eta^{\mu \nu} \eta^{\rho_{1} (\kappa_1} \eta^{\lambda_1) \rho_{2}} \eta^{\kappa_2 \lambda_2}
\\
&\quad
- \frac{1}{4} \eta^{\mu (\kappa_1} \eta^{\lambda_1) (\kappa_2} \eta^{\lambda_2) \nu} \eta^{\rho_{1} \rho_{2}}
- \frac{1}{4} \eta^{\rho_{2} (\kappa_1} \eta^{\lambda_1) (\mu } \eta^{\nu) (\kappa_2} \eta^{\lambda_2) \rho_{1}}
- \frac{1}{8} \eta^{\rho_{1} (\mu} \eta^{\nu) \rho_{2}} \eta^{\kappa_1 (\kappa_2} \eta^{\lambda_2) \lambda_1}
\\
&\quad
+\frac{1}{4}\eta^{\mu (\kappa_2} \eta^{\lambda_2) \nu} \eta^{\rho_{1} (\kappa_1} \eta^{\lambda_1) \rho_{2}}
+\frac{1}{2} \eta^{\rho_{1} (\mu} \eta^{\nu) ({\kappa_{2}}} \eta^{{\lambda_{2}}) ({\kappa_{1}}} \eta^{{\lambda_{1}}) \rho_{2}} \,,
\end{aligned}\label{}
\end{equation}}
and
{\small
\begin{equation}
\begin{aligned}
  &\bar{\mathcal{I}}_{1,2}^{\mu \nu, \kappa_1 \lambda_1, \rho_{1} \rho_{2} \kappa_2 \lambda_2}
  \\
  &=
  \eta^{\mu (\kappa_1} \eta^{\lambda_1) \nu} \bigg[
  	  \frac{1}{8}\eta^{\rho_{1} \rho_{2}} \eta^{\kappa_2 \lambda_2}
  	+ \frac{1}{4} \eta^{\rho_{1} ({\kappa_{1}}} \eta^{{\lambda_{1}}) \rho_{2}}
  \bigg]
  -\frac{\eta^{\mu \nu}}{8}  \bigg[\eta^{\kappa_2 \lambda_2} \eta^{\kappa_1 (\rho_{1}} \eta^{\rho_{2}) \lambda_1}
  -\eta^{\kappa_1 (\kappa_2} \eta^{\lambda_2) \lambda_2} \eta^{\rho_{1} \rho_{2}}\bigg]
  \\&\quad
  - \frac{1}{2}\eta^{\mu ({\kappa_{1}}} \eta^{{\lambda_{1}}) ({\kappa_{2}}} \eta^{{\lambda_{2}}) \nu} \eta^{\rho_{1} \rho_{2}}
  +\frac{1}{2}\eta^{\mu ({\kappa_{1}}} \eta^{{\lambda_{1}}) ({\lambda_{2}}} \eta^{{\kappa_{2}}) (\rho_{1}} \eta^{\rho_{2}) \nu} \,.
\end{aligned}\label{}
\end{equation}}

Similarly, the $\mathcal{G}^{\mu\nu}_{2}$ is defined in \eqref{G_tensor} as follows:
\begin{equation}
\begin{aligned}
\mathcal{G}^{\mu\nu}_{2} &=
  \mathcal{I}^{\mu\nu,\kappa_{1} \lambda_{1},\rho_{1} \kappa_{2} \lambda_{2},\rho_{2} \kappa_{3} \lambda_{3}}_{2,1} h^{\kappa_{1} \lambda_{1}} \partial_{\rho_{1}} h^{\kappa_{2} \lambda_{2}} \partial_{\rho_{2}} h^{\kappa_{3} \lambda_{3}}
  \\&\quad
  + \mathcal{I}^{\mu\nu,\kappa_{1} \lambda_{1},\kappa_{2} \lambda_{2},\rho_{1} \rho_{2} \kappa_{3} \lambda_{3}}_{2,2} h^{\kappa_{1} \lambda_{1}} h^{\kappa_{2} \lambda_{2}} \partial_{\rho_{1}} \partial_{\rho_{2}} h^{\kappa_{3} \lambda_{3}}\,.
\end{aligned}\label{}
\end{equation}
Again, we decompose the $\mathcal{I}$-tensors as the
\begin{equation}
\begin{aligned}
  \mathcal{I}^{\mu\nu,\kappa_{1} \lambda_{1},\rho_{1} \kappa_{2} \lambda_{2},\rho_{2} \kappa_{3} \lambda_{3}}_{2,1} &=
  	\bar{\mathcal{I}}^{\mu\nu,\kappa_{1} \lambda_{1},\rho_{1} \kappa_{2} \lambda_{2},\rho_{2} \kappa_{3} \lambda_{3}}_{2,1}
  + \bar{\mathcal{I}}^{\mu\nu,\kappa_{1} \lambda_{1},\rho_{2} \kappa_{3} \lambda_{3},\rho_{1} \kappa_{2} \lambda_{2}}_{2,1}
  \\
  \mathcal{I}^{\mu\nu,\kappa_{1} \lambda_{1},\kappa_{2} \lambda_{2},\rho_{1} \rho_{2} \kappa_{3} \lambda_{3}}_{2,2} &=
  	\bar{\mathcal{I}}^{\mu\nu,\kappa_{1} \lambda_{1},\kappa_{2} \lambda_{2},\rho_{1} \rho_{2} \kappa_{3} \lambda_{3}}_{2,2}
  + \bar{\mathcal{I}}^{\nu\mu,\kappa_{1} \lambda_{1},\kappa_{2} \lambda_{2},\rho_{1} \rho_{2} \kappa_{3} \lambda_{3}}_{2,2}\,,
\end{aligned}\label{I_12_length4}
\end{equation}
where
{\small
\begin{equation}
\begin{aligned}
&\bar{\mathcal{I}}_{2,1}^{\mu \nu, \kappa_1 \lambda_1, \rho_{1} \lambda_2 \kappa_2, \rho_{2} \kappa_3 \lambda_3}
\\
&=
  \frac{\eta^{\mu \nu} }{8} \Big[2\eta^{\rho_{1} \rho_{2}} \eta^{{\kappa_{1}} ({\kappa_{2}}} \eta^{{\lambda_{2}}) ({\kappa_{3}}} \eta^{{\lambda_{3}}) {\lambda_{1}}}
	- \eta^{\rho_{1} ({\kappa_{1}}} \eta^{{\lambda_{1}}) \rho_{2}} \eta^{{\kappa_{2}} ({\kappa_{3}}} \eta^{{\lambda_{3}}) {\lambda_{2}}}
    - \eta^{\rho_{1} ({\kappa_{2}}} \eta^{{\lambda_{2}}) \rho_{2}} \eta^{{\kappa_{1}} ({\kappa_{3}}} \eta^{{\lambda_{3}}) {\lambda_{1}}}\Big]
\\&\quad
  - \frac{1}{8} \eta^{\rho_{1} (\mu} \eta^{\nu) \rho_{2}}\Big[ 2\eta^{{\kappa_{1}} ({\kappa_{2}}} \eta^{{\lambda_{2}}) ({\kappa_{3}}} \eta^{{\lambda_{3}}) {\lambda_{1}}}
	-\eta^{{\kappa_{1}} ({\kappa_{2}}} \eta^{{\lambda_{2}}) {\lambda_{1}}} \eta^{{\kappa_{3}} {\lambda_{3}}}\Big]
\\&\quad
+\frac{1}{8}\eta^{\mu ({\kappa_{1}}} \eta^{{\lambda_{1}}) \nu} \Big[
	  \eta^{{\kappa_{2}} ({\kappa_{3}}} \eta^{{\lambda_{3}}) {\lambda_{2}}} \eta^{\rho_{1} \rho_{2}}
	+ \eta^{\rho_{1} ({\kappa_{3}}} \eta^{{\lambda_{3}}) \rho_{2}} \eta^{{\kappa_{2}} {\lambda_{2}}}\Big]
\\&\quad
- \frac{\eta^{\rho_{1} \rho_{2}}}{4} \Big[
	  \eta^{\mu ({\kappa_{1}}} \eta^{{\lambda_{1}}) ({\kappa_{2}}} \eta^{{\lambda_{2}}) ({\kappa_{3}}} \eta^{{\lambda_{3}}) \nu}
	+ \eta^{\mu ({\kappa_{2}}} \eta^{{\lambda_{2}}) ({\kappa_{3}}} \eta^{{\lambda_{3}}) ({\kappa_{1}}} \eta^{{\lambda_{1}}) \nu}
	+ \eta^{\mu ({\kappa_{3}}} \eta^{{\lambda_{3}}) ({\kappa_{1}}} \eta^{{\lambda_{1}}) ({\kappa_{2}}} \eta^{{\lambda_{2}}) \nu}
  \Big]
\\&\quad
+\frac{1}{4} \Big[\eta^{\mu ({\kappa_{1}}} \eta^{{\lambda_{1}}) ({\kappa_{2}}} \eta^{{\lambda_{2}}) \nu}
  + \eta^{\nu ({\kappa_{1}}} \eta^{{\lambda_{1}}) ({\kappa_{2}}} \eta^{{\lambda_{2}}) \mu} \Big] \eta^{\rho_{1} ({\kappa_{3}}} \eta^{{\lambda_{3}}) \rho_{2}}
+\frac{1}{4} \eta^{\mu ({\kappa_{2}}} \eta^{{\lambda_{2}}) ({\kappa_{3}}} \eta^{{\lambda_{3}}) \nu} \eta^{\rho_{1} ({\kappa_{1}}} \eta^{{\lambda_{1}}) \rho_{2}}
\\&\quad
+\frac{1}{2} \eta^{\rho_{1} (\mu} \eta^{\nu) ({\kappa_{3}}} \eta^{{\lambda_{3}}) ({\kappa_{1}}} \eta^{{\lambda_{1}}) ({\kappa_{2}}} \eta^{{\lambda_{2}}) \rho_{2}}
+\frac{1}{2} \eta^{\rho_{1} (\mu} \eta^{\nu) ({\kappa_{1}}} \eta^{{\lambda_{1}}) ({\kappa_{3}}} \eta^{{\lambda_{3}}) ({\kappa_{2}}} \eta^{{\lambda_{2}}) \rho_{2}}
\\ &\quad
- \frac{1}{2} \eta^{\rho_{1} ({\kappa_{3}}} \eta^{{\lambda_{3}}) (\mu} \eta^{\nu) ({\kappa_{1}}} \eta^{{\lambda_{1})} {(\kappa_{2}}} \eta^{{\lambda_{2}}) \rho_{2}}\,,
\end{aligned}\label{}
\end{equation}}
and
{\small
\begin{equation}
\begin{aligned}
&\bar{\mathcal{I}}_{2,2}^{\mu \nu, \kappa_1 \lambda_1, \kappa_2 \lambda_2, \rho_{1} \rho_{2} \kappa_3 \lambda_3}
\\
&= \frac{\eta^{\mu \nu} \eta^{\rho_{1} \rho_{2}} }{16} \Big[
	  \eta^{{\kappa_{1}} ({\kappa_{2}}} \eta^{{\lambda_{2})} {(\kappa_{3}}} \eta^{{\lambda_{3}}) {\lambda_{1}}}
	+ \eta^{\lambda_{1} ({\kappa_{2}}} \eta^{{\lambda_{2})} {(\kappa_{3}}} \eta^{{\lambda_{3}}) {{\kappa_{1}}}} \Big]
	+\frac{1}{8} \eta^{\mu ({\kappa_{1}}} \eta^{{\lambda_{1}}) ({\kappa_{2}}} \eta^{{\lambda_{2}}) \nu} \eta^{{\kappa_{3}} {\lambda_{3}}} \eta^{\rho_{1} \rho_{2}}
  \\&\quad
	- \frac{\eta^{\mu \nu}}{16} \Big[\eta^{\rho_{1} ({\kappa_{2}}} \eta^{{\lambda_{2}}) \rho_{2}} \eta^{{\kappa_{1}} ({\kappa_{3}}} \eta^{{\lambda_{3}}) {\lambda_{1}}}
	+\eta^{\rho_{1} ({\kappa_{1}}} \eta^{{\lambda_{1}}) \rho_{2}} \eta^{{\kappa_{2}} ({\kappa_{3}}} \eta^{{\lambda_{3}}) {\lambda_{2}}}
  \Big]
  \\&\quad
  +\frac{\eta^{\mu ({\kappa_{1}}} \eta^{{\lambda_{1}}) \nu}}{16} \Big[
	  \eta^{{\kappa_{2}} ({\kappa_{3}}} \eta^{{\lambda_{3}}) {\lambda_{2}} } \eta^{\rho_{1} \rho_{2}}
	- \eta^{\rho_{1} ({\kappa_{2}}} \eta^{{\lambda_{2}}) \rho_{2}} \eta^{{\kappa_{3}} {\lambda_{3}}}\Big]
  \\&\quad
  +\frac{\eta^{\mu ({\kappa_{2}}} \eta^{{\lambda_{2}}) \nu} }{16} \Big[
	  \eta^{{\kappa_{1}} ({\kappa_{3}}} \eta^{{\lambda_{3}}) {\lambda_{1}}} \eta^{\rho_{1} \rho_{2}}
	+ \eta^{\rho_{1} ({\kappa_{1}}} \eta^{{\lambda_{1}}) \rho_{2}} \eta^{{\kappa_{3}} {\lambda_{3}}}
	\Big]
  \\&\quad
  - \frac{\eta^{\rho_{1} \rho_{2}} }{4} \Big[
	  \eta^{\mu ({\kappa_{1}}} \eta^{{\lambda_{1}}) ({\kappa_{2}}} \eta^{{\lambda_{2}}) ({\kappa_{3}}} \eta^{{\lambda_{3}}) \nu}
	+ \eta^{\mu ({\kappa_{2}}} \eta^{{\lambda_{2}}) ({\kappa_{3}}} \eta^{{\lambda_{3}}) ({\kappa_{1}}} \eta^{{\lambda_{1}}) \nu}
	+ \eta^{\mu ({\kappa_{3}}} \eta^{{\lambda_{3}}) ({\kappa_{1}}} \eta^{{\lambda_{1}}) ({\kappa_{2}}} \eta^{{\lambda_{2}}) \nu}
  \Big]
  \\&\quad
+\frac{1}{4}\eta^{\mu ({\kappa_{1}}} \eta^{{\lambda_{1}}) ({\kappa_{3}}} \eta^{{\lambda_{3}}) \nu} \eta^{\rho_{1} ({\kappa_{2}}} \eta^{{\lambda_{2}}) \rho_{2}}
+\frac{1}{4} \eta^{\mu ({\kappa_{2}}} \eta^{{\lambda_{2}}) ({\kappa_{3}}} \eta^{{\lambda_{3}}) \nu} \eta^{\rho_{1} ({\kappa_{1}}} \eta^{{\lambda_{1}}) \rho_{2}}
\\
&\quad
+\frac{1}{4} \eta^{\mu ({\kappa_{2}}} \eta^{{\lambda_{2}}) ({\kappa_{1}}} \eta^{{\lambda_{1}}) ({\kappa_{3}}} \eta^{{\lambda_{3}}) (\rho_{1}} \eta^{\rho_{2}) \nu}
+\frac{1}{8} \eta^{\mu (\rho_{1}} \eta^{\rho_{2}) ({\kappa_{3}}} \eta^{{\lambda_{3}}) ({\kappa_{2}}} \eta^{{\lambda_{2}}) ({\kappa_{1}}} \eta^{{\lambda_{1}}) \nu} \,.
\end{aligned}\label{}
\end{equation}}

Let us consider the energy-momentum tensor part. Since the $\mathcal{T}^{M}_{1}$ does not involve the $\mathcal{I}$-tensor, we focus on the $\mathcal{T}^{M}_{2}$ and $\mathcal{T}^{M}_{3}$,
\begin{equation}
  \mathcal{T}^{M}_{2} = \mathcal{I}_{\rm EM}^{M,N} \Theta^{N}_{2}\,,
  \qquad
  \mathcal{T}^{M}_{3} = \mathcal{I}_{\rm EM}^{M,N,P} \Theta^{N,P}_{3}\,,
\label{}\end{equation}
where
{\small
\begin{equation}
\begin{aligned}
  \mathcal{I}^{\mu \nu, \kappa_{1} \lambda_{1}}_{\rm EM} =
  \frac{1}{2} \eta^{\mu (\kappa_{1}} \eta^{\lambda_{1}) \nu} - \frac{1}{4} \eta^{\mu \nu} \eta^{\kappa_{1} \lambda_{1}} \,,
\end{aligned}\label{I_EM1}
\end{equation}}
and
{\small
\begin{equation}
\begin{aligned}
 \mathcal{I}^{\mu \nu, \kappa_{1} \lambda_{1}, \kappa_{2} \lambda_{2}}_{\rm EM} &=
  	  \frac{1}{16} \eta^{\mu \nu} \eta^{\kappa_{1} \lambda_{1}} \eta^{\kappa_{2} \lambda_{2}}
  	- \frac{1}{8} \eta^{\mu \nu} \eta^{\kappa_{2} (\kappa_{1}} \eta^{\lambda_{1}) \lambda_{2}}
  	- \frac{1}{8} \eta^{\mu (\kappa_{1}} \eta^{\lambda_{1}) \nu} \eta^{\kappa_{2} \lambda_{2}}
  \\&\quad
  	- \frac{1}{8} \eta^{\mu (\kappa_{2}} \eta^{\lambda_{2}) \nu} \eta^{\kappa_{1} \lambda_{1}}
  	+ \frac{1}{4} \eta^{\mu (\kappa_{1}} \eta^{\lambda_{1}) (\kappa_{2}} \eta^{\lambda_{2}) \nu}
  	+ \frac{1}{4} \eta^{\nu (\kappa_{1}} \eta^{\lambda_{1}) (\kappa_{2}} \eta^{\lambda_{2}) \mu} \,.
\end{aligned}\label{I_EM2}
\end{equation}}

Finally, $\mathcal{I}$-tensors that appear in the EoM of $\phi_{I}$ \eqref{EoM_phi} are
{\small
\begin{equation}
\begin{aligned}
  \mathcal{I}^{\kappa_{1} \lambda_{1}, \kappa_{2} \lambda_{2}}_{\phi} &=
  \frac{1}{8} \eta^{\kappa_{1} \lambda_{1}} \eta^{\kappa_{2} \lambda_{2}} - \frac{1}{4} \eta^{\kappa_{1} (\kappa_{2}} \eta^{\lambda_{2}) \lambda_{1}}\,,
  \\
  \mathcal{I}^{\kappa_{1} \lambda_{1}, \kappa_{2} \lambda_{2}, \kappa_{3} \lambda_{3}}_{\phi} &=
  - \frac{1}{48} \eta^{\kappa_{1} \lambda_{1}} \eta^{\kappa_{2} \lambda_{2}} \eta^{\kappa_{3} \lambda_{3}}
  + \frac{1}{8} \eta^{\kappa_{1} \lambda_{1}} \eta^{\kappa_{2} (\kappa_{3}} \eta^{\lambda_{3}) \lambda_{2}}
  \\&\quad
  + \frac{1}{8} \eta^{\kappa_{2} \lambda_{2}} \eta^{\kappa_{3} (\kappa_{1}} \eta^{\lambda_{1}) \lambda_{3}}
  + \frac{1}{8} \eta^{\kappa_{3} \lambda_{3}} \eta^{\kappa_{1} (\kappa_{2}} \eta^{\lambda_{2}) \lambda_{1}}
  \\&\quad
  - \frac{1}{12} \eta^{\kappa_{1} (\kappa_{2}} \eta^{\lambda_{2}) (\kappa_{3}} \eta^{\lambda_{3}) \lambda_{1}}
  - \frac{1}{12} \eta^{\lambda_{1} (\kappa_{2}} \eta^{\lambda_{2}) (\kappa_{3}} \eta^{\lambda_{3}) \kappa_{1}}\,.
\end{aligned}\label{I_phi}
\end{equation}}

\section{Descendant equations} \label{App:B}

\subsection{Deformation rule for the product of multiple fields} \label{Sec:B.1}

We explain the general structure of the deformation, which is a key step in deriving the DS equation. We first consider arbitrary fields $A_{x},B_{x},C_{x},D_{x}$ and their external sources $a_{x},b_{x},c_{x},d_{x}$ respectively. The deformation is defined as follows:
\begin{equation}
\begin{aligned}
  A_{x} \mapsto A_{x} + \frac{\hbar}{i} \frac{\delta}{\delta a_{x}}\,,
  \qquad
  B_{x} \mapsto B_{x} + \frac{\hbar}{i} \frac{\delta}{\delta b_{x}}\,,
  \\
  C_{x} \mapsto C_{x} + \frac{\hbar}{i} \frac{\delta}{\delta c_{x}}\,,
  \qquad
  D_{x} \mapsto D_{x} + \frac{\hbar}{i} \frac{\delta}{\delta d_{x}}\,.
\end{aligned}\label{}
\end{equation}
We introduce a shorthand notation for the functional derivatives as
\begin{equation}
\begin{aligned}
  \psi^{ab}_{xy} &= \frac{\delta A_{x}}{\delta b_{y}}\,,
  &\quad
  \psi^{bc}_{yz} &= \frac{\delta B_{y}}{\delta c_{z}}\,,
  &\quad
  \psi^{ca}_{zx} &= \frac{\delta C_{z}}{\delta a_{x}}\,,
  \\
  \psi^{abc}_{xyz} &= \frac{\delta^{2} A_{x}}{\delta b_{y} \delta c_{z}}\,,
  &\quad
  \psi^{bca}_{yzx} &= \frac{\delta^{2} B_{y}}{\delta c_{z} \delta a_{x}}\,,
  &\quad
  \psi^{cab}_{zxy} &= \frac{\delta^{2} C_{z}}{\delta a_{x} \delta b_{y}}\,,
\end{aligned}\label{}
\end{equation}
and their permutations as
\begin{equation}
\begin{aligned}
  \hat{\psi}^{ab}_{xy} &= \frac{1}{2} \Big(\psi^{ab}_{xy} + \psi^{ba}_{yx}\Big)\,,
  \\
  \hat{\psi}^{abc}_{xyz} &= \frac{1}{3!} \Big(\psi_{xyz}^{abc} +\psi_{xzy}^{acb} +\psi_{yxz}^{bac} +\psi_{yzx}^{bca} +\psi_{zxy}^{cab} +\psi_{zyx}^{cba}\Big)\,.
\end{aligned}\label{}
\end{equation}
Under the deformation, we derive the following rules:
\begin{equation}
\begin{aligned}
  & A_{x} B_{y} \mapsto A_{x} B_{y} + \frac{\hbar}{i} \hat{\psi}^{ab}_{x,y}\,,
  \\
  & A_{x} B_{y} C_{z} \mapsto A_{x} B_{y} C_{z}
  +\frac{\hbar}{i}\left[ A_{x} \hat{\psi}_{yz}^{bc} + B_{y} \hat{\psi}_{zx}^{ca} + C_{z} \hat{\psi}_{x y}^{ab}
  \right]
  +\Big(\frac{\hbar}{i}\Big)^{2} \hat{\psi}_{xyz}^{abc}\,,
  \\
  & A_{x} B_{y} C_{z} D_{w} \mapsto A_{x} B_{y} C_{z} D_{w}
  \\&\quad~
  + \frac{\hbar}{i}\bigg[
	 \hat{\psi}^{ab}_{xy} C_{z} D_{w}	
	+ \hat{\psi}^{ac}_{xz} B_{y} D_{w}
	+ \hat{\psi}^{ad}_{xw} B_{y} C_{z}
  	+ \hat{\psi}^{bc}_{yz} A_{x} D_{w}
  	+ \hat{\psi}^{bd}_{yw} A_{x} C_{z}
	+ \hat{\psi}^{cd}_{zw} A_{x} B_{y}
  \bigg]
  \\&\quad~
  + \Big(\frac{\hbar}{i}\Big)^{2} \bigg[
    A_{x} \hat{\psi}_{yzw}^{bcd}
  + B_{y} \hat{\psi}_{xzw}^{acd}
  + C_{z} \hat{\psi}_{xyw}^{abd}
  + D_{w} \hat{\psi}_{xyz}^{abc}
    + \hat{\psi}^{ab}_{xy} \hat{\psi}^{cd}_{zw}
  	+ \hat{\psi}^{ac}_{xz} \hat{\psi}^{bd}_{yw}
  	+ \hat{\psi}^{ad}_{xw} \hat{\psi}^{bc}_{yz} \bigg]
  \\&\quad~
  + \Big(\frac{\hbar}{i}\Big)^{3}
    \hat{\psi}_{xyzw}^{abcd}\,.
\end{aligned}\label{}
\end{equation}
\subsection{Explicit form of the DS equations} \label{App:B.2}
The DS equation for $\varphi^{I}_{x}$ is given by
\begin{equation}
\begin{aligned}
  &\left(\square-m_{I}^{2}\right) \varphi^{I}_{x} =
  	- \rho_{x}^{I}
  	+ \kappa \Omega^{I}_{1}
  	+ \kappa^{2} \mathcal{I}^{I,M,N}_{\phi} \Omega^{IMN}_{2}
  	+ \kappa^{3} \mathcal{I}^{I,M,N,P}_{\phi} \Omega^{IMNP}_{3}\,,
\end{aligned}\label{DSeq_scalar2}
\end{equation}
where
\begin{equation}
\begin{aligned}
  \Omega^{I}_{1} &= \partial_{\kappa} \big(h_{x}^{\kappa\lambda} \partial_{\lambda} \varphi_{x}^{I}\big)
  	- \frac{ m_{I}^{2}}{2} h^{\kappa\kappa}_{x} \varphi_{x}^{I}
  	  + \left(\partial_{\kappa}^{y} \partial_{\lambda}^{x} + \partial_{\kappa}^{x} \partial_{\lambda}^{x}\right) \hat{\psi}_{xy}^{I \kappa\lambda}\Big|_{y\to x}
  	- \frac{m_{I}^{2}}{2} \hat{\psi}^{\kappa\kappa, I}_{xx}\,,
  \\
  \Omega^{IMN}_{2} &= h_{x}^{M} h_{x}^{N} \varphi_{x}^{I}
    + \hat{\psi}_{xx}^{MN} \varphi^{I}_{x} + 2 h^{(M}_{x} \hat{\psi}_{xx}^{N)I}
  	+ \hat{\psi}_{xxx}^{MNI}\,,
  \\
  \Omega^{IMNP}_{3} &= h_{x}^{M} h_{x}^{N} h_{x}^{P} \varphi_{x}^{I}
  	+ 3\Big[h^{(M}_{x} \hat{\psi}^{NP)}_{xx} \varphi^{I}_{x} + h^{(M}_{x} h^{N}_{x} \hat{\psi}^{P)I}_{x,x} \Big]
  \\&\quad
  + \Big[\hat{\psi}^{MNP}_{xxx} \varphi^{I}_{x} + 3h^{(M} \hat{\psi}_{xxx}^{NP)I}
  		+3\hat{\psi}^{(MN}_{xx} \hat{\psi}^{P)I}_{xx} \Big]\,,
\end{aligned}\label{Omega_fields}
\end{equation}
where
\begin{equation}
\begin{aligned}
  \hat{\psi}^{AB}_{xy} &= \frac{1}{2} \frac{\hbar}{i} \left( \psi^{AB}_{xy} +\psi^{BA}_{yx} \right)\,,
  \\
  \hat{\psi}_{xyz}^{ABC} &= \frac{1}{3!} \bigg(\frac{\hbar}{i}\bigg)^{2} \left(
		  \psi_{xyz}^{ABC} + \psi_{xzy}^{ACB} + \psi_{yzx}^{BCA} + \psi_{yxz}^{BAC}
		+ \psi_{zxy}^{CAB} + \psi_{zyx}^{CBA}
  		\right)\,.
\end{aligned}\label{}
\end{equation}
The DS equation for $h^{M}$ is given by
\begin{equation}
  \Box h_{x}^{M} = 2 P^{M,N}\bigg(\sum_{n=1}^{2}\kappa^{n}\mathcal{G}^{N}_{n} - \sum_{n=1}^{3} \kappa^{n} \mathcal{T}^{N}_{n} -2 j^{N}_{x}\bigg)\,,
\label{DSeq_h2}\end{equation}
where $\mathcal{G}^{M}_{x}$ represents the curvature perturbations
\begin{equation}
\begin{aligned}
  \mathcal{G}^{M}_{1} &=
  	  \mathcal{I}^{M,\rho N,\sigma P}_{1,1} \Lambda^{\rho N,\sigma P}_{1,1}
  	+ \mathcal{I}^{M,N,\rho \sigma P}_{1,2} \Lambda^{N,\rho \sigma P}_{1,2}\,,
  	\\
    \mathcal{G}^{M}_{2} &=
      \mathcal{I}^{M,N,\rho P,\sigma Q}_{2,1} \Lambda^{N,\rho P,\sigma Q}_{2,1}
    + \mathcal{I}^{M,N,P,\rho \sigma Q}_{2,2} \Lambda^{N,P,\rho \sigma Q}_{2,2} \,,
\end{aligned}\label{}
\end{equation}
and
\begin{equation}
\begin{aligned}
  \mathcal{T}^{M}_{1} = \Theta^{M}_{1}\,,
  \qquad
  \mathcal{T}^{M}_{2} = \mathcal{I}_{\rm EM}^{M,N} \Theta^{N}_{2}\,,
  \qquad
  \mathcal{T}^{M}_{3} = \mathcal{I}_{\rm EM}^{M,N,P} \Theta^{N,P}_{3}\,.
\end{aligned}\label{}
\end{equation}
The explicit structure of $\Lambda$ and $\Theta$ are given by
\begin{equation}
\begin{aligned}
  \Lambda^{\rho N,\sigma P}_{1,1} &=
  	  \partial_{\rho} h^{N}_{x} \partial_{\sigma} h^{P}_{x}
  	+ \frac{\hbar}{i}\partial^{x}_{\rho} \partial^{y}_{\sigma}\hat{\psi}_{xy}^{NP}\Big|_{y\to x}\,,
  \\
  \Lambda^{N,\rho\sigma P}_{1,2} &=
 	  h^{N}_{x} \partial_{\rho} \partial_{\sigma} h^{P}_{x}
  	+ \frac{\hbar}{i}\partial^{y}_{\rho} \partial^{y}_{\sigma} \hat{\psi}_{xy}^{NP}\Big|_{y\to x} \,,
  \\
  \Lambda^{N,\rho P,\sigma Q}_{2,1} &=
  	  h^{N}_{x} \partial_{\rho} h^{P}_{x} \partial_{\sigma} h^{Q}_{x}
  	+ \frac{\hbar}{i} \Big[ \partial_{\rho}^{x} \partial_{\sigma}^{y} \hat{\psi}_{xy}^{PQ} h^{N}_{x}
    + \partial_{\rho}^{y} \hat{\psi}^{NP}_{xy} \partial^{x}_{\sigma} h_{x}^{Q}
    + \partial_{\sigma}^{y} \hat{\psi}^{NQ}_{xy} \partial^{x}_{\rho} h_{x}^{P}
    \Big]_{y \to x}
  \\&\quad
    + \left(\frac{\hbar}{i}\right)^{2} \partial_{\rho}^{y} \partial_{\sigma}^{z} \hat{\psi}_{xyz}^{NPQ}\Big|_{y,z\to x}
  \\
  \Lambda^{N,P,\rho \sigma Q}_{2,2} &=
 	h^{N}_{x} h^{P}_{x} \partial_{\rho} \partial_{\sigma} h^{Q}_{x}
 	+ \frac{\hbar}{i}\Big[\hat{\psi}_{xy}^{NP} \partial_{\rho}^{x} \partial_{\sigma}^{x} h_{x}^{Q}
 	+ h^{N}_{x} \partial_{\rho}^{y} \partial_{\sigma}^{y} \hat{\psi}_{xy}^{PQ}
 	+ h^{P}_{x} \partial_{\rho}^{x} \partial_{\sigma}^{x} \hat{\psi}_{xy}^{QN} \Big]_{y\to x}
  \\&\quad
    + \left(\frac{\hbar}{i}\right)^{2} \partial_{\rho}^{z} \partial_{\sigma}^{z} \hat{\psi}_{xyz}^{NPQ}\Big|_{y,z\to x}
\end{aligned}\label{Lambda_fields}
\end{equation}
where
\begin{equation}
\begin{aligned}
  \Theta^{\mu\nu}_{1} &=
  \sum_{I=1}^{2} \Bigg[
  	  \partial_{\mu} \varphi^{I}_{x} \partial_{\nu} \varphi^{I}_{x}
  	+ \frac{m_{I}^{2}}{2} \eta_{\mu\nu} \varphi^{I}_{x} \varphi^{I}_{x}
  	+ \frac{\hbar}{i}\bigg(\partial^{x}_{\mu}\partial^{y}_{\nu} + \frac{m_{I}^{2}}{2} \eta_{\mu\nu} \bigg)\hat{\psi}^{I,I}_{x,y}\Big|_{y\to x}
  \Bigg]\,,
  \\
  \Theta^{N}_{2} &=
  \sum_{I=1}^{2} m^{2}_{I}\left[
  	  h^{N}_{x} (\varphi^{I}_{x})^{2}
  	+ \frac{\hbar}{i}\Big(2 \hat{\psi}^{N,I}_{x,x} \varphi^{I}_{x} + h^{N}_{x} \hat{\psi}^{I,I}_{x,x}\Big)
	+ \left(\frac{\hbar}{i}\right)^{2} \hat{\psi}^{N,I,I}_{x,x,x}
  \right]\,,
  \\
  \Theta^{N,P}_{3} &=
  \sum_{I=1}^{2} m^{2}_{I} \left[
      h^{N}_{x} h^{P}_{x} (\varphi^{I}_{x})^{2}
    + \frac{\hbar}{i}\Big( \hat{\psi}_{x,x}^{N,P} (\varphi^{I}_{x})^{2}
		+ h^{(N}_{x} h^{P)}_{x} \hat{\psi}^{I,I}_{x,x}
		+ 4h^{(N}_{x} \hat{\psi}^{P), I}_{x,x} \varphi^I_{x}
	\Big)\right.
  \\&\qquad\qquad\quad
  \left.
  + \left(\frac{\hbar}{i}\right)^{2}\Big( 3 \hat{\psi}^{(N,P}_{x,x}\hat{\psi}^{I),I}_{x,x}
  	+ 2 \hat{\psi}^{N,P,I}_{x,x,x} \varphi^{I}_{x}
  	+ 2 h^{(N}_{x} \hat{\psi}^{P), I,I}_{x,x,x}\Big)
  \right]\,.
\end{aligned}\label{Theta_fields}
\end{equation}
%

\subsection{DS equations for the descendant fields} \label{App:B.3}
To solve the DS equations, we need to derive the equations for the descendant fields. It is straightforward to achieve by acting the functional derivative $\frac{\delta}{\delta \mathbf{j}^{A}_{x}}$ on the DS equations for $h^{\mu \nu}$ and $\varphi^{I}$.

Let us consider the scalar field sector. The DS equation for the first descendant fields $\psi^{I,A}$ are given by
\begin{equation}
\begin{aligned}
  \left(\square-m_{I}^2\right) \psi^{IA}_{xy} &=
    - \hat{\delta}_{xy}\delta^{IA}
  	+ \kappa\Delta^{A}_{y}\big[\Omega^{I}_{1} \big]
  	+ \kappa^2\mathcal{I}^{M,N}_{\varphi,I} \Delta^{A}_{y} \big[\Omega^{IMN}_{2}\big]
  	+ \kappa^3\mathcal{I}^{M,N,P}_{\varphi,I} \Delta^{A}_{y} \big[\Omega^{IMNP}_{3}\big]\,,
\end{aligned}\label{PsiIA}
\end{equation}
where
\begin{equation}
\begin{aligned}
  \Delta^{A}_{y}\big[\Omega^{I}_{1} \big]  &=
  	\partial^{x}_{\kappa} \Big( \psi^{\kappa\lambda A}_{xy}\partial_{\lambda} \varphi_{x}^{I}
  + h^{\kappa\lambda}_{x} \partial^{x}_{\lambda} \psi^{I A}_{xy} \Big)
  - \frac{m^{2}_I}{2} \Big(
  	 \psi^{\lambda\lambda A}_{xy} \varphi^{I}_{x} + h_{x}^{\lambda\lambda} \psi^{I A}_{x y}\Big)
  \\&\quad
  + \bigg[\left(\partial_{\kappa}^x \partial_{\lambda}^x +\partial_{\kappa}^z \partial_{\lambda}^x\right)\hat{\psi}_{xz|y}^{I \kappa\lambda|A}
  		- \frac{m_{I}^{2}}{2} \hat{\psi}^{\lambda\lambda I|A}_{xz|y} \bigg]_{z\to x}\,,
  \\
  \Delta^{A}_{y} \big[\Omega^{IMN}_{2}\big] &=
  	  2h^{(M}_{x}\psi^{N)A}_{xy} \varphi^{I}_{x}
  	+ h^{M}_{x}h^{N}_{x} \psi^{I,A}_{x,y}
  \\&\quad
  +  \Big[
	  \hat{\psi}^{MN|A}_{xz|y} \varphi^{I}_{x}
  	+ \hat{\psi}^{MN}_{xz} \psi^{IA}_{xy}
	+ 2 \psi^{(M|A|}_{xy} \hat{\psi}^{N)I}_{xz}
  	+ 2 h^{(M}_{x} \hat{\psi}^{N)I|A}_{xz|y} \Big]_{z\to x}\,,
  \\
  \Delta^{A}_{y} \big[\Omega^{IMNP}_{3}\big] &= 3h^{(M}_{x} h^{N}_{x} \psi^{P)A}_{xy} \varphi^I_{x} + h^{M}_{x} h^{N}_{x} h^{P}_{x} \psi^{IA}_{xy}
  \\&\quad
  	+ 3\Big[ \psi^{(M|A|}_{xy} \hat{\psi}^{NP)}_{xz} \varphi^{I}_{x}
  		+ h^{(M}_{x} \hat{\psi}^{NP)|A}_{xz|y} \varphi^{I}_{x}
  		+ h^{(M}_{x} \hat{\psi}^{NP)}_{xz} \psi^{IA}_{xy}
  		\\&\qquad\quad
  		+ 2\psi^{(M|A|}_{xy} h^{N}_{x} \hat{\psi}^{P)I}_{xz}
  		+ h^{(M}_{x} h^{N}_{x} \hat{\psi}^{P)I|A}_{xz|y}\Big]_{z\to x}\,,
\end{aligned}\label{}
\end{equation}
Here we denote the functional derivative to the first descendant fields as
\begin{equation}
\begin{aligned}
  \hat{\psi}^{AB|C}_{xy|z} := \frac{\delta \hat{\psi}^{AB}_{xy}}{\delta \mathbf{j}^{C}_{z}} = \frac{1}{2}\Big( \psi_{xyz}^{ABC}+\psi_{yxz}^{BAC}\Big)\,.
\end{aligned}\label{}
\end{equation}
Next, the DS equation for second descendant fields $\psi^{I,AB}_{x,y,z}$
\begin{equation}
\begin{aligned}
  &\left(\square-m_{I}^{2}\right) \psi^{IAB}_{xyz} =
  	 \kappa \Delta^{A,B}_{y,z} \big[\Omega^{I}_{1} \big]
  	+ \kappa^2\mathcal{I}^{M,N}_{\varphi,I} \Delta^{A,B}_{y,z} \big[\Omega^{IMN}_{2}\big]
  	+ \kappa^3\mathcal{I}^{M,N,P}_{\varphi,I} \Delta^{A,B}_{y,z} \big[\Omega^{IMNP}_{3}\big]\,,
\end{aligned}
\label{PsiIAB}\end{equation}
where
\begin{equation}
\begin{aligned}
  \Delta^{A,B}_{y,z} \big[\Omega^{I}_{1} \big] &= \partial_{\kappa} \left(
  	\psi^{\kappa\lambda AB}_{xyz} \partial_{\lambda} \varphi^{I}_{x}
  + \psi_{xy}^{\kappa\lambda A} \partial_{\lambda} \psi^{IB}_{xz}
  + \psi_{xz}^{\kappa\lambda B} \partial_{\lambda} \psi^{IA}_{xy}
  + h_{x}^{\kappa\lambda} \partial_{\lambda} \psi^{IAB}_{xyz}
  \right)
  \\&\quad
  - \frac{m^{2}_I}{2}  \Big(
  	  \psi^{\lambda\lambda AB}_{xyz} \varphi^{I}_{x}
  	+ \psi^{\lambda\lambda A}_{xy} \psi^{I B}_{xz}
  	+ \psi_{xz}^{\lambda\lambda B} \psi^{IA}_{xy}
  	+ h_{x}^{\lambda\lambda} \psi^{IAB}_{xyz}\Big)\,,
  \\
  \Delta^{A,B}_{y,z} \big[\Omega^{IMN}_{2}\big] &=
  	  2 \Big( \psi^{(M|B|}_{x,z} \psi^{N)A}_{xy} + h^{(M}_{x} \psi^{N)AB}_{xyz}\Big) \varphi^{I}_{x}
  	+ 2 h^{(M}_{x} \psi^{N)A}_{xy} \psi^{IB}_{xz}
  \\&\quad
  + 2 h^{(M}_{x} \psi^{N)B}_{xz} \psi^{IA}_{xy} + h^{M}_{x} h^{N}_{x} \psi^{IAB}_{xyz}\,,
  \\
  \Delta^{A,B}_{y,z} \big[\Omega^{IMNP}_{3}\big] &=
    6 h^{(M}_{x} \psi^{N|B|}_{xz} \psi^{P)A}_{xy} \varphi^{I}_{x}
  + 3 h^{(M}_{x} h^{N}_{x} \psi^{P)AB}_{xyz} \varphi^{I}_{x}
  \\&\quad
  + 3 h^{(M}_{x} h^{N}_{x} \psi^{P)A}_{xy} \psi^{IB}_{xz}
  + 3 h^{(M}_{x} h^{N}_{x} \psi^{P)B}_{xz} \psi^{IA}_{xy}
  + h^{M}_{x} h^{N}_{x} h^{P}_{x} \psi^{IAB}_{xyz}\,.
\end{aligned}
\label{}\end{equation}
We then consider the graviton sector. The first descendent equation for the EoM of $h^{\mu\nu}$ is given by a functional derivative with respect to $\mathbf{j}^{A}$ to the EoM
\begin{equation}
  \Box\psi_{xy}^{MA} = 2 P^{M,N} \Big( \sum_{n=1}^{2}\kappa^n \Delta^{A}_{y}\big[\mathcal{G}^{N}_{n}\big] - \sum_{n=1}^{3} \kappa^n \Delta^{A}_{y}\big[{\mathcal{T}}^{N}_{n}\big] -2 \hat{\delta}_{x,y} \delta^{N,A}\Big)\,,
\label{PsiMA}\end{equation}
where the $\Delta$ operation on $\mathcal{G}$
\begin{equation}
\begin{aligned}
  \Delta^{A}_{y}\big[\mathcal{G}^{M}_{1} \big] &=
  	  \mathcal{I}^{M,\rho N,\sigma P}_{1,1} \Delta^{A}_{y}\big[\Lambda^{\rho N,\sigma P}_{1,1} \big]
  	+ \mathcal{I}^{M,N,\rho \sigma P}_{1,2} \Delta^{A}_{y}\big[\Lambda^{N,\rho \sigma P}_{1,2} \big]\,,
  	\\
   \Delta^{A}_{y}\big[ \mathcal{G}^{M}_{2} \big] &=
      \mathcal{I}^{M,N,\rho P,\sigma Q}_{2,1} \Delta^{A}_{y}\big[\Lambda^{N,\rho P,\sigma Q}_{2,1} \big]
    + \mathcal{I}^{M,N,P,\rho \sigma Q}_{2,2}\Delta^{A}_{y}\big[ \Lambda^{N,P,\rho \sigma Q}_{2,2}  \big]\,, \\
\end{aligned}\label{}
\end{equation}
and $\Delta$ operation on $\mathcal{T}$
\begin{equation}
\begin{aligned}
&     \Delta^{A}_{y}\big[ \mathcal{T}^{M}_{1} \big]= \Delta^{A}_{y}\big[\Theta^{M}_{1}\big]\,,
  \\
  &
\Delta^{A}_{y}\big[  \mathcal{T}^{M}_{2} \big] = \mathcal{I}_{\rm EM}^{M,N} \Delta^{A}_{y}\big[\Theta^{N}_{2} \big]\,,
  \\
  &
\Delta^{A}_{y}\big[  \mathcal{T}^{M}_{3} \big]= \mathcal{I}_{\rm EM}^{M,N,P} \Delta^{A}_{y}\big[\Theta^{N,P}_{3} \big]\,.
\end{aligned}\label{}
\end{equation}
with
\begin{equation}
\begin{aligned}
  \Delta^{A}_{y}\big[\Lambda^{\rho N,\sigma P}_{1,1}\big] &=
  	  \partial^{x}_{\rho} \psi^{NA}_{xy} \partial^{x}_{\sigma} h^{P}_{xy}
  	+ \partial^{x}_{\rho} h^{N}_{x} \partial^{x}_{\sigma} \psi^{PA}_{xy}
  	+ \partial^{x}_{\rho} \partial^{z}_{\sigma}\hat{\psi}_{xz|y}^{NP|A}\Big|_{z\to x}\,,
  \\
   \Delta^{A}_{y}\big[\Lambda^{N,\rho\sigma P}_{1,2}\big] &=
 	  \psi^{NA}_{xy} \partial_{\rho} \partial_{\sigma} h^{P}_{x}
 	+ h^{N}_{x} \partial^{x}_{\rho} \partial^{x}_{\sigma} \psi^{PA}_{xy}
  	+ \partial^{z}_{\rho} \partial^{z}_{\sigma} \hat{\psi}_{xz|y}^{NP|A}\Big|_{z\to x}\,,
  \\
   \Delta^{A}_{y}\big[\Lambda^{N,\rho P,\sigma Q}_{2,1}\big] &=
  	  \psi^{NA}_{xy} \partial_{\rho} h^{P}_{x} \partial_{\sigma} h^{Q}_{x}
  	+ h^{N}_{x} \partial^{x}_{\rho} \psi^{PA}_{xy} \partial^{x}_{\sigma} h^{Q}_{x}
  	+ h^{N}_{x} \partial^{x}_{\rho} h^{P}_{x} \partial^{x}_{\sigma} \psi^{QA}_{xy}
  \\&\quad
  	+ \Big[
  	  \partial_{\rho}^{x} \partial_{\sigma}^{z} \hat{\psi}_{xz|y}^{PQ|A} h^{N}_{x}
  	+ \partial_{\rho}^{x} \partial_{\sigma}^{z} \hat{\psi}_{xz}^{PQ} \psi^{NA}_{xy}
  	+ \partial_{\rho}^{y} \hat{\psi}^{NP|A}_{xz|y} \partial^{x}_{\sigma} h_{x}^{Q}
  \\&\qquad\quad
    + \partial_{\rho}^{y} \hat{\psi}^{NP}_{xz} \partial^{x}_{\sigma} \psi_{xy}^{QA}
    + \partial_{\sigma}^{y} \hat{\psi}^{NQ|A}_{xz|y} \partial^{x}_{\rho} h_{x}^{P}
    + \partial_{\sigma}^{y} \hat{\psi}^{NQ}_{xy} \partial^{x}_{\rho} \psi_{xy}^{PA}\Big]_{z\to x} \,,
  \\
   \Delta^{A}_{y}\big[\Lambda^{N,P,\rho \sigma Q}_{2,2}\big] &=
 	  \psi^{NA}_{xy} h^{P}_{x} \partial_{\rho} \partial_{\sigma} h^{Q}_{x}
 	+ h^{N}_{x} \psi^{PA}_{xy} \partial_{\rho} \partial_{\sigma} h^{Q}_{x}
 	+ h^{N}_{x} h^{P}_{x} \partial_{\rho} \partial_{\sigma} \psi^{QA}_{xy}
  \\&\quad
 	+ \Big[\hat{\psi}_{xzy}^{NPA} \partial_{\rho}^{x} \partial_{\sigma}^{x} h_{x}^{Q}
 	+ \hat{\psi}_{xz}^{NP} \partial_{\rho}^{x} \partial_{\sigma}^{x} \psi_{xy}^{QA}
 	+ \psi^{NA}_{xy} \partial_{\rho}^{z} \partial_{\sigma}^{z} \hat{\psi}_{xz}^{PQ}
  \\&\qquad\quad
 	+ h^{N}_{x} \partial_{\rho}^{z} \partial_{\sigma}^{z} \hat{\psi}_{xz|y}^{PQ|A}
 	+ \psi^{PA}_{xy} \partial_{\rho}^{x} \partial_{\sigma}^{x} \hat{\psi}_{xz}^{QN}
 	+ h^{P}_{x} \partial_{\rho}^{x} \partial_{\sigma}^{x} \hat{\psi}_{xz|y}^{QN|A} \Big]_{z\to x} \,,
\end{aligned}\label{}
\end{equation}
and
\begin{equation}
\begin{aligned}
  \Delta^{A}_{y}\big[\Theta^{\kappa\lambda}_{1}\big] &=
    \partial^{x}_{\kappa} \psi^{IA}_{xy} \partial^{x}_{\lambda} \varphi^{I}_{x}
  + \partial^{x}_{\kappa} \varphi^{I}_{x} \partial^{x}_{\lambda} \psi^{IA}_{xy}
  + m_{I}^{2} \eta_{\kappa\lambda} \psi^{IA}_{xy} \varphi^{I}_{x}
  + \Big(\partial^{x}_{\kappa}\partial^{y}_{\lambda} + \frac{m_{I}^{2}}{2} \eta_{\kappa\lambda} \Big)\hat{\psi}^{II|A}_{xz|y}\Big|_{z\to x} \,,
  \\
  \Delta^{A}_{y}\big[\Theta^{N,I}_{2}\big] &=
    \psi^{NA}_{xy} (\varphi^{I}_{x})^{2}
  + 2 h^{N}_{x} \varphi^{I}_{x} \psi^{IA}_{xy}
  \\&\quad
  + \Big[ 2 \hat{\psi}^{NI|A}_{xx|y} \varphi^{I}_{x}
    + 2 \hat{\psi}^{NI}_{xx} \varphi^{IA}_{xy}
  	+ \psi^{NA}_{xy} \hat{\psi}^{II}_{xx}
  	+ h^{N}_{x} \hat{\psi}^{II|A}_{xx|y} \Big] \,,
  \\
  \Delta^{A}_{y}\big[\Theta^{N,P,I}_{3}\big] &=
      2\psi^{(N|A|}_{xy} h^{P)}_{x} (\varphi^{I}_{x})^{2}
    + 2 h^{N}_{x} h^{P}_{x} \varphi^{I}_{x} \psi^{IA}_{xy}
  \\&\quad
    + \Big[
    	  \hat{\psi}_{xx|y}^{NP|A} (\varphi^{I}_{x})^{2}
    	+ 2\hat{\psi}_{xx}^{N,P} \varphi^{I}_{x} \psi^{IA}_{xy}
		+ 2 \psi^{(N|A|}_{xy} h^{P)}_{x} \hat{\psi}^{II}_{xx}
		+ h^{N}_{x}h^{P}_{x} \hat{\psi}^{II|A}_{xx|y}
  \\&\qquad\quad
		+ 4\psi^{(N|A|}_{xy} \hat{\psi}^{P)I}_{xx} \varphi^I_{x}
		+ 4h^{(N}_{x} \hat{\psi}^{P)I|A}_{xx|y} \varphi^I_{x}
		+ 4h^{(N}_{x} \hat{\psi}^{P)I}_{xx} \psi^{IA}_{x}\, .
	\Big]
\end{aligned}\label{}
\end{equation}
Next, the DS equation for the second descendent field up to $\hbar^{2}$ order is given by
\begin{equation}
  \Box\psi_{xy}^{MAB} = 2 P^{M,N}\Big( \sum_{n=1}^{2} \kappa^{n} \Delta^{A,B}_{y,z}\big[\mathcal{G}^{N}_{n}\big] - \sum_{n=1}^{3} \kappa^{n} \Delta^{A,B}_{y,z}\big[{\mathcal{T}}^{N}_{n}\big] \Big)\,,
\label{PsiMAB}\end{equation}
where the $\Delta$ operations on $\mathcal{G}$
\begin{equation}
\begin{aligned}
  \Delta^{A,B}_{y,z}\big[\mathcal{G}^{M}_{1} \big] &=
  	  \mathcal{I}^{M,\rho N,\sigma P}_{1,1} \Delta^{A,B}_{y,z}\big[\Lambda^{\rho N,\sigma P}_{1,1} \big]
  	+ \mathcal{I}^{M,N,\rho \sigma P}_{1,2} \Delta^{A,B}_{y,z}\big[\Lambda^{N,\rho \sigma P}_{1,2} \big]\,,
  	\\
   \Delta^{A,B}_{y,z}\big[ \mathcal{G}^{M}_{2} \big] &=
      \mathcal{I}^{M,N,\rho P,\sigma Q}_{2,1} \Delta^{A,B}_{y,z}\big[\Lambda^{N,\rho P,\sigma Q}_{2,1} \big]
    + \mathcal{I}^{M,N,P,\rho \sigma Q}_{2,2}\Delta^{A,B}_{y,z}\big[ \Lambda^{N,P,\rho \sigma Q}_{2,2}  \big]\,, \\
\end{aligned}\label{}
\end{equation}
and $\Delta$ operation on $\mathcal{T}$
\begin{equation}
\begin{aligned}
&     \Delta^{A,B}_{y,z}\big[ \mathcal{T}^{M}_{1} \big]= \Delta^{A,B}_{y,z}\big[\Theta^{M}_{1}\big]\,,
  \\
  &
\Delta^{A,B}_{y,z}\big[  \mathcal{T}^{M}_{2} \big] = \mathcal{I}_{\rm EM}^{M,N} \Delta^{A,B}_{y,z}\big[\Theta^{N}_{2} \big]\,,
  \\
  &
\Delta^{A,B}_{y,z}\big[  \mathcal{T}^{M}_{3} \big]= \mathcal{I}_{\rm EM}^{M,N,P} \Delta^{A,B}_{y,z}\big[\Theta^{N,P}_{3} \big]\,.
\end{aligned}\label{}
\end{equation}
with
\begin{equation}
\begin{aligned}
  \Delta^{A,B}_{y,z}\big[\Lambda^{\rho N,\sigma P}_{1,1}\big] &=  \partial_{\rho}^{x} \psi_{xyz}^{NAB} \partial_{\sigma}^{x} h_{x}^{P} + 2 \partial_{\rho}^{x} \psi^{N(A}_{xy} \partial_{\sigma}^{x} \psi_{xz}^{|P|B)} + \partial_{\rho}^{x} h_{x}^{N} \partial_{\sigma}^{x} \psi_{xyz}^{PAB} \,,
  \\
   \Delta^{A,B}_{y,z}\big[\Lambda^{N,\rho\sigma P}_{1,2}\big] &=  \psi^{NAB}_{xyz} \partial_{\rho}^{x} \partial_{\sigma}^{x} h_{x}^{P} + \psi^{NA}_{xy} \partial_{\rho}^{x} \partial_{\sigma}^{x} \psi_{xz}^{PB} + \psi^{NB}_{xz} \partial^{x}_{\rho}\partial^{x}_{\sigma} \psi^{PA}_{xy} + h_{x}^{N} \partial^{x}_{\rho} \partial^{x}_{\sigma} \psi^{NAB}_{xyz}\,,
  \\
   \Delta^{A,B}_{y,z}\big[\Lambda^{N,\rho P,\sigma Q}_{2,1}\big] &= \psi^{NAB}_{xyz} \partial_{\rho}^{x} h_{x}^{P} \partial_{\sigma}^{x} h_{x}^{Q}
   + h_{x}^{N} \partial_{\rho}^{x} \psi^{PAB}_{xyz} \partial_{\sigma}^{x} h_{x}^{Q} + h_{x}^{N} \partial_{\rho}^{x} h_{x}^{P} \partial_{\sigma}^{x} \psi^{QAB}_{xyz}
   \\
   & + \psi^{NA}_{xy} ( \partial_{\rho}^{x} \psi_{xz}^{PB} \partial_{\sigma}^{x} h_{x}^{Q} + \partial_{\rho}^{x} h_{x}^{P} \partial_{\sigma}^{x} \psi^{QB}_{xz})
   + \psi^{NB}_{xz} ( \partial_{\rho}^{x} \psi^{PA}_{xy} \partial^{x}_{\sigma} h_{x}^{Q} + \partial_{\rho}^{x} h_{x}^{P} \partial_{\sigma}^{x} \psi^{QA}_{xy})
   \\
   & + h_{x}^{N} ( \partial_{\rho}^{x} \psi^{PA}_{xy} \partial^{x}_{\sigma} \psi^{QB}_{xz} + \partial^{x}_{\rho} \psi^{PB}_{xz} \partial_{\sigma}^{x} \psi^{QA}_{xy}) \, ,\\
   \Delta^{A,B}_{y,z}\big[\Lambda^{N,P,\rho \sigma Q}_{2,2}\big] &=  2 h_{x}^{(N} \psi^{P)AB}_{xyz} \partial^{x}_{\rho} \partial^{x}_{\sigma} h_{x}^{Q} + h_{x}^{N} h_{x}^{P} \partial_{\rho}^{x} \partial_{\sigma}^{x} \psi_{xyz}^{QAB}
   + (\psi^{NA}_{xy} \psi^{PB}_{xz} + \psi^{NB}_{xz} \psi^{PA}_{xy}) \partial_{\rho}^{x} \partial_{\sigma}^{x} h_{x}^{Q} \\
   & + 2 h_{x}^{(N} \psi^{P)A}_{xy} \partial_{\rho}^{x} \partial_{\sigma}^{x} \psi^{QB}_{xz}
   + 2 h_{x}^{(N} \psi^{P)B}_{xz} \partial_{\rho}^{x} \partial_{\sigma}^{x} \psi^{QA}_{xy}\,,
\end{aligned}\label{}
\end{equation}
and
\begin{equation}
\begin{aligned}
  \Delta^{A,B}_{y,z}\big[\Theta^{\kappa\lambda}_{1}\big] &=
      2\partial^{x}_{(\kappa} \psi^{IAB}_{xyz} \partial^{x}_{\lambda)} \varphi^{I}_{x}
    + 2 \partial^{x}_{(\kappa} \psi^{IA}_{xy} \partial_{\lambda)}^{x} \psi^{IB}_{xz} 
    + m_I^2 \eta_{\kappa \lambda} ( \psi^{IAB}_{xyz} \varphi^{I}_{x} + \psi^{IA}_{xy} \psi^{IB}_{xz} )\,,
  \\
  \Delta^{A,B}_{y,z}\big[\Theta^{N,I}_{2}\big] &=  \psi^{NAB}_{xyz} (\varphi^{I}_{x})^2 + 2 \psi^{NA}_{xy} \psi^{IB}_{xz} \varphi_{x}^{I}
  + 2 \psi_{xz}^{NB} \psi^{IA}_{xy} \varphi_{x}^{I} + 2 h_{x}^{N} \psi^{IAB}_{xyz} \varphi_{x}^{I} + 2 h_{x}^N \psi^{IA}_{xy}\psi^{IB}_{xz}\,,
  \\
  \Delta^{A,B}_{y,z}\big[\Theta^{N,P,I}_{3}\big] &=  2  h^{(N}_{x}\psi^{P)AB}_{xyz} (\varphi_{x}^{I})^2
  + 2 \psi^{(N|A|}_{xy} \psi^{P)B}_{xz} (\varphi_{x}^I)^2 + 2 \varphi_{x}^{N} \psi^{PA}_{xy} \psi^{IB}_{xz} \varphi_{x}^{I} + 4 \psi^{(N|A|}_{xy} \psi^{I)B}_{xz} h_{x}^{P} \varphi_{x}^{I} \\
  & \quad + 2 \psi_{xz}^{PB} \psi^{IA}_{xy} h^{N}_{x} \varphi_{x}^{I} + 2 h_{x}^{N} h_{x}^{P} \psi^{IAB}_{xyz} \varphi_{x}^{I} + 2 h_{x}^{N} h_{x}^{P} \psi^{IA}_{xy} \psi^{IB}_{xz} .
   \end{aligned}\label{}
\end{equation}

\section{Explicit Form of the Quantum Recursion Relations}
Here we collect the $\Omega, \Lambda, \Theta$ currents for the quantum recursion relation of scalar field \eqref{Phi_recursion} and graviton \eqref{recursion_h}.
The $\Omega$ currents are
\begin{equation}
\begin{aligned}
  \Omega^{\ord{n}}\big|^{I}_{\hat{\mathcal{P}}} &= k^{\kappa}_{\hat{\mathcal{P}}} \Big\lceil J^{\kappa\lambda},k^{\lambda}\Phi^{I} \Big\rfloor^{\ord{n}}_{\hat{\mathcal{P}}}
  + \frac{m_{I}^{2}}{2} \Big\lceil J^{\lambda\lambda}, \Phi^{I} \Big\rfloor^{\ord{n}}_{\hat{\mathcal{P}}}
  \\&\quad
  + \int_{\ell} \bigg[ \frac{k^{\kappa}_{\hat{\mathcal{P}}}}{2}\Big(
  	  k_{\hat{\mathcal{P}},\ell}^{\lambda} \Psi^{\ord{n-1}}\big|_{\ell| \hat{\mathcal{P}}}^{I,\kappa\lambda}
  	+ \ell^{\lambda} \Psi^{\ord{n-1}}\big|_{\ell| \hat{\mathcal{P}}}^{\kappa\lambda,I}
  	\Big)
  + \frac{m_{I}^{2}}{2} \hat{\Psi}^{\ord{n-1}}\big|^{I,\lambda\lambda}_{\ell| \hat{\mathcal{P}}}\bigg]\,,
  \\
  \Omega^{\ord{n}}\big|^{IMN}_{\hat{\mathcal{P}}} &=
  \Big\lceil J^{M}{,} J^{N}{,} \Phi^{I} \Big\rfloor^{\ord{n}}_{\hat{\mathcal{P}}}
  + \int_{\ell} \Big[2\big\lceil J^{M}{,} \hat{\Psi}_{\ell}^{IN} \big\rfloor^{\ord{n-1}}_{\hat{\mathcal{P}}}
  + \big\lceil \hat{\Psi}_{\ell}^{MN} {,} \Phi^{I} \big\rfloor^{\ord{n-1}}_{\hat{\mathcal{P}}} \Big]
  \\&\quad
  + \int_{\ell_{1} \ell_{2}} \hat{\Psi}^{\ord{n-2}}\big|_{\ell_{1}\ell_{2}| \hat{\mathcal{P}}}^{MNI}\,,
  \\
  \Omega^{\ord{n}}\big|^{IMNP}_{\hat{\mathcal{P}}} &=
    \Big\lceil J^{M}{,} J^{N}{,} J^{P}{,} \Phi^{I} \Big\rfloor^{\ord{n}}_{\hat{\mathcal{P}}}
  \\&\quad
  + \int_{\ell} \left[ 3\big\lceil J^{(M}{,} J^{N} {,} \hat{\Psi}^{P),I}_{\ell} \big\rfloor^{\ord{n-1}}_{\hat{\mathcal{P}}}
  + \frac{3}{2} \Big\lceil \Delta^{(P}_{\ell}\big\lceil J^{M}{,}J^{N)} \big\rfloor{,}\Phi^{I} \Big\rfloor^{\ord{n-1}}_{\hat{\mathcal{P}}} \right]
  \\&\quad
  + \int_{\ell_{1}\ell_{2}} \left[ \big\lceil \hat{\Psi}_{\ell_{1}\ell_{2}}^{MNP}, \Phi^{I} \big\rfloor^{\ord{n-2}}_{\hat{\mathcal{P}}}
  + 3 \big\lceil \hat{\Psi}_{\ell_{1}\ell_{2}}^{I(MN}, J^{P)} \big\rfloor^{\ord{n-2}}_{\hat{\mathcal{P}}}
  + 3 \big\lceil \Psi^{(MN}_{\ell_{2}}, \hat{\Psi}_{\ell_{1}}^{P)I} \big\rfloor^{\ord{n-2}}_{\hat{\mathcal{P}}} \right]\,.
\end{aligned}\label{Omega_currents}
\end{equation}

The $\Lambda$ currents are
\begin{equation}
\begin{aligned}
  \Lambda^\ord{n}_{1,1}\big|^{\rho N, \sigma P}_{\hat{\mathcal{P}}}&=
    \big\lceil k^{\rho} J^{N}, k^{\sigma} J^{P} \big\rfloor^\ord{n}_{\hat{\mathcal{P}}}
  + \int_{\ell} k^{\rho}_{\ell} \ell^{\sigma} \Psi^\ord{n-1}\big|_{\ell | \hat{\mathcal{P}}}^{N,P}\,,
  \\
  \Lambda^\ord{n}_{1,2}\big|^{N, \rho \sigma, P}_{\hat{\mathcal{P}}} &=
    \big\lceil J^{N}, k^{\rho} k^{\sigma} J^{P} \big\rfloor^\ord{n}_{\hat{\mathcal{P}}}
  + \frac{1}{2}\int_{\ell} \Big( k^{\rho}_{\ell} k^{\sigma}_{\ell} \Psi^\ord{n-1}\big|^{P,N}_{\ell |\hat{\mathcal{P}}}
  	+ \ell^{\rho} \ell^{\sigma} \Psi^\ord{n-1}\big|^{N,P}_{\ell |\hat{\mathcal{P}}} \Big)\,,
  \\
  \Lambda^\ord{n}_{2,1}\big|^{N, \rho P, \sigma Q}_{\hat{\mathcal{P}}} &=
      \big\lceil J^{N}, k^{\rho} J^{P}, k^{\sigma} J^{Q} \big\rfloor^\ord{n}_{\hat{\mathcal{P}}}
    + \frac{1}{2} \int_{\ell} \Delta^{N}_{\ell}\big\lceil k^{\rho} J^{P}, k^{\sigma} J^{Q} \big\rfloor^{\ord{n-1}}_{\hat{\mathcal{P}}}
  \\&
    + \int_{\ell} \ell^{\sigma} \Big[ \big\lceil k^{\rho} J^{P}, \Psi_{\ell}^{NQ} \big\rfloor^\ord{n-1}_{\hat{\mathcal{P}}}
  	+ \big\lceil J^{N}, k^{\rho}_{\ell} \Psi_{\ell }^{PQ}  \big\rfloor^\ord{n-1}_{\hat{\mathcal{P}}}\Big]
  \\&
  	+ \frac{1}{3} \int_{\ell_{1}\ell_{2}} \bigg[ k^{\rho}_{\hat{\mathcal{P}},\ell_{12}} \ell_{1}^{\sigma} \Psi^\ord{n-2}\big|_{\ell_{1}\ell_{2}| \hat{\mathcal{P}}}^{PQN}
  	+ \ell_{2}^{\rho} k^{\sigma}_{\hat{\mathcal{P}},\ell_{12}} \Psi^\ord{n-2}\big|_{\ell_{1}\ell_{2}| \hat{\mathcal{P}}}^{QNP}
  	+ \ell_{1}^{\rho} \ell_{2}^{\sigma} \Psi^\ord{n-2}\big|^{NPQ}_{\ell_{1}\ell_{2}| \hat{\mathcal{P}}} \bigg]\,,
  \\
  \Lambda^\ord{n}_{2,2}\big|^{N,P, \rho \sigma, Q}_{\hat{\mathcal{P}}} &=
  	\big\lceil J^{N},J^{P}, k^{\rho} k^{\sigma} J^{Q} \big\rfloor^\ord{n}_{\hat{\mathcal{P}}}
  \\&
  + \int_{\ell} \Big[
  	\big\lceil k^{\rho} k^{\sigma} J^{Q}, \Psi_{\ell}^{NP} \big\rfloor^\ord{n-1}_{\hat{\mathcal{P}}}
  + \big\lceil J^{N}, k^{\rho}_{\ell} k^{\sigma}_{\ell} \Psi_{\ell}^{QP} + \ell^{\rho} \ell^{\sigma} \Psi_{\ell}^{PQ} \big\rfloor^\ord{n-1}_{\hat{\mathcal{P}}}\Big]
  \\&
  + \frac{1}{3} \int_{\ell_{1}\ell_{2}} \bigg[
    \ell_{1}^{\rho} \ell_{1}^{\sigma} \Psi^\ord{n-2}\big|_{\ell_{1}\ell_{2}|\hat{\mathcal{P}}}^{PQN}
  + k^{\rho}_{\hat{\mathcal{P}},\ell_{12}} k^{\sigma}_{\hat{\mathcal{P}},\ell_{12}} \Psi^\ord{n-2}\big|_{\ell_{1}\ell_{2}| \hat{\mathcal{P}}}^{QNP}
  + \ell_{2}^{\rho} \ell_{2}^{\sigma} \Psi^\ord{n-2}\big|_{\ell_{1}\ell_{2}|\hat{\mathcal{P}}}^{NPQ}\bigg]\,.
\end{aligned}\label{Lambda_currents}
\end{equation}

The $\Theta$ currents are
\begin{equation}
\begin{aligned}
  \Theta^\ord{n}_{1}\big|^{\rho \sigma}_{\hat{\mathcal{P}}} &= \sum_{I=1}^{2} \bigg[
    \big\lceil k^{\rho} \Phi^{I}, k^{\sigma} \Phi^{I}\big\rfloor^{\ord{n}}_{\hat{\mathcal{P}}}
  - \frac{m_{I}^{2}}{2}\eta^{\rho\sigma} \big[\Phi^{I},\Phi^{I}\big]^{\ord{n}}_{\hat{\mathcal{P}}}
  + \int_{\ell} \Big(k^{(\rho}_{\ell} \ell^{\sigma)} - \frac{m_{I}^{2}}{2} \eta^{\rho\sigma}  \Big) \Psi^{\ord{n}}\big|_{\ell | \hat{\mathcal{P}}}^{II}\bigg]\,,
  \\
  \Theta^\ord{n}_{2}\big|^{N}_{\hat{\mathcal{P}}} &= \sum_{I=1}^{2} m_{I}^{2} \Bigg[
  	  \big\lceil J^{N}, \Phi^{I},\Phi^{I} \big\rfloor^{\ord{n}}_{\hat{\mathcal{P}}}
  	+ \int_{\ell} \left( \big\lceil J^{N}{,} \Psi^{II}_{\ell} \big\rfloor^{\ord{n-1}}_{\hat{\mathcal{P}}}
  	+ 2 \big\lceil \Phi^{I}{,} \hat{\Psi}^{IN}_{\ell} \big\rfloor^{\ord{n-1}}_{\hat{\mathcal{P}}} \right)
  \\&\qquad\qquad
  	+ \int_{\ell_1 \ell_2} \hat{\Psi}^{\ord{n-2}}_{\hat{\mathcal{P}}}\big|^{NII}_{\ell_{1}\ell_{2}} 
  \Bigg]\,,
  \\
  \Theta^\ord{n}_{3}\big|^{N,P}_{\hat{\mathcal{P}}} &= \sum_{I=1}^{2} m_{I}^{2} \Bigg[
    \big\lceil \Phi^{I}{,}\Phi^{I} {,} J^{N} {,} J^{P} \big\rfloor^{\ord{n}}_{\hat{\mathcal{P}}}
  + \int_{\ell} \Big(\big\lceil \Phi^{I}{,}\Phi^{I}{,}\Psi_{\ell}^{NP} \big\rfloor^{\ord{n-1}}_{\hat{\mathcal{P}}} + \big\lceil J^{N}{,}J^{P}{,} \Psi^{II}_{\ell} \big\rfloor^{\ord{n-1}}_{\hat{\mathcal{P}}}\Big)
  \\&\qquad\qquad
  + \int_{\ell} \left( \Big\lceil \Delta^{I}_{\ell}\big\lceil J^{N}{,}J^{P} \big\rfloor, \Phi^{I} \Big\rfloor^{\ord{n-1}}_{\hat{\mathcal{P}}}
  + \Big\lceil \Delta^{P}_{\ell}\big[\Phi^{I},\Phi^{I}\big],J^{N} \Big\rfloor^{\ord{n-1}}_{\hat{\mathcal{P}}} \right)
  \\&\qquad\qquad
  + \int_{\ell_{1}\ell_{2}} \bigg[
      \big\lceil \Psi_{\ell_{1}}^{NP}, \Psi_{\ell_{2}}^{II} \big\rfloor^{\ord{n-2}}_{\hat{\mathcal{P}}}
  + 2 \big\lceil \hat{\Psi}_{\ell_{1}}^{PI}, \hat{\Psi}_{\ell_{2}}^{NI} \big\rfloor^{\ord{n-2}}_{\hat{\mathcal{P}}}
  \\&\qquad\qquad\qquad\quad
  + 2 \big\lceil \Phi^{I}, \hat{\Psi}^{NPI}_{\ell_{1}\ell_{2}} \big\rfloor^{\ord{n-2}}_{\hat{\mathcal{P}}}
  + 2 \big\lceil J^{N}, \hat{\Psi}^{PII}_{\ell_{1}\ell_{2}} \big\rfloor^{\ord{n-2}}_{\hat{\mathcal{P}}} \bigg] \Bigg]\,.
\end{aligned}\label{Theta_currents}
\end{equation}
%

\section{IBP reduction of the Two loop integrals}

In Section \ref{Sec:6}, the classical amplitudes in each sector are represented by linear combinations of the master integrals through IBP reduction. Here, we denote the explicit forms of the IBP reduction of the two-loop integrands.

\subsection{$\Box\!\Box$ class}

The IBP reduction for each $I^{\rm class}_{i}$

\subsubsection{III sector}\label{App:C.1.1}
The coefficients are
{\small
\begin{equation}
\begin{aligned}
c^{\text{III}}_{1}&=\frac{m_{1}^2 m_{2}^2 \left(2 \sigma ^2-1\right) }{3 \left(\sigma ^2-1\right)^2 \epsilon ^3 (2 \epsilon -1)}
	\Big[c^{\rm III}_{1,1} (m_{1}^2+m_{2}^2) + c^{\rm III}_{1,2} m_{1} m_{2} \Big]\,,
  \\
  c^{\text{III}}_{2}&=\frac{8 m_{1}^2 m_{2}^2}{\left(\sigma ^2-1\right)^{3/2} \epsilon ^4 (2 \epsilon -1)}
  	\Big[c^{\rm III}_{2,1} (m_{1}^2+m_{2}^2) + c^{\rm III}_{2,2} m_{1} m_{2}. \Big]\,,
  \\
  c^{\text{III}}_{3}&=\frac{m_{1}^2 m_{2}^2 \left(2 \sigma ^2-1\right)^3 \left(\sigma  \left(m_{1}^2+m_{2}^2\right)+2 m_{1} m_{2}\right)}{\left(\sigma ^2-1\right)^{3/2} \epsilon ^3}
  \\
  c^{\text{III}}_{4}&=\frac{2 m_{1}^2 m_{2}^2 \left(2 \sigma ^2-1\right)}{3 \left(\sigma. ^2-1\right)^2 \epsilon ^3 (2 \epsilon -1)}
	\Big[c^{\rm III}_{4,1} (m_{1}^2+m_{2}^2) + c^{\rm III}_{4,2} m_{1} m_{2} \Big]\,,
  \\
  c^{\text{III}}_{5}&=-\frac{2 m_{1}^2 m_{2}^2 \left(2 \sigma ^2-1\right)}{3 \left(\sigma ^2-1\right)^{3/2} \epsilon ^3 (2 \epsilon -1)}
	\Big[c^{\rm III}_{5,1} (m_{1}^2+m_{2}^2) + c^{\rm III}_{5,2} m_{1} m_{2} \Big]\,,
  \\
  c^{\text{III}}_{6}&=\frac{2 m_{1}^2 m_{2}^2}{3 \left(\sigma ^2-1\right)^2 \epsilon ^3 (6 \epsilon -5) (6 \epsilon -1)}
	\Big[c^{\rm III}_{6,1} (m_{1}^2+m_{2}^2) + c^{\rm III}_{6,2} m_{1} m_{2} \Big]\,,
  \\
  c^{\text{III}}_{7}&=8 m_{1}^2 m_{2}^2 \left(2 \sigma ^2-1\right)^3 \left(\frac{m_{1}^2+2 m_{1} m_{2} \sigma +m_{2}^2}{3 \left(\sigma ^2-1\right)^2 \epsilon ^3}-\frac{m_{1}^2 m_{2}^2}{\left(\sigma ^2-1\right) \epsilon ^4 q^2}\right)\,,
  \\
  c^{\text{III}}_{8}&=0\,,
  \\
  c^{\text{III}}_{9}&=0\,,
  \\
  c^{\text{III}}_{10}&=-\frac{16 m_{1}^3 m_{2}^3 (2 \sigma^2-1) (m_{1}+m_{2}) \left((1-2 \sigma ^2)^2 \epsilon -(\sigma -1) (4 \sigma ^3-2 \sigma +1)\right)}{(\sigma -1)^{\frac{3}{2}} \epsilon ^4 q}\,,
\end{aligned}\label{}
\end{equation}
}

where
{\small
\begin{equation}
\begin{aligned}
  c^{\rm III}_{1,1} &= 6 \epsilon ^2+12 \sigma ^6 (\epsilon -1) (2 \epsilon -3)+16 \sigma ^4 (5 \epsilon -4)+\sigma ^2 (34-\epsilon  (18 \epsilon +35))+11 \epsilon -7 \,,
  \\
  c^{\rm III}_{1,2} & = 2 \sigma  (2 \epsilon -1) \left(-20 \sigma ^4 +26\sigma^2 -5 +6 \left(1-2 \sigma^2\right)^2 \epsilon \right)\,,
  \\
  c^{\rm III}_{2,1} &= \sigma \left(2 (\sigma ^2-1)+2(2 \sigma ^2-1)^3 \epsilon ^3-\left(1-2 \sigma ^2\right)^2 \left(10 \sigma ^2-9\right) \epsilon ^2+4 \sigma^2 \left(6 \sigma ^4-11 \sigma ^2+5\right) \epsilon \right)\,,
  \\
  c^{\rm III}_{2,2} & = 2  \left(2 \sigma ^2-1\right) (2 \epsilon -1) \left(\sigma ^4 \left(4 \epsilon ^2-3\right)-2 \sigma ^2 \left(2 \epsilon ^2+\epsilon -2\right)+\epsilon ^2+2 \epsilon -1\right)\,,
  \\
  c^{\rm III}_{4,1} &=  6 \epsilon ^2+12 \sigma ^6 (\epsilon -1) (2 \epsilon -3)+16 \sigma ^4 (5 \epsilon -4)+\sigma ^2 (34-\epsilon  (18 \epsilon +35))+11\epsilon -7 \,,
  \\
  c^{\rm III}_{4,2} &= 2 \sigma  (2 \epsilon -1) \left(-20 \sigma ^4+26 \sigma^2 -5 +6 \left(1-2 \sigma ^2\right)^2 \epsilon \right)\,,
  \\
  c^{\rm III}_{5,1} &= 4\sigma \left(\sigma^4 (4 (\epsilon -1) \epsilon +3)+\sigma ^2 (-4 (\epsilon -2) \epsilon -5)+(\epsilon -3) \epsilon \right)+5\sigma\,,
  \\
  c^{\rm III}_{5,2} &= 2 \left(20 \sigma ^2+4 \sigma ^4 (4 \epsilon  (\epsilon +1)-5)-8 \sigma ^2 \epsilon  (2 \epsilon +3)+4 \epsilon  (\epsilon +3)-3\right)\,,
  \\
  c^{\rm III}_{6,1} &= -108 \epsilon ^3-144 \epsilon ^2+453 \epsilon -191 +3 \sigma ^2 (\epsilon  (36 \epsilon  (5 \epsilon +7)-935)+476)
  \\&\quad
  -6 \sigma ^4 (3 \epsilon  (12 \epsilon  (3 \epsilon +11)-343)+508)
  +4 \sigma ^6 (664-3 \epsilon  (12 \epsilon  (3 \epsilon -29)+533))
  \\&\quad
  +24 \sigma ^8 (\epsilon -1) (6 \epsilon -7) (6 \epsilon -5)\,,
  \\
  c^{\rm III}_{6,2} &= 2\sigma (6 \epsilon -5) \Big(8 \sigma ^6 (6 \epsilon -5) (6 \epsilon -1) -12\sigma^{2}(4\sigma^{2}-1)
  \\&\qquad\qquad\qquad
  - 36 \sigma^{2}(2\sigma ^2-1) \epsilon(6 \epsilon -7)-36 (\epsilon -1) \epsilon +19\Big)\,.
\end{aligned}\label{}
\end{equation}
}
\subsubsection{$\overline{\rm III}$}\label{App:C.1.2}

The coefficients are
{\small
\begin{equation}
\begin{aligned}
c^{\overline{\rm III}}_{1}&=\frac{m_{1}^2 m_{2}^2 \left(2 \sigma ^2-1\right) }{3 \left(\sigma ^2-1\right)^2 \epsilon ^3 (2 \epsilon -1)}
\Big[c^{\overline{\rm III}}_{1,1} (m_{1}^2+m_{2}^2) + c^{\overline{\rm III}}_{1,2} m_{1} m_{2} \Big]\,,
\\
c^{\overline{\rm III}}_{2}&=-\frac{8 m_{1}^2 m_{2}^2}{\left(\sigma ^2-1\right)^{3/2} \epsilon ^4 (2 \epsilon -1)}
\Big[c^{\overline{\rm III}}_{2,1} (m_{1}^2+m_{2}^2) + c^{\overline{\rm III}}_{2,2} m_{1} m_{2}. \Big]\,,
\\
c^{\overline{\rm III}}_{3}&=\frac{m_{1}^2 m_{2}^2 \left(2 \sigma ^2-1\right)^3 \left(\sigma \left(m_{1}^2+m_{2}^2\right)-2 m_{1} m_{2}\right)}{\left(\sigma ^2-1\right)^{3/2} \epsilon ^3}
\\
c^{\overline{\rm III}}_{4}&=\frac{2 m_{1}^2 m_{2}^2 \left(2 \sigma ^2-1\right)}{3 \left(\sigma^2-1\right)^2 \epsilon ^3 (2 \epsilon -1)}
\Big[c^{\overline{\rm III}}_{4,1} (m_{1}^2+m_{2}^2) + c^{\overline{\rm III}}_{4,2} m_{1} m_{2} \Big]\,,
\\
c^{\overline{\rm III}}_{5}&=\frac{2 m_{1}^2 m_{2}^2 \left(2 \sigma ^2-1\right)}{3 \left(\sigma ^2-1\right)^{3/2} \epsilon ^3 (2 \epsilon -1)}
\Big[c^{\overline{\rm III}}_{5,1} (m_{1}^2+m_{2}^2) + c^{\overline{\rm III}}_{5,2} m_{1} m_{2} \Big]\,,
\\
c^{\overline{\rm III}}_{6}&=\frac{2 m_{1}^2 m_{2}^2}{3 \left(\sigma ^2-1\right)^2 \epsilon ^3 (6 \epsilon -5) (6 \epsilon -1)}
\Big[c^{\overline{\rm III}}_{6,1} (m_{1}^2+m_{2}^2) + c^{\overline{\rm III}}_{6,2} m_{1} m_{2} \Big]\,,
\\
c^{\overline{\rm III}}_{7}&=8 m_{1}^2 m_{2}^2 \left(2 \sigma ^2-1\right)^3 \left(\frac{m_{1}^2-2 m_{1} m_{2} \sigma +m_{2}^2}{3 \left(\sigma ^2-1\right)^2 \epsilon ^3}-\frac{m_{1}^2 m_{2}^2}{\left(\sigma ^2-1\right) \epsilon ^4 q^2}\right)\,,
\\
c^{\overline{\rm III}}_{8}&=0\,,
\\
c^{\overline{\rm III}}_{9}&=0\,,
\\
c^{\overline{\rm III}}_{10}&=\frac{16 m_{1}^3 m_{2}^3 (2 \sigma^2-1) (m_{1}+m_{2}) \left((1-2 \sigma ^2)^2 \epsilon -(\sigma +1) (4 \sigma ^3-2 \sigma -1)\right)}{(\sigma -1)^{\frac{3}{2}} \epsilon ^4 q}\,,
\end{aligned}\label{}
\end{equation}
}

where
{\small
\begin{equation}
\begin{aligned}
c^{\overline{\rm III}}_{1,1} &= 6 \left(4 \sigma ^6-3 \sigma ^2+1\right) \epsilon ^2+\left(-60 \sigma ^6+80 \sigma ^4-35 \sigma ^2+11\right) \epsilon + 36 \sigma ^6-64 \sigma ^4+34 \sigma ^2 -7 \,,
\\
c^{\overline{\rm III}}_{1,2} & = -2 \sigma (2 \epsilon -1) \left(6 \left(1-2 \sigma^2\right)^2 \epsilon -20 \sigma ^4 +26\sigma^2 -5 \right)\,,
\\
c^{\overline{\rm III}}_{2,1} &= -\sigma \left[2(2 \sigma^2{-}1)^3 \epsilon^3-\left(2\sigma^2{-}1\right)^2 \left(10 \sigma^2 {-}9\right) \epsilon ^2+4 \sigma^2 \left(6 \sigma ^4-11 \sigma ^2+5\right) \epsilon + 2 (\sigma ^2-1) \right]\,,
\\
c^{\overline{\rm III}}_{2,2} & = 2 \left(2 \sigma ^2-1\right) (2 \epsilon -1) \left(\left(1-2 \sigma ^2\right)^2 \epsilon ^2+\left(2-2 \sigma ^2\right) \epsilon -3 \sigma ^4+4 \sigma ^2 -1\right)\,,
\\
c^{\overline{\rm III}}_{4,1} &= 6 \left(4 \sigma ^6-3 \sigma ^2+1\right) \epsilon ^2 -\left(60 \sigma^6 -80 \sigma^4 +35 \sigma^2 -11\right) \epsilon + 36 \sigma ^6-64 \sigma ^4+34 \sigma ^2 -7 \,,
\\
c^{\overline{\rm III}}_{4,2} &= -2 \sigma (2 \epsilon -1) \left(6 \left(1-2 \sigma ^2\right)^2 \epsilon -20 \sigma ^4+26 \sigma^2 -5 \right)\,,
\\
c^{\overline{\rm III}}_{5,1} &=-\sigma \big( 4 \left(1-2 \sigma ^2\right)^2 \epsilon ^2-4 \left(4 \sigma ^4-8 \sigma ^2+3\right) \epsilon + \left(12 \sigma ^4-20 \sigma ^2+5\right) \big) \,,
\\
c^{\overline{\rm III}}_{5,2} &=8 \left(1-2 \sigma ^2\right)^2 \epsilon ^2+8 \left(4 \sigma ^4-6 \sigma ^2+3\right) \epsilon -40 \sigma ^4+40 \sigma ^2 -6\,,
\\
c^{\overline{\rm III}}_{6,1} &= 108 \left(\sigma ^2+1\right) \left(2 \sigma ^2-1\right)^3 \epsilon ^3
-36 \left(72 \sigma ^8-116 \sigma ^6+66 \sigma ^4-21 \sigma ^2+4\right) \epsilon ^2
\\
&\quad
+3 \left(856 \sigma ^8-2132 \sigma ^6+2058 \sigma ^4-935 \sigma ^2+151\right) \epsilon
\\
&\quad
+4 \sigma ^2 \left(-210 \sigma ^6+664 \sigma ^4-762 \sigma ^2+357\right)-191 \,,
\\
c^{\overline{\rm III}}_{6,2} &=-2 \sigma (6 \epsilon -5) \Big[ 36(2 \sigma^2{-}1)^3 \epsilon^2-36 \left(8 \sigma^6 {-}14 \sigma^4 {+}7\sigma ^2{-}1\right) \epsilon + 40 \sigma^6 -48 \sigma^4 -12 \sigma^2 +19\Big]\,.
\end{aligned}\label{}
\end{equation}
}

\subsubsection{IX}\label{App:C.1.3}
The coefficients are
{\small
\begin{equation}
\begin{aligned}
c^{\text{IX}}_{1}&=-\frac{m_{1}^2 m_{2}^2 \left(2 \sigma ^2-1\right)}{6 \left(\sigma ^2-1\right)^2 \epsilon ^4 (2 \epsilon -1)}
\Big[c^{\rm IX}_{1,1} (m_{1}^2+m_{2}^2) + c^{\rm IX}_{1,2} m_{1} m_{2} \Big]\,,
\\
c^{\text{IX}}_{2}&=\frac{2 m_{1}^2 m_{2}^2}{\left(\sigma ^2-1\right)^{3/2} \epsilon ^4 (2 \epsilon -1)}
\Big[c^{\rm IX}_{2,1} (m_{1}^2+m_{2}^2) + c^{\rm IX}_{2,2} m_{1} m_{2}. \Big]\,,
\\
c^{\text{IX}}_{3}&=-\frac{m_{1}^2 m_{2}^2\left(2 \sigma ^2-1\right)}{2 \left(\sigma ^2-1\right)^{3/2} \epsilon ^3 (2 \epsilon -1)}
\Big[c^{\rm IX}_{3,1} (m_{1}^2+m_{2}^2) + c^{\rm IX}_{3,2} m_{1} m_{2}. \Big]
\\
c^{\text{IX}}_{4}&=\frac{m_{1}^2 m_{2}^2}{6 \left(\sigma ^2-1\right)^2 \epsilon ^4 (2 \epsilon -1)}
\Big[c^{\rm IX}_{4,1} (m_{1}^2+m_{2}^2) + c^{\rm IX}_{4,2} m_{1} m_{2} \Big]\,,
\\
c^{\text{IX}}_{5}&=\frac{2 m_{1}^2 m_{2}^2}{\left(\sigma ^2-1\right)^{3/2} \epsilon ^4 (2 \epsilon -1)}
\Big[c^{\rm IX}_{5,1} (m_{1}^2+m_{2}^2) + c^{\rm IX}_{5,2} m_{1} m_{2} \Big]\,,
\\
c^{\text{IX}}_{6}&=-\frac{m_{1}^2 m_{2}^2 \left(2 \sigma ^2-1\right)}{\left(\sigma ^2-1\right)^{3/2} \epsilon ^3 (2 \epsilon -1)}
\Big[c^{\rm IX}_{6,1} (m_{1}^2+m_{2}^2) + c^{\rm IX}_{6,2} m_{1} m_{2} \Big]\,,
\\
c^{\text{IX}}_{7}&= \frac{m_{1}^2 m_{2}^2}{6 \left(\sigma ^2-1\right)^2 \epsilon ^4 (2 \epsilon -1)}
\Big[c^{\rm IX}_{7,1} (m_{1}^2+m_{2}^2) + c^{\rm IX}_{7,2} m_{1} m_{2} \Big]\,,
\\
c^{\text{IX}}_{8}&=-\frac{m_{1}^2 m_{2}^2 \left(2 \sigma ^2-1\right)}{6 \left(\sigma ^2-1\right)^2 \epsilon ^4 (6 \epsilon -1)}
\Big[c^{\rm IX}_{8,1} (m_{1}^2+m_{2}^2) + c^{\rm IX}_{8,2} m_{1} m_{2} \Big]\,,
\\
c^{\text{IX}}_{9}&=\frac{m_{1}^2 m_{2}^2 \left(2 \sigma ^2-1\right)}{3 \left(\sigma ^2-1\right)^{3/2} \epsilon ^3 (2 \epsilon -1)}
\Big[c^{\rm IX}_{9,1} (m_{1}^2+m_{2}^2) + c^{\rm IX}_{9,2} m_{1} m_{2} \Big]\,,
\\
c^{\text{IX}}_{10}&=4 m_{1}^2 m_{2}^2 \left(2  \sigma ^2 - 1 \right)^2
\Big[ \frac{2 m_{1}^2 m_{2}^2 \left(1-2 \sigma ^2\right)}{ \left(\sigma ^2-1\right) \epsilon ^4 q^2} + \frac{c^{\rm IX}_{10,1}  \left(m_{1}^2+m_{2}^2\right)+c^{\rm IX}_{10,2} m_{1} m_{2}}{3 \left(\sigma ^2-1\right)^2 \epsilon ^4} \Big]\,,
\\
c^{\text{IX}}_{11}&=0\,,
\\
c^{\text{IX}}_{12}&=\frac{4 m_{1}^3 m_{2}^3 \left(2 \sigma ^2-1\right)^3 (m_{1}+m_{2})}{\left(\sigma ^2-1\right) \epsilon ^3 q}\,,
\\
c^{\text{IX}}_{13}&=-\frac{4 m_{1}^3 m_{2}^3 \left(2 \sigma ^2-1\right)^3 (m_{1}+m_{2})}{\left(\sigma ^2-1\right) \epsilon ^3 q}\,,
\\
c^{\text{IX}}_{14}&=-\frac{4 m_{1}^3 m_{2}^3 \left(2 \sigma ^2-1\right) (m_{1}+m_{2})}{\left(\sigma ^2-1\right)^{3/2} \epsilon ^4 q}
c^{\rm IX}_{14,1} \,,
\\
c^{\text{IX}}_{15}&=\frac{8 m_{1}^3 m_{2}^3 \left(1-2 \sigma ^2\right)^2 (m_{1}+m_{2}) \left(\left(2 \sigma ^2-1\right) \epsilon -2 \sigma  (\sigma +1)\right)}{(\sigma +1) \sqrt{\sigma ^2-1} \epsilon ^4 q}\,.
\end{aligned}\label{}
\end{equation}
}
\newpage
where
{\small
\begin{equation}
\begin{aligned}
c^{\rm IX}_{1,1} &= \epsilon  \Big[6 \left(4 \sigma ^6-3 \sigma ^2+1\right) \epsilon ^2 -\left(60 \sigma^6-80 \sigma^4 +35 \sigma^2 -11\right) \epsilon + 36 \sigma ^6-64 \sigma ^4+34 \sigma ^2 -7\Big] \,,
\\
c^{\rm IX}_{1,2} & = 4 \sigma  \left(2 \sigma ^2-1\right) (2 \epsilon -1) \big(\left(12 \sigma ^2-6\right) \epsilon ^2+\left(23-16 \sigma ^2\right) \epsilon -6 \sigma ^2 +7\big) \,,
\\
c^{\rm IX}_{2,1} &=\sigma  \big(2 \left(2 \sigma ^2-1\right)^3 \epsilon ^3+\left(14 \sigma ^2-15\right) \left(1-2 \sigma ^2\right)^2 \epsilon ^2
\\
&\qquad\;\;
+\left(-96 \sigma ^6+152 \sigma ^4-64 \sigma ^2+8\right) \epsilon + 2 \left(\sigma ^2-1\right) \big) \,,
\\
c^{\rm IX}_{2,2} & =4 \big(2 \left(2 \sigma ^2-1\right)^3 \epsilon ^3+\left(8 \sigma ^4-10 \sigma ^2+3\right) \sigma ^2 \epsilon ^2
\\
&\qquad\;\;
+\left(-60 \sigma ^6+94 \sigma ^4-40 \sigma ^2+5\right) \epsilon + 26 \sigma ^6-43 \sigma ^4+19 \sigma ^2 -2\big) \,,
\\
c^{\rm IX}_{3,1} &=\sigma  \left(1-2 \sigma ^2\right)^2 \left(\left(24 \sigma ^2-12\right) \epsilon ^2+\left(28-24 \sigma ^2\right) \epsilon -2 \sigma ^2 +1\right)  \,,
  \\
  c^{\rm IX}_{3,2} & =2 \left(2 \sigma ^2-1\right) \left[\left(8 \sigma ^4-12 \sigma ^2+6\right) \epsilon -\left(1-2 \sigma ^2\right)^2\right] \,,
  \\
  c^{\rm IX}_{4,1} &= \epsilon \left(2 \sigma ^2-1\right)   \Big[12 \left(2 \sigma ^2-1\right)^3 \epsilon^2 -4 \left(4 \sigma^4 -\sigma^2 -2\right) \epsilon -144 \sigma ^6+296 \sigma ^4-158 \sigma ^2 +5\Big] \,,
  \\
  c^{\rm IX}_{4,2} &= 4 \sigma  \Big[12 \left(2 \sigma ^2-1\right)^3 \epsilon ^3-8 \left(1-2 \sigma ^2\right)^2 \left(4 \sigma ^2-5\right) \epsilon ^2
  \\&\qquad\;\;
  +\left(-8 \sigma ^6-16 \sigma ^4+34 \sigma ^2-15\right) \epsilon + \left(6 \sigma ^2-7\right) \left(1-2 \sigma ^2\right)^2 \Big] \,,
  \\
  c^{\rm IX}_{5,1} &= \sigma  \Big[-2 \left(2 \sigma ^2-1\right)^3 \epsilon ^3+\left(34 \sigma ^2-33\right) \left(1-2 \sigma ^2\right)^2 \epsilon ^2
  \\&\qquad\;\;
  -8\left( 18\sigma^6 -30 \sigma^4 +13 \sigma^2 -1\right) \epsilon -2 \sigma^2 +2\Big] \,,
  \\
  c^{\rm IX}_{5,2} &= 4 \Big[-2 \left(2 \sigma ^2-1\right)^3 \epsilon ^3+\left(-8 \sigma ^6+14 \sigma ^4-9 \sigma ^2+2\right) \epsilon ^2
  \\&\qquad\;\;
  +\left(64 \sigma ^6-104 \sigma ^4+44 \sigma ^2-3\right) \epsilon -28 \sigma ^6+44 \sigma ^4-17 \sigma ^2 +1\Big] \,,
\\
c^{\rm IX}_{6,1} &=\sigma  \Big[\left(12 \sigma ^2-6\right) \epsilon ^2+\left(13-10 \sigma ^2\right) \epsilon -2 \sigma ^2 +1\Big] \,,
\\
c^{\rm IX}_{6,2} &=8 \left(1-2 \sigma ^2\right)^2 \epsilon ^2+\left(4-4 \sigma ^2\right) \epsilon -2 \left(1-2 \sigma ^2\right)^2 \,,
\\
  c^{\rm IX}_{7,1} &= \epsilon (2 \sigma^2-1)  \Big[-12 \left(1-2 \sigma ^2\right)^2 \left(\sigma ^2-2\right) \epsilon ^2
  \\
  &\qquad\qquad\qquad\
    -2 \left(60 \sigma^6 -88 \sigma^4+37 \sigma^2 -7\right) \epsilon + 216 \sigma ^6-424 \sigma ^4+226 \sigma ^2 -19\Big] \,,
  \\
  c^{\rm IX}_{7,2} &= 4 \sigma  \Big[12 \left(2 \sigma ^2-1\right)^3 \epsilon ^3-8 \left(1-2 \sigma ^2\right)^2 \left(7 \sigma ^2-8\right) \epsilon ^2
  \\
  &\qquad\quad\;\;
  +\left(40 \sigma ^6-88 \sigma ^4+46 \sigma ^2-3\right) \epsilon + \left(6 \sigma ^2-7\right) \left(1-2 \sigma ^2\right)^2 \Big] \,,
  \\
  c^{\rm IX}_{8,1} &= \epsilon \Big[36 \left(\sigma ^2-2\right) \left(1-2 \sigma ^2\right)^2 \epsilon ^2
  \\
  &\qquad\;\;
  -6 \left(52 \sigma ^6-80 \sigma ^4+23 \sigma ^2+5\right) \epsilon + 168 \sigma ^6-320 \sigma ^4+158 \sigma ^2-5\Big] \,,
  \\
  c^{\rm IX}_{8,2} &=-4 \sigma  \left(2 \sigma ^2-1\right) (6 \epsilon -1) \Big[\left(12 \sigma ^2-6\right) \epsilon ^2+\left(2 \sigma ^2+5\right) \epsilon -6 \sigma ^2 +7\Big]  \,,
  \\
  c^{\rm IX}_{9,1} &= \sigma  \left[4 \left(1-2 \sigma ^2\right)^2 \epsilon ^2-4 \left(4 \sigma ^4-8 \sigma ^2+3\right) \epsilon + 12 \sigma ^4-20 \sigma ^2+5\right] \,,
  \\
  c^{\rm IX}_{9,2} &= -8 \left(1-2 \sigma ^2\right)^2 \epsilon ^2+8 \left(1-2 \sigma ^2\right)^2 \epsilon + 8 \sigma ^4-8 \sigma ^2 -2 \,,
  \\
  c^{\rm IX}_{10,1} &= 2 \left(2 \sigma ^2-1\right) \epsilon \,,
  \\
  c^{\rm IX}_{10,2} &= \sigma  \left[\left(4 \sigma ^2-2\right) \epsilon +6 \sigma ^2 -7\right] \,,
  \\
  c^{\rm IX}_{14,1} &= 2 (2 \sigma +1) \left(1-2 \sigma ^2\right)^2 \epsilon - 12 \sigma ^5+20 \sigma ^3 -\sigma^2 -7 \sigma+1 \,.
\end{aligned}\label{}
\end{equation}
}

\newpage
\subsubsection{$\overline{\rm IX}$}\label{App:C.1.4}

The coefficients are
{\small
\begin{equation}
\begin{aligned}
c^{\overline{\rm IX}}_{1}&=-\frac{m_{1}^2 m_{2}^2 \left(2 \sigma ^2-1\right)}{6 \left(\sigma ^2-1\right)^2 \epsilon ^4 (2 \epsilon -1)}
\Big[c^{\overline{\rm IX}}_{1,1} (m_{1}^2+m_{2}^2) + c^{\overline{\rm IX}}_{1,2} m_{1} m_{2} \Big]\,,
\\
c^{\overline{\rm IX}}_{2}&=-\frac{2 m_{1}^2 m_{2}^2}{\left(\sigma ^2-1\right)^{3/2} \epsilon ^4 (2 \epsilon -1)}
\Big[c^{\overline{\rm IX}}_{2,1} (m_{1}^2+m_{2}^2) + c^{\overline{\rm IX}}_{2,2} m_{1} m_{2}. \Big]\,,
\\
c^{\overline{\rm IX}}_{3}&=\frac{m_{1}^2 m_{2}^2\left(2 \sigma ^2-1\right)}{2 \left(\sigma ^2-1\right)^{3/2} \epsilon ^3 (2 \epsilon -1)}
\Big[c^{\overline{\rm IX}}_{3,1} (m_{1}^2+m_{2}^2) + c^{\overline{\rm IX}}_{3,2} m_{1} m_{2}. \Big]
\\
c^{\overline{\rm IX}}_{4}&=\frac{m_{1}^2 m_{2}^2}{6 \left(\sigma ^2-1\right)^2 \epsilon ^4 (2 \epsilon -1)}
\Big[c^{\overline{\rm IX}}_{4,1} (m_{1}^2+m_{2}^2) + c^{\overline{\rm IX}}_{4,2} m_{1} m_{2} \Big]\,,
\\
c^{\overline{\rm IX}}_{5}&=-\frac{2 m_{1}^2 m_{2}^2}{\left(\sigma ^2-1\right)^{3/2} \epsilon ^4 (2 \epsilon -1)}
\Big[c^{\overline{\rm IX}}_{5,1} (m_{1}^2+m_{2}^2) + c^{\overline{\rm IX}}_{5,2} m_{1} m_{2} \Big]\,,
\\
c^{\overline{\rm IX}}_{6}&=\frac{m_{1}^2 m_{2}^2 \left(2 \sigma ^2-1\right)}{\left(\sigma ^2-1\right)^{3/2} \epsilon ^3 (2 \epsilon -1)}
\Big[c^{\overline{\rm IX}}_{6,1} (m_{1}^2+m_{2}^2) + c^{\overline{\rm IX}}_{6,2} m_{1} m_{2} \Big]\,,
\\
c^{\overline{\rm IX}}_{7}&= \frac{m_{1}^2 m_{2}^2}{6 \left(\sigma ^2-1\right)^2 \epsilon ^4 (2 \epsilon -1)}
\Big[c^{\overline{\rm IX}}_{7,1} (m_{1}^2+m_{2}^2) + c^{\overline{\rm IX}}_{7,2} m_{1} m_{2} \Big]\,,
\\
c^{\overline{\rm IX}}_{8}&=-\frac{m_{1}^2 m_{2}^2 \left(2 \sigma ^2-1\right)}{6 \left(\sigma ^2-1\right)^2 \epsilon ^4 (6 \epsilon -1)}
\Big[c^{\overline{\rm IX}}_{8,1} (m_{1}^2+m_{2}^2) + c^{\overline{\rm IX}}_{8,2} m_{1} m_{2} \Big]\,,
\\
c^{\overline{\rm IX}}_{9}&=-\frac{m_{1}^2 m_{2}^2 \left(2 \sigma ^2-1\right)}{3 \left(\sigma ^2-1\right)^{3/2} \epsilon ^3 (2 \epsilon -1)}
\Big[c^{\overline{\rm IX}}_{9,1} (m_{1}^2+m_{2}^2) + c^{\overline{\rm IX}}_{9,2} m_{1} m_{2} \Big]\,,
\\
c^{\overline{\rm IX}}_{10}&=4 m_{1}^2 m_{2}^2 \left(2 \sigma ^2 - 1 \right)^2
\Big[ \frac{2 m_{1}^2 m_{2}^2 \left(1-2 \sigma ^2\right)}{ \left(\sigma ^2-1\right) \epsilon ^4 q^2} + \frac{c^{\overline{\rm IX}}_{10,1} \left(m_{1}^2+m_{2}^2\right)+c^{\overline{\rm IX}}_{10,2} m_{1} m_{2}}{3 \left(\sigma ^2-1\right)^2 \epsilon ^4} \Big]\,,
\\
c^{\overline{\rm IX}}_{11}&=0\,,
\\
c^{\overline{\rm IX}}_{12}&=\frac{4 m_{1}^3 m_{2}^3 \left(2 \sigma ^2-1\right)^3 (m_{1}+m_{2})}{\left(\sigma ^2-1\right) \epsilon ^3 q}\,,
\\
c^{\overline{\rm IX}}_{13}&=-\frac{4 m_{1}^3 m_{2}^3 \left(2 \sigma ^2-1\right)^3 (m_{1}+m_{2})}{\left(\sigma ^2-1\right) \epsilon ^3 q}\,,
\\
c^{\overline{\rm IX}}_{14}&=-\frac{4 m_{1}^3 m_{2}^3 \left(2 \sigma ^2-1\right) (m_{1}+m_{2})}{\left(\sigma ^2-1\right)^{3/2} \epsilon ^4 q}
c^{\overline{\rm IX}}_{14,1} \,,
\\
c^{\overline{\rm IX}}_{15}&=\frac{8 m_{1}^3 m_{2}^3 \left(1-2 \sigma ^2\right)^2 (m_{1}+m_{2}) \left(\left(2 \sigma ^2-1\right) \epsilon -2 \sigma (\sigma -1)\right)}{(\sigma -1) \sqrt{\sigma ^2-1} \epsilon ^4 q}\,.
\end{aligned}\label{}
\end{equation}
}
\newpage

where
{\small
\begin{equation}
\begin{aligned}
c^{\overline{\rm IX}}_{1,1} &= \epsilon \left(6 \left(4 \sigma ^6-3 \sigma ^2+1\right) \epsilon ^2+\left(-60 \sigma ^6+80 \sigma ^4-35 \sigma ^2+11\right) \epsilon + 36 \sigma ^6-64 \sigma ^4+34 \sigma ^2 -7\right) \,,
\\
c^{\overline{\rm IX}}_{1,2} & = -4 \sigma \left(2 \sigma ^2-1\right) (2 \epsilon -1) \big(\left(12 \sigma ^2-6\right) \epsilon ^2+\left(23-16 \sigma ^2\right) \epsilon -6 \sigma ^2 +7\big) \,,
\\
c^{\overline{\rm IX}}_{2,1} &=-\sigma \big(2 \left(2 \sigma ^2-1\right)^3 \epsilon ^3+\left(14 \sigma ^2-15\right) \left(1-2 \sigma ^2\right)^2 \epsilon ^2
\\
&\qquad\;\;
+\left(-96 \sigma ^6+152 \sigma ^4-64 \sigma ^2+8\right) \epsilon + 2 \left(\sigma ^2-1\right) \big) \,,
\\
c^{\overline{\rm IX}}_{2,2} & =4 \big(2 \left(2 \sigma ^2-1\right)^3 \epsilon ^3+\left(8 \sigma ^4-10 \sigma ^2+3\right) \sigma ^2 \epsilon ^2
\\
&\qquad\;\;
+\left(-60 \sigma ^6+94 \sigma ^4-40 \sigma ^2+5\right) \epsilon + 26 \sigma ^6-43 \sigma ^4+19 \sigma ^2 -2\big) \,,
\\
c^{\overline{\rm IX}}_{3,1} &=-\sigma \left(1-2 \sigma ^2\right)^2 \left(\left(24 \sigma ^2-12\right) \epsilon ^2+\left(28-24 \sigma ^2\right) \epsilon -2 \sigma ^2 +1\right) \,,
\\
c^{\overline{\rm IX}}_{3,2} & =2 \left(2 \sigma ^2-1\right) \left(\left(8 \sigma ^4-12 \sigma ^2+6\right) \epsilon -\left(1-2 \sigma ^2\right)^2\right) \,,
\\
c^{\overline{\rm IX}}_{4,1} &= \epsilon \left(2 \sigma ^2-1\right) \big(12 \left(2 \sigma ^2-1\right)^3 \epsilon ^2+4 \left(-4 \sigma ^4+\sigma ^2+2\right) \epsilon -144 \sigma ^6+296 \sigma ^4-158 \sigma ^2 +5\big) \,,
\\
c^{\overline{\rm IX}}_{4,2} &= -4 \sigma \big(12 \left(2 \sigma ^2-1\right)^3 \epsilon ^3-8 \left(1-2 \sigma ^2\right)^2 \left(4 \sigma ^2-5\right) \epsilon ^2
\\
&\qquad\;\;
+\left(-8 \sigma ^6-16 \sigma ^4+34 \sigma ^2-15\right) \epsilon + \left(6 \sigma ^2-7\right) \left(1-2 \sigma ^2\right)^2 \big) \,,
\\
c^{\overline{\rm IX}}_{5,1} &= -\sigma \big(-2 \left(2 \sigma ^2-1\right)^3 \epsilon ^3+\left(34 \sigma ^2-33\right) \left(1-2 \sigma ^2\right)^2 \epsilon ^2
\\
&\qquad\;\;
+\left(-144 \sigma ^6+240 \sigma ^4-104 \sigma ^2+8\right) \epsilon -2 \sigma ^2 +2\big) \,,
\\
c^{\overline{\rm IX}}_{5,2} &= 4 \big(-2 \left(2 \sigma ^2-1\right)^3 \epsilon ^3+\left(-8 \sigma ^6+14 \sigma ^4-9 \sigma ^2+2\right) \epsilon ^2
\\
&\qquad\;\;
+\left(64 \sigma ^6-104 \sigma ^4+44 \sigma ^2-3\right) \epsilon -28 \sigma ^6+44 \sigma ^4-17 \sigma ^2 +1\big) \,,
\\
c^{\overline{\rm IX}}_{6,1} &=-\sigma \left(\left(12 \sigma ^2-6\right) \epsilon ^2+\left(13-10 \sigma ^2\right) \epsilon -2 \sigma ^2 +1\right) \,,
\\
c^{\overline{\rm IX}}_{6,2} &=8 \left(1-2 \sigma ^2\right)^2 \epsilon ^2+\left(4-4 \sigma ^2\right) \epsilon -2 \left(1-2 \sigma ^2\right)^2 \,,
\\
c^{\overline{\rm IX}}_{7,1} &= \epsilon \left(2 \sigma ^2-1\right) \big(-12 \left(1-2 \sigma ^2\right)^2 \left(\sigma ^2-2\right) \epsilon ^2
\\
&\qquad\qquad\qquad\quad
+2 \left(-60 \sigma ^6+88 \sigma ^4-37 \sigma ^2+7\right) \epsilon + 216 \sigma ^6-424 \sigma ^4+226 \sigma ^2 -19\big) \,,
\\
c^{\overline{\rm IX}}_{7,2} &= -4 \sigma \big(12 \left(2 \sigma ^2-1\right)^3 \epsilon ^3-8 \left(1-2 \sigma ^2\right)^2 \left(7 \sigma ^2-8\right) \epsilon ^2
\\
&\qquad\quad\;\;
+\left(40 \sigma ^6-88 \sigma ^4+46 \sigma ^2-3\right) \epsilon + \left(6 \sigma ^2-7\right) \left(1-2 \sigma ^2\right)^2 \big) \,.
\\
c^{\overline{\rm IX}}_{8,1} &= \epsilon \big(36 \left(\sigma ^2-2\right) \left(1-2 \sigma ^2\right)^2 \epsilon ^2
\\
&\qquad\;\;
-6 \left(52 \sigma ^6-80 \sigma ^4+23 \sigma ^2+5\right) \epsilon + 168 \sigma ^6-320 \sigma ^4+158 \sigma ^2-5\big) \,,
\\
c^{\overline{\rm IX}}_{8,2} &=4 \sigma \left(2 \sigma ^2-1\right) (6 \epsilon -1) \left(\left(12 \sigma ^2-6\right) \epsilon ^2+\left(2 \sigma ^2+5\right) \epsilon -6 \sigma ^2 +7\right) \,,
\\
c^{\overline{\rm IX}}_{9,1} &=- \sigma \left(4 \left(1-2 \sigma ^2\right)^2 \epsilon ^2-4 \left(4 \sigma ^4-8 \sigma ^2+3\right) \epsilon + 12 \sigma ^4-20 \sigma ^2+5\right) \,,
\\
c^{\overline{\rm IX}}_{9,2} &= -8 \left(1-2 \sigma ^2\right)^2 \epsilon ^2+8 \left(1-2 \sigma ^2\right)^2 \epsilon + 8 \sigma ^4-8 \sigma ^2 -2 \,,
\\
c^{\overline{\rm IX}}_{10,1} &= 2 \left(2 \sigma ^2-1\right) \epsilon \,,
\\
c^{\overline{\rm IX}}_{10,2} &= -\sigma \left(\left(4 \sigma ^2-2\right) \epsilon +6 \sigma ^2 -7\right) \,,
\\
c^{\overline{\rm IX}}_{14,1} &= 2 (2 \sigma +1) \left(1-2 \sigma ^2\right)^2 \epsilon + 12 \sigma ^5-20 \sigma ^3 -\sigma^2 +7 \sigma+1 \,.
\end{aligned}\label{}
\end{equation}
}
%

\subsection{\textbf{H}-class}

The \textbf{H} class consists of two sectors, H and $\overline{\rm H}$. We present their coefficients of the IBP reduction for the loop integrands. The master integrals are decomposed into the even and the odd terms in $q$, and the odd terms do not contribute to the amplitudes. Thus we focus on the even terms only.

\subsubsection{{\rm H} sector}\label{App:C.2.1}

The coefficients of the IBP reduction of the amplitude for the H sector defined in \eqref{IBP_coeff_H} are given by
{\small
\begin{equation}
\begin{aligned}
c^{\text{H}}_{1}&=\frac{2 m_{1}^3 m_{2}^3 \sigma }{(\epsilon -1) \epsilon ^4 (4 \epsilon ^2-8 \epsilon +3)^2 (9 \epsilon ^2-9 \epsilon +2)}
c^{\rm H}_{1,1}\,,
\\
c^{\text{H}}_{2}&=-\frac{8 m_{1}^3 m_{2}^3 \sigma }{(\epsilon -1) \epsilon ^3 (2 \epsilon -3) (2 \epsilon -1) (4 \epsilon -1)}
c^{\rm H}_{2,1}\,,
\\
c^{\text{H}}_{3}&=-\frac{4 m_{1}^3 m_{2}^3 \sigma }{\epsilon ^4 (4 \epsilon ^2-8 \epsilon +3)^2}
c^{\rm H}_{3,1}
\\
c^{\text{H}}_{4}&=\frac{m_{1}^3 m_{2}^3 }{8 (\epsilon -1)^2 \epsilon ^4 (4 \epsilon ^2-8 \epsilon +3)}
c^{\rm H}_{4,1}\,,
\\
c^{\text{H}}_{5}&=\frac{32 m_{1}^3 m_{2}^3 }{\sqrt{\sigma ^2-1} (\epsilon -1) \epsilon ^4 (2 \epsilon -3) (2 \epsilon -1)}
c^{\rm H}_{5,1}\,,
\\
c^{\text{H}}_{6}&=\frac{8 m_{1}^3 m_{2}^3 }{\sqrt{\sigma ^2-1} (\epsilon -1) \epsilon ^4 (2 \epsilon -3) (2 \epsilon -1)}
c^{\rm H}_{6,1}\,,
\\
c^{\text{H}}_{7}&=\frac{32 m_{1}^3 m_{2}^3 \sigma  \left(\left(4 \sigma ^2-10\right) \epsilon ^2+\left(15-8 \sigma ^2\right) \epsilon -6\right)}{(\epsilon -1) \epsilon ^3 (2 \epsilon -3) (2 \epsilon -1)}\,,
\\
c^{\text{H}}_{8}&=\frac{8 m_{1}^3 m_{2}^3}{\sqrt{\sigma ^2-1} (\epsilon -1) \epsilon ^4 (2 \epsilon -3) (2 \epsilon -1) (4 \epsilon -1)} c^{\rm H}_{8,1}\,,
\\
c^{\text{H}}_{9}&=\frac{4 m_{1}^3 m_{2}^3 \left(-3 \left(1-2 \sigma ^2\right)^2+4 \left(4 \sigma ^4-3 \sigma ^2-2\right) \epsilon ^2+\left(-12 \sigma ^4+8 \sigma ^2+12\right) \epsilon\right)}{\sqrt{\sigma ^2-1} \epsilon ^4 (4 \epsilon ^2-8 \epsilon +3)}\,,
\\
c^{\text{H}}_{10}&=-\frac{8 m_{1}^3 m_{2}^3 \sigma  \left(4 \left(4 \sigma ^2+3\right) \epsilon ^2+\left(-22 \sigma ^2-19\right) \epsilon +6\right)}{\epsilon ^4 (4 \epsilon ^2-8 \epsilon +3)}\,.
\end{aligned}\label{}
\end{equation}
}
where
{\small
\begin{equation}
\begin{aligned}
c^{\rm H}_{1,1} &=-96 \left(3 \sigma ^2+37\right) \epsilon ^7+8 \left(187 \sigma ^2+2117\right) \epsilon ^6-60 \left(56 \sigma ^2+549\right) \epsilon ^5+38 \left(97 \sigma ^2+897\right) \epsilon ^4
\\
&\quad
+\left(-1564 \sigma ^2-20197\right) \epsilon ^3+\left(6773-18 \sigma ^2\right) \epsilon ^2+6 \left(18 \sigma ^2-193\right) \epsilon +72 \,,
\\
c^{\rm H}_{2,1} &= \left(36 \sigma ^4+60 \sigma ^2-4\right) \epsilon ^3+\left(-92 \sigma ^4-146 \sigma ^2+2\right) \epsilon ^2
\\
&\quad
+\left(56 \sigma ^4+142 \sigma ^2-9\right) \epsilon -54 \sigma ^2 +9\,,
\\
c^{\rm H}_{3,1} &=8 \left(9 \sigma ^2+4\right) \epsilon ^4-4 \left(68 \sigma ^2+41\right) \epsilon ^3+\left(258 \sigma ^2+224\right) \epsilon ^2-3 \left(18 \sigma ^2+35\right) \epsilon +18 \,,
\\
c^{\rm H}_{4,1} &=\left(3328 \sigma ^2+368\right) \epsilon ^4-8 \left(1424 \sigma ^2+185\right) \epsilon ^3
\\
&\quad
+4 \left(3377 \sigma ^2+502\right) \epsilon ^2-2 \left(3060 \sigma ^2+539\right) \epsilon +  675 \sigma ^2 +183\,,
\\
c^{\rm H}_{5,1} &=\left(4 \sigma ^4-23 \sigma ^2-21\right) \epsilon ^3+\left(14 \sigma ^2+46\right) \epsilon ^2+\left(-22 \sigma ^4-\sigma ^2-31\right) \epsilon + 6 \sigma ^4-3 \sigma ^2 +6 \,,
\\
c^{\rm H}_{6,1} &= 3 \left(1-2 \sigma ^2\right)^2+4 \left(6 \sigma ^4-\sigma ^2-6\right) \epsilon ^3+\left(-50 \sigma ^4+12 \sigma ^2+50\right) \epsilon ^2+\left(22 \sigma ^4-4 \sigma ^2-29\right) \epsilon\,,
\\
c^{\rm H}_{8,1} &=-4 \left( 9 \sigma ^6-45 \sigma ^4+27 \sigma ^2  +13\right) \epsilon ^4+4 \left(23 \sigma ^6-112 \sigma ^4+68 \sigma ^2+34\right) \epsilon ^3
\\
&\quad
-2 \left(28 \sigma ^6-129 \sigma ^4+68 \sigma ^2+61\right) \epsilon ^2+\left(26 \sigma ^4-44 \sigma ^2+41\right) \epsilon -3 \left(1-2 \sigma ^2\right)^2 \,.
\end{aligned}\label{}
\end{equation}
}
%

\subsubsection{$\overline{\rm H}$ sector}\label{App:C.2.2}

Similarly, the coefficients of the IBP reduction of the amplitude for the $\overline{\rm H}$ sector defined in \eqref{IBP_coeff_bH} are given by
{\small
\begin{equation}
\begin{aligned}
c^{\overline{\rm H}}_{1}&=-\frac{2 m_{1}^3 m_{2}^3 \sigma }{(\epsilon -1) \epsilon ^4 (4 \epsilon ^2-8 \epsilon +3)^2 (9 \epsilon ^2-9 \epsilon +2)}
c^{\overline{\rm H}}_{1,1}\,,
\\
c^{\overline{\rm H}}_{2}&=\frac{8 m_{1}^3 m_{2}^3 \sigma }{(\epsilon -1) \epsilon ^3 (2 \epsilon -3) (2 \epsilon -1) (4 \epsilon -1)}
c^{\overline{\rm H}}_{2,1}\,,
\\
c^{\overline{\rm H}}_{3}&=\frac{4 m_{1}^3 m_{2}^3 \sigma }{\epsilon ^4 (4 \epsilon ^2-8 \epsilon +3)^2}
c^{\overline{\rm H}}_{3,1}
\\
c^{\overline{\rm H}}_{4}&=\frac{m_{1}^3 m_{2}^3 }{8 (\epsilon -1)^2 \epsilon ^4 (4 \epsilon ^2-8 \epsilon +3)}
c^{\overline{\rm H}}_{4,1}\,,
\\
c^{\overline{\rm H}}_{5}&=-\frac{32 m_{1}^3 m_{2}^3 }{\sqrt{\sigma ^2-1} (\epsilon -1) \epsilon ^4 (2 \epsilon -3) (2 \epsilon -1)}
c^{\overline{\rm H}}_{5,1}\,,
\\
c^{\overline{\rm H}}_{6}&=-\frac{8 m_{1}^3 m_{2}^3 }{\sqrt{\sigma ^2-1} (\epsilon -1) \epsilon ^4 (2 \epsilon -3) (2 \epsilon -1)}
c^{\overline{\rm H}}_{6,1}\,,
\\
c^{\overline{\rm H}}_{7}&=-\frac{32 m_{1}^3 m_{2}^3 \sigma \left(4 \sigma ^2 \epsilon ^2-10 \epsilon ^2-8 \sigma ^2 \epsilon +15 \epsilon -6\right)}{(\epsilon -1) \epsilon ^3 (2 \epsilon -3) (2 \epsilon -1)}\,,
\\
c^{\overline{\rm H}}_{8}&=-\frac{8 m_{1}^3 m_{2}^3}{\sqrt{\sigma ^2-1} (\epsilon -1) \epsilon ^4 (2 \epsilon -3) (2 \epsilon -1) (4 \epsilon -1)} c^{\overline{\rm H}}_{8,1}\,,
\\
c^{\overline{\rm H}}_{9}&=-\frac{4 m_{1}^3 m_{2}^3 \left(-12 \sigma ^4+12 \sigma ^2+16 \sigma ^4 \epsilon ^2-12 \sigma ^2 \epsilon ^2-8 \epsilon ^2-12 \sigma ^4 \epsilon +8 \sigma ^2 \epsilon +12 \epsilon -3\right)}{\sqrt{\sigma ^2-1} \epsilon ^4 (4 \epsilon ^2-8 \epsilon +3)}\,,
\\
c^{\overline{\rm H}}_{10}&=-\frac{8 m_{1}^3 m_{2}^3 \sigma \left(16 \sigma ^2 \epsilon ^2+12 \epsilon ^2-22 \sigma ^2 \epsilon -19 \epsilon +6\right)}{\epsilon ^4 (4 \epsilon ^2-8 \epsilon +3)}\,.
\end{aligned}\label{}
\end{equation}
}
where
{\small
\begin{equation}
\begin{aligned}
c^{\overline{\rm H}}_{1,1} &=-96 \left(3 \sigma ^2+37\right) \epsilon ^7+8 \left(187 \sigma ^2+2117\right) \epsilon ^6-60 \left(56 \sigma ^2+549\right) \epsilon ^5+38 \left(97 \sigma ^2+897\right) \epsilon ^4
\\
&\quad
+\left(-1564 \sigma ^2-20197\right) \epsilon ^3+\left(6773-18 \sigma ^2\right) \epsilon ^2+6 \left(18 \sigma ^2-193\right) \epsilon +72 \,,
\\
c^{\overline{\rm H}}_{2,1} &= \left(36 \sigma ^4+60 \sigma ^2-4\right) \epsilon ^3+\left(-92 \sigma ^4-146 \sigma ^2+2\right) \epsilon ^2
\\
&\quad
+\left(56 \sigma ^4+142 \sigma ^2-9\right) \epsilon -54 \sigma ^2 +9\,,
\\
c^{\overline{\rm H}}_{3,1} &=8 \left(9 \sigma ^2+4\right) \epsilon ^4-4 \left(68 \sigma ^2+41\right) \epsilon ^3+\left(258 \sigma ^2+224\right) \epsilon ^2-3 \left(18 \sigma ^2+35\right) \epsilon +18 \,,
\\
c^{\overline{\rm H}}_{4,1} &=\left(3328 \sigma ^2+368\right) \epsilon ^4-8 \left(1424 \sigma ^2+185\right) \epsilon ^3
\\
&\quad
+4 \left(3377 \sigma ^2+502\right) \epsilon ^2-2 \left(3060 \sigma ^2+539\right) \epsilon + 675 \sigma ^2 +183\,,
\\
c^{\overline{\rm H}}_{5,1} &=\left(4 \sigma ^4-23 \sigma ^2-21\right) \epsilon ^3+\left(14 \sigma ^2+46\right) \epsilon ^2+\left(-22 \sigma ^4-\sigma ^2-31\right) \epsilon + 6 \sigma ^4-3 \sigma ^2 +6 \,,
\\
c^{\overline{\rm H}}_{6,1} &= 3 \left(1-2 \sigma ^2\right)^2+4 \left(6 \sigma ^4-\sigma ^2-6\right) \epsilon ^3+\left(-50 \sigma ^4+12 \sigma ^2+50\right) \epsilon ^2+\left(22 \sigma ^4-4 \sigma ^2-29\right) \epsilon\,,
\\
c^{\overline{\rm H}}_{8,1} &=-4 \left( 9 \sigma ^6-45 \sigma ^4+27 \sigma ^2  +13\right) \epsilon ^4+4 \left(23 \sigma ^6-112 \sigma ^4+68 \sigma ^2+34\right) \epsilon ^3
\\
&\quad
-2 \left(28 \sigma ^6-129 \sigma ^4+68 \sigma ^2+61\right) \epsilon ^2+\left(26 \sigma ^4-44 \sigma ^2+41\right) \epsilon -3 \left(1-2 \sigma ^2\right)^2 \,.
\end{aligned}\label{}
\end{equation}
}
%

\subsection{\textbf{IY}-class}

We now consider the coefficients of the IBP reduction of the amplitude in the \textbf{IY}-class. There are the horizontal mirror dual pairs, and we focus on the three independent sectors: IY, $\overline{\rm IY}$ and /\!\!\!Y. Note that the coefficients for the \flip{\textbf{IY}} class can be derived from the coefficients in the \textbf{IY}-class by the vertical mirror dual relations \eqref{eq:mirror_duals}.

\subsubsection{IY sector} \label{App:C.3.1}

The coefficients of the IBP reduction of the amplitude for the IY sector defined in \eqref{IBP_coeff_IY} are given by
{\small
\begin{equation}
\begin{aligned}
c^{\text{IY}}_{1}&=\frac{m_{1}^3 m_{2}^2}{12 \left(\sigma ^2-1\right) (1-2 \epsilon )^2 (\epsilon -1) \epsilon ^3 (2 \epsilon -3) (2 \epsilon +1) (9 \epsilon ^2-9 \epsilon +2)}
\Big[c^{\rm IY}_{1,1} m_{1} + c^{\rm IY}_{1,2} m_{2} \Big]\,,
\\
c^{\text{IY}}_{2}&=\frac{4 m_{1}^3 m_{2}^2}{\sqrt{\sigma ^2-1} (\epsilon -1) \epsilon ^4 (2 \epsilon -1)}
\Big[c^{\rm IY}_{2,1} m_{1} + c^{\rm IY}_{2,2} m_{2} \Big]\,,
\\
c^{\text{IY}}_{3}&=\frac{2 m_{1}^3 m_{2}^2}{\sqrt{\sigma ^2-1} (\epsilon -1) \epsilon ^3 (2 \epsilon -1)}
\Big[c^{\rm IY}_{3,1} m_{1} + c^{\rm IY}_{3,2} m_{2} \Big]\,,
\\
c^{\text{IY}}_{4}&=\frac{2 m_{1}^3 m_{2}^2}{\left(\sigma ^2-1\right) (\epsilon -1) \epsilon ^3 (2 \epsilon -1)}
\Big[c^{\rm IY}_{4,1} m_{1} + c^{\rm IY}_{4,2} m_{2} \Big]\,,
\\
c^{\text{IY}}_{5}&=\frac{2 m_{1}^3 m_{2}^3 \left(2 \sigma ^2-1\right) \left(\left(32 \sigma ^2-16\right) \epsilon ^2+\left(38-84 \sigma ^2\right) \epsilon + 54 \sigma ^2 -17\right)}{3 \sqrt{\sigma ^2-1} \epsilon ^3 (4 \epsilon ^2-8 \epsilon +3)}\,,
\\
c^{\text{IY}}_{6}&=\frac{2 m_{1}^3 m_{2}^2}{3 \left(\sigma ^2-1\right) (\epsilon -1) \epsilon ^3 (2 \epsilon +1) (6 \epsilon -5) (6 \epsilon -1)}
\Big[c^{\rm IY}_{6,1} m_{1} + c^{\rm IY}_{6,2} m_{2} \Big]\,,
\\
c^{\text{IY}}_{7}&=-\frac{2 m_{1}^4 m_{2}^3 \left(2 \sigma ^2-1\right) \left(\left(4 \sigma ^2-6 \sigma -1\right) \epsilon ^2+(10 \sigma^2 - 8 \sigma +6) \epsilon + 4 (\sigma -1) \sigma -2\right)}{(\epsilon -1) \epsilon ^4 (2 \epsilon -1) q}\,,
\\
c^{\text{IY}}_{8}&=-\frac{4 m_{1}^4 m_{2}^3 \left(2 \sigma ^2-1\right) \left(4 \sigma  \epsilon ^2+(3-7 \sigma ) \epsilon + 2 \sigma -1\right)}{(\epsilon -1) \epsilon ^4 (2 \epsilon -1) q}\,,
\\
c^{\text{IY}}_{9}&= \frac{2 m_{1}^4 m_{2}^3 \left(2 \sigma ^2-1\right) \left(\left(16 \sigma ^2-8\right) \epsilon -15 \sigma ^2 +7\right)}{\sqrt{\sigma ^2-1} (\epsilon -1) \epsilon ^4 q}\,,
\\
c^{\text{IY}}_{10}&=\frac{m_{1}^3 m_{2}^2}{ (\epsilon -1) \epsilon ^3 (2 \epsilon -1) (2 \epsilon +1) (4 \epsilon -3) (4 \epsilon -1)}\Big[c^{\rm IY}_{10,1} m_{1} + c^{\rm IY}_{10,2} m_{2} \Big]\,,
\\
c^{\text{IY}}_{11}&=\frac{2 m_{1}^3 m_{2}^2}{\sqrt{\sigma ^2-1} (\epsilon -1) \epsilon ^3 \left(8 \epsilon ^2-6 \epsilon +1\right)}\Big[c^{\rm IY}_{11,1} m_{1} + c^{\rm IY}_{11,2} m_{2} \Big]\,,
\\
c^{\text{IY}}_{12}&=-\frac{4 m_{1}^4 m_{2}^3 \sigma  \left(2 \sigma ^2-1\right) (1-2 \epsilon )^2 (3 \epsilon -4)}{(\epsilon -1) \epsilon ^4 (2 \epsilon -1)^2 q}\,,
\\
c^{\text{IY}}_{13}&=\frac{4 m_{1}^4 m_{2}^3 \left(2 \sigma ^2-1\right) \left((10 \sigma -2) \epsilon ^2+(4-12 \sigma ) \epsilon + 2 \sigma -1\right) }{(\epsilon -1) \epsilon ^4 (2 \epsilon -1) q} \,.
\end{aligned}\label{}
\end{equation}
}
\newpage
where
{\small
\begin{equation}
\begin{aligned}
c^{\rm IY}_{1,1} &=-432 \left(32 \sigma ^6-18 \sigma ^4-13 \sigma ^2+7\right) \epsilon ^7+24 \left(2904 \sigma ^6-2366 \sigma ^4-297 \sigma ^2+335\right) \epsilon ^6
\\
&\quad
-4 \left(32760 \sigma ^6-34268 \sigma ^4+5187 \sigma ^2+1265\right) \epsilon ^5
\\
&\quad
+2 \left(52920 \sigma ^6-67802 \sigma ^4+21173 \sigma ^2-819\right) \epsilon ^4
\\
&\quad+\left(-19656 \sigma ^6+37910 \sigma ^4-20080 \sigma ^2+2282\right) \epsilon ^3
\\
&\quad
+\left(-22176 \sigma ^6+24662 \sigma ^4-5245 \sigma ^2-265\right) \epsilon ^2
\\
&\quad
+\left(13320 \sigma ^6-18176 \sigma ^4+6241 \sigma ^2-209\right) \epsilon
\\
&\quad
 -6 \left(360 \sigma ^6-532 \sigma ^4+203 \sigma ^2-7\right)\,,
\\
c^{\rm IY}_{1,2} & =-864 \left(10 \sigma ^5-7 \sigma ^3+\sigma \right) \epsilon ^7+864 \sigma ^3 \left(41 \sigma ^2-25\right) \epsilon ^6+24 \sigma  \left(-1894 \sigma ^4+621 \sigma ^2+449\right) \epsilon ^5
\\
&\quad
+8 \sigma  \left(1621 \sigma ^4+2293 \sigma ^2-2546\right) \epsilon ^4+8 \sigma  \left(1444 \sigma ^4-2696 \sigma ^2+1309\right) \epsilon ^3
\\
&\quad
+24 \sigma  \left(-298 \sigma ^4+85 \sigma ^2+87\right) \epsilon ^2
+24 \sigma  \left(27 \sigma ^4+144 \sigma ^2-122\right) \epsilon
+ 144 \sigma  \left(\sigma ^4-6 \sigma ^2+4\right) \,,
\\
c^{\rm IY}_{2,1} &= \sigma  \left(-8 \sigma ^2+\left(12 \sigma ^2-24 \sigma ^4\right) \epsilon ^3+\left(60 \sigma ^4-44 \sigma ^2+16\right) \epsilon ^2+\left(-36 \sigma ^4+40 \sigma ^2-13\right) \epsilon +4\right) \,,
\\
c^{\rm IY}_{2,2} & =\left(-8 \sigma ^4+20 \sigma ^2-8\right) \epsilon ^3+2 \left(6 \sigma ^4-8 \sigma ^2+5\right) \epsilon ^2+\left(12 \sigma ^4+6 \sigma ^2-11\right) \epsilon -4 \left(\sigma ^4+\sigma ^2\right) +3 \,,
\\
c^{\rm IY}_{3,1} &= \left(8 \sigma ^4-4 \sigma ^2\right) \epsilon ^2-4 \left(5 \sigma ^4-5 \sigma ^2+1\right) \epsilon + 12 \sigma ^4-16 \sigma ^2 +5 \,,
\\
c^{\rm IY}_{3,2} & = \left(8 \sigma ^4-4 \sigma ^2\right) \epsilon ^2-2 \left(8 \sigma ^4-7 \sigma ^2+1\right) \epsilon + 4 \sigma ^4-2 \sigma ^2 -1 \,,
\\
c^{\rm IY}_{4,1} &= -4 \left(2 \sigma ^2-1\right) \left(2 \sigma ^4-1\right) \epsilon ^2+\left(40 \sigma ^6-40 \sigma ^4+4 \sigma ^2+2\right) \epsilon -2 \sigma ^2 \left(12 \sigma ^4-16 \sigma ^2+5\right) \,,
\\
c^{\rm IY}_{4,2} &=\sigma  \left(-4 \left(1-2 \sigma ^2\right)^2 \epsilon ^2+\left(30 \sigma ^4-29 \sigma ^2+5\right) \epsilon + -6 \sigma ^4+\sigma ^2 +3\right)\,,
\\
c^{\rm IY}_{6,1} &=-432 \left(1-2 \sigma ^2\right)^2 \epsilon ^4+24 \left(132 \sigma ^4-136 \sigma ^2+31\right) \epsilon ^3-4 \left(260 \sigma ^4-294 \sigma ^2+49\right) \epsilon ^2
\\
&\quad
+\left(-784 \sigma ^4+792 \sigma ^2-194\right) + 6 \left(62 \sigma ^4-69 \sigma ^2+12\right)\epsilon \,,
\\
c^{\rm IY}_{6,2} &= -432 \sigma  \left(1-2 \sigma ^2\right)^2 \epsilon ^4+108 \sigma  \left(34 \sigma ^4-39 \sigma ^2+11\right) \epsilon ^3-12 \sigma  \left(144 \sigma ^4-205 \sigma ^2+66\right) \epsilon ^2
\\
&\quad
+\left(-770 \sigma ^5+781 \sigma ^3-197 \sigma \right) \epsilon + 5 \sigma  \left(110 \sigma ^4-151 \sigma ^2+47\right) \,,
\\
c^{\rm IY}_{10,1} &= 128 \left(10 \sigma ^4-7 \sigma ^2+1\right) \epsilon ^5-8 \left(352 \sigma ^4-303 \sigma ^2+46\right) \epsilon ^4+8 \left(194 \sigma ^4-203 \sigma ^2+27\right) \epsilon ^3
\\
&\quad
+\left(392 \sigma ^4-278 \sigma ^2+66\right) \epsilon ^2+\left(-480 \sigma ^4+474 \sigma ^2-61\right) \epsilon + 72 \sigma ^4-76 \sigma ^2 +7 \,,
\\
c^{\rm IY}_{10,2} &= 4 \sigma  (4 \epsilon -3) \big(\left(40 \sigma ^4-52 \sigma ^2+16\right) \epsilon ^4+\left(-84 \sigma ^4+142 \sigma ^2-50\right) \epsilon ^3
\\
&\qquad\qquad\qquad\;
+12 \left(\sigma ^4-4 \sigma ^2+2\right) \epsilon ^2+\left(32 \sigma ^4-46 \sigma ^2+14\right) \epsilon + 8 \sigma ^2 -5\big) \,,
\\
c^{\rm IY}_{11,1} &=  \sigma  \big(\left(80 \sigma ^4-56 \sigma ^2+8\right) \epsilon ^3-4 \left(39 \sigma ^4-37 \sigma ^2+8\right) \epsilon ^2
\\
&\qquad\;\;
+4 \left(22 \sigma ^4-26 \sigma ^2+7\right) \epsilon -12 \sigma ^4+16 \sigma ^2 -5\big)\,,
\\
c^{\rm IY}_{11,2} & =4 \left(10 \sigma ^6+3 \sigma ^4-6 \sigma ^2+1\right) \epsilon ^3+\left(-104 \sigma ^6+48 \sigma ^4+26 \sigma ^2-10\right) \epsilon ^2
\\
&\quad
+2 \left(32 \sigma ^6-28 \sigma ^4+\sigma ^2+1\right) \epsilon  -4 \sigma ^4+2 \sigma ^2 +1 \,.
\end{aligned}\label{}
\end{equation}
}

\newpage
\subsubsection{$\overline{\rm IY}$ sector} \label{App:C.3.2}

The coefficients of the IBP reduction of the amplitude for the $\overline{\rm IY}$ sector defined in \eqref{IBP_coeff_IY} are given by
{\small
\begin{equation}
\begin{aligned}
c^{\overline{\text{IY}}}_{1}&=\frac{m_{1}^3 m_{2}^2}{12 \left(\sigma ^2-1\right) (1-2 \epsilon )^2 (\epsilon -1) \epsilon ^3 (2 \epsilon -3) (2 \epsilon +1) (9 \epsilon ^2-9 \epsilon +2)}
\Big[c^{\overline{\rm IY}}_{1,1} m_{1} + c^{\overline{\rm IY}}_{1,2} m_{2} \Big]\,,
\\
c^{\overline{\text{IY}}}_{2}&=-\frac{4 m_{1}^3 m_{2}^2}{\sqrt{\sigma ^2-1} (\epsilon -1) \epsilon ^4 (2 \epsilon -1)}
\Big[c^{\overline{\rm IY}}_{2,1} m_{1} + c^{\overline{\rm IY}}_{2,2} m_{2} \Big]\,,
\\
c^{\overline{\text{IY}}}_{3}&=-\frac{2 m_{1}^3 m_{2}^2}{\sqrt{\sigma ^2-1} (\epsilon -1) \epsilon ^3 (2 \epsilon -1)}
\Big[c^{\overline{\rm IY}}_{3,1} m_{1} + c^{\overline{\rm IY}}_{3,2} m_{2} \Big]\,,
\\
c^{\overline{\text{IY}}}_{4}&=\frac{2 m_{1}^3 m_{2}^2}{\left(\sigma ^2-1\right) (\epsilon -1) \epsilon ^3 (2 \epsilon -1)}
\Big[c^{\overline{\rm IY}}_{4,1} m_{1} + c^{\overline{\rm IY}}_{4,2} m_{2} \Big]\,,
\\
c^{\overline{\text{IY}}}_{5}&=-\frac{2 m_{1}^3 m_{2}^3 \left(2 \sigma ^2-1\right) \left(\left(32 \sigma ^2-16\right) \epsilon ^2+\left(38-84 \sigma ^2\right) \epsilon + 54 \sigma ^2 -17\right)}{3 \sqrt{\sigma ^2-1} \epsilon ^3 (4 \epsilon ^2-8 \epsilon +3)}\,,
\\
c^{\overline{\text{IY}}}_{6}&=\frac{2 m_{1}^3 m_{2}^2}{3 \left(\sigma ^2-1\right) (\epsilon -1) \epsilon ^3 (2 \epsilon +1) (6 \epsilon -5) (6 \epsilon -1)}
\Big[c^{\overline{\rm IY}}_{6,1} m_{1} + c^{\overline{\rm IY}}_{6,2} m_{2} \Big]\,,
\\
c^{\overline{\text{IY}}}_{7}&=-\frac{2 m_{1}^4 m_{2}^3 \left(2 \sigma ^2-1\right) \left(\left(4 \sigma ^2-6 \sigma -1\right) \epsilon ^2+(10 \sigma^2 - 8 \sigma +6) \epsilon + 4 (\sigma -1) \sigma -2\right)}{(\epsilon -1) \epsilon ^4 (2 \epsilon -1) q}\,,
\\
c^{\overline{\text{IY}}}_{8}&=-\frac{4 m_{1}^4 m_{2}^3 \left(2 \sigma ^2-1\right) \left(4 \sigma \epsilon ^2+(3-7 \sigma ) \epsilon + 2 \sigma -1\right)}{(\epsilon -1) \epsilon ^4 (2 \epsilon -1) q}\,,
\\
c^{\overline{\text{IY}}}_{9}&= -\frac{2 m_{1}^4 m_{2}^3 \left(2 \sigma ^2-1\right) \left(\left(16 \sigma ^2-8\right) \epsilon -15 \sigma ^2 +7\right)}{\sqrt{\sigma ^2-1} (\epsilon -1) \epsilon ^4 q}\,,
\\
c^{\overline{\text{IY}}}_{10}&=\frac{m_{1}^3 m_{2}^2}{ (\epsilon -1) \epsilon ^3 (2 \epsilon -1) (2 \epsilon +1) (4 \epsilon -3) (4 \epsilon -1)}\Big[c^{\overline{\rm IY}}_{10,1} m_{1} + c^{\overline{\rm IY}}_{10,2} m_{2} \Big]\,,
\\
c^{\overline{\text{IY}}}_{11}&=-\frac{2 m_{1}^3 m_{2}^2}{\sqrt{\sigma ^2-1} (\epsilon -1) \epsilon ^3 \left(8 \epsilon ^2-6 \epsilon +1\right)}\Big[c^{\overline{\rm IY}}_{11,1} m_{1} + c^{\overline{\rm IY}}_{11,2} m_{2} \Big]\,,
\\
c^{\overline{\text{IY}}}_{12}&=-\frac{4 m_{1}^4 m_{2}^3 \sigma \left(2 \sigma ^2-1\right) (1-2 \epsilon )^2 (3 \epsilon -4)}{(\epsilon -1) \epsilon ^4 (2 \epsilon -1)^2 q}\,,
\\
c^{\overline{\text{IY}}}_{13}&=\frac{4 m_{1}^4 m_{2}^3 \left(2 \sigma ^2-1\right) \left((10 \sigma -2) \epsilon ^2+(4-12 \sigma ) \epsilon + 2 \sigma -1\right) }{(\epsilon -1) \epsilon ^4 (2 \epsilon -1) q} \,.
\end{aligned}\label{}
\end{equation}
}
\newpage
where
{\small
\begin{equation}
\begin{aligned}
c^{\overline{\rm IY}}_{1,1} &=-432 \left(32 \sigma ^6-18 \sigma ^4-13 \sigma ^2+7\right) \epsilon ^7+24 \left(2904 \sigma ^6-2366 \sigma ^4-297 \sigma ^2+335\right) \epsilon ^6
\\
&\quad
-4 \left(32760 \sigma ^6-34268 \sigma ^4+5187 \sigma ^2+1265\right) \epsilon ^5
\\
&\quad
+2 \left(52920 \sigma ^6-67802 \sigma ^4+21173 \sigma ^2-819\right) \epsilon ^4
\\
&\quad+\left(-19656 \sigma ^6+37910 \sigma ^4-20080 \sigma ^2+2282\right) \epsilon ^3
\\
&\quad
+\left(-22176 \sigma ^6+24662 \sigma ^4-5245 \sigma ^2-265\right) \epsilon ^2
\\
&\quad
+\left(13320 \sigma ^6-18176 \sigma ^4+6241 \sigma ^2-209\right) \epsilon
\\
&\quad
-6 \left(360 \sigma ^6-532 \sigma ^4+203 \sigma ^2-7\right)\,,
\\
c^{\overline{\rm IY}}_{1,2} & =864 \left(10 \sigma ^5-7 \sigma ^3+\sigma \right) \epsilon ^7-864 \sigma ^3 \left(41 \sigma ^2-25\right) \epsilon ^6-24 \sigma \left(-1894 \sigma ^4+621 \sigma ^2+449\right) \epsilon ^5
\\
&\quad
-8 \sigma \left(1621 \sigma ^4+2293 \sigma ^2-2546\right) \epsilon ^4-8 \sigma \left(1444 \sigma ^4-2696 \sigma ^2+1309\right) \epsilon ^3
\\
&\quad
-24 \sigma \left(-298 \sigma ^4+85 \sigma ^2+87\right) \epsilon ^2
-24 \sigma \left(27 \sigma ^4+144 \sigma ^2-122\right) \epsilon
- 144 \sigma \left(\sigma ^4-6 \sigma ^2+4\right) \,,
\\
c^{\overline{\rm IY}}_{2,1} &= -\sigma \left(-8 \sigma ^2+\left(12 \sigma ^2-24 \sigma ^4\right) \epsilon ^3+\left(60 \sigma ^4-44 \sigma ^2+16\right) \epsilon ^2+\left(-36 \sigma ^4+40 \sigma ^2-13\right) \epsilon +4\right) \,,
\\
c^{\overline{\rm IY}}_{2,2} & =\left(-8 \sigma ^4+20 \sigma ^2-8\right) \epsilon ^3+2 \left(6 \sigma ^4-8 \sigma ^2+5\right) \epsilon ^2+\left(12 \sigma ^4+6 \sigma ^2-11\right) \epsilon -4 \sigma ^4-4\sigma ^2 +3 \,,
\\
c^{\overline{\rm IY}}_{3,1} &= \left(8 \sigma ^4-4 \sigma ^2\right) \epsilon ^2-4 \left(5 \sigma ^4-5 \sigma ^2+1\right) \epsilon + 12 \sigma ^4-16 \sigma ^2 +5 \,,
\\
c^{\overline{\rm IY}}_{3,2} & = \left(8 \sigma ^4-4 \sigma ^2\right) \epsilon ^2-2 \left(8 \sigma ^4-7 \sigma ^2+1\right) \epsilon + 4 \sigma ^4-2 \sigma ^2 -1 \,,
\\
c^{\overline{\rm IY}}_{4,1} &= -4 \left(2 \sigma ^2-1\right) \left(2 \sigma ^4-1\right) \epsilon ^2+\left(40 \sigma ^6-40 \sigma ^4+4 \sigma ^2+2\right) \epsilon -2 \sigma ^2 \left(12 \sigma ^4-16 \sigma ^2+5\right) \,,
\\
c^{\overline{\rm IY}}_{4,2} &=-\sigma \left(-4 \left(1-2 \sigma ^2\right)^2 \epsilon ^2+\left(30 \sigma ^4-29 \sigma ^2+5\right) \epsilon + -6 \sigma ^4+\sigma ^2 +3\right)\,,
\\
c^{\overline{\rm IY}}_{6,1} &=-432 \left(1-2 \sigma ^2\right)^2 \epsilon ^4+24 \left(132 \sigma ^4-136 \sigma ^2+31\right) \epsilon ^3-4 \left(260 \sigma ^4-294 \sigma ^2+49\right) \epsilon ^2
\\
&\quad
+\left(-784 \sigma ^4+792 \sigma ^2-194\right) + 6 \left(62 \sigma ^4-69 \sigma ^2+12\right)\epsilon \,,
\\
c^{\overline{\rm IY}}_{6,2} &= 432 \sigma \left(1-2 \sigma ^2\right)^2 \epsilon ^4-108 \sigma \left(34 \sigma ^4-39 \sigma ^2+11\right) \epsilon ^3+12 \sigma \left(144 \sigma ^4-205 \sigma ^2+66\right) \epsilon ^2
\\
&\quad
+\left(770 \sigma ^5-781 \sigma ^3+197 \sigma \right) \epsilon - 5 \sigma \left(110 \sigma ^4-151 \sigma ^2+47\right) \,,
\\
c^{\overline{\rm IY}}_{10,1} &= 128 \left(10 \sigma ^4-7 \sigma ^2+1\right) \epsilon ^5-8 \left(352 \sigma ^4-303 \sigma ^2+46\right) \epsilon ^4+8 \left(194 \sigma ^4-203 \sigma ^2+27\right) \epsilon ^3
\\
&\quad
+\left(392 \sigma ^4-278 \sigma ^2+66\right) \epsilon ^2+\left(-480 \sigma ^4+474 \sigma ^2-61\right) \epsilon + 72 \sigma ^4-76 \sigma ^2 +7 \,,
\\
c^{\overline{\rm IY}}_{10,2} &=-4 \sigma (4 \epsilon -3) \big(\left(40 \sigma ^4-52 \sigma ^2+16\right) \epsilon ^4+\left(-84 \sigma ^4+142 \sigma ^2-50\right) \epsilon ^3
\\
&\qquad\qquad\qquad\;
+12 \left(\sigma ^4-4 \sigma ^2+2\right) \epsilon ^2+\left(32 \sigma ^4-46 \sigma ^2+14\right) \epsilon + 8 \sigma ^2 -5\big) \,,
\\
c^{\overline{\rm IY}}_{11,1} &= - \sigma \big(\left(80 \sigma ^4-56 \sigma ^2+8\right) \epsilon ^3-4 \left(39 \sigma ^4-37 \sigma ^2+8\right) \epsilon ^2
\\
&\qquad\;\;
+4 \left(22 \sigma ^4-26 \sigma ^2+7\right) \epsilon -12 \sigma ^4+16 \sigma ^2 -5\big)\,,
\\
c^{\overline{\rm IY}}_{11,2} & =4 \left(10 \sigma ^6+3 \sigma ^4-6 \sigma ^2+1\right) \epsilon ^3+\left(-104 \sigma ^6+48 \sigma ^4+26 \sigma ^2-10\right) \epsilon ^2
\\
&\quad
+2 \left(32 \sigma ^6-28 \sigma ^4+\sigma ^2+1\right) \epsilon -4 \sigma ^4+2 \sigma ^2 +1 \,.
\end{aligned}\label{}
\end{equation}
}
\newpage
\subsubsection{/\!\!\!Y sector} \label{App:C.3.3}

The coefficients are
{\small
\begin{equation}
\begin{aligned}
c^{\text{/\!\!\!Y}}_{1}&=\frac{m_{1}^4 m_{2}^2}{6 \left(\sigma ^2-1\right) (1-2 \epsilon )^2 (\epsilon -1) \epsilon ^3 (2 \epsilon -3) (9 \epsilon ^2-9 \epsilon +2)}c^{\text{/\!\!\!Y}}_{1,1}  \,,
\\
c^{\text{/\!\!\!Y}}_{2}&=\frac{4 m_{1}^3 m_{2}^2}{\sqrt{\sigma ^2-1} (\epsilon -1) \epsilon ^4 (2 \epsilon -3) (2 \epsilon -1)}
\Big[c^{\text{/\!\!\!Y}}_{2,1} m_{1} + c^{\text{/\!\!\!Y}}_{2,2} m_{2} \Big]\,,
\\
c^{\text{/\!\!\!Y}}_{3}&=\frac{2 m_{1}^3 m_{2}^2}{\sqrt{\sigma ^2-1} (\epsilon -1) \epsilon ^4 (2 \epsilon -3) (2 \epsilon -1)}
\Big[c^{\text{/\!\!\!Y}}_{3,1} m_{1} + c^{\text{/\!\!\!Y}}_{3,2} m_{2} \Big]\,,
\\
c^{\text{/\!\!\!Y}}_{4}&=\frac{4 m_{1}^3 m_{2}^2}{\left(\sigma ^2-1\right) (\epsilon -1) \epsilon ^3 (2 \epsilon -3) (2 \epsilon -1)}
\Big[c^{\text{/\!\!\!Y}}_{4,1} m_{1} + c^{\text{/\!\!\!Y}}_{4,2} m_{2} \Big]\,,
\\
c^{\text{/\!\!\!Y}}_{5}&=\frac{4 m_{1}^3 m_{2}^2}{\sqrt{\sigma ^2-1} (\epsilon -1) \epsilon ^4 (2 \epsilon -3) (2 \epsilon -1)}
\Big[c^{\text{/\!\!\!Y}}_{5,1} m_{1} + c^{\text{/\!\!\!Y}}_{5,2} m_{2} \Big]\,,
\\
c^{\text{/\!\!\!Y}}_{6}&=\frac{2 m_{1}^3 m_{2}^2}{\sqrt{\sigma ^2-1} (\epsilon -1) \epsilon ^4 (2 \epsilon -3) (2 \epsilon -1)}
\Big[c^{\text{/\!\!\!Y}}_{6,1} m_{1} + c^{\text{/\!\!\!Y}}_{6,2} m_{2} \Big]\,,
\\
c^{\text{/\!\!\!Y}}_{7}&=\frac{4 m_{1}^3 m_{2}^2}{\left(\sigma ^2-1\right) (\epsilon -1) \epsilon ^3 (2 \epsilon -3) (2 \epsilon -1)}
\Big[c^{\text{/\!\!\!Y}}_{7,1} m_{1} + c^{\text{/\!\!\!Y}}_{7,2} m_{2} \Big]\,,
\\
c^{\text{/\!\!\!Y}}_{8}&=\frac{4 m_{1}^4 m_{2}^2}{3 \left(\sigma ^2-1\right) (\epsilon -1) \epsilon ^3 (6 \epsilon -5) (6 \epsilon -1)}c^{\text{/\!\!\!Y}}_{8,1}  \,,
\\
c^{\text{/\!\!\!Y}}_{9}&= \frac{4 m_{1}^4 m_{2}^3 \left(2 \sigma ^2-1\right) \left(\left(4 \sigma ^2-1\right) \epsilon ^2+\left(6-8 \sigma ^2\right) \epsilon + 4 \sigma ^2-2\right)}{(\epsilon -1) \epsilon ^4 (2 \epsilon -1) q}\,,
\\
c^{\text{/\!\!\!Y}}_{10}&=\frac{4 m_{1}^4 m_{2}^3 \left(2 \sigma ^2-1\right) \left(4 \sigma ^2 \epsilon ^2+\left(-6 \sigma ^2-\sigma +3\right) \epsilon + 2 \sigma ^2 -1\right)}{(\epsilon -1) \epsilon ^4 (2 \epsilon -1) q}\,,
\\
c^{\text{/\!\!\!Y}}_{11}&=\frac{4 m_{1}^4 m_{2}^3 \left(2 \sigma ^2-1\right) \left(4 \sigma ^2 \epsilon ^2+\left(-6 \sigma ^2+\sigma +3\right) \epsilon + 2 \sigma ^2 -1\right)}{(\epsilon -1) \epsilon ^4 (2 \epsilon -1) q}\,,
\\
c^{\text{/\!\!\!Y}}_{12}&=\frac{2 m_{1}^4 m_{2}^3 \left(2 \sigma ^2-1\right) \left(\left(16 \sigma ^2-8\right) \epsilon -15 \sigma ^2+7\right)}{\sqrt{\sigma ^2-1} (\epsilon -1) \epsilon ^4 q}\,,
\\
c^{\text{/\!\!\!Y}}_{13}&=-\frac{2 m_{1}^4 m_{2}^2}{(\epsilon -1) \epsilon ^3 (2 \epsilon -1) (4 \epsilon -3) (4 \epsilon -1)}c^{\text{/\!\!\!Y}}_{13,1} \,,
\\
c^{\text{/\!\!\!Y}}_{14}&=\frac{2 m_{1}^3 m_{2}^2}{\sqrt{\sigma ^2-1} (\epsilon -1) \epsilon ^4 (2 \epsilon -3) \left(8 \epsilon ^2-6 \epsilon +1\right)}
\Big[c^{\text{/\!\!\!Y}}_{14,1} m_{1} + c^{\text{/\!\!\!Y}}_{14,2} m_{2} \Big] \,,
\\
c^{\text{/\!\!\!Y}}_{15}&=\frac{2 m_{1}^3 m_{2}^2}{\sqrt{\sigma ^2-1} (\epsilon -1) \epsilon ^4 (2 \epsilon -3) \left(8 \epsilon ^2-6 \epsilon +1\right)}
\Big[c^{\text{/\!\!\!Y}}_{15,1} m_{1} + c^{\text{/\!\!\!Y}}_{15,2} m_{2} \Big] \,,
\\
c^{\text{/\!\!\!Y}}_{16}&=0 \,,
\\
c^{\text{/\!\!\!Y}}_{17}&=-\frac{4 m_{1}^4 m_{2}^3 \left(2 \sigma ^2-1\right) \left(\left(10 \sigma ^2-2\right) \epsilon ^2+\left(4-12 \sigma ^2\right) \epsilon + 2 \sigma ^2-1\right)}{(\epsilon -1) \epsilon ^4 (2 \epsilon -1) q} \,,
\\
c^{\text{/\!\!\!Y}}_{18}&=-\frac{4 m_{1}^4 m_{2}^3 \left(2 \sigma ^2-1\right) \left(\left(10 \sigma ^2-2\right) \epsilon ^2+\left(4-12 \sigma ^2\right) \epsilon + 2 \sigma ^2 -1\right)}{(\epsilon -1) \epsilon ^4 (2 \epsilon -1) q} \,.
\end{aligned}\label{}
\end{equation}
}
\newpage
where
{\small
\begin{equation}
\begin{aligned}
c^{\text{/\!\!\!Y}}_{1,1} &= 216 \left(32 \sigma ^6-18 \sigma ^4-13 \sigma ^2+7\right) \epsilon ^6+24 \left(-1596 \sigma ^6+1264 \sigma ^4+207 \sigma ^2-199\right) \epsilon ^5
\\
&\quad
+\left(84672 \sigma ^6-83704 \sigma ^4+7890 \sigma ^2+4918\right) \epsilon ^4
-2 \left(47628 \sigma ^6-54827 \sigma ^4+12559 \sigma ^2+820\right) \epsilon ^3
\\
&\quad
+\left(9 \sigma ^2 \left(6384 \sigma ^4-8198 \sigma ^2+2511\right)-321\right) \epsilon ^2+\left(-17640 \sigma ^6+24560 \sigma ^4-8677 \sigma ^2+293\right) \epsilon
\\
&\quad
+ 6 \left(360 \sigma ^6-532 \sigma ^4+203 \sigma ^2-7\right)\,,
\\
c^{\text{/\!\!\!Y}}_{2,1} &= \sigma  \big(\left(24 \sigma ^2-48 \sigma ^4\right) \epsilon ^4+4 \left(48 \sigma ^4-31 \sigma ^2+8\right) \epsilon ^3
\\
&\qquad\;\;
+\left(-252 \sigma ^4+212 \sigma ^2-74\right) \epsilon ^2+\left(108 \sigma ^4-136 \sigma ^2+47\right) \epsilon + 24 \sigma ^2 -12\big) \,,
\\
c^{\text{/\!\!\!Y}}_{2,2} & =-8 \left(2 \sigma ^4+\sigma ^2-1\right) \epsilon ^4+2 \left(18 \sigma ^4+5 \sigma ^2+5\right) \epsilon ^3
\\
&\quad
+\left(-80 \sigma ^4+116 \sigma ^2-34\right) \epsilon ^2+\left(132 \sigma ^4-104 \sigma ^2+15\right) \epsilon -36 \sigma ^4+24 \sigma ^2 -3 \,,
\\
c^{\text{/\!\!\!Y}}_{3,1} &= 8 \left(2 \sigma ^2-1\right) \sigma ^2 \epsilon ^3+\left(-64 \sigma ^4+52 \sigma ^2-8\right) \epsilon ^2+\left(84 \sigma ^4-92 \sigma ^2+22\right) \epsilon -36 \sigma ^4+48 \sigma ^2 -15\,,
\\
c^{\text{/\!\!\!Y}}_{3,2} & =-3 \left(1-2 \sigma ^2\right)^2+8 \left(2 \sigma ^4-3 \sigma ^2+1\right) \epsilon ^4+\left(-28 \sigma ^4+34 \sigma ^2-10\right) \epsilon ^3+4 \sigma ^2 \epsilon ^2+3 \left(1-2 \sigma ^2\right)^2 \epsilon  \,,
\\
c^{\text{/\!\!\!Y}}_{4,1} &= -4 \left(2 \sigma ^2-1\right) \left(2 \sigma ^4-1\right) \epsilon ^3+\left(64 \sigma ^6-52 \sigma ^4-8 \sigma ^2+8\right) \epsilon ^2
\\
&\quad
+\left(-84 \sigma ^6+92 \sigma ^4-16 \sigma ^2-3\right) \epsilon + 3 \sigma ^2 \left(12 \sigma ^4-16 \sigma ^2+5\right)\,,
\\
c^{\text{/\!\!\!Y}}_{4,2} &= -2 (\sigma -1) \sigma  (\sigma +1) \left(-\sigma ^2+8 \left(\sigma ^2-1\right) \epsilon ^2+\left(6-13 \sigma ^2\right) \epsilon +1\right) \,,
\\
c^{\text{/\!\!\!Y}}_{5,1} &= \sigma  \big(24 \sigma ^2 \left(2 \sigma ^2-1\right) \epsilon ^4-4 \left(48 \sigma ^4-31 \sigma ^2+8\right) \epsilon ^3
\\
&\qquad\;\;
+\left(252 \sigma ^4-212 \sigma ^2+74\right) \epsilon ^2+\left(-108 \sigma ^4+136 \sigma ^2-47\right) \epsilon -24 \sigma ^2 +12\big) \,,
\\
c^{\text{/\!\!\!Y}}_{5,2} &=24 \sigma ^2-8 \left(2 \sigma ^4+\sigma ^2-1\right) \epsilon ^4+2 \left(18 \sigma ^4+5 \sigma ^2+5\right) \epsilon ^3
\\
&\quad
+\left(-80 \sigma ^4+116 \sigma ^2-34\right) \epsilon ^2+\left(132 \sigma ^4-104 \sigma ^2+15\right) \epsilon  -36 \sigma ^4 -3 \,,
\\
c^{\text{/\!\!\!Y}}_{6,1} &=\sigma  \epsilon  \left(\left(8 \sigma ^2-16 \sigma ^4\right) \epsilon ^3+\left(64 \sigma ^4-52 \sigma ^2+8\right) \epsilon ^2+\left(-84 \sigma ^4+92 \sigma ^2-22\right) \epsilon + 36 \sigma ^4-48 \sigma ^2 +15\right)\,,
\\
c^{\text{/\!\!\!Y}}_{6,2} &= 8 \left(2 \sigma ^4-3 \sigma ^2+1\right) \epsilon ^4+\left(-28 \sigma ^4+34 \sigma ^2-10\right) \epsilon ^3+4 \sigma ^2 \epsilon ^2+3 \left(1-2 \sigma ^2\right)^2 \epsilon-3 \left(1-2 \sigma ^2\right)^2\,,
\\
c^{\text{/\!\!\!Y}}_{7,1} &=-4 \left(2 \sigma ^2-1\right) \left(2 \sigma ^4-1\right) \epsilon ^3+\left(64 \sigma ^6-52 \sigma ^4-8 \sigma ^2+8\right) \epsilon ^2
\\
&\quad
+\left(-84 \sigma ^6+92 \sigma ^4-16 \sigma ^2-3\right) \epsilon + 3 \sigma ^2 \left(12 \sigma ^4-16 \sigma ^2+5\right)\,,
\\
c^{\text{/\!\!\!Y}}_{7,2} &= 2 \sigma  \left(\sigma ^2-1\right) \left(8 \left(\sigma ^2-1\right) \epsilon ^2+\left(6-13 \sigma ^2\right) \epsilon -\sigma ^2 +1\right)\,,
\\
c^{\text{/\!\!\!Y}}_{8,1} &=-108 \left(1-2 \sigma ^2\right)^2 \epsilon ^3+24 \left(42 \sigma ^4-43 \sigma ^2+10\right) \epsilon ^2
\\
&\quad
+\left(-764 \sigma ^4+810 \sigma ^2-169\right) \epsilon + 3 \left(62 \sigma ^4-69 \sigma ^2+12\right)\,,
\\
c^{\text{/\!\!\!Y}}_{13,1} &=64 \left(10 \sigma ^4-7 \sigma ^2+1\right) \epsilon ^4-4 \left(432 \sigma ^4-359 \sigma ^2+54\right) \epsilon ^3
\\
&\quad
+2 \left(820 \sigma ^4-765 \sigma ^2+108\right) \epsilon ^2+\left(-624 \sigma ^4+626 \sigma ^2-75\right) \epsilon + 72 \sigma ^4-76 \sigma ^2 +7
\\
c^{\text{/\!\!\!Y}}_{14,1} &=\sigma  \epsilon  \big(-16 \left(10 \sigma ^4-7 \sigma ^2+1\right) \epsilon ^4+8 \left(69 \sigma ^4-58 \sigma ^2+11\right) \epsilon ^3
\\
&\qquad\quad
+\left(-644 \sigma ^4+652 \sigma ^2-152\right) \epsilon ^2+\left(288 \sigma ^4-344 \sigma ^2+94\right) \epsilon -36 \sigma ^4+48 \sigma ^2 -15\big) \,,
\\
c^{\text{/\!\!\!Y}}_{14,2} &= 8 \left(10 \sigma ^6+3 \sigma ^4-6 \sigma ^2+1\right) \epsilon ^5+4 \left(96 \sigma ^6-48 \sigma ^4-13 \sigma ^2+5\right) \epsilon ^4
\\
&\quad
-2 \left(53 \sigma ^2+292 \left(\sigma ^2-1\right) \sigma ^4+5\right) \epsilon ^3+4 \left(70 \sigma ^6-81 \sigma ^4+17 \sigma ^2+2\right) \epsilon ^2
\\
&\quad
-15 \left(1-2 \sigma ^2\right)^2 \epsilon + 3 \left(1-2 \sigma ^2\right)^2\,,
\\
c^{\text{/\!\!\!Y}}_{15,1} &=\sigma  \epsilon  \big(-16 \left(10 \sigma ^4-7 \sigma ^2+1\right) \epsilon ^4+8 \left(69 \sigma ^4-58 \sigma ^2+11\right) \epsilon ^3
\\
&\qquad\quad
+\left(-644 \sigma ^4+652 \sigma ^2-152\right) \epsilon ^2+\left(288 \sigma ^4-344 \sigma ^2+94\right) \epsilon -36 \sigma ^4+48 \sigma ^2 -15\big)\,,
\\
c^{\text{/\!\!\!Y}}_{15,2} &=8 \left(10 \sigma ^6+3 \sigma ^4-6 \sigma ^2+1\right) \epsilon ^5+4 \left(-96 \sigma ^6+48 \sigma ^4+13 \sigma ^2-5\right) \epsilon ^4
\\
&\quad
+2 \left(53 \sigma ^2+292 \left(\sigma ^2-1\right) \sigma ^4+5\right) \epsilon ^3-4 \left(70 \sigma ^6-81 \sigma ^4+17 \sigma ^2+2\right) \epsilon ^2
\\
&\quad
+15 \left(1-2 \sigma ^2\right)^2 \epsilon -3 \left(1-2 \sigma ^2\right)^2 \,.
\end{aligned}\label{}
\end{equation}
}
\newpage
\section{Master Integrals}
In this section, we present the explicit form of the master integrals at two loops derived in \cite{Parra-Martinez:2020dzs} by solving the differential equation with the static boundary conditions in the potential region. Here, we modified the results in our convention. We have shown in Section \ref{Sec:6.3} that the master integrals in the \textbf{IY} and \flip{\textbf{IY}} classes can be represented by the master integrals in the $\Box\!\Box$ and $\mathbf{H}$ classes. Thus, we omit the master integrals for the \textbf{IY} and \flip{\textbf{IY}} classes here.

We first consider the $\Box\!\Box$ class, which consists of six sectors: III, $\overline{\rm III}$, IX, $\overline{\rm IX}$, XI and $\overline{\rm XI}$ sectors. The XI and $\overline{\rm XI}$ are the horizontal mirror dual of IX, $\overline{\rm IX}$, and we also omit them. The non-vanishing master integrals for the III sector are given by
\begin{equation}
\begin{aligned}
  f^{\text{III}}_{2} &= \frac{\epsilon^{2} \pi^{2}}{(q^{2})^{2\epsilon}} \bigg[\frac{\epsilon}{3} \log x - \epsilon^2 \left(\text{Li}_2\left(1-x^2\right)+\log^2 x\right)\bigg]\,,
  \\
  f^{\text{III}}_{3} &= \frac{\epsilon^{2} \pi^{2}}{(q^{2})^{2\epsilon}} \bigg[
  	\frac{2\epsilon}{3} \log x+\frac{2 \epsilon^2}{3} \left(\text{Li}_2\left(1-x^2\right)+\log^2 x \right)\bigg]\,,
  \\
  f^{\text{III}}_{4} &= \frac{\epsilon^{2} \pi^{2}}{(q^{2})^{2\epsilon}} \bigg[ -\frac{1}{3} + \frac{\epsilon^4}{18} \left(48 \log^2 x + 7 \pi ^2\right)\bigg]\,,
  \\
  f^{\text{III}}_{6} &= \frac{\epsilon^{2} \pi^{2}}{(q^{2})^{2\epsilon}} \bigg[-\frac{1}{6}+ \frac{7 \pi^2 \epsilon^2}{36}\bigg]\,,
  \\
  f^{\text{III},}_{7} &= \frac{\epsilon^{2} \pi^{2}}{(q^{2})^{2\epsilon}} \bigg[\frac{1}{2}-\frac{ \epsilon^2}{12} \left(\pi ^2-4 \log^2 x\right)\bigg]\,,
  \\
  f^{\text{III}}_{10} &= \frac{\epsilon^{2} \pi^{2}}{(q^{2})^{2\epsilon}} \bigg[\frac{i \epsilon\pi}{4} -\frac{i \pi \epsilon^{2}}{2} \log 2 \bigg]\,,
\end{aligned}\label{MasterIII}
\end{equation}
The non-vanishing master integrals for the $\overline{\rm III}$ sector are given by
\begin{equation}
\begin{aligned}
  f^{\overline{\text{III}}}_{2} &= \frac{\epsilon^{2} \pi^{2}}{(q^{2})^{2\epsilon}} \bigg[ -\frac{\epsilon}{6} \log x + \frac{\epsilon^{2}}{2} \left(\text{Li}_2\left(1-x^2\right)+\log^2 x\right) \bigg] \,,
  \\
  f^{\overline{\text{III}}}_{3} &= \frac{\epsilon^{2} \pi^{2}}{(q^{2})^{2\epsilon}} \bigg[ -\frac{\epsilon}{3} \log x -\frac{\epsilon^2}{3}  \left(\text{Li}_2\left(1-x^2\right)+\log^2 x \right) \bigg]\,,
  \\
  f^{\overline{\text{III}}}_{4} &= \frac{\epsilon^{2} \pi^{2}}{(q^{2})^{2\epsilon}} \bigg[ \frac{1}{6}-\frac{\epsilon^2}{36} \left(7 \pi^2 +48 \pi ^2 \log^2 x\right) \bigg]\,,
  \\
  f^{\overline{\text{III}}}_{6} &= \frac{\epsilon^{2} \pi^{2}}{(q^{2})^{2\epsilon}} \bigg[ -\frac{1}{6} + \frac{7 \pi^2 \epsilon^2}{36} \bigg]\,,
  \\
  f^{\overline{\text{III}}}_{7} &= \frac{\epsilon^{2} \pi^{2}}{(q^{2})^{2\epsilon}} \bigg[-\frac{\epsilon^2}{6} \log^2 x\bigg]\,,
\end{aligned}\label{MasterxIII}
\end{equation}
The non-vanishing master integrals for the IX sector are given by
\begin{equation}
\begin{aligned}
f_{\mathrm{IX},2}&= \frac{\epsilon^{2} \pi^{2}}{(q^{2})^{2\epsilon}} \bigg[ \frac 1 6 \epsilon \log x -\frac{1}{2} \epsilon^2 \left( \operatorname{Li}_2(1-x^2) + \log^2(x) \right) \bigg]\,,
\\
f_{\mathrm{IX},3}&= \frac{\epsilon^{2} \pi^{2}}{(q^{2})^{2\epsilon}} \bigg[ \frac 1 3 \epsilon \log x +\frac 1 3 \epsilon^2 \left( \operatorname{Li}_2(1-x^2) + \log^2(x) \right) \bigg]\,,
\\
f_{\mathrm{IX},4}&=\frac{\epsilon^{2} \pi^{2}}{(q^{2})^{2\epsilon}} \bigg[-\frac 1 6 + \frac 1 {36} \epsilon^2 \left( 7 \pi^2 +48 \log^2(x) \right) \bigg]\,,
\\
f_{\mathrm{IX},5}&=\frac{\epsilon^{2} \pi^{2}}{(q^{2})^{2\epsilon}} \bigg[ \frac 1 3\epsilon\log x -\epsilon^2 \left( \operatorname{Li}_2(1-x^2) + \log^2(x) \right) \bigg]\,,
\\
f_{\mathrm{IX},6}&=\frac{\epsilon^{2} \pi^{2}}{(q^{2})^{2\epsilon}} \bigg[ \frac 2 3\epsilon\log x +\frac{2}{3}\epsilon^2 \left( \operatorname{Li}_2(1-x^2) + \log^2(x) \right) \bigg]\,,
\\
f_{\mathrm{IX},7}&=\frac{\epsilon^{2} \pi^{2}}{(q^{2})^{2\epsilon}} \bigg[\frac{1}{3}-\frac{1}{18}\epsilon^2 \left( 7 \pi^2 +48 \log^2(x) \right)\bigg]\,,
\\
f_{\mathrm{IX},8}&=\frac{\epsilon^{2} \pi^{2}}{(q^{2})^{2\epsilon}} \bigg[ -\frac{1}{6}+\frac{7\pi^2\epsilon^2}{36}\bigg]\,,
\\
f_{\mathrm{IX},10}&= \frac{\epsilon^{2} \pi^{2}}{(q^{2})^{2\epsilon}} \bigg[ -\frac{5}{12}\epsilon^2\log^2(x)\bigg]\,,
\\
f_{\mathrm{IX},15}&=\frac{\epsilon^{2} \pi^{2}}{(q^{2})^{2\epsilon}} \bigg[ \frac {i\pi \epsilon }{4} -\frac{i \pi \log(2) \epsilon^2}{2} \bigg]
\,.
\end{aligned}\label{MasterIX}
\end{equation}
The non-vanishing master integrals for the $\overline{\rm IX}$ sector are given by
\begin{equation}
\begin{aligned}
f_{\overline{\mathrm{IX}},2}&= \frac{\epsilon^{2} \pi^{2}}{(q^{2})^{2\epsilon}} \bigg[-\frac {1}{3} \epsilon \log x +\epsilon^2 \left( \operatorname{Li}_2(1-x^2) + \log^2(x) \right) \bigg]\,,
\\
f_{\overline{\mathrm{IX}},3}&=\frac{\epsilon^{2} \pi^{2}}{(q^{2})^{2\epsilon}} \bigg[ -\frac {2}{3} \epsilon \log x -\frac{2}{3} \epsilon^2 \left( \operatorname{Li}_2(1-x^2) + \log^2(x) \right) \bigg]\,,
\\
f_{\overline{\mathrm{IX}},4}&=\frac{\epsilon^{2} \pi^{2}}{(q^{2})^{2\epsilon}} \bigg[ \frac 1 3 - \frac 1 {18} \epsilon^2 \left( 7 \pi^2 +48 \log^2(x) \right) \bigg]\,,
\\
f_{\overline{\mathrm{IX}},5}&= \frac{\epsilon^{2} \pi^{2}}{(q^{2})^{2\epsilon}} \bigg[ -\frac {1}{6} \epsilon \log x +\frac{1}{2} \epsilon^2 \left( \operatorname{Li}_2(1-x^2) + \log^2(x) \right) \bigg]\,,
\\
f_{\overline{\mathrm{IX}},6}&= \frac{\epsilon^{2} \pi^{2}}{(q^{2})^{2\epsilon}} \bigg[ -\frac {1}{3} \epsilon \log x -\frac{1}{3} \epsilon^2 \left( \operatorname{Li}_2(1-x^2) + \log^2(x) \right) \bigg]\,,
\\
f_{\overline{\mathrm{IX}},7}&= \frac{\epsilon^{2} \pi^{2}}{(q^{2})^{2\epsilon}} \bigg[ -\frac 1 6 + \frac 1 {36} \epsilon^2 \left( 7 \pi^2 +48 \log^2(x) \right) \bigg]\,,
\\
f_{\overline{\mathrm{IX}},8}&= \frac{\epsilon^{2} \pi^{2}}{(q^{2})^{2\epsilon}} \bigg[ -\frac 1 6 + \frac{7\pi^2\epsilon^2} {36} \bigg]\,,
\\
f_{\overline{\mathrm{IX}},10}&= \frac{\epsilon^{2} \pi^{2}}{(q^{2})^{2\epsilon}} \bigg[ \frac 1 3\epsilon^2\log^2(x) \bigg]\,.
\end{aligned}\label{MasterxIX}
\end{equation}
Next, we consider the $\mathbf{H}$ class. As discussed in Section \ref{Sec:6.2}, the master integrals in $\mathbf{H}$ class are decomposed into the even and odd powers in $|q|$. Since the odd terms do not contribute to the total amplitude, we only focus on the even terms.

The non-vanishing master integrals for the H sector are given by
\begin{equation}
\begin{aligned}
f_{\rm H,4} &= \frac{\epsilon^{2} \pi^{2}}{(q^{2})^{2\epsilon}} \bigg[ \frac{\epsilon^2 \pi^2} 4 \bigg]\,,
\\
f_{\rm H,5} &=\frac{\epsilon^{2} \pi^{2}}{(q^{2})^{2\epsilon}} \bigg[ \frac 1 3 \epsilon \log x -  \epsilon^2 \left( \operatorname{Li}_2(1-x^2) + \log^2 x \right) \bigg]\,,
\\
f_{\rm H,6} &= \frac{\epsilon^{2} \pi^{2}}{(q^{2})^{2\epsilon}} \bigg[ \frac{2}{3} \epsilon \log x + \frac{2}{3}  \epsilon^2 \left( \operatorname{Li}_{2}(1-x^2) + \log^2 x \right) \bigg] \,,
\\
f_{\rm H,7} &= \frac{\epsilon^{2} \pi^{2}}{(q^{2})^{2\epsilon}} \bigg[ -\frac 1 3 + \epsilon^2 \left( \frac{7\pi^2} {18} + \frac{8}{3} \log^2 x \right) \bigg]\,,
\\
f_{\rm H,9} &= \frac{\epsilon^{2} \pi^{2}}{(q^{2})^{2\epsilon}} \bigg[ - \frac{2}{3} \epsilon \log x + \frac{2}{3} \epsilon^2 \left( \operatorname{Li}_{2}(1-x^2) + \log^2 x \right) \bigg]\,,
\\
f_{\rm H,10} &= \frac{\epsilon^{2} \pi^{2}}{(q^{2})^{2\epsilon}} \bigg[\frac{1}{2} \epsilon^2 \left( \pi^2 + 8 \log^2 x \right) \bigg]\,.
\end{aligned}\label{MasterH}
\end{equation}
Similarly, the non-vanishing master integrals for the $\overline{\rm H}$ sector are given by
\begin{equation}
\begin{aligned}
f_{\overline{\rm H},4} &= \frac{\epsilon^{2} \pi^{2}}{(q^{2})^{2\epsilon}} \bigg[ \frac{\epsilon^2 \pi^2} 4 \bigg]\,,
\\
f_{\overline{\rm H},5} &=\frac{\epsilon^{2} \pi^{2}}{(q^{2})^{2\epsilon}} \bigg[ - \frac 1 6 \epsilon \log x + \frac 1 2 \epsilon^2 \left( \operatorname{Li}_2(1-x^2) + \log^2 x \right) \bigg]\,,
\\
f_{\overline{\rm H},6} &= \frac{\epsilon^{2} \pi^{2}}{(q^{2})^{2\epsilon}} \bigg[ - \frac{1}{3} \epsilon \log x + \frac{1}{3} \epsilon^2 \left( \operatorname{Li}_{2}(1-x^2) + \log^2 x \right) \bigg]\,,
\\
f_{\overline{\rm H},7} &= \frac{\epsilon^{2} \pi^{2}}{(q^{2})^{2\epsilon}} \bigg[ \frac 1 6 - \epsilon^2 \left( \frac{7\pi^2} {36} + \frac{4}{3} \log^2 x \right) \bigg]\,,
\\
f_{\overline{\rm H},9} &= \frac{\epsilon^{2} \pi^{2}}{(q^{2})^{2\epsilon}} \bigg[ \frac{1}{3} \epsilon \log x - \frac{1}{3} \epsilon^2 \left( \operatorname{Li}_{2}(1-x^2) + \log^2 x \right) \bigg]\,,
\\
f_{\overline{\rm H},10} &= \frac{\epsilon^{2} \pi^{2}}{(q^{2})^{2\epsilon}} \bigg[ - \epsilon^2 \left( \frac{1}{2} \pi^2 + 2 \log^2 x  \right) \bigg]\,.
\end{aligned}\label{MasterxH}
\end{equation}

\bibliography{references}

\bibliographystyle{JHEP}

\end{document}